\pdfoutput=1

\documentclass[a4paper,twoside,openright,12pt]{memoir}

\DoubleSpacing
\usepackage[greek, english]{babel}
\usepackage{amsmath}
\usepackage{amsfonts}
\usepackage{amssymb}
\usepackage{bigfoot}
\usepackage{bm}
\usepackage{caption}
\usepackage{soul}
\usepackage{color}
\usepackage{graphicx}
\usepackage{indentfirst}
\usepackage{textgreek}
\usepackage{setspace}
\usepackage[utf8]{inputenc}
\usepackage{libertine}
\usepackage{lipsum}
\usepackage{oxfordthesis}
\usepackage{xr}
\usepackage{xr-hyper}

\thetitle{Kinetic treatment of magnetized and collisionless plasma near a wall}
\theauthor{Alessandro Geraldini}
\degreedate{Jul 2018}
\degree{Doctor of Philosophy}
\college{Merton College}
\university{University of Oxford}

\usepackage[
	pdftitle={\oxfthetitle},
	pdfauthor={\oxftheauthor},
	pdfsubject={Thesis for the Degree of \oxfdegree, \oxfdegreedate},
	pdfborder=0,
	bookmarks=true,
	bookmarksnumbered=true,
	bookmarksopen=true,
	bookmarksopenlevel=1,
	plainpages=false,
	pdfpagelabels=true,
	colorlinks=false,
    citecolor=blue
]{hyperref}
\usepackage{memhfixc}

\begin{document}

\titlepage

\frontmatter

\OnehalfSpacing

\begin{abstract}

Charged particles gyrate around magnetic field lines, a property that is exploited to confine plasma in magnetic confinement fusion devices. 
Typically, the gyroradius is small compared to the system size and thus the gyromotion can be averaged out.
The resulting charged particle motion closely follows a magnetic field line.
At the edge of fusion devices, the magnetic field usually impinges on a wall at a shallow angle.
A boundary layer forms in which the plasma density changes over a characteristic distance from the wall of the order of the ion gyroradius, as ions are absorbed during their gyromotion.
This boundary layer is called magnetic presheath, and is typically collisionless and quasineutral. 
Importantly, the electric field in this region distorts the ion gyro-orbits, making them non-circular and thus affecting the ion density profile.
Solving the magnetic presheath amounts to obtaining the self-consistent electric field for which the net charge density is zero.

In this thesis, I assume a small magnetic field angle and small gradients parallel to the wall to develop an asymptotic theory for the magnetic presheath, which is used to obtain the ion density.
The small, yet crucial, contribution of the part of the orbit near the wall is included. 
To demonstrate the theory for a case without any gradients parallel to the wall, I calculate numerically the self-consistent electrostatic potential by assuming the electron density to be a Boltzmann distribution.
The model is used to study the dependence of magnetic presheath characteristics on magnetic field angle and ion temperature.
The distribution function of ions that have traversed the magnetic presheath is obtained, which is important to predict the amount of sputtering and erosion at the wall of a fusion device.

\end{abstract}

\cleardoublepage

\begin{acknowledgements}

I cannot thank enough my PhD supervisor, Felix Parra: in these four years, he has been able to simultaneously motivate, encourage and teach me in the best possible way.
I also thank my co-supervisor at Culham Centre for Fusion Energy (CCFE), Fulvio Militello, for his contribution and support.

Several people have given me feedback about my work. I thank Greg Hammett, Paolo Ricci, Dmitri Ryutov, Paul Dellar, Chris Ham, John Omotani and Ian Abel for discussions and comments. I am especially grateful to Paolo for reading this thesis with genuine curiosity and for giving me valuable advice.

Together with Felix, Alex Schekochihin, Michael Barnes and Steve Cowley have created the perfect research environment in the Oxford Plasma Theory group; I thank them for this and for their continuous feedback.
Especially Alex, who not only read this thesis but has also been like a mentor.

During my time in the group, I am lucky to have bridged two “generations” of students and postdocs: the “Lads on Tour” Justin, Michael F, Ferdinand and Marek; the “Plasmaniacs” Nick, Michael H, Adwiteey, Valerian, Plamen, Jason, Ollie, Mantas, Yohei and Adnane.
I have enjoyed spending my days with them, as well as conferences (tours) and ping-pong games.

A number of people have made my time at Oxford special. 
I thank all of those who kept me company in these years, be it with dinner nights, pints, coffees, house-sharing, tennis or simply an occasional life talk.
I especially thank Luca, Francesco, Jutta, Karl and Richard, who were always there when I needed them.
And Andriana, who taught me the meaning of \textgreek{ερωτευμένος}.

I am grateful to my family for their wholehearted love and support throughout my life, and for encouraging me to do what I wanted.

This work has received funding from the RCUK Energy Programme [grant number EP/P012450/1].
I am grateful to CCFE for supporting me financially. 
I have also received financial support for conferences from Merton College and the Wolfgang Pauli Institute.

\end{acknowledgements}

\cleardoublepage

\DoubleSpacing
\tableofcontents
\listoffigures

\mainmatter
\cleardoublepage

\chapter{Introduction}

Harnessing the energy released during nuclear fusion reactions is the objective of nuclear fusion research, with a collective effort from physicists, engineers and materials scientists.
Confinement of the hot plasma required for fusion reactions can be achieved by using magnetic fields.
Magnetic forces cause the charged particles in a plasma to gyrate, thus confining them within a gyroradius in the two dimensions perpendicular to the magnetic field.
Existing concepts for magnetic confinement fusion include tokamaks \cite{Mukhovatov-1971}, stellarators \cite{Spitzer-1958}, magnetic mirrors \cite{Post-1987} and reversed-field pinches \cite{Bodin-1980}.
Most of these concepts \cite{Mukhovatov-1971, Spitzer-1958, Bodin-1980} rely on a magnetic field with magnetic field lines that do not reach the wall in order to achieve confinement parallel to the magnetic field.
Even in such devices, particle and energy confinement is far from perfect for several reasons, including collisions, orbit drifts, plasma instabilities and turbulence.
Thus, particles and energy slowly travel perpendicular to the magnetic field and eventually leave the confinement region.

The plasma and the closed field lines must be contained in a finite volume with boundaries, because the magnets generating most of the confining field must be protected from the plasma and from the neutrons generated by the fusion reactions.
Hence, at the boundary of a fusion device there are regions in which the magnetic field lines are open and terminate at a wall.
These walls are subject to a constant particle and energy flux from the plasma leaving the confinement region.
Therefore, it is crucial to understand and predict how the particles and energy leaving the plasma affect the device walls, and in turn how the plasma-wall interaction affects device performance and confinement.

Typically, the interaction between the confined plasma and the wall of the magnetic fusion device happens at locations specified by design, which in tokamaks are called divertor or limiter targets \cite{Stangeby-book}.
The magnetic field usually makes a shallow angle with these targets in order to minimize the heat flux onto the wall materials \cite{Loarte-2007}. 
Motivated by this observation, in this thesis I study the plasma-wall boundary assuming that the magnetic field makes a small angle $\alpha \ll 1$ with the wall.
With this magnetic field configuration, I focus on the boundary layer with characteristic thickness of the order of the ion gyroradius that forms near the wall, called the magnetic presheath, and assume that ions are collisionless in this thin region.
This study is applicable to other systems: near spacecraft \cite{Laframboise-1997}, plasma thrusters \cite{Martinez-1998}, probes \cite{Hutchinson-book} and magnetic filters \cite{Anders-1995-filters}. 

The rest of this introductory chapter is structured as follows.
In Section \ref{sec-plasmawall}, I discuss the basic physics of plasma-wall interaction and introduce the boundary layers present next to the wall.
In Section \ref{sec-intro-MPS} I describe the magnetic presheath and explain the structure of the thesis.

\section{Plasma-wall interaction} \label{sec-plasmawall}

When a steady-state plasma is in contact with a wall, a potential difference between the bulk plasma and the wall is present which depends on the density and temperature of the plasma and on the current flowing from the plasma to the wall. 
This potential drop forms due to the difference in mobility between ions and electrons.
When quasineutral plasma is placed in contact with a wall, electrons usually reach the wall faster and thus charge it negatively \cite{Cipolla-1981}, leaving a thin layer of plasma next to the wall, called the ``Debye sheath'', to be positively charged.
The Debye sheath has a thickness of several Debye lengths $\lambda_{\text{D}} = \sqrt{e^2 n_{\text{e}}/\epsilon_0 T_{\text{e}}}$, where $e$ is the proton charge, $n_{\text{e}}$ is the number density of electrons in the plasma, $\epsilon_0$ is the permittivity of free space and $T_{\text{e}}$ is the electron temperature. 
Most of the wall charge is shielded from the bulk plasma by the Debye sheath, allowing a steady-state in which most of the electrons are repelled from the wall while the ions are accelerated towards the wall.

Some of the electrostatic potential difference between wall and plasma occurs in a quasineutral ``presheath'', of size $\lambda_{\text{ps}} \gg \lambda_{\text{D}}$. 
Usually $\lambda_{\text{ps}} \ll a$, where $a$ is the scale of the device (for example, the minor radius of a tokamak), which implies that the presheath can be treated as a thin boundary layer with respect to the bulk plasma. 
The validity of two-scale theories that exploit the limit $\lambda_{\text{D}} / \lambda_{\text{ps}} \rightarrow 0$ has been justified by Riemann \cite{Riemann-2003, Riemann-2005-matching}. 
In unmagnetized plasmas, or magnetized plasmas in which the magnetic field is normal to the wall, the size of the presheath is determined by the ion collisional mean free path, $\lambda_{\text{mfp}}$.
When a magnetic field is present, the ion motion perpendicular to the magnetic field is confined within a gyroradius $\rho_{\text{i}}$, while ions are free to move parallel to the magnetic field.
The ions are considered magnetized if $ \rho_{\text{i}} \ll \lambda_{\text{mfp}} $, and thus travel in the direction parallel to the magnetic field for a distance of the order of a mean free path before colliding.
If the magnetic field makes an oblique angle $\alpha$ with the wall, the ion gyro-orbit touches the wall when it is a gyroradius $\rho_{\text{i}}$ away from it. 
The mean displacement of an ion between collisions ($\lambda_{\text{mfp}}$ parallel to the magnetic field) projected in the direction normal to the wall is $\lambda_{\text{mfp}} \sin \alpha $.
This length scale sets the size of the boundary layer in which the transition from a collisional to a collisionless plasma occurs.
Thus, there are two presheath length scales: $ \lambda_{\text{mfp}} \sin \alpha $ and $\rho_{\text{i}}$.

\begin{figure}
\centering
\includegraphics[width = 0.7\textwidth]{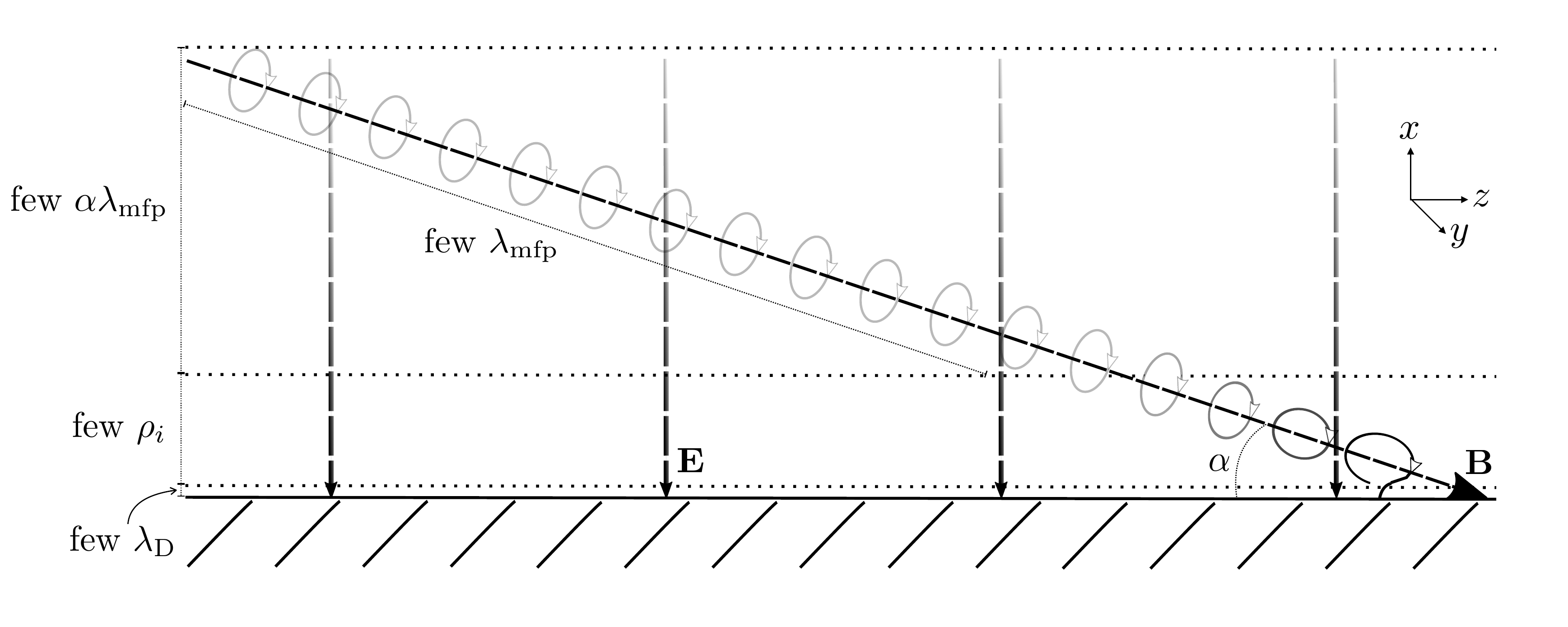}
\caption[Boundary layers of magnetized plasma]{Cartoon of ion orbits in the neighbourhood of the divertor target of a tokamak plasma, with $\lambda_{\text{D}} \ll \rho_{\text{i}} \ll \alpha \lambda_{\text{mfp}}$. The orbits have a size $\rho_{\text{i}}$ and are tied to a dashed line representing the magnetic field $\vec{B}$, which is inclined at an angle $\alpha$ with the wall. The electric field $\vec{E}$ is shown as a dashed vertical line, and is shaded darker nearer to the wall, where it is stronger. Highly distorted orbits in the magnetic presheath are black, while circular orbits in the collisional presheath are light grey.  }
\label{figure-boundary-layers}
\end{figure}

I consider the fusion-relevant case of a magnetic field impinging on the wall at a shallow angle, $\alpha \ll 1$.
The size of the collisional length scale at such a small angle is $\lambda_{\text{mfp}} \sin \alpha  \simeq \alpha \lambda_{\text{mfp}} $.
Moreover, I consider a plasma-wall boundary that satisfies the scale separation
\begin{align} \label{scale-sep}
\lambda_{\text{D}} \ll \rho_{\text{i}} \ll  \alpha \lambda_{\text{mfp}} \text{.}
\end{align}
As shown in Figure \ref{figure-boundary-layers}, with this scale separation, the presheath can be split into two separate layers: a ``collisional presheath'' of size $\alpha \lambda_{\text{mfp}}$ and a collisionless ``magnetic presheath'' of size $\rho_{\text{i}}$ \cite{Loizu-2012}. 
The ion motion in the two layers has a very different nature: in the collisional layer, ions are magnetized in circular gyro-orbits and stream parallel to the magnetic field; in the magnetic presheath, ion gyro-orbits are distorted by increasingly strong electric fields.
In the Debye sheath, ions are accelerated towards the wall by an electric force much larger than the magnetic force.
In Appendix \ref{appendix:widths} I justify the ordering (\ref{scale-sep}) in the context of a typical tokamak plasma.
A cartoon of the ion motion across all boundary layers is shown in Figure \ref{figure-boundary-layers}.

There is a vast literature that treats the plasma-wall boundary using fluid equations \cite{Chodura-1982, Riemann-1994, Hutchinson-2008, Stangeby-1995, Stanojevic-2005, Chankin-1994}.
Although, with due care, fluid equations may capture well most of the underlying physics, it is widely accepted that a kinetic treatment should be carried out to describe the plasma-wall boundary, which is kinetic in nature from the collisional layer to the wall. 
For example, Siddiqui \emph{et al} \cite{Siddiqui-Hershkowitz-2016} argue, using experimental measurements of ion flows in three dimensions near the plasma-wall boundary, that a kinetic theory of ions and neutrals is necessary in the boundary layer in order to accurately predict the location and intensity of ion and charge exchanged neutral fluxes to the wall.
Most of the existing kinetic solutions to plasma-wall boundary layers are numerical and use Particle-in-Cell (PIC) codes \cite{Tskhakaya-2004, Tskhakaya-2003, Kovacic-2009, Khaziev-Curreli-2015}, although some use Eulerian-Vlasov advection schemes \cite{Gerhauser-1998, Devaux-2006, Coulette-2014, Coulette-Manfredi-2016}. 
Analytical kinetic treatments of the plasma-wall boundary are more rare \cite{Harrison-Thompson-1959, Riemann-review, Daube-Riemann-1999, Cohen-Ryutov-1998, Holland-Fried-Morales-1993, Daybelge-Bein-1981, Sato-1994, Tskhakaya-2014}. 
They often provide more insight, although generally suffer from the limitation of making oversimplifying assumptions and/or solving the problem only partially.
A combination of robust analytical work supported by numerical solutions is often the best compromise.

A kinetic treatment of the collisional presheath must describe the transition from collisional to collisionless ions and electrons, while retaining the effect of magnetic and electric forces on each individual particle.
A conventional drift-kinetic or gyrokinetic ordering \cite{Shi-Hammett-2017, Parra-Catto-2008, Jorge-2017} is expected to hold for both ions and electrons in the collisional presheath, which can be used to substantially simplify the particle trajectories.
Apart from this simplification, a kinetic theory of this layer remains complicated and dependent on the dominant collision process. 

Kinetic treatments of the collisionless layers (magnetic presheath and Debye sheath) are simplified by the absence of collisions.
However, compared with the Debye sheath, an ion kinetic treatment in the magnetic presheath is complicated by the role of the oblique magnetic field, which exerts dynamically important forces on the ions.
Electric forces perpendicular to the wall dominate the ion dynamics in the Debye sheath; hence, the change in electrostatic potential energy of an ion is equal to the change in kinetic energy of the component of the ion velocity normal to the wall.
In the magnetic presheath, the typical size of electric and magnetic forces acting on an ion are similar; hence, the kinetic energy of each ion is continuously transferred between the three velocity components and the electrostatic potential.
From this discussion, it is clear that two co-ordinates are sufficient to describe ion motion in the Debye sheath: the distance from the wall, $x$, and the velocity component normal to the wall, $v_x$.
Conversely, a kinetic treatment of ions in the magnetic presheath requires two additional velocity co-ordinates to describe the ion dynamics in the plane parallel to the wall (even when the system is translationally invariant in the directions parallel to the wall).
Hence, a kinetic treatment of the magnetic presheath offers a substantial analytical and numerical challenge, which is addressed by this thesis.


\section{The magnetic presheath} \label{sec-intro-MPS}

The magnetic presheath was first studied by Chodura \cite{Chodura-1982}. 
For this reason, it is often referred to as the ``Chodura sheath'' in the literature, although it has also been referred to as the ``gyrosheath'' \cite{Cohen-Ryutov-2004}.
Chodura used fluid equations for the electrons and ions, which are valid provided ions are much colder than electrons. 
He thus found a solution for the electrostatic potential and ion flow across the magnetic presheath. 
Chodura also found that, for cold ions, the ion flow parallel to the magnetic field at the presheath entrance must at least be equal to the Bohm speed
\begin{align} \label{vB}
v_{\text{B}} = \sqrt{\frac{ZT_{\text{e}}}{m_{\text{i}}}} \text{,}
\end{align} 
which is known as the Chodura (or Bohm-Chodura) condition \cite{Chodura-1982, Riemann-1994}. 
In equation (\ref{vB}), $Z$ is the ion charge as a multiple of the proton charge $e$, and $m_{\text{i}}$ is the ion mass.
Bohm \cite{Bohm-1949} first obtained a similar condition at the entrance of the Debye sheath, where he concluded that the ions enter with a normal velocity component at least equal to $v_{\text{B}}$.
Hence, the ion fluid velocity in the magnetic presheath is turned by the strong electric field from being parallel to the magnetic field to being normal to the wall. 
Chodura also predicted that the electrostatic potential drop across the magnetic presheath is approximately $\left( T_{\text{e}} / e \right) \ln\left( 1 / \sin \alpha \right)$.
His predictions were supported by PIC simulations. 

The problem of how an oblique magnetic field influences the plasma-wall interaction and the plasma-wall boundary conditions has been of interest and is yet to be fully resolved.
After Chodura, many other studies of the magnetic presheath that also used fluid equations to model the ion species have been carried out \cite{Ahedo-1997, Ahedo-2009, Riemann-1994, Chankin-1994}.
The main results and the physical picture provided by Chodura survived.
After the work of Stangeby and Chankin \cite{Stangeby-1995}, who incorporated the $\vec{E} \times \vec{B}$ drifts expected in the magnetic presheath (due to, for example, turbulence in the scrape-off-layer) and thus generalized Chodura's condition, Loizu \cite{Loizu-2012} obtained a set of boundary conditions for fluid models of the Scrape-Off-Layer.
The problem of an ion fluid model is that it cannot accurately describe fluid quantities across the magnetic presheath, since kinetic effects due to the finite ion Larmor radius and due to the width of the ion distribution function are neglected.
For example, an ion that is initially at a distance $\rho_{\text{i}}$ from the wall may or may not reach the wall before completing a gyration depending on its initial velocity, which in turn also affects the velocity of the ion at the wall.
Moreover, ion Larmor orbits have a characteristic radius of the order of the length scale of variation of the electric field, and are thus highly distorted \cite{Cohen-Ryutov-2004}. 
A fluid description can only be used when $T_{\text{i}} \ll T_{\text{e}}$, where $T_{\text{i}}$ is the ion temperature, because such ions can be treated as mono-energetic and with zero Larmor radius. 
Several kinetic treatments of the ions in the magnetic presheath have been carried out \cite{Daube-Riemann-1999, Cohen-Ryutov-1998, Holland-Fried-Morales-1993, Daybelge-Bein-1981, Sato-1994, Tskhakaya-2014, Claassen-Gerhauser-1996-gyrocooling}, all of which have made significant progress in our understanding of the magnetic presheath.
With this thesis, I hope to make further progress by providing a complete kinetic theory of ions in the magnetic presheath based on an asymptotic expansion and a numerical procedure that is efficient and requires very little computational power.

In this thesis, I do not consider the effect of kinetic electrons and take a Boltzmann electron distribution across the magnetic presheath, as is done in most other studies.
This relies on the assumptions that the electron distribution function entering the magnetic presheath is Maxwellian, and that the potential drop across the Debye sheath repels most of the electrons away from the wall.
I derive the electron-repelling magnetic presheath equations valid for general $T_{\text{i}} \sim T_{\text{e}}$ by retaining the full ion distribution function and exploiting an asymptotic expansion in the small angle $\alpha$.
This work is consistent with the qualitative picture of fluid models of the electron-repelling magnetic presheath, with a quantitative modification of the electrostatic potential and ion flow profiles. 
Additionally, the ion distribution function at the entrance of the Debye sheath is obtained from the ion distribution function entering the magnetic presheath.
The distribution function at the entrance of the Debye sheath is found to be narrower when the angle $\alpha$ is smaller.
A kinetic solution of the Debye sheath, as shown in reference \cite{Riemann-1981}, 
can be used to obtain the ion distribution function at the wall from the distribution function at the Debye sheath entrance.
The distribution of ion velocities at the wall is important because the sputtering yield of each ion depends on its kinetic energy and its angle of impact with the wall, and thus sputtering --- which is a source of impurities in the plasma --- can be predicted from the ion distribution function.


When the angle between the magnetic field and the wall is sufficiently small and/or there is a sufficiently large electron current to the wall, the strongly electron-repelling model of the magnetic presheath is invalid (see the discussion in section \ref{sec-KMPS-Boltzmann}) \cite{Stangeby-2012}. 
In order to further improve the models of plasma-wall interaction at grazing angles, a kinetic description of the electrons is therefore necessary.
The strongly electron-repelling assumption can only break down if $T_{\text{i}} \neq 0$, and thus an essential prerequisite of any improved model is to have a kinetic description of the ions.
Therefore, the theory presented in this thesis is crucial for future models of the collisionless boundary layers that do not assume an electron-repelling Debye sheath.
With the inclusion of a kinetic description of the electrons, and a solution of the Debye sheath \cite{Riemann-1981} and magnetic presheath, one could reliably obtain the relationship between current through the plasma and electrostatic potential drop across the collisionless sheath and presheath at small angles.
This relationship could provide a simple boundary condition to drift-kinetic models describing the plasma in the Scrape-Off-Layer.
The equations derived herein for the ion density in the magnetic presheath can also be used to obtain the electron density in the Debye sheath when the electron finite Larmor orbit effects are important $\rho_{\text{e}} \gtrsim \lambda_{\text{D}}$.
In previous studies, finite Larmor orbit effects are often neglected $\rho_{\text{e}} \ll \lambda_{\text{D}}$ \cite{Cohen-Ryutov-1998, Loizu-2012} or considered using the assumption that $\rho_{\text{e}} \gg \lambda_{\text{D}}$ \cite{Cohen-Ryutov-1995-spreading}. 


The thesis is structured as follows.
In chapter \ref{chap-traj}, I solve for the ion trajectories by using an asymptotic expansion for $\alpha \sim \delta \ll 1$, which is equivalent to a gyrokinetic separation of timescales.
In chapter \ref{chap-dens}, I solve for the ion distribution function and density in the magnetic presheath.
Up to this point, my equations also include the effect of the small turbulent electric fields parallel to the wall, ordered smaller than the electric field normal to the wall by a factor $\delta \ll 1$ as was done in reference \cite{Loizu-2012}.
In chapter \ref{chap-KMPS}, I close the system of equations derived in the previous chapters using the quasineutrality equation, and subsequently consider a magnetic presheath with no gradients parallel to the wall ($\delta = 0$).
After a detailed kinetic analysis of the Debye sheath and magnetic presheath entrance conditions, I present numerical solutions for different values of $\alpha$.
In chapter \ref{chap-Tdep}, I study the dependence of the magnetic presheath results on the ion temperature. 
Finally, in chapter \ref{chap-conc} I conclude by summarizing and discussing my work\footnote{Much of the material presented in this thesis appears in references \cite{Geraldini-2017,Geraldini-2018}.}.


\chapter{Ion trajectories}
\label{chap-traj}

The ion trajectories are necessary in order to solve the ion kinetic equation in the magnetic presheath and thus obtain the ion density.
In this chapter I solve for the ion motion in the magnetic presheath assuming that:
\begin{itemize}
\item the angle between the magnetic field and the wall is small, $\alpha \ll 1$; 
\item variations parallel to the wall are in the cross-field direction only, with a characteristic length scale $l$ much longer than the ion gyroradius $\rho_{\text{i}}$, such that $\delta = \rho_{\text{i}} / l \ll 1$.
\end{itemize}
In Section \ref{sec-traj-orderings} I introduce and explain the orderings of all quantities.
In Section \ref{sec-traj-periodic} I solve for periodic ion trajectories in a system in which the magnetic field is exactly parallel to the wall ($\alpha = 0$) with no gradients in the direction parallel to the wall ($\delta = 0$).
Finally, in Section \ref{sec-traj-real}, I use the asymptotic expansion in $\alpha \sim \delta \ll 1$ to solve for the approximate ion trajectories in a grazing angle magnetic presheath with weak turbulent gradients parallel to the wall.

\section{Orderings} \label{sec-traj-orderings}

I denote the electric and magnetic fields in the magnetic presheath as $\vec{E}$ and $\vec{B}$ respectively, and use the coordinate axes in Figure \ref{figure-boundary-layers}. 
The magnetic field is
\begin{align}
\vec{B} = B \cos \alpha \hat{\vec{z}} - B \sin \alpha \hat{\vec{x}}  \text{,}
\end{align}
where $\hat{\vec{x}}$ is the unit vector in the $x$ direction, normal to the wall, and $\hat{\vec{z}}$ is a unit vector parallel to the wall.
Throughout this work I assume a negatively charged, electron repelling wall, valid if the time it takes for the electrons to reach the wall is shorter than the time it takes for the ions. 
The criterion required for assuming an electron-repelling wall is quantified in chapter \ref{chap-KMPS}.

The gyrofrequency of an ion orbit in a constant and uniform magnetic field is
\begin{align} \label{Omega}
\Omega = \frac{ZeB}{m_{\text{i}}} \text{,}
\end{align}
where $Z \sim 1$ is the ion atomic number and $e$ is the proton charge. 
As we will see, orbit distortion in the magnetic presheath changes the exact value of the gyrofrequency, whose size nonetheless remains of order $\Omega$.
The individual ion velocity is ordered
\begin{align} \label{v-order}
\left| \vec{v} \right| \sim v_x \sim v_y \sim v_z  \sim v_{\text{t,i}} = \sqrt{ \frac{2T_{\text{i}}}{m_{\text{i}}} }  \text{,}
\end{align}
where $v_{\text{t,i}}$ is the thermal velocity.
The thermal gyroradius, which is the typical size of an ion orbit, is defined as
\begin{align} \label{rhoi}
\rho_{\text{i}} = \frac{v_{\text{t,i}}}{\Omega} \text{.}
\end{align}
As we will see in chapter \ref{chap-Tdep}, the magnetic presheath has a characteristic thickness of the order of the ion sound Larmor radius 
\begin{align} \label{rho-s-firstdef}
\rho_{\text{s}} = \sqrt{ \frac{ZT_{\text{e}} + T_{\text{i}} }{m_{\text{i}}} } \text{.}
\end{align}
In this chapter, and in chapters \ref{chap-dens} and \ref{chap-KMPS}, I will consider magnetic presheaths with $T_{\text{e}} \sim T_{\text{i}}$, and thus with $\rho_{\text{s}} \sim \rho_{\text{i}}$.
Such magnetic presheaths have a characteristic thickness of the order of the thermal ion Larmor radius $\rho_{\text{i}}$.

The characteristic length perpendicular to the wall is set by the ion gyroradius $\sim \rho_{\text{i}}$. 
The characteristic lengths parallel to the wall are constrained by the size of the turbulent structures in the scrape-off-layer (SOL) \cite{DIppolito-2004}, which are assumed much larger than $\rho_{\text{i}}$.
The $z$ direction is mostly along the magnetic field, a direction in which turbulent structures are elongated, while the $y$ direction is mostly across the magnetic field. 
From this, I argue in Appendix \ref{appendix:turbulence} that gradients in the $z$ direction must be ordered smaller than in the $y$ direction.
I thus take the orderings 
\begin{align} \label{xyz-order}
x \sim \rho_{\text{i}} \ll y \sim l \ll z \sim l/\alpha  \sim  l / \delta  \text{}
\end{align}
for the length scales associated with the different coordinate directions, where $l$ is the characteristic cross-field size of turbulent structures in the SOL and $\delta = \rho_{\text{i}} / l \ll 1$ is a small parameter relating the different length scales. 
From reference \cite{Carralero-2015}, I estimate $l \sim 10 \text{ mm}$ in a typical tokamak, which leads to $\delta \sim 0.07$, which is indeed small.  
I take the maximal ordering 
\begin{align} \label{max-ord}
\alpha \sim \delta
\end{align}
throughout this chapter and chapter \ref{chap-dens}. 
From section \ref{sec-KMPS-quasi} of chapter \ref{chap-KMPS} and throughout chapter \ref{chap-Tdep}, I take the ordering $\delta \ll \alpha$, effectively setting $\delta = 0$ and ignoring gradients parallel to the wall.

The external magnetic field $\vec{B}$ is assumed constant in space and time. 
This is justified in the context of a fusion device because the length scale of $\vec{B}$ is set by the curvature of the device, which is typically much larger than $l$, and the time variations of this field are also expected to be negligible.
I make two further assumptions which are justified in the last few paragraphs of this section. 
Firstly, I assume that the magnetic fields produced by currents in the magnetic presheath plasma are so small compared to the external magnetic field that they can be neglected. 
Secondly, I assume that the plasma in the magnetic presheath is electrostatic, $\vec{E} = - \nabla \phi$, where $\phi$ is the electrostatic potential.

The electrostatic potential changes in the magnetic presheath are ordered
 \begin{align} \label{order-pot}
\phi \sim \frac{T_{\text{e}}}{e} \text{.}
 \end{align} 
The ordering (\ref{order-pot}) is expected from a magnetic presheath in which electrons are being repelled from the wall, and is consistent with the potential drop across the magnetic presheath being $\sim (T_{\text{e}}/e) \ln (1/\alpha )$ as predicted by fluid treatments \cite{Chodura-1982}. 
The ion and electron temperatures are ordered to be of similar size, $T_{\text{i}} \sim T_{\text{e}} \sim T$.
From the ordering of the potential (\ref{order-pot}) and the length scales (\ref{xyz-order}), the size of the electric field components is
\begin{align} \label{order-E}
\frac{\partial \phi}{\partial z}  \sim \frac{\delta T}{el} \sim \frac{\alpha T}{el} \ll  \frac{\partial \phi}{\partial y} \sim \frac{T}{el} \ll \frac{\partial \phi}{\partial x} \sim \frac{T}{e \rho_{\text{i}}} \text{.}
\end{align}

The strong electric field normal to the wall leads to an $\vec{E} \times \vec{B}$ drift in the $y$ direction comparable with the ion thermal velocity, $(1 / B) \partial \phi / \partial x \sim v_{\text{t,i}}$. 
Because potential gradients in this direction are small, the drifting particles will be exposed to significant potential changes only after a timescale much longer than the period of gyration. 
This means that the effect of this drift is unimportant to lowest order.
The electric field component $\partial \phi / \partial y$ parallel to the wall leads to an $\vec{E} \times \vec{B}$ drift in the $x$ direction, normal to the wall, that is first order in $\delta$. 
Therefore, ions drift towards or away from the wall at a speed $\sim \delta v_{\text{t,i}}$. 
This drift competes with the projection of the parallel flow towards the wall $\sim \alpha v_{\text{t,i}}$ when we take the maximal ordering $\delta \sim \alpha$, consistent with reference \cite{Loizu-2012}. 
Parallel streaming and the presence of an absorbing wall leads to an expected ion flow $\sim v_{\text{t,i}}$ in the $z$ direction. 
From (\ref{v-order}) and (\ref{order-E}), potential changes due to motion in the $z$ direction happen over a timescale so much longer than the orbital timescale that their effect is small even to first order in $\delta \sim \alpha$.

The magnetic presheath has a size $\rho_{\text{i}}$. 
Considering that the drift in the $x$ direction is of order $\delta v_{\text{t,i}} \sim \alpha v_{\text{t,i}}$, the characteristic time $t_{\text{MPS}}$ that it takes for an ion to reach the wall after having entered the magnetic presheath is expected to be $\rho_{\text{i}} / \delta v_{\text{t,i}} \sim \rho_{\text{i}} / \alpha v_{\text{t,i}} $, which becomes 
\begin{align} \label{tMPS}
t_{\text{MPS}} \sim \frac{1}{\Omega \delta} \sim \frac{1}{\Omega \alpha} \text{.}
\end{align}
The size of the time derivative $\partial / \partial t$ is set by the turbulence in the SOL, and is given by (see Appendix \ref{appendix:turbulence}) 
\begin{align} \label{t-dep}
 \frac{\partial}{\partial t} \sim  \delta^2 \Omega \text{.}
 \end{align}
Because this partial derivative is higher order compared to $1/t_{\text{MPS}}$, it is negligible. 
From (\ref{v-order}) and (\ref{xyz-order}), the gradients in the $z$ direction are also negligible because
\begin{align}
v_z \frac{ \partial }{ \partial z } \sim  \alpha^2 \Omega \sim \delta^2 \Omega \text{.}
\end{align}
    
Ions and electrons $\vec{E} \times \vec{B}$ drift in the same direction, so their contributions to the current partially cancel each other. 
However, because ions have a large Larmor orbit, they experience a strongly varying electric field over an orbit. 
Therefore, their $\vec{E}\times \vec{B}$ drift in the $y$ direction can differ from the electron one substantially, which leads to a large current density $j^D_y \sim en_{\text{i}}v_{\text{t,i}}$ in this direction.
Here, $n_{\text{i}}$ is the ion density. 
The ``D'' superscript denotes current that is produced by the particle drifts in the plasma.
This estimate for $j_y^D$ also follows from analyzing the size of diamagnetic ion and electron flows parallel to the wall \cite{Chankin-1994}. 
The order of magnitude of the diamagnetic current in the $y$ direction is $\left(1/ B^2\right) \left( \vec{B} \times \nabla p \right)_y \sim \left(1/B\right)\partial p / \partial x \sim en_{\text{i}} v_{\text{t,i}}$, where $p \sim m_{\text{i}} n_{\text{i}} v_{\text{t,i}}^2$ is the plasma pressure. 
From the ordering of the plasma flow in the $x$ direction, the size of the current normal to the wall is expected to be $j_x^D \sim \delta en_{\text{i}}v_{\text{t,i}} \sim \alpha e n_{\text{i}} v_{\text{t,i}}$. 
Again, this also follows by considering the component of the diamagnetic current in the $x$ direction, $\left(1/ B^2\right) \left( \vec{B} \times \nabla p \right)_x \sim \left(1/B\right)\partial p / \partial y \sim \delta en_{\text{i}} v_{\text{t,i}}$.

I proceed to demonstrate that the neglect of magnetic fields produced by magnetic presheath currents and the electrostatic assumption are both justified.
For the remainder of this section (and in Appendix \ref{appendix:largejz}), I refer to the constant externally produced field as $\vec{B}^{c}$. 
From my choice of axes (see Figure \ref{figure-boundary-layers}), $B_x^{c} = - B^{c} \sin \alpha \sim \alpha B^c$, $B_y^{c} = 0$, $B_z^{c} = B^c \cos \alpha \sim B^c$. 
The plasma current $\vec{j}^D$ in the boundary layer can produce a magnetic field $\vec{B}^p$.
Using (\ref{xyz-order}), $\nabla \cdot \vec{B^{p}} = 0$ gives
\begin{align} \label{ordering-Bplasma}
B^{p}_x  \sim \delta^2 B^{p} \sim \delta \alpha B^{p} \ll B_y^{p} \sim \delta B^{p} \sim \alpha B^{p} \ll B_z^{p} \sim B^{p} \text{.}
\end{align}
Amp\`ere's law is
\begin{align} \label{Ampere}
\mu_0 \vec{j}^D = \nabla \times \vec{B^{p}} \text{,}
\end{align}
where $\mu_0$ is the vacuum permeability. 
Using (\ref{xyz-order}) and (\ref{ordering-Bplasma}) to order the right hand side of (\ref{Ampere}), the orderings
\begin{align} \label{j-orderings}
j_x^D \sim j_z^D \sim \frac{B^p}{\mu_0 l} \ll j_y^D \sim \frac{B^p}{\mu_0 \rho_{\text{i}}} \text{}
\end{align}
are obtained.
The earlier orderings for the current deduced from particle motion ($j_x^D \sim \delta en_{\text{i}} v_{\text{t,i}} \sim \alpha en_{\text{i}} v_{\text{t,i}}$ and $j_y^D \sim en_{\text{i}} v_{\text{t,i}}$) are consistent with equation (\ref{j-orderings}) if we take $j_z^D \sim j_x^D \sim \delta en_{\text{i}} v_{\text{t,i}} \sim \alpha en_{\text{i}} v_{\text{t,i}}$. 
This ordering is consistent with what is expected from the piece of the parallel current produced in response to the perpendicular currents resulting from particle drifts (an analogue of the Pfirsch-Schl\"uter current \cite{Hirshman-1978-Pfirsch-Schluter}).\footnote{This does not imply, as discussed in the last paragraph of Section 2 and in Appendix \ref{appendix:largejz}, that larger parallel currents cannot be present in the magnetic presheath.}
From these estimates of the currents, it follows that 
\begin{align} \label{ordering-beta}
\frac{B^{p}}{B^c} \sim \beta \ll 1 \text{,}
\end{align}
where $\beta = 2 \mu_0 p / (B^{c})^2$ is the plasma beta parameter. 
This parameter is typically small in the core and is even smaller in the SOL ($\beta \sim 0.004$ inferred from reference \cite{Militello-Fundamenski-2011}), so that the field produced by the plasma in the magnetic presheath is much smaller than the externally generated one. 

In order to neglect the plasma produced magnetic field, I require each component of it to be negligible compared to either the respective component or the smallest retained component of the external magnetic field $\vec{B}^c$. 
Considering the non-zero components of $\vec{B}^c$ (the $z$ and $x$ components), the orderings $B^p_z \sim B^p \ll B_z^c \sim B^c$ and $B_x^p \sim \delta \alpha B^p \ll B_x^c \sim \alpha B^c$ are required, which are both satisfied if the inequality (\ref{ordering-beta}) holds. 
In addition to this we require that $B_y^p \ll B_x^c$ (because $B^c_x$ is the smallest retained component of the external magnetic field), which is satisfied if (\ref{ordering-beta}) holds. 
This discussion justifies taking $\vec{B} = \vec{B}^c = \text{constant}$ in my equations and hence neglecting all plasma produced magnetic field components.

The electrostatic approximation is valid if each component of the \emph{non-electrostatic} piece, $\vec{E}^p$, of the electric field (which is induced by the plasma produced magnetic fields) is negligible compared to either the respective component or the smallest retained component of the \emph{electrostatic} piece, $-\nabla \phi$, of the electric field. 
The smallest retained component of the electric field is $\partial \phi / \partial y \sim T_{\text{e}} /el$ because $\partial \phi / \partial z$ will be neglected, as discussed earlier. 
With this consideration and using (\ref{order-E}), $E_x^p \ll T / e\rho_{\text{i}} \sim v_{\text{t,i}} B^c $, $E_y^p \ll T / el \sim \delta v_{\text{t,i}} B^c $ and $E_z^p \ll T / el \sim \delta v_{\text{t,i}} B^c $ are required in order to justify the electrostatic approximation. 
The induction equation is
\begin{align} \label{induction}
\frac{\partial \vec{B}^{p}}{\partial t} = - \nabla \times \vec{E}^{p} \text{.}
\end{align}
Using (\ref{t-dep}) and (\ref{ordering-Bplasma}) to order the left hand side, and (\ref{xyz-order}) to order the partial derivatives on the right hand side of (\ref{induction}), I obtain an ordering for the induced electric field components,
\begin{align} \label{eq:ordering-Einduced}
E_z^p \sim \delta^2 E^p \sim \delta \alpha E^p \ll E_y^p \sim \delta E^p \ll E_x^p \sim E^p \sim \delta v_{\text{t,i}} B^p \text{.}
\end{align}
In order to neglect $E_x^p$ and $E_y^p$ compared to their electrostatic counterparts I require $\delta B^{p} \ll B^{c}$, which is automatically satisfied if (\ref{ordering-beta}) holds. 
It follows that $E_z^p$ can also be neglected, because $E_z^p \ll E_y^p$ (from (\ref{eq:ordering-Einduced})) and the neglect of $E_y^p$ has been justified. 
This discussion justifies the electrostatic approximation and hence the use of $\vec{E} = - \nabla \phi$ in the equations of this paper.

Note that my orderings do not preclude a larger parallel current $\vec{j}^L$ (e.g. due to divertor target potential bias, Edge Localized Mode disruptions \cite{Kirk-2006}, etc.) with $j_x^L = - j^L \sin\alpha $, $j^L_y = 0$ and $j_z^L = j^L \cos \alpha$, provided that the magnetic field produced by the plasma in the magnetic presheath remains much smaller than the external one and that the electrostatic assumption remains valid. 
In order for our approximations to be valid, such a current would have to satisfy $\nabla \cdot \vec{j}^L = 0$ independently of $\vec{j}^{\text{D}}$. 
In Appendix \ref{appendix:largejz}, I show that my equations allow for a parallel current density of size $j^L \ll \left( \alpha/\beta \right) \delta en_{\text{i}}v_{\text{t,i}}$. 
This current density can be large, $j^L \gg j_z^D \sim \delta n_{\text{i}} e v_{\text{t,i}}$, because $\alpha \sim 0.1 \gg \beta \sim 0.004$ in the magnetic presheath.  

To conclude, in this section I have justified and introduced the following assumptions and orderings:
\begin{itemize}
\item The magnetic field $\vec{B}$ is uniform and constant in time, making a small angle $\alpha \ll 1$ with the wall;
\item the electric field is described by an electrostatic potential $\phi(x,y)$;
\item the scale length of variation perpendicular to the wall is set by the size of the ion gyro-orbits, $\rho_{\text{i}}$; 
\item turbulent variations in the plane parallel to the wall are constrained to the direction perpendicular to the magnetic field, with a typical length scale $l$ longer than the ion gyroradius $\rho_{\text{i}}$ such that $\delta = \rho_{\text{i}} / l \ll 1$.
\end{itemize}

\section{Periodic orbits} \label{sec-traj-periodic}

When $\alpha = \delta = 0$, the uniform magnetic field is parallel to the wall and the system is translationally invariant in the plane parallel to the wall.
I proceed to study ion trajectories in this system, as a lowest order solution to the ion trajectories in the magnetic presheath, where $\alpha \sim \delta \ll 1$.
Section \ref{subsec-traj-orbitparameters} is devoted to obtaining constants of the ion motion in such a system.
Using these constants of the motion, the ion velocity is expressed in terms of the instantaneous position and the orbit parameters, using an ``effective potential''.
In Section \ref{subsec-traj-effpottypes} I introduce two distinct types of effective potential curves.
In Section \ref{subsec-traj-closed} I briefly discuss periodic solutions to the equations of motion with $\alpha = \delta = 0$.

\subsection{Orbit parameters} \label{subsec-traj-orbitparameters}

The equations of motion for an ion moving in the collisionless magnetic presheath are \cite{Geraldini-2017}
\begin{align}
\label{x-EOM-exact}
\dot{x} = v_x \text{,}
\end{align}
\begin{align}
\label{y-EOM-exact}
\dot{y} = v_y \text{,}
\end{align}
\begin{align}
\label{z-EOM-exact}
\dot{z} = v_z \text{,}
\end{align}
\begin{align}
\label{vx-EOM-exact}
\dot{v}_x = -\frac{\Omega}{B} \frac{\partial \phi}{\partial x} + \Omega v_{y}\cos\alpha \text{,}
\end{align}
\begin{align}
\label{vy-EOM-exact}
\dot{v}_y = -\frac{\Omega}{B} \frac{\partial \phi}{\partial y} - \Omega v_{x}\cos\alpha - \Omega v_{z}\sin\alpha \text{,}
\end{align}
\begin{align}\label{vz-EOM-exact}
\dot{v}_z = -\frac{\Omega}{B} \frac{\partial \phi}{\partial z} + \Omega v_{y}\sin\alpha \text{,}
\end{align}
where a dot $\dot{}$ denotes a time derivative, $d / dt$.
Setting $\alpha = 0$ and $\delta = 0$, equations (\ref{vx-EOM-exact})-(\ref{vz-EOM-exact}) become
\begin{align}
\label{vx-EOM-zero}
\dot{v}_x = -\frac{\Omega}{B} \frac{\partial \phi}{\partial x} + \Omega v_{y} \text{,}
\end{align}
\begin{align}
\label{vy-EOM-zero}
\dot{v}_y =  - \Omega v_{x}  \text{,}
\end{align}
\begin{align}\label{vz-EOM-zero}
\dot{v}_z = 0 \text{.}
\end{align}
Using (\ref{x-EOM-exact}), direct integration of (\ref{vy-EOM-zero}) leads to
\begin{align} \label{xbar-def}
 \bar{x} = \frac{v_y}{\Omega} +  x \sim \rho_{\text{i}} \text{,}
\end{align}
where $\bar{x}$ is the constant of integration which represents the position of an ion orbit. 
The statement that $\bar{x} \sim \rho_{\text{i}}$ is understood to mean ``the typical changes of the value of $\bar{x}$ (that occur as ions move in the magnetic presheath) are of the order of $\rho_{\text{i}}$''.
Multiplying (\ref{vx-EOM-zero}) by $v_x$ and adding it to (\ref{vy-EOM-zero}) multiplied by $v_y$, we obtain $\dot{U}_{\perp} = 0$, where
\begin{align} \label{Uperp-def}
U_{\perp} = \frac{1}{2} v_x^2 + \frac{1}{2} v_y^2 + \frac{\Omega \phi}{B} \sim v_{\text{t,i}}^2
\end{align}
is the perpendicular energy.
From (\ref{vz-EOM-zero}), the parallel velocity $v_z$ of the ion is a constant of the motion. Adding the parallel kinetic energy $v_z^2 / 2 $ to the perpendicular energy, we obtain the total energy, 
\begin{align} \label{U-def}
U = \frac{1}{2} v_x^2 + \frac{1}{2} v_y^2 + \frac{1}{2} v_z^2 + \frac{\Omega \phi }{B}  \sim v_{\text{t,i}}^2 \text{.}
\end{align}
The quantities $\bar{x}, ~ U_{\perp} \text{ and } U$ constitute the orbit parameters of ion motion. 
When $\alpha = \delta = 0$ they are exactly conserved, and when $\alpha \sim \delta \ll 1$ they change slowly (except for $U$ which remains constant).

The ion velocity components $v_x, ~ v_y \text{ and } v_z$ can be expressed in terms of the orbit parameters and the instantaneous ion position $x$. Inserting (\ref{xbar-def}) into (\ref{Uperp-def}) and rearranging gives
\begin{align} \label{vx-Uperp-xbar-x}
v_x = \sigma_x V_x \left( x, y, \bar{x}, U_{\perp} \right)  \text{ with } V_x \left( x, y, \bar{x}, U_{\perp} \right) = \sqrt{ 2\left( U_{\perp} - \chi\left( x; y, \bar{x} \right) \right) } \text{,}
\end{align}
where I introduced $\sigma_x = \pm 1$ to account for the two possible signs of $v_x$, and an effective potential function
\begin{align} \label{chi-def}
\chi \left(x; y, \bar{x} \right) = \frac{1}{2} \Omega^2 \left(x - \bar{x} \right)^2 + \frac{\Omega \phi(x,y)}{B} \text{.}
\end{align}
The dependence on $z$ is negligible even to first order in $\delta$, and therefore it is omitted.
The $y$-component of the velocity is obtained by rearranging equation (\ref{xbar-def}),
\begin{align} \label{vy-xbar-x}
\dot{y} = v_y = \Omega \left( \bar{x} - x \right) \text{.}
\end{align} 
The $z$-component of the velocity is obtained by subtracting equation (\ref{Uperp-def}) from (\ref{U-def}), multiplying by $2$ and taking a square root,
\begin{align} \label{vz-U-Uperp}
v_z = \sigma_{\parallel} V_{\parallel} \left( U_{\perp}, U \right) \text{ with } V_{\parallel} \left( U_{\perp}, U \right) = \sqrt{2\left( U - U_{\perp} \right) } \text{,}
\end{align}
where $\sigma_{\parallel} = \pm 1$ is the sign of $v_z$.

\subsection{Types of effective potential curves} \label{subsec-traj-effpottypes}

\begin{figure}
\centering
\includegraphics[width= 0.47\textwidth]{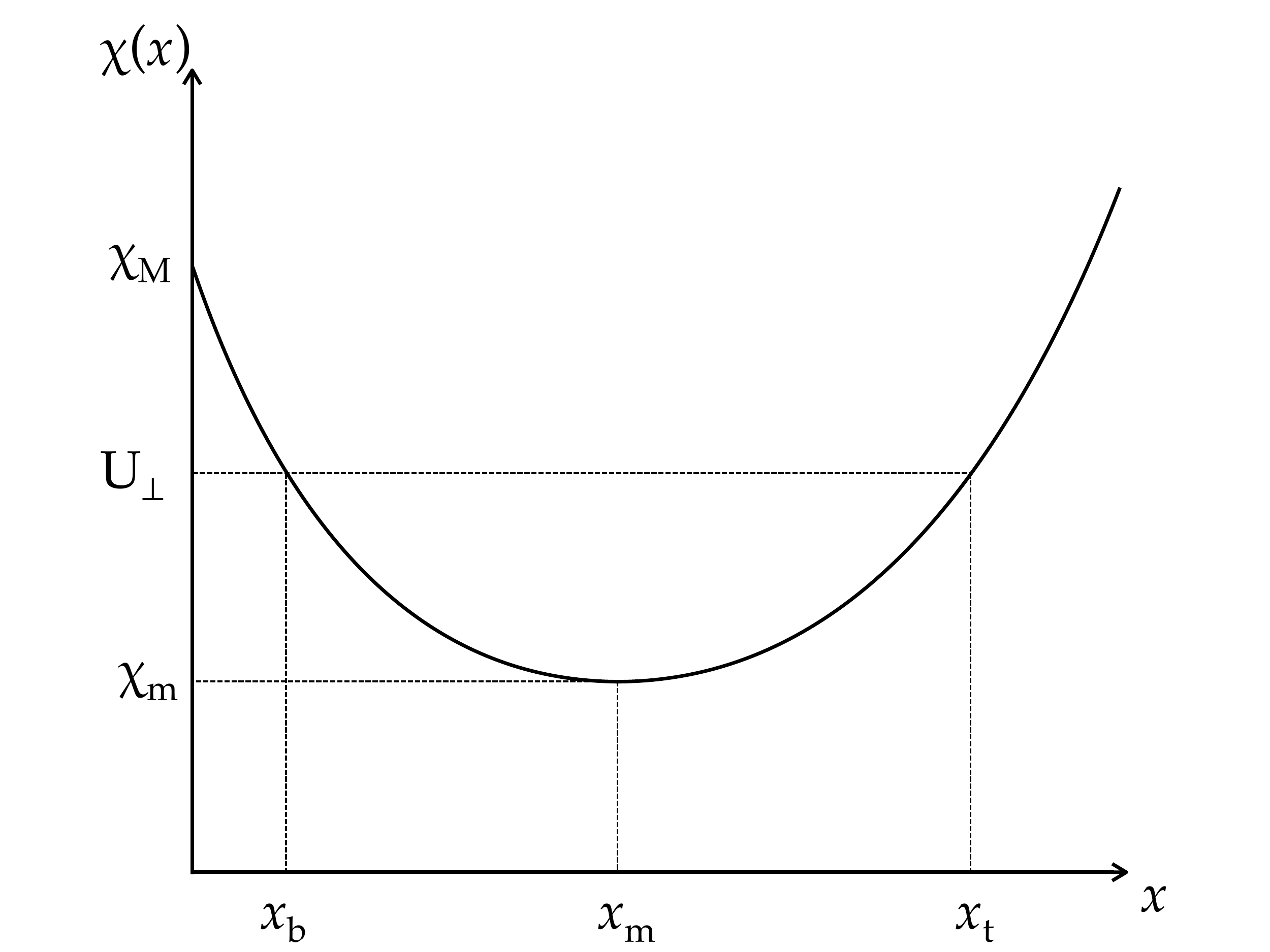}
\includegraphics[width= 0.47\textwidth]{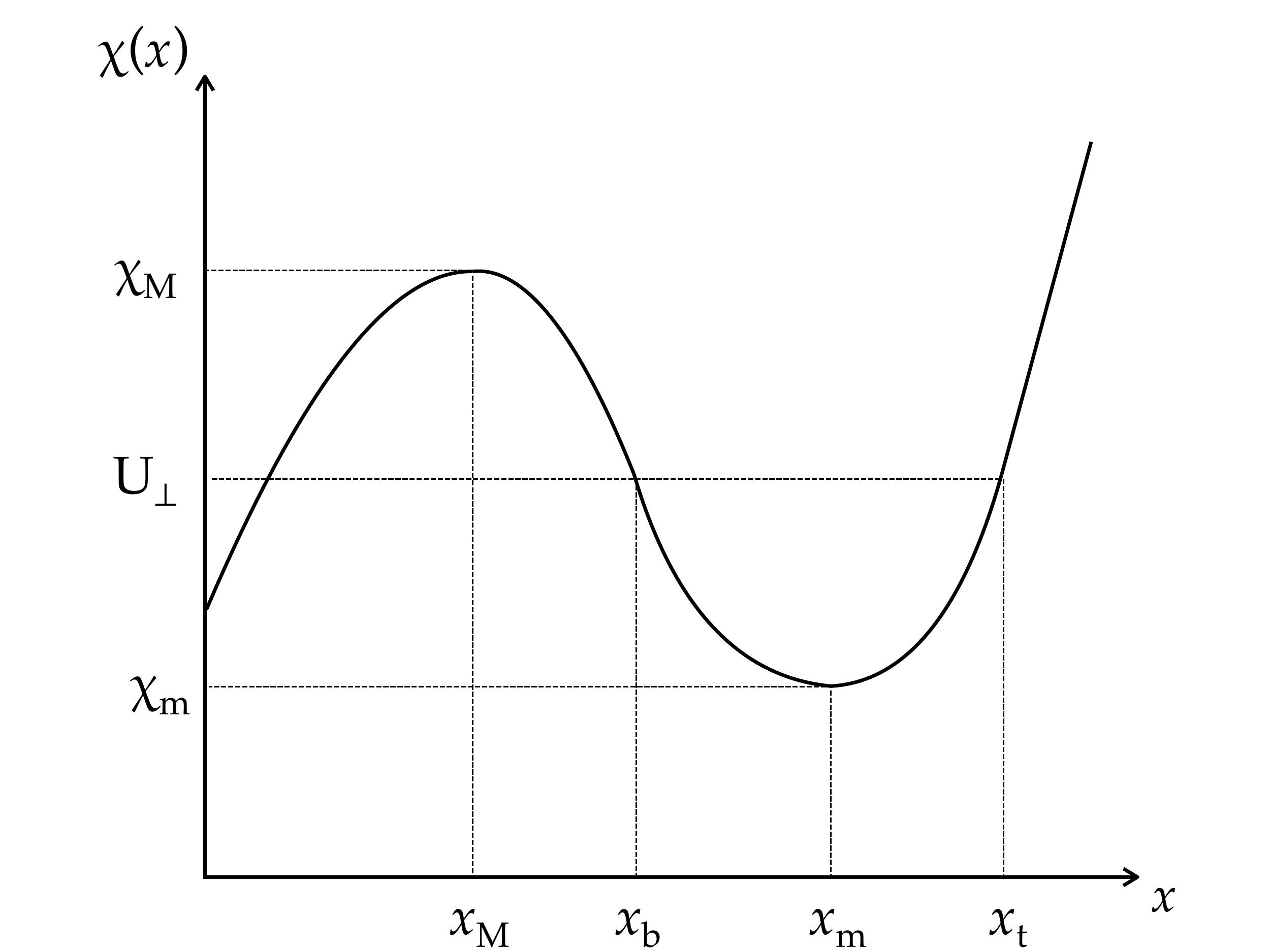}
\caption[Effective potential types]{Type I (left) and II (right) effective potential curves,
both with a stationary point corresponding to a minimum at $x=x_{\text{m}}$. 
A type II curve is characterized by a stationary point corresponding to a maximum at $x=x_{\text{M}}$. 
These curves allow closed orbits for any value of $U_{\perp}$ in the range $\chi_{\text{m}} \left(y, \bar{x} \right) \leqslant U_{\perp} \leqslant \chi_{\text{M}} \left(y, \bar{x} \right) $  with bottom and top bounce points at positions $x_{\text{b}}$ and $x_{\text{t}}$.
}
\label{fig-effpotclosed}
\end{figure}

\begin{figure}
\centering
\includegraphics[width= 0.6\textwidth]{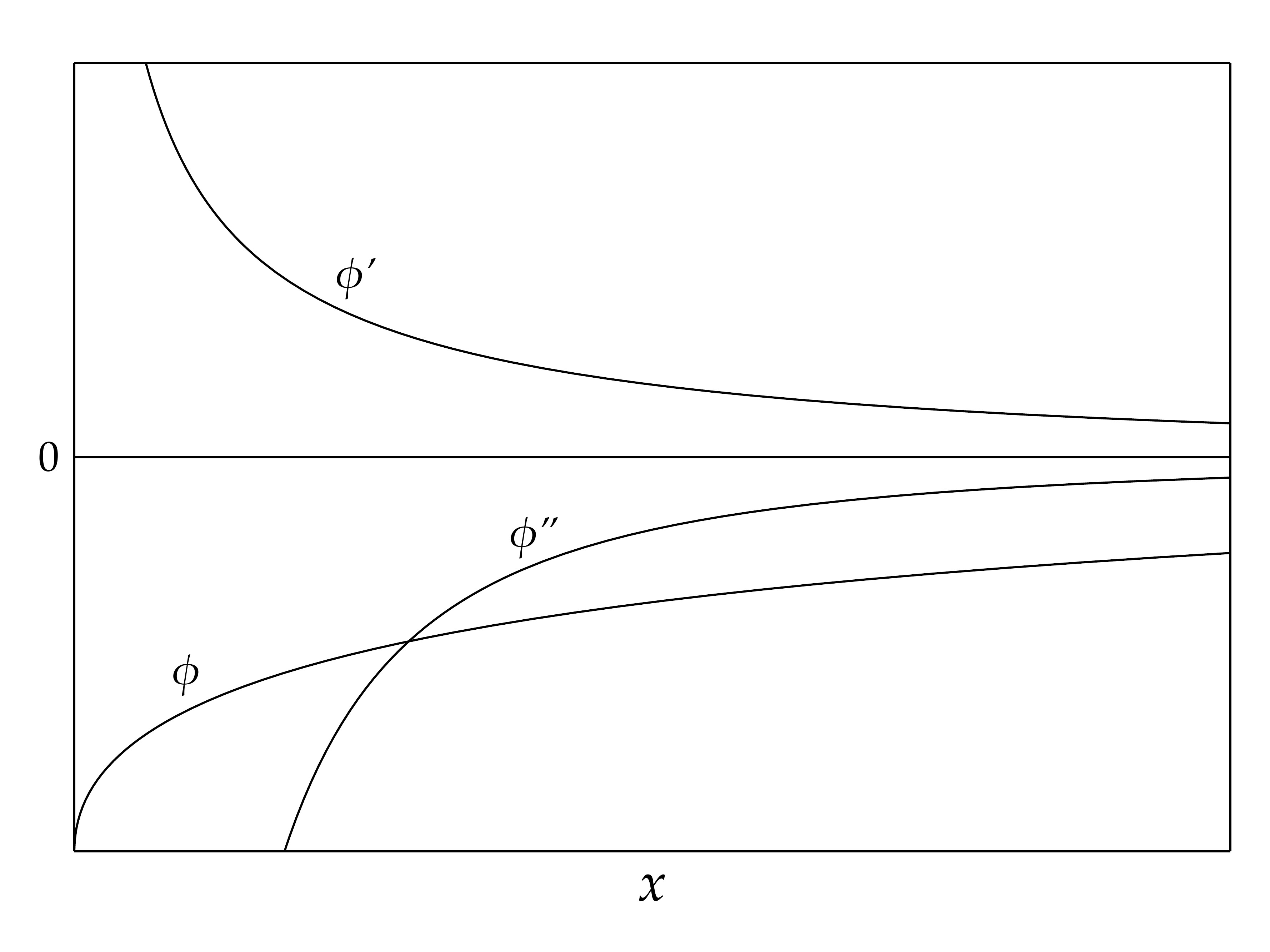}
\caption[Electrostatic potential example and its derivatives]{An example of a monotonic electrostatic potential profile $\phi(x)$ and its monotonic first and second derivatives $\phi'(x)$ and $\phi''(x)$.
}
\label{fig-phiandderivatives}
\end{figure}

By imposing that $v_x$ be real in equation (\ref{vx-Uperp-xbar-x}), I find that the allowed ion positions must satisfy $U_{\perp} \geqslant \chi \left(x;  y, \bar{x} \right)$.   
A particle moves periodically if, for given values of $U_{\perp}$ and $\bar{x}$, it is trapped around a minimum (with respect to $x$) of the effective potential $\chi(x; y, \bar{x})$. 
Then, the ion motion is confined between bounce points $x_{\text{b}}$ (bottom) and $x_{\text{t}}$ (top) defined by (see Figure \ref{fig-effpotclosed})
\begin{align} \label{closed-orbit-bounce}
U_{\perp} = \chi \left(x_b; y, \bar{x} \right) =  \chi \left(x_t; y, \bar{x} \right) \text{ with }  x_{\text{b}} \leqslant x_{\text{t}} \text{.} 
\end{align}
Throughout this work, the electrostatic potential across the magnetic presheath is assumed to be such that $\phi (x,y)$, $\partial \phi(x,y) / \partial x $ and $\partial^2 \phi(x,y) / \partial x^2$ are all monotonic as a function of $x$, as shown in Figure \ref{fig-phiandderivatives}.
My numerical results, presented in chapters \ref{chap-KMPS} and \ref{chap-Tdep} and obtained with no $y$-dependence ($\delta = 0$), satisfy these monotonicity conditions. 
Then, for values of $\bar{x}$ and $y$ for which the effective potential has a minimum, there are two possible types of effective potential $\chi(x; y, \bar{x})$:
\begin{itemize}
\item a type I effective potential has one stationary point, corresponding to a minimum at $x_{\text{m}}$, such that $\chi_{\text{m}} \left(\bar{x}, y \right)  \equiv \chi \left( x_{\text{m}}; y, \bar{x} \right) $, and \emph{no} other stationary point --- in this case, it is important to consider the local maximum at position $x_{\text{M}} = 0$ (which is \emph{not} a stationary point) with $\chi_{\text{M}} \left( \bar{x}, y \right) = \chi \left( 0, \bar{x}, y \right) $;
\item a type II effective potential has two stationary points: one at position $x_{\text{m}}$ which corresponds to a minimum $\chi_{\text{m}} \left(\bar{x} , y\right)$, and one at position $x_{\text{M}}$ which corresponds to a maximum $\chi_{\text{M}} \left( y, \bar{x} \right) \equiv \chi \left( x_{\text{M}}; y, \bar{x} \right)$.
\end{itemize}
These two types of effective potential are shown in Figure \ref{fig-effpotclosed}. 
I will refer to the ion trajectories arising due to each curve type as type I and type II orbits.

I proceed to obtain the range of values of $\bar{x}$ for which the effective potential is of either type, and to give a physical explanation of the difference between the two types of curves.
Differentiating equation (\ref{chi-def}) with respect to $x$, we obtain
\begin{align} \label{chi'}
\frac{\partial \chi}{\partial x} (x; y, \bar{x}) = \Omega^2 (x - \bar{x} ) + \frac{\Omega }{B} \frac{\partial \phi}{\partial x} \left(x,y\right) \text{.}
\end{align} 
Substituting $v_y = \Omega \left( \bar{x} - x \right)$ into equation (\ref{vx-EOM-zero}), which is valid to lowest order in $\alpha$ and $\delta$, one sees that $-\partial \chi (x; y, \bar{x}) / \partial x $ coincides with the acceleration of an ion at that point, in the $x$ direction. 
Indeed, the magnetic force in the $x$ direction is, to lowest order in $\alpha$ and $\delta$, given by $m_{\text{i}} \Omega v_y = m_{\text{i}} \Omega^2 \left( \bar{x} - x \right) $ and the electric force is $-m_{\text{i}} \Omega \phi'(x,y)/B$.
Consider an ion that reaches the point $x=0$.
For type I curves the gradient of the effective potential at $x=0$ must be negative. 
Hence, from equation (\ref{chi'}), we obtain $-\Omega^2 \bar{x}  + \Omega \phi'(0,y)/ B < 0$ which leads to the requirement that $\bar{x} > \bar{x}_{\text{m,I}}(y)$ with
\begin{align} \label{xbarmI}
\bar{x}_{\text{m,I}} (y) = \frac{ \phi'(0,y) }{ \Omega B } \text{.} 
\end{align} 
Physically, the difference between type I and type II curves only manifests itself only when the ion reaches the region $x \leqslant x_{\text{M}}$, and therefore always manifests itself at $x=0$. 
For type I curves, the magnetic force directed away from the wall at $x=0$ is larger than the electric force towards the wall.
For type II curves, the opposite is true and the electric force towards the wall is larger than the magnetic force turning the ions away from the wall at $x=0$.

\begin{figure}
\centering
\includegraphics[width= 0.7\textwidth]{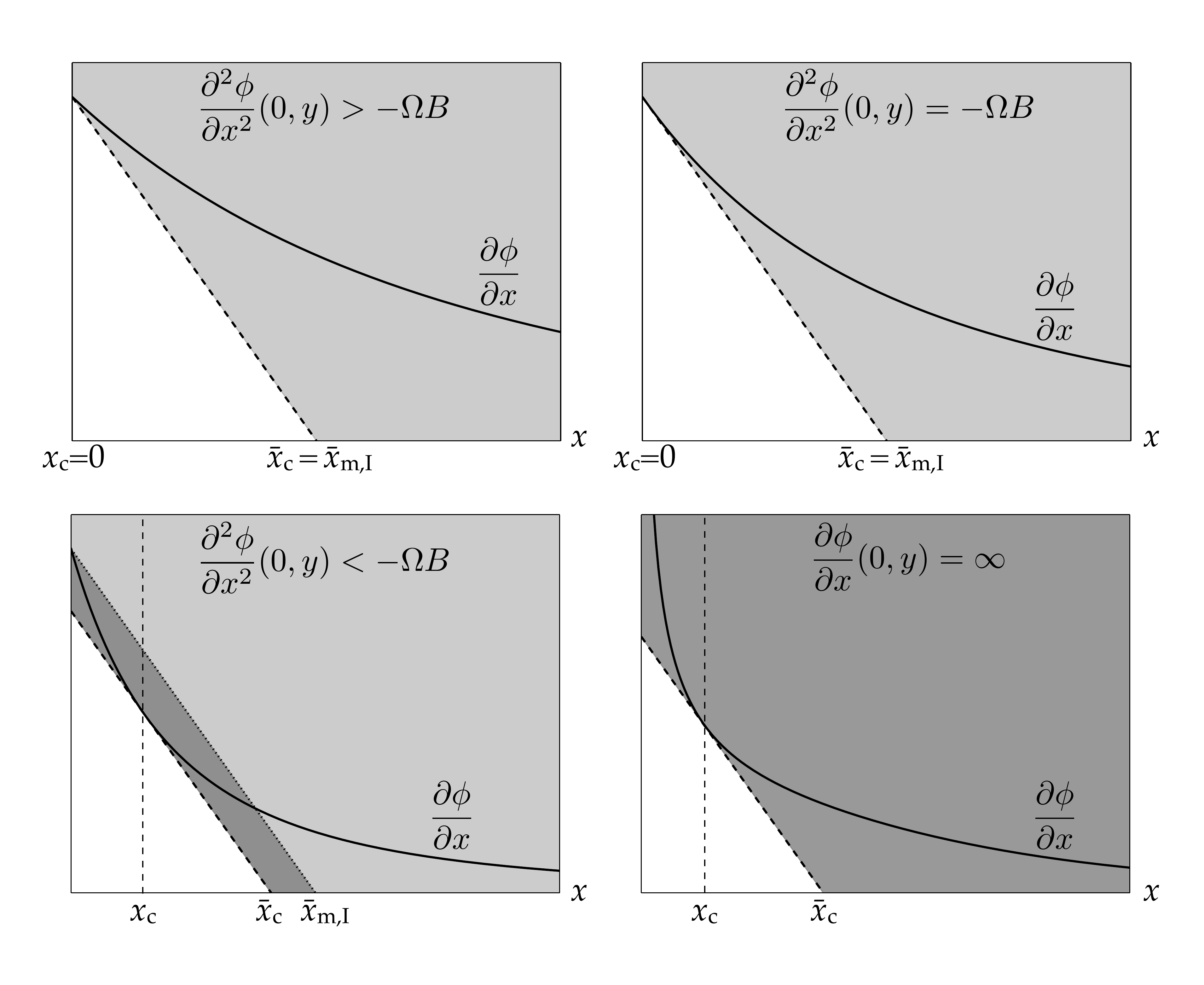}
\caption[Values of orbit position for which the effective potential is type I or type II]{The stationary points of the effective potential satisfy equation (\ref{stationary-points}), $\partial  \phi /\partial x = \Omega B \left( \bar{x} - x \right)$. 
In each of the four diagrams the solid curves represent $\partial \phi  / \partial x$, the oblique dashed lines are $\Omega B \left( \bar{x} - x_{\text{c}} \right)$, while the additional oblique dotted line in the bottom-left diagram is $\Omega B \left( \bar{x} - x_{\text{m,I}} \right)$.
For given values of $y$ and $\bar{x}$, equation (\ref{stationary-points}) can have two solutions (dark grey region, $\chi$ is type II), one solution (light grey region, $\chi$ is type I) or no solution (unshaded region, $\chi$ has no minimum).
The smallest value of $\bar{x}$ for which a stationary point exists, at position $x_{\text{c}}(y)$, is $\bar{x}_{\text{c}}(y)$.
The value of $\bar{x}$ which corresponds to a stationary point at $x=0$ is $\bar{x}_{\text{m,I}}(y)$.
}
\label{fig-typesofcurves}
\end{figure}

Setting equation (\ref{chi'}) to zero gives an equation for the stationary points of $\chi$, which can be rearranged to
\begin{align} \label{stationary-points}
\frac{ \partial  \phi }{\partial x} \left( x,y \right) = \Omega B \left( \bar{x}  - x \right)  \text{.}
\end{align}
The stationary points are minima if the second derivative of $\chi$ is positive.
This condition is equivalent to the gradient of $\partial \phi (x,y) / \partial x$ with respect to $x$ being larger than the gradient of the line $\Omega B ( \bar{x } - x)$.
By rearranging equation (\ref{stationary-points}) to an equation for $\bar{x}$ as a function of $x$ and then minimizing it with respect to $x$, we obtain the minimum value of the orbit position $\bar{x}_{\text{c}}(y)$ for which the effective potential has a stationary point,
\begin{align} \label{xbarc}
\bar{x}_{\text{c}} (y) \equiv \min_{x \in [0,\infty]} \left( x  + \frac{\phi'( x, y )}{\Omega B} \right) \equiv x_{\text{c}}(y) + \frac{\phi'(x_{\text{c}}(y),y)}{\Omega B}  \text{.}
\end{align}
Note that, in equation (\ref{xbarc}), I also defined the position $x_{\text{c}}(y)$ of the stationary point of the effective potential $\chi$ when $\bar{x} = \bar{x}_{\text{c}}(y)$.
In Figure \ref{fig-typesofcurves}, $\bar{x}_{\text{c}}(y)$ is the smallest value of $\bar{x}$ for which the straight line $\Omega B \left( \bar{x} - x \right) $ touches the curve $\partial \phi (x,y) / \partial x$, and $x_{\text{c}}(y)$ is the value of $x$ at which they intersect. 
From Figure \ref{fig-typesofcurves}, $\bar{x}_{\text{c}}$ and $\bar{x}_{\text{m,I}}$ coincide if $\partial^2 \phi (0,y) / \partial x^2 \geqslant - \Omega B$. 
Then, all effective potential curves are type I for $\bar{x} > \bar{x}_{\text{c}}(y) = \bar{x}_{\text{m,I}} (y) $.
If $\partial^2 \phi (0,y) / \partial x^2 < - \Omega B$, $\bar{x} = \bar{x}_{\text{c}}(y)$ is the orbit parameter value corresponding to when the straight line $\Omega B (\bar{x} - x )$ touches the curve $\partial \phi (x,y) / \partial x$ tangentially.
Then, for orbit parameter values in the range $\bar{x}_{\text{c}} (y) \leqslant \bar{x} \leqslant \bar{x}_{\text{m,I}} (y)$, there are two stationary points (a minimum in the region $x>x_{\text{c}}(y)$ and a maximum in the region $0 \leqslant x < x_{\text{c}}(y)$), corresponding to type II curves, while for $\bar{x} > \bar{x}_{\text{m,I}}(y)$ there is only one minimum, corresponding to type I curves.
Summarizing these observations with the aid of Figure \ref{fig-typesofcurves}:
\begin{itemize}
\item if $  \partial^2  \phi (0,y) / \partial x^2  \geqslant - \Omega B $, $\chi$ is a type I curve for $\bar{x} > \bar{x}_{\text{c}}(y) = \bar{x}_{\text{m,I}}(y)$;
\item if $ \partial^2  \phi (0,y) / \partial x^2    < - \Omega B $, $\chi$ is a type II curve for $ \bar{x}_{\text{c}}(y) < \bar{x} < \bar{x}_{\text{m,I}}(y)$ and a type I curve for $ \bar{x} > \bar{x}_{\text{m,I}}(y)$.
\end{itemize}

In chapter \ref{chap-KMPS} I set $\delta = 0$ and thus solve a magnetic presheath with no $y$-dependence.
I will show that a self-consistent solution $\phi(x)$ of such a magnetic presheath requires the electric field to diverge at $x=0$, $\partial \phi (0) / \partial x  \rightarrow \infty$, which is confirmed numerically.
Hence, the effective potential curves are type II for all values of $\bar{x}$ larger than $\bar{x}_{\text{c}}$ because $\bar{x}_{\text{m,I}} = \left( \partial \phi(0) / \partial x \right) / \Omega B  \rightarrow \infty$ (see Figure \ref{fig-typesofcurves}, bottom-right diagram). 
It is nonetheless useful to consider also type I curves because I obtain my numerical solution by iterating over possible electrostatic potential profiles starting from the initial guess of a flat potential, $\phi(x) =0$ for all $x$.

\subsection{Periodicity} \label{subsec-traj-closed}

The ion motion for $\alpha =0$ and $\delta = 0$ is a periodic (closed) orbit provided that an effective potential minimum exists, $\bar{x}> \bar{x}_{\text{c}}(y)$, and that a pair of bounce points $x_{\text{b}}$ and $x_{\text{t}}$ exist, $U_{\perp} < \chi_{\text{M}} (y, \bar{x})$ (see Figure \ref{fig-effpotclosed}). 
For a closed orbit, the position $x$ can be written as a function of a gyrophase angle which parameterizes the particular point of the orbit in which the particle lies. 
The period of the orbit, $2\pi/ \overline{\Omega}$, where $\overline{\Omega}$ is the generalized gyrofrequency, is the integral of all the time elements $dt = dx/v_x$ over a whole orbit,
\begin{align} \label{Omegabar-def}
\frac{2\pi}{\overline{\Omega}} = 2 \int_{x_{\text{b}}}^{x_{\text{t}}} \frac{dx}{V_x \left( x, y, \bar{x}, U_{\perp} \right)} \text{.}
\end{align}
The gyrophase angle $\varphi$ of the orbit is defined as $\overline{\Omega} t$, where $t$ is defined in the interval $-\pi / \overline{\Omega} < t < \pi / \overline{\Omega} $ and is (when positive) the time elapsed since the particle last reached the top bounce point,
\begin{align} \label{varphi-def}
\varphi = \sigma_x \overline{\Omega} \int_{x_{\text{t}}}^x \frac{ds}{V_x \left( s, y, \bar{x}, U_{\perp} \right) } \text{.}
\end{align}
The time derivative of the gyrophase is the gyrofrequency, 
\begin{align} \label{varphidot}
\dot{\varphi} = \overline{\Omega} \text{.}
\end{align}
To obtain this, I have used $\partial \varphi / \partial x = \sigma_x \overline{\Omega} /V_x$ and $\sigma_x \dot{x} = V_x$ to simplify $\dot{\varphi} = \sigma_x \overline{\Omega} \dot{x} \partial \varphi / \partial x $.
Note that inverting (\ref{varphi-def}) gives $x(\varphi, y, \bar{x}, U_{\perp})$.
It will turn out that, for $\alpha \ll 1$, the ion distribution function is independent of $\varphi$.

It will be useful to define the gyroaveraging operation as a time average over possible values of gyrophase, or equivalently as an average over the period of a closed orbit,
\begin{align} \label{gyroaverage}
\left\langle \ldots \right\rangle_{\varphi} = \frac{1}{2\pi} \int_{-\pi}^{\pi} \left( \ldots \right) d\varphi = \sum_{\sigma_x = \pm 1} \frac{\overline{\Omega}}{2\pi} \int_{x_{\text{b}}}^{x_{\text{t}}} \frac{\left( \ldots \right) dx}{V_x \left( x,y, \bar{x}, U_{\perp} \right)} \text{.}
\end{align}
The second equality in (\ref{gyroaverage}) is obtained using (\ref{varphi-def}). 
The closed orbit has an $\vec{E}\times \vec{B}$ drift in the $y$ direction (parallel to the wall), with drift velocity $V_{\vec{E} \times \vec{B} }$ defined as the gyroaverage of $v_y$,
\begin{align} \label{V-ExB}
V_{\vec{E} \times \vec{B} } \left( y, \bar{x}, U_{\perp} \right) = \frac{ \overline{\Omega} }{\pi} \int_{x_{\text{b}}}^{x_{\text{t}}} \frac{\Omega \left( \bar{x} - x \right)}{V_x \left( x, y, \bar{x}, U_{\perp} \right)} dx = \frac{ \overline{\Omega} }{\pi} \int_{x_{\text{b}}}^{x_{\text{t}}} \frac{ \left(1/B\right) \partial \phi (x,y) / \partial x  }{V_x \left( x, y, \bar{x}, U_{\perp} \right)}  dx \text{.}
\end{align}
The second equality in (\ref{V-ExB}) comes from using equation (\ref{chi'}) and the result
\begin{align} \label{gyroaveragevy-trick}
\int_{x_{\text{b}}}^{x_{\text{t}}} \frac{\partial \chi (x;y, \bar{x})/ \partial x }{V_x \left( x, y, \bar{x}, U_{\perp} \right)} dx & = - \int_{x_{\text{b}}}^{x_{\text{t}}}   \frac{ \partial V_x }{\partial x}  \left( x, y, \bar{x}, U_{\perp} \right)  dx \nonumber \\
& = V_x \left( x_{\text{b}}, y, \bar{x}, U_{\perp} \right)  - V_x \left( x_{\text{t}}, \bar{x}, U_{\perp} \right) = 0 \text{,}
\end{align}
where I used $V_x \left( x_{\text{b}}; y, \bar{x}, U_{\perp} \right)  = V_x \left( x_{\text{t}}; y, \bar{x}, U_{\perp} \right) = 0$. The first equality in (\ref{gyroaveragevy-trick}) comes from differentiating equation (\ref{vx-Uperp-xbar-x}).

\section{Magnetic presheath trajectories} \label{sec-traj-real}

The main effect of non-zero $\alpha$ and $\delta$ is to break the exact periodicity by making the orbit parameters vary over a characteristic time $1/\alpha \Omega \sim 1/\delta \Omega \gg 1/ \Omega$.
Using the results of Section \ref{sec-traj-periodic}, in Section \ref{subsec-traj-appclosed} I explain how the ion trajectories consist of a sequence of approximately closed orbits.
I quantify the variation of the orbit parameters to first order in $\alpha$ and $\delta$, and obtain two quantities that are conserved to lowest order over the long timescale $1/\alpha \Omega  \sim 1/\delta \Omega$: the adiabatic invariant and the $z$-component of the canonical momentum. 
A time $\sim 1/\Omega$ before the ion reaches the wall, the ion is considered to be in an ``open'' orbit. 
In Section \ref{subsec-traj-open}, I define an open orbit and obtain the conditions that orbit parameters must satisfy for an ion to be in an open orbit.

\subsection{Approximately closed orbits} \label{subsec-traj-appclosed}

When $\alpha = \delta = 0$ an ion moves in a closed orbit which $\vec{E}\times \vec{B}$ drifts in the $y$ direction (equation (\ref{V-ExB})) and streams parallel to the magnetic field in the $z$ direction (equation (\ref{vz-U-Uperp})). 
When $\alpha \sim \delta \ll 1$, the motion is \emph{approximately} periodic because the orbit parameters vary over a timescale $1/\alpha \Omega$ that is much longer than the typical gyroperiod $1/ \Omega$.
The equations of motion (\ref{vx-EOM-exact})-(\ref{vz-EOM-exact}) are approximately given by
\begin{align}
\label{vx-EOM-first}
\dot{v}_x = -\frac{\Omega}{B} \frac{\partial \phi}{\partial x} + \Omega v_{y} + O\left( \alpha^2 \Omega v_{\text{t,i}} \right)  \text{,}
\end{align}
\begin{align}
\label{vy-EOM-first}
\dot{v}_y =  - \Omega v_{x} -\frac{\Omega}{B} \frac{\partial \phi}{\partial y}  - \alpha \Omega v_{z} + O\left( \alpha^2 \Omega v_{\text{t,i}} \right)  \text{,}
\end{align}
\begin{align}\label{vz-EOM-first}
\dot{v}_z = \alpha \Omega v_{y}  + O\left( \alpha^2 \Omega v_{\text{t,i}} \right)  \text{.}
\end{align}

The time variation of the total energy is given by
\begin{align} \label{Udot}
\dot{U} = O\left( \delta^2 \Omega v_{\text{t,i}}^2 \right) \text{}
\end{align}
because energy is exactly conserved up to explicit time dependence, which is second order in $\delta$ (equation (\ref{t-dep})). 
Differentiating (\ref{xbar-def}) with respect to time and using (\ref{vy-EOM-first}), we find
\begin{align} \label{xbardot}
\dot{\bar{x}} = - \sigma_{\parallel} \alpha V_{\parallel} \left( U_{\perp}, U \right)  - \frac{1}{B} \frac{\partial \phi}{\partial y} (x,y) + O \left( \alpha^2 v_{\text{t,i}} \right) \text{.}
\end{align}
Physically, this drift is a combination of the small $x$-component of the parallel motion, and the small $\vec{E} \times \vec{B}$ drift due to the weak turbulent electric fields \cite{Loizu-2012}.
Differentiating (\ref{Uperp-def}) and using (\ref{vx-EOM-first}) and (\ref{vy-EOM-first}) we get
\begin{align} \label{Uperpdot}
\dot{U}_{\perp}  = - \sigma_{\parallel} \alpha \Omega^2 V_{\parallel} \left( U_{\perp}, U \right) \left( \bar{x} - x\right)  + O \left( \alpha^2 \Omega v_{t,i}^2 \right) \text{.}
\end{align}

\begin{figure}
\centering
\includegraphics[width= 0.8\textwidth]{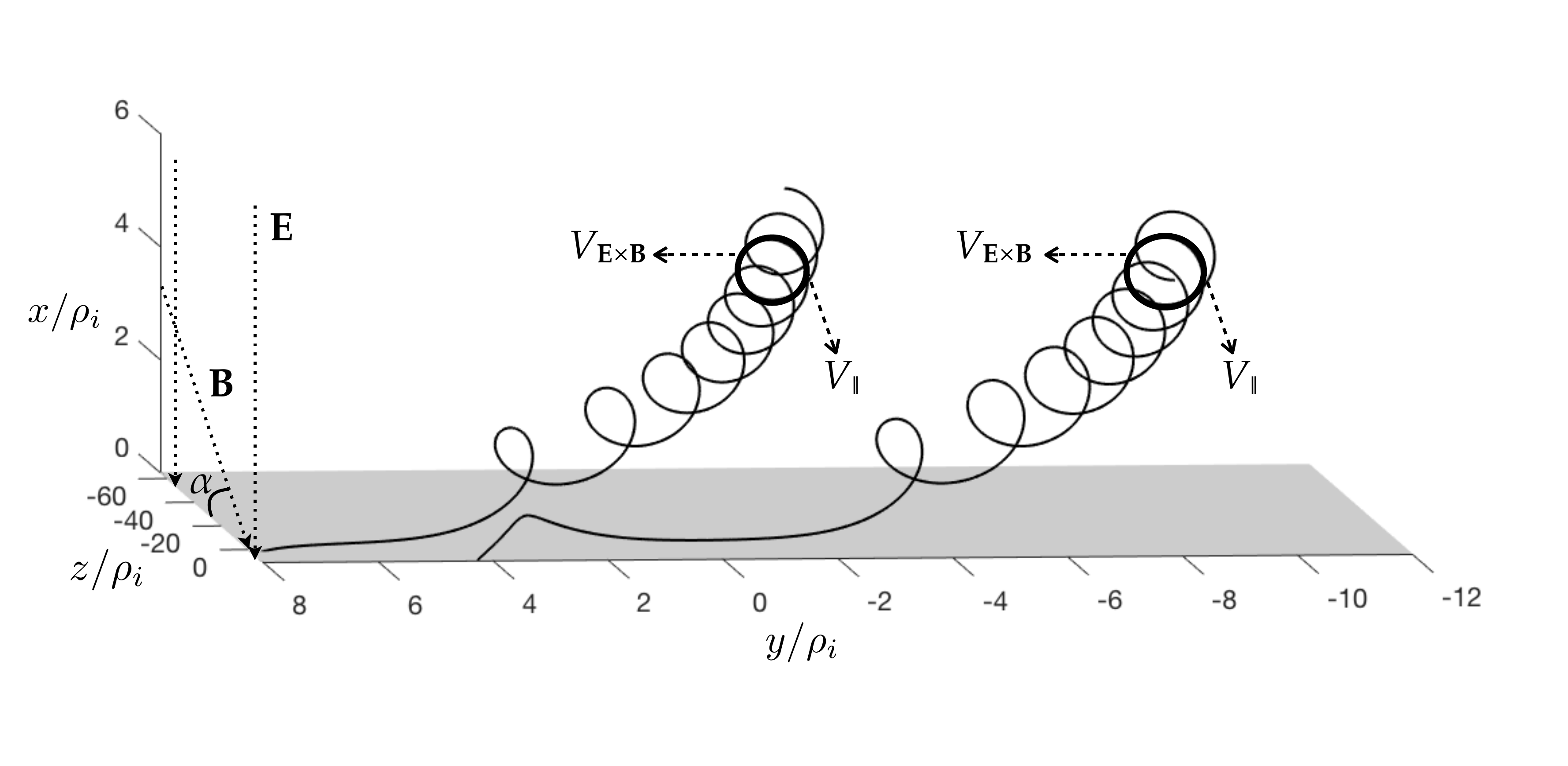}
\caption[Ion trajectories in the magnetic presheath]{Two ion trajectories approaching the wall, represented as a grey surface at $x=0$, are shown as black lines. 
The electric and magnetic fields are marked with dotted arrows. 
The angle between the magnetic field and wall is $\alpha = 0.05 \text{ radians}$, although it looks large because the $z$ direction has been squashed in order to draw the 3 dimensional ion trajectories. 
The electrostatic potential used to obtain the trajectories is the solution presented in chapter \ref{chap-KMPS}. 
Most of the ion motion is locally approximated by closed orbits, represented as superimposed rings. 
Ions stream along the magnetic field $\vec{B}$ at velocity $V_{\parallel} \left( U_{\perp}, U \right)$, and the strong electric field towards the wall causes the approximately closed orbits to $\vec{E} \times \vec{B}$ drift at velocity $V_{\vec{E} \times \vec{B} } \left( y, \bar{x}, U_{\perp} \right) $ in the $y$ direction. 
The increasing electric field as the orbits approach $x=0$ causes the  $\vec{E} \times \vec{B}$ velocity to noticeably increase (see equation (\ref{V-ExB})). 
Since these are type II orbits, there is a point ($x_{\text{M}}$) in which the electric and magnetic forces on the ion are equal and opposite, where the trajectories spend a long time moving almost tangential to the wall. 
The trajectory on the left crosses $x_{\text{M}}$ with near-zero $v_x$, while the trajectory on the right crosses $x_{\text{M}}$ with the largest velocity possible.
The two trajectories have, to lowest order, equal values of $\mu$ and $U$; the only difference between them is the gyrophase $\varphi$.
}
\label{fig-iontraj}
\end{figure}

Both $\dot{\bar{x}}$ and $\dot{U}_{\perp}$ depend on the instantaneous particle position $x$, and therefore on the gyrophase $\varphi$. 
Since the orbit parameters are varying over the long timescale $t_{\text{MPS}}$, they are approximately constant over a single orbit, and hence $\dot{\bar{x}}$ and $\dot{U}_{\perp}$ are approximately periodic at small timescales (because $x$ is approximately periodic). 
Hence, the gyroaveraged time derivatives of $\bar{x}$ and $U_{\perp}$ determine the behaviour of $\bar{x}$ and $U_{\perp}$ at long timescales. 
From (\ref{xbardot}) the gyroaveraged time derivative of $\bar{x}$ is
\begin{align} \label{xbardot-gyro}
\left\langle \dot{\bar{x}} \right\rangle_{\varphi} = - \sigma_{\parallel} \alpha V_{\parallel} \left( U_{\perp}, U \right)  - \frac{1}{B} \left\langle \frac{\partial \phi}{\partial y} (x,y) \right\rangle_{\varphi} + O \left( \alpha^2 v_{\text{t,i}} \right) \text{.}
\end{align}
Exploiting (\ref{gyroaverage}) and (\ref{V-ExB}), the gyroaverage of (\ref{Uperpdot}) is
\begin{align} \label{Uperpdot-gyro}
\left\langle \dot{U}_{\perp} \right\rangle_{\varphi} = - \frac{ \sigma_{\parallel} \alpha \Omega V_{\parallel} \left( U_{\perp}, U \right) }{B} \left\langle \frac{\partial \phi}{\partial x} (x,y) \right\rangle_{\varphi}  + O \left( \alpha^2 \Omega v_{t,i}^2 \right) \text{.}
\end{align}
Note that the overall motion in the $y$ direction is determined by the gyroaveraged time derivative of $y$, which is the $\vec{E} \times \vec{B}$ drift of equation (\ref{V-ExB}),
\begin{align} \label{ydot-gyro}
\left\langle \dot{y} \right\rangle_{\varphi} = \left\langle v_y \right\rangle_{\varphi}  = \Omega  \left\langle \bar{x} - x \right\rangle_{\varphi} = V_{\vec{E} \times \vec{B} } \left( y, \bar{x}, U_{\perp} \right) = \frac{1}{B} \left\langle \frac{ \partial \phi}{ \partial x } \left( x,y \right) \right\rangle_{\varphi} \text{.}
\end{align}
The variation of $z$ is second order, $\dot{z}/z \sim \alpha^2 \Omega$. 
Two ion trajectories, which were obtained by varying the orbit parameters according to equations (\ref{Udot})-(\ref{Uperpdot}), are shown in Figure \ref{fig-iontraj}.

In a Hamiltonian system, when the parameters of periodic motion change over a timescale much longer than the period of the motion, an \emph{adiabatic invariant} exists. 
Here, it is given by the definite integral
\begin{align} \label{mu-Uperp-xbar}
\mu = \frac{1}{\pi} \int_{x_b}^{x_t} V_x \left( x, y, \bar{x}, U_{\perp} \right) dx \sim \frac{v_{\text{t,i}}^2}{\Omega} \text{.}
\end{align}
This is derived in Appendix \ref{appendix:mucalc}.
Equation (\ref{mu-Uperp-xbar}) is a generalization of the usual magnetic moment to the grazing angle presheath geometry which I study in this thesis. It was derived by Cohen and Ryutov in reference \cite{Cohen-Ryutov-1998}.
Note that we can use $\mu$ as an orbit parameter instead of $U_{\perp}$.
Unlike $U_{\perp}$, the adiabatic invariant (\ref{mu-Uperp-xbar}) is conserved to lowest order over the much longer timescale $t_{\text{MPS}}$,
\begin{align} \label{mudot-gyro}
\left \langle \dot{\mu} \right \rangle_{\varphi} = O \left( \alpha^2 v_{\text{t,i}}^2 \right) \simeq 0 \text{.}
\end{align}
I proceed to prove equation (\ref{mudot-gyro}).
Differentiating (\ref{mu-Uperp-xbar}) with respect to $\bar{x}$ gives, using $\partial \chi / \partial \bar{x} = \Omega^2 \left( \bar{x} - x \right) $ and equation (\ref{V-ExB}),
\begin{align} \label{dmudxbar}
\frac{\partial \mu}{\partial \bar{x}} = \frac{1}{\pi} \int_{x_{\text{b}}}^{x_{\text{t}}} \frac{\Omega^2 \left( x - \bar{x} \right)}{\sqrt{2\left( U_{\perp} - \chi\left( x; y, \bar{x} \right) \right)}}dx = -\frac{\Omega}{B \overline{\Omega}} \left\langle \frac{\partial \phi}{\partial x} \left( x, y \right) \right\rangle_{\varphi} \text{.}
\end{align}
Differentiating (\ref{mu-Uperp-xbar}) with respect to $y$, I obtain, using $\partial \chi / \partial y = \left( \Omega /B \right) \partial \phi / \partial y$,
\begin{align} \label{dmudy}
\frac{\partial \mu}{\partial y} = - \frac{1}{\pi} \frac{\Omega}{B} \int_{x_{\text{b}}}^{x_{\text{t}}} \frac{\partial \phi / \partial y \left(x, y \right)}{\sqrt{2\left( U_{\perp} - \chi\left( x; y, \bar{x} \right) \right)}} dx = -\frac{\Omega}{B \overline{\Omega} } \left\langle \frac{\partial \phi}{\partial y} \left( x, y \right) \right\rangle_{\varphi} \text{.}
\end{align}
Finally, differentiating (\ref{mu-Uperp-xbar}) with respect to $U_{\perp}$ and using (\ref{Omegabar-def}), I have
\begin{align} \label{dmudUperp}
\frac{\partial \mu}{\partial U_{\perp}} = \frac{1}{\pi} \int_{x_{\text{b}}}^{x_{\text{t}}} \frac{1}{\sqrt{2\left( U_{\perp} - \chi\left( x; y, \bar{x} \right) \right)}} dx = \frac{1}{\overline{\Omega}} \text{.}
\end{align}
Using the chain rule to take the time derivative $\dot{\mu}$ gives, to first order in $\alpha$ and $\delta$,
\begin{align}\label{mudot}
\dot{\mu} = \frac{\partial \mu}{\partial \bar{x}} \dot{\bar{x}}  + \frac{\partial \mu}{\partial y}  \dot{y}  + \frac{\partial \mu}{\partial U_{\perp}}  \dot{U}_{\perp} \text{.}
\end{align}
Gyroaveraging equation (\ref{mudot}), the first order gyroaveraged total derivative of the magnetic moment with respect to time is 
\begin{align}
\left\langle \dot{\mu} \right\rangle_{\varphi} = \frac{\partial \mu}{\partial \bar{x}} \left\langle \dot{\bar{x}} \right\rangle_{\varphi} + \frac{\partial \mu}{\partial y} \left\langle \dot{y} \right\rangle_{\varphi}  + \frac{\partial \mu}{\partial U_{\perp}} \left\langle \dot{U}_{\perp} \right\rangle_{\varphi} \text{.}
\end{align}
Upon using (\ref{dmudxbar})-(\ref{dmudUperp}) and (\ref{xbardot-gyro})-(\ref{ydot-gyro}), I obtain (\ref{mudot-gyro}).

The variation of $y$ with time is first order in my ordering, $\dot{y}/y \sim \alpha \Omega$.
In what follows, I introduce another orbit parameter $y_{\star} \sim l$ which varies over a timescale much longer than $t_{\text{MPS}}$, $\dot{y}_{\star} / y_{\star} \sim \alpha^2 \Omega \ll 1/t_{\text{MPS}}$.
The equation of motion in the $z$ direction is approximately (\ref{vz-EOM-first}).
Integrating this in time and introducing the constant of integration $y_{\star}$ gives
\begin{align}
y_{\star} = y - \frac{v_z}{\alpha \Omega} \sim l \text{.}
\end{align}
This quantity is proportional to the canonical momentum in the $z$ direction \cite{Holland-Fried-Morales-1993}, if the magnetic vector potential is written so that it has no $z$ dependence. 
Such a choice for the vector potential is, for example,
\begin{align} \label{mag-vec-pot}
\vec{A} = \begin{pmatrix}
0 \\ xB\cos\alpha  \\ -yB \sin \alpha 
\end{pmatrix} \text{.}
\end{align}
This vector potential can be checked by calculating the magnetic field that corresponds to it, 
\begin{align}
\vec{B} = \nabla \times \vec{A} = \begin{pmatrix}
-B \sin \alpha  \\ 0 \\ B\cos\alpha
\end{pmatrix} \text{,}
\end{align}
which is exactly the magnetic field present in the magnetic presheath. 
Using $\sin \alpha \simeq \alpha$, the canonical momentum in the $z$ direction, $p_z$, is proportional to $y_{\star}$, $p_z = m_{\text{i}} v_z + ZeA_z = m_{\text{i}} \left( v_z -  \Omega y \sin\alpha \right) \simeq -m_{\text{i}} \alpha \Omega y_{\star} $. 
Because the magnetic vector potential is written such that it is independent of $z$ and the electrostatic potential depends on $z$ only to second order (see equation (\ref{order-E})), the canonical momentum that we have just calculated is a constant of motion to first order, $\dot{p}_z / p_z \sim \alpha^2 \Omega$. 
Note that the orbit position $\bar{x}$ is proportional to the canonical momentum in the $y$ direction \cite{Gerver-Parker-Theilhaber-1990}, $p_y = m_{\text{i}} v_y + ZeA_y = m_{\text{i}} \left( v_y +  \Omega x \cos\alpha \right) \simeq m_{\text{i}} \Omega \bar{x} $. 
Because both the magnetic vector potential in (\ref{mag-vec-pot}) and the electrostatic potential have a first order dependence on $y$, we have $\dot{p}_y / p_y \sim \alpha \Omega$ as expected.
Using (\ref{vz-U-Uperp}) $y_{\star}$ is re-expressed in terms of $y$, $U_{\perp}$ and $U$,
\begin{align} \label{ystar-y-vpar}
y_{\star} = y - \frac{1}{\alpha \Omega}v_{\parallel}\left(U_{\perp}, U, \sigma_{\parallel} \right) \text{.}
\end{align}

The picture that emerges of the ion trajectory in a grazing angle magnetic presheath is that of a sequence of approximately closed orbits whose parallel streaming brings them slowly towards the wall, as shown in Figure \ref{fig-iontraj}.
As the ion approaches the wall, its $\vec{E}\times \vec{B}$ drift increases.
The adiabatic invariant $\mu$, the $z$ component of the canonical momentum $y_{\star}$, and total energy $U$ are conserved as the ion traverses the magnetic presheath.
The final piece of the ion trajectory terminates at the wall, a characteristic that is clearly incompatible with periodicity.
Hence, a time $1/\Omega$ before the ion reaches the wall, I consider the ion's trajectory to be an open orbit, as it can no longer be approximated by a closed orbit.
The gyroaveraged time derivatives (equations (\ref{xbardot-gyro})-(\ref{ydot-gyro})) are not an accurate description of the open orbit, and the instantaneous variation of the orbit parameters (equations (\ref{xbardot})-(\ref{Uperpdot})) must be considered in order to study open orbits.


\subsection{Open orbits} \label{subsec-traj-open}

When the ion reaches values of the orbit parameters for which its lowest order motion intersects the wall (and is therefore no longer periodic), it reaches the wall and is lost from the system over the fast timescale $1/\Omega$ (as I will show).
In this short period of time, the ion is in an open orbit.
The number of ions in open orbits is small (higher order in $\alpha$) compared with the number of ions in closed orbits because open orbits exist for a much shorter time. 
However, the number of ions in closed orbits that cross a point arbitrarily close to the wall is small because it only includes those ions in type I orbits that are near the bottom bounce point of their orbit (and thus, from equation (\ref{varphi-def}), it only includes ions with a small range of gyrophases around $\varphi = \pm \pi$). 
Therefore, it is essential to obtain the contribution to the density due to ions in open orbits.

\begin{figure}[h]
\centering
\includegraphics[width=0.75\textwidth]{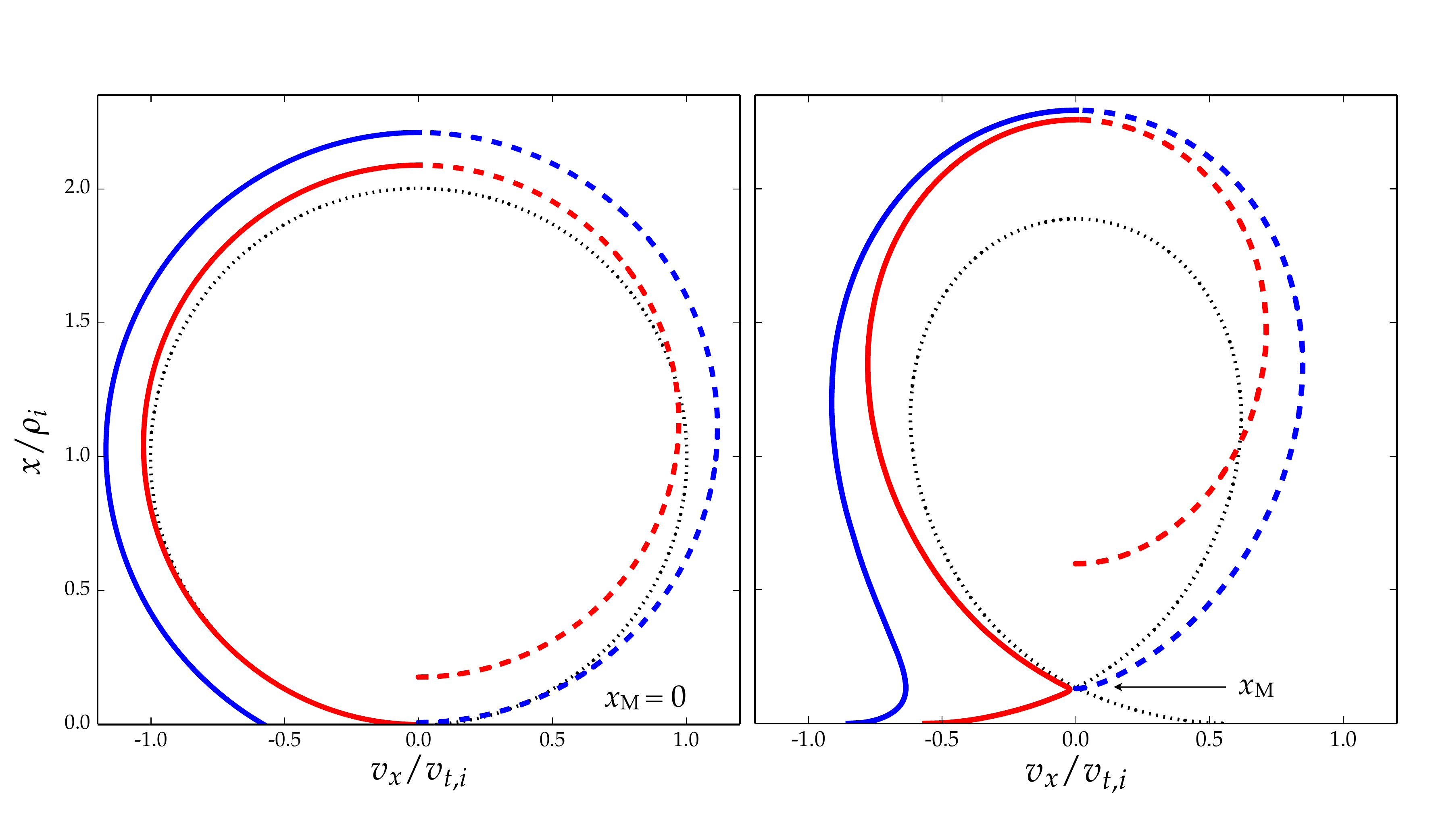}
\caption[Phase space ion trajectories in the transition from a closed to an open orbit]{Two sets of phase space trajectories corresponding to type I (left diagram) and type II (right diagram) orbits.
The type I trajectories are obtained using $\phi =0$ everywhere, while the type II trajectories are evaluated using the electrostatic potential solution of chapter \ref{chap-KMPS} for $\alpha = 0.02$, which has no $y$ dependence. 
The dotted lines are trajectories of motion with $\alpha =0$ when $U_{\perp} =\chi_{\text{M}} $, with $\bar{x} = \rho_{\text{i}}$ (type I) and $\bar{x} = 1.6 \rho_{\text{i}}$ (type II). 
The solid and dashed lines are trajectories calculated by integrating equations (\ref{x-EOM-exact}), (\ref{vx-Uperp-xbar-x}) and (\ref{xbardot})-(\ref{Uperpdot}) backwards in time from $x=0$ with $\alpha = 0.02$, starting with the same value of $\bar{x}$ used to obtain the dotted trajectories and with $U - \chi_{\text{M}} = v_{\text{t,i}}^2$. 
The solid lines are the open orbit pieces of the trajectories, while the dashed lines are approximately closed orbits according to my definition.
In each diagram, the red trajectory corresponds to the ion crossing $x_{\text{M}}$ with $v_x \simeq 0$, while the blue trajectory corresponds to the ion crossing $x_{\text{M}}$ with the largest possible value of $|v_x|$. 
}
\label{fig-phase}
\end{figure}

It is clear that an ion is in an open orbit when $x \leqslant x_{\text{M}}$, because a closed orbit cannot access this region by definition (see Figure \ref{fig-effpotclosed}).
For the ion to reach $x \leqslant x_{\text{M}}$, it must have crossed the maximum of the effective potential $\chi$ from the region $x>x_{\text{M}}$.
The exact point $x>x_{\text{M}}$ at which we consider its orbit to be open is arbitrary, but this arbitrariness does not matter because the ion density for $x>x_{\text{M}}$ is dominated by closed orbits.
We exploit this to generalize the open orbit definition in a way that includes all ions at $x\leqslant x_{\text{M}}$ and smoothly extends the open orbit density to $x> x_{\text{M}}$. 
We consider an ion to be in an open orbit if:
\begin{align}
& \text{~~~at future times, its trajectory has no bounce points;} \label{cond-i} \\ 
& \text{~~~at past times, its trajectory has several bounce points.} \label{cond-ii}
\end{align}
Note that the criterion (\ref{cond-ii}) corresponds to the past trajectory becoming an approximately closed orbit. 
Both criteria (\ref{cond-i}) and (\ref{cond-ii}) rely on the wall being electron-repelling, because they assume that any ion reaching $x=0$ (the Debye sheath entrance) does not bounce back.
Examples of pieces of trajectories considered to be open orbits are shown in Figure \ref{fig-phase} by solid lines.
We consider open orbit the part of a trajectory between the wall and the top bounce point.

To study open orbits, it will be useful to consider the difference between the perpendicular energy and the effective potential maximum as a separate quantity $D$, 
\begin{align} \label{D-def}
D = U_{\perp} - \chi_{\text{M}} \left(y, \bar{x} \right) \text{.}
\end{align}
The velocity component $v_x$, given by equation (\ref{vx-Uperp-xbar-x}), is
\begin{align} \label{vx-open}
v_x = \sigma_x V_x \left( x; y, \bar{x}, D + \chi_{\text{M}} \left( \bar{x}\right)  \right) = \sigma_x \sqrt{ 2\left( D + \chi_{\text{M}} \left( y, \bar{x}\right) - \chi \left( x; y, \bar{x} \right)\right) } \text{.}
\end{align}
When $x=x_{\text{M}}$ is reached from $x > x_{\text{M}}$, the velocity is given by $v_x = -\sqrt{2D}$, hence only ions with $D > 0$ cross the effective potential maximum and reach $x \leqslant x_{\text{M}}$.
To obtain the rate of change of $D$, we calculate the rate of change of $\chi_{\text{M}} \left( y, \bar{x} \right)$,
\begin{align} \label{chiMdot1}
\dot{\chi}_{\text{M}} \left( y, \bar{x} \right) = \frac{\partial \chi}{\partial \bar{x}} \left( x_{\text{M}}, y, \bar{x} \right) \dot{\bar{x}} +  \frac{ \partial  \chi}{\partial x} \left( x_{\text{M}}; y, \bar{x}  \right) \left( \frac{\partial x_{\text{M}}}{\partial \bar{x}}  \dot{\bar{x}} + \frac{\partial x_{\text{M}}}{\partial y}  \dot{y} \right) +  \frac{ \partial  \chi}{\partial y} \left( x_{\text{M}}; y, \bar{x} \right) \dot{y}  \text{.}
\end{align} 
For both type I and type II orbits, the second term in (\ref{chiMdot1}) vanishes (type I curves have $\partial x_{\text{M}}  / \partial \bar{x} = \partial x_{\text{M}}  / \partial y = 0$, while type II curves have $ \partial  \chi   / \partial x \left( x_{\text{M}}; y, \bar{x} \right)   = 0$) and, using (\ref{vy-xbar-x}) and (\ref{xbardot}), we find
\begin{align} \label{chiMdot}
\dot{\chi}_{\text{M}} \left( \bar{x} \right) = & ~\sigma_{\parallel} \alpha \Omega^2 V_{\parallel} \left( U_{\perp}, U \right) \left( x_{\text{M}} - \bar{x} \right)  
+ \frac{\Omega^2}{B} \left(  x_{\text{M}} - \bar{x}  \right) \frac{\partial \phi}{\partial y} (x, y)  \nonumber \\ 
& ~ + \frac{\Omega^2}{B} \left( \bar{x} - x \right) \frac{\partial \phi}{\partial y} (x_{\text{M}}, y) + O\left( \alpha^2 \Omega v_{\text{t,i}}^2 \right)  \text{.}
\end{align}
Combining (\ref{chiMdot}) with the result for $\dot{U}_{\perp}$ in (\ref{Uperpdot}), we get
\begin{align} \label{D-dot-general}
\dot{D} ( x, y, \bar{x}, U_{\perp}, U )   = & ~  \sigma_{\parallel} \alpha \Omega^2 V_{\parallel} ( U_{\perp} , U ) \left( x - x_{\text{M}} \right) 
- \frac{\Omega^2}{B} \left(  x_{\text{M}} - \bar{x} \right) \frac{\partial \phi}{\partial y} (x, y)  \nonumber \\ 
& ~ + \frac{\Omega^2}{B} \left(  x - \bar{x} \right) \frac{\partial \phi}{\partial y} (x_{\text{M}}, y)  + O\left( \alpha^2 \Omega v_{\text{t,i}}^2 \right) \text{.}
\end{align}

Consider an ion that reaches $U_{\perp} = \chi_{\text{M}} (y,\bar{x})$ at a position $x' > x_{\text{M}}$ and is travelling towards the maximum ($\sigma_x = -1$).
We use the relation
\begin{align} \label{dt-open-naive}
dt = \frac{dx}{v_x} \simeq \frac{dx}{ \sigma_x V_x \left( x; y, \bar{x}, U_{\perp} \right) }
\end{align}
to estimate the time taken for the ion to reach the effective potential maximum,
\begin{align} \label{t-open}
\delta t_{\text{M}} = \int dt \simeq \int^{x'}_{x_{\text{M}}}  \frac{ds}{ V_x \left( s; y, \bar{x}, U_{\perp} \right) } \text{.}
\end{align}
We assume that the difference between $U_{\perp}$ and $\chi_{\text{M}} $ stays small and that the change in $\bar{x}$ and $y$ during the time $\delta t_{\text{M}}$ is small (which we will show to be true), so that $U_{\perp} \simeq \chi_{\text{M}} (y,\bar{x})$.
If the effective potential curve is of type I, $\delta t_{\text{M}}^{\text{I}} \sim 1/\Omega $, whereas for type II curves, $\delta t_{\text{M}}^{\text{II}}$ diverges according to equation (\ref{t-open}).
We show this by expanding $V_x \left( x, y, \bar{x}, U_{\perp} \right) $ near $x \simeq x_{\text{M}}$ for a type II curve, using $U_{\perp} \simeq \chi_{\text{M}}$ and defining $ \chi_{\text{M}}'' \left( y, \bar{x} \right) \equiv \partial^2 \chi / \partial x^2 \left( x_{\text{M}}, y, \bar{x} \right)$ to obtain 
\begin{align} \label{VxII}
V_x^{\text{II}} \left( x; y, \bar{x}, U_{\perp} \right)  \simeq V_x^{\text{II}} \left( x, y, \bar{x},  \chi_{\text{M}} \left( y, \bar{x} \right) \right) \simeq \sqrt{\left| \chi''_{\text{M}} \right| } \left| x - x_{\text{M}} \right| \text{.}
\end{align}
The time $\delta t_{\text{M}}^{\text{II}}$ is then
\begin{align}
\delta t_{\text{M}}^{\text{II}} \simeq \int^{x'}_{x_{\text{M}}} \frac{ds}{ \sqrt{\left| \chi''_{\text{M}} \right| } \left( s - x_{\text{M}} \right)  } \rightarrow \infty \text{.}
\end{align}
Despite this apparent divergence, the variation of $D$ \emph{during} the time $\delta t_{\text{M}}$ can be evaluated using (\ref{dt-open-naive}).
Using $U_{\perp} \simeq \chi_{\text{M}} (y, \bar{x})$, equation (\ref{D-dot-general}) becomes
\begin{align} \label{D-dot}
\dot{D} ( x, y, \bar{x}, U_{\perp}, U )  = & ~ \alpha \Omega^2 V_{\parallel} ( \chi_{\text{M}} (y, \bar{x}) , U ) \left( x - x_{\text{M}} \right) 
- \frac{\Omega^2}{B} \left(  x_{\text{M}} - \bar{x} \right) \frac{\partial \phi}{\partial y} (x, y)  \nonumber \\ 
& + \frac{\Omega^2}{B} \left(  x - \bar{x} \right) \frac{\partial \phi}{\partial y} (x_{\text{M}}, y)  + O\left( \alpha^2 \Omega v_{\text{t,i}}^2 \right) \text{.}
\end{align} 
Note that $\dot{D} = 0$ at $x=x_{\text{M}}$.
Thus, equations (\ref{VxII}) and (\ref{D-dot}) imply that $\dot{D}/ V_x \left( x, y, \bar{x}, \chi_{\text{M}} \right) $ is not divergent at $x = x_{\text{M}}$.
Integrating equation (\ref{D-dot}) in time using (\ref{dt-open-naive}), we have
\begin{equation} \label{D-estimate}
D = \int \dot{D} dt \simeq  \int_{x_{\text{M}}}^{x'}  \frac{ \dot{D} ( s, y, \bar{x}, U_{\perp}, U )   }{V_x \left( s, y, \bar{x}, \chi_{\text{M}}  \right)}  ds \text{;}
\end{equation}
hence, we expect $D \sim \alpha v_{\text{t,i}}^2 \sim  \delta v_{\text{t,i}}^2$ for both orbit types, justifying $U_{\perp} \simeq \chi_{\text{M}} (y, \bar{x})$ a posteriori.
Using $U_{\perp} = \chi_{\text{M}} (y,\bar{x}) + D$ with $D \sim \alpha v_{\text{t,i}}^2$, equation (\ref{t-open}) can be used to obtain the more accurate estimate $\delta t_{\text{M}}^{\text{II}} \sim \ln \left( 1/\alpha \right) /\Omega$. 
Putting together the estimates for both orbit types, I have 
\begin{align} \label{deltat_M}
\Omega \delta t_{\text{M}} \sim \begin{cases}
1 & \text{ for type I orbits,} \\
 \ln \left( \frac{1}{\alpha} \right) & \text{ for type II orbits.}
\end{cases}
\end{align}
During the time $\delta t_{\text{M}}$, the change in $\bar{x}$ and $y$ is small.

I proceed to find the possible values of $D$ which satisfy the open orbit criteria that I have defined.
If $x < x_{\text{M}}$, the particle has already crossed the effective potential maximum and one has to integrate backwards in time to obtain the value of $D$ at the moment $x_{\text{M}}$ was crossed, denoted $D_{\text{X}}$, and further back to obtain the value of $D$ during the last bounce from the bottom bounce point $x_{\text{b}} \simeq x_{\text{M}}$, denoted $D_{\text{B}}$. 
If $x>x_{\text{M}}$, one must integrate $\dot{D}$ forwards in time to obtain $D_{\text{X}}$ (because by definition the particle trajectory must cross $x_{\text{M}}$ when it next reaches it, otherwise it would not be an open orbit), and backwards in time to obtain $D_{\text{B}}$. 

I first obtain $D_{\text{X}} - D$ in terms of $x$, $\bar{x}$ and $U$. 
If $x>x_{\text{M}}$, $\dot{D} $ is integrated \emph{forwards} in time (so $dt>0$), and if $x<x_{\text{M}}$, $\dot{D} $ is integrated \emph{backwards} in time (so $dt<0$).
From equation (\ref{D-estimate}), I obtain
\begin{align} \label{Delta+}
D_{\text{X}} - D \simeq  \Delta_+ \left(x,  y, \bar{x}, U \right)  \equiv \int_{x_{\text{M}}}^{x} \frac{ \dot{D} ( s, y, \bar{x}, U_{\perp}, U )  }{ V_x \left( s, y, \bar{x}, \chi_{\text{M}} \left( \bar{x} \right) \right)   } ds   \sim \alpha v_{\text{t,i}}^2 \text{;}
\end{align}
therefore, $D_{\text{X}}$ is
\begin{align} \label{D-X}
D_{\text{X}} = D + \Delta_+ \left(x, y, \bar{x}, U \right)  + O\left( \alpha^{1+p} v_{\text{t,i}}^2 \right)   \text{.}
\end{align}
The power $p$ used to quantify the error is given by 
\begin{align} \label{p-def}
p = \begin{cases}
1 & \text{ for type I orbits,} \\
\frac{1}{2} & \text{ for type II orbits.}
\end{cases}
\end{align}
The larger error from type II orbits comes from the fact that $D \sim \alpha v_{\text{t,i}}^2 $ is neglected when using $dt \simeq ds / V_x \left(s, y, \bar{x}, \chi_{\text{M}}(\bar{x})  \right)$. Estimating $|v_x|$ more accurately in the region near the maximum, I have
\begin{align} \label{VxII-modified}
V_x^{\text{II}} \left( x, y, \bar{x},  U_{\perp} \right)  = V_x^{\text{II}} \left( x, y, \bar{x},  \chi_{\text{M}}  \left( y, \bar{x} \right) + D \right) \simeq \sqrt{\left| \chi''_{\text{M}} \right| \left( x - x_{\text{M}} \right)^2 + 2D } \text{.}
\end{align}
Hence, there is a region of size $| x-x_{\text{M}}| \sim \alpha^{1/2} \rho_{\text{i}}$ 
where the estimate (\ref{VxII}) is incorrect. 
The contribution from this region to the integral (\ref{Delta+}) is therefore incorrect, and the size of this contribution is the size of the error in equation (\ref{D-X}).
Indeed, multiplying the size of the region ($\int_{x_{\text{M}}}^{x} ds \sim \alpha^{1/2} \rho_{\text{i}}$) by the size of the integrand ($| x-x_{\text{M}}|/V_x^{\text{II}} \sim 1/\Omega$) and by the prefactor ($\alpha \Omega^2 v_{\text{t,i}}$), I obtain an error of $\alpha^{3/2} v_{\text{t,i}}^2$, in accordance with equation (\ref{D-X}) with $p=1/2$.

I proceed to obtain $D_{\text{B}} - D_{\text{X}}$ by integrating $\dot{D} $ \emph{backwards} in time (so $dt < 0$) from the point at which the maximum is crossed. 
The backwards integration is identical to the forwards one; hence, using equation (\ref{D-estimate}), I obtain
\begin{align} \label{DeltaM}
D_{\text{X}} - D_{\text{B}} \simeq \Delta_{\text{M}}  \left( y, \bar{x}, U \right) = 2 \int_{x_{\text{M}}}^{x_{\text{t,M}}} \frac{ \dot{D} ( s, y, \bar{x}, \chi_{\text{M}}, U )  }{ V_x \left( x, y, \bar{x}, \chi_{\text{M}} \left( y, \bar{x} \right) \right)  } ds  \sim 2\pi \alpha v_{\text{t,i}}^2 \text{,}
\end{align}
where $x_{\text{t,M}}$ is the top bounce point corresponding to $U_{\perp} =  \chi_{\text{M}} (y,\bar{x})$. 
The factor of $2\pi$ in the final scaling of (\ref{DeltaM}) is due to having integrated in time over a gyroperiod $\sim 2\pi / \Omega$.
Then, $D_{\text{B}}$ is
\begin{align} \label{D-turn}
D_{\text{B}} = D_{\text{X}} -  \Delta_{\text{M}}  \left( y, \bar{x}, U \right)   + O\left( \alpha^{1+p} v_{\text{t,i}}^2 \right)   \text{.}
\end{align}

The criteria (\ref{cond-i}) and (\ref{cond-ii}) used to determine whether an ion is in an open orbit can be used to obtain the possible values of $D_{\text{B}}$ and $D_{\text{X}}$ for an ion in an open orbit.
We only consider $\sigma_x = -1$, as previously argued.
Then, for an ion at position $x<x_{\text{M}}$, condition (\ref{cond-i}) is automatically satisfied as the ion's fate is to fall down the effective potential towards $x=0$. 
However, condition (\ref{cond-ii}) is satisfied provided the ion's past trajectory crosses the maximum $x_{\text{M}}$, so that $D_{\text{X}}  > 0$, and then bounces back from the bottom bounce point (which, to lowest order, coincides with $x_{\text{M}}$) so that $D_{\text{B}} < 0$.
For an ion at position $x>x_{\text{M}}$, condition (\ref{cond-i}) implies that $D_{\text{X}}  > 0$ while condition (\ref{cond-ii}) implies that $D_{\text{B}} < 0$.
In both cases, imposing both conditions implies that both $D_{\text{X}} > 0$ and $D_{\text{B}} < 0$ must be satisfied.
Note that the argument can be applied in the reverse direction as well, and hence conditions (\ref{cond-i}) and (\ref{cond-ii}) are satisfied if and only if the following inequalities are satisfied:
\begin{align} \label{cond-open-DX}
& D_{\text{X}} > O  \left( \alpha^{1+p} v_{\text{t,i}}^2 \right) \text{;}  \\
& D_{\text{B}} < O \left( \alpha^{1+p} v_{\text{t,i}}^2 \right) \text{.}  \label{cond-open-DB}  
\end{align}
The limited accuracy in the evaluation of $D_{\text{X}}$ and $D_{\text{B}}$ leads to the  $O \left( \alpha^{1+p} v_{\text{t,i}}^2 \right) $ error in the inequality.
Using conditions (\ref{cond-open-DX}) and (\ref{cond-open-DB}), and equations (\ref{D-X}) and (\ref{D-turn}), the inequality
\begin{align} \label{D-range}
 -  \Delta_+ \left(x, y,  \bar{x}, U \right)  + O \left( \alpha^{1+p} v_{\text{t,i}}^2 \right)  <  D  <  \Delta_{\text{M}} \left( y, \bar{x}, U \right)  -  \Delta_+  \left(x, y,  \bar{x}, U \right) + O \left( \alpha^{1+p} v_{\text{t,i}}^2 \right)   \text{}
\end{align}
gives the values of $D$ that an open orbit can have.
From (\ref{D-range}), for an open orbit to exist at some value of $\bar{x}$, $U$ and $y$, the integral $\Delta_{\text{M}}$ must be positive.
From equations (\ref{vx-open}) and (\ref{D-range}), there is a range of possible particle velocities $v_x$ for open orbits, with maximum given by $-V_{x+} \left( x, \bar{x}, U \right)$, where
\begin{align} \label{Vx+}
V_{x+} \left( x, \bar{x}, U \right) = \sqrt{ 2 \left(  -  \Delta_+ \left(x, y, \bar{x}, U \right) + \chi_{\text{M}} \left( y, \bar{x} \right) - \chi \left( x, y, \bar{x} \right) \right)  + O \left( \alpha^{1+p} v_{\text{t,i}}^2 \right)  } \text{,}
\end{align}
and with range of values given by
\begin{align} \label{Deltavx}
\Delta v_x = &  \sqrt{ 2 \left(   \Delta_{\text{M}} \left( y, \bar{x}, U \right) -  \Delta_+ \left(x, y, \bar{x}, U \right) + \chi_{\text{M}} \left( y, \bar{x} \right) - \chi \left( x, y, \bar{x} \right) \right) + O \left( \alpha^{1+p} v_{\text{t,i}}^2 \right)   } \nonumber \\
 & - \sqrt{ 2 \left(  -  \Delta_+ \left(x, y, \bar{x}, U \right) + \chi_{\text{M}} \left( y, \bar{x} \right) - \chi \left( x, y, \bar{x} \right) \right)  + O \left( \alpha^{1+p} v_{\text{t,i}}^2 \right)  } \text{,}
\end{align}
such that
\begin{align} \label{vx-range}
- V_{x+} \left( x, y, \bar{x}, U \right) - \Delta v_x <  v_x <  - V_{x+} \left( x, y, \bar{x}, U \right)  \text{.}
\end{align}
Note that equations (\ref{Vx+})-(\ref{vx-range}) are defined, for a given $\bar{x}$ and $U$, in the region $0 \leqslant x \leqslant x_{\text{t,M}+}$, where $ |x_{\text{t,M}} - x_{\text{t,M}+} | \sim \alpha \rho_{\text{i}}$ and $ x_{\text{t,M}+}$ is obtained by setting $V_{x+}  \left( x_{\text{t,M}+}, \bar{x}, U \right)  $ to zero, $\chi_{\text{M}} (\bar{x} ) - \chi  \left( x_{\text{t,M+}} ,  \bar{x} \right)  - \Delta_+ \left(x_{\text{t,M+}} ,  \bar{x}, U \right) = 0$.
In section \ref{subsec-iondens-open} of chapter \ref{chap-dens}, I will obtain a useful approximation to equations (\ref{Vx+})-(\ref{vx-range}) which eliminates the dependence on $\Delta_+$ and is defined in the region $0 \leqslant x \leqslant x_{\text{t,M}}$ (instead of $0 \leqslant x \leqslant x_{\text{t,M}+}$).

\chapter{Ion distribution function and density}
\label{chap-dens}

The magnetic presheath solution requires a treatment of the behaviour of a collection of ions in this system, which is the subject of this chapter.
The ion distribution function $f(t, x,y,z, v_x, v_y, v_z)$ describes the probability of an ion having a certain position $(x,y,z)$ and a certain velocity $(v_x, v_y, v_z)$ at a time $t$, such that integrating the distribution function in velocity space at fixed time and position gives the number density $n_{\text{i}}$ at that position and time,
\begin{align}
n_{\text{i}} (t, x, y, z) = \int f(t, x,y,z, v_x, v_y, v_z) d^3v \text{.}
\end{align}
This chapter is devoted to obtaining $f$ and $n_{\text{i}}$ in the magnetic presheath, with the assumptions and approximations presented in chapter \ref{chap-traj}.

The structure of this chapter is as follows.
First, in section \ref{sec-iondens-distfunc}, I obtain the ion distribution function by simplifying the kinetic equation using the asymptotic expansion $\alpha \sim \delta \ll 1$.
Then, in section \ref{sec-iondens}, I obtain the ion density by treating the contribution due to approximately closed orbits (described in subsection \ref{subsec-traj-appclosed}) and due to open orbits (described in subsection \ref{subsec-traj-open}) separately.

\section{Ion distribution function} \label{sec-iondens-distfunc}

The Vlasov equation, with an ion distribution function $ f \left( t, x,y,z,v_x,v_y,v_z \right)$ is
\begin{align} \label{kinetic-start}
\frac{\partial f}{\partial t} + \dot{x} \frac{\partial f}{\partial x} + \dot{y} \frac{\partial f}{\partial y} + \dot{z} \frac{\partial f}{\partial z} + \dot{v}_x \frac{\partial f}{\partial v_x} + \dot{v}_y \frac{\partial f}{\partial v_y} + \dot{v}_z \frac{\partial f}{\partial v_z} = 0 \text{.}
\end{align}
In this paragraph, it will be useful to introduce vectors to represent the two possible sets of variables in which the kinetic equation can be written.
By denoting $\vec{H} = \left( x,y,z,v_x,v_y,v_z \right)$, the kinetic equation is 
\begin{align} \label{kinetic-vector}
\frac{\partial f}{\partial t} + \dot{\vec{H}} \cdot \frac{\partial f}{\partial \vec{H}} = 0 \text{.}
\end{align}
Applying the change of variables $\left( x,y,z,v_x,v_y,v_z \right) \rightarrow \left( \varphi, y_{\star}, z, \bar{x}, \mu, U, \sigma_{\parallel} \right) $, the distribution function has the form $ F \left( t, \varphi, y_{\star}, z, \bar{x}, \mu, U, \sigma_{\parallel} \right) $.
By denoting $\vec{G} =  \left( \varphi, y_{\star}, z, \bar{x}, \mu, U \right)$ and appying the chain rule on \ref{kinetic-vector}, the kinetic equation is re-expressed to
\begin{align}
\frac{\partial F}{\partial t} + \left[ \frac{\partial \vec{G}}{\partial t} +  \dot{\vec{H}} \cdot \frac{\partial \vec{G}}{\partial \vec{H}} \right] \cdot \frac{\partial F}{\partial \vec{G}} = 0 \text{.}
\end{align}
The expression in the square bracket is a vector, and can be identified as the time derivative of the set of variables $\vec{G}$
\begin{align}
\dot{ \vec{G} } = \frac{\partial \vec{G}}{\partial t} + \dot{\vec{H}} \cdot \frac{\partial \vec{G}}{\partial \vec{H}} \text{.}
\end{align}
Hence, the kinetic equation (\ref{kinetic-start}) is, when expressed in the set of variables $\vec{G}$,
\begin{align} \label{kinetic-gyro}
\frac{\partial F}{\partial t} + \dot{\varphi} \frac{\partial F}{\partial \varphi} + \dot{y}_{\star} \frac{\partial F}{\partial y_{\star}} + \dot{z} \frac{\partial F}{\partial z} + \dot{\bar{x}} \frac{\partial F}{\partial \bar{x}} + \dot{\mu} \frac{\partial F}{\partial \mu} + \dot{U} \frac{\partial F}{\partial U}  = 0 \text{.}
\end{align}

I expand the distribution function for $\alpha \sim \delta \ll 1$,
\begin{align}
F = F_0 + F_1 + \ldots \text{,}
\end{align}
where $F_0 \sim F$, $F_1 \sim \alpha F \sim \delta F$.
Taking $O(\Omega F)$ terms only in (\ref{kinetic-gyro}) and using  $\dot{\varphi} \simeq \overline{\Omega}$ (from equation (\ref{varphidot})) leads to the lowest order equation 
\begin{align} \label{gyro1}
\overline{\Omega} \frac{\partial F_0}{\partial \varphi} = 0 \text{.}
\end{align}
To obtain this, I used $\overline{\Omega} \sim \Omega$, $\dot{\bar{x}} / \bar{x} \sim  \dot{\mu}  / \mu  \sim \alpha \Omega $ and $\partial / \partial t \sim \dot{y}_{\star} / y_{\star}  \sim \dot{z}/z \sim \dot{U} / U  \sim \alpha^2 \Omega $. 
From (\ref{gyro1}),  the lowest order distribution function is gyrophase independent, $F_0 = F_0(t, y_{\star}, z, \bar{x}, \mu, U )$.

The first order of (\ref{kinetic-gyro}) is
\begin{align} \label{kinetic-star}
\dot{\bar{x}}  \frac{\partial F_0}{\partial \bar{x}}  + \dot{\mu} \frac{\partial F_0}{\partial \mu} + \overline{\Omega} \frac{\partial F_1}{\partial \varphi} = 0   \text{.}
\end{align}
where we used $\partial / \partial t \sim \dot{y}_{\star} / y_{\star}  \sim \dot{z}/z \sim \dot{U} / U  \sim \alpha^2 \Omega $. 
Taking the gyroaverage of (\ref{kinetic-star}) and, using $ \left\langle \dot{\mu} \right\rangle_{\varphi} / \mu \sim \alpha^2 \Omega \sim \delta^2 \Omega$ and $\left\langle \partial F_1 / \partial \varphi \right\rangle_{\varphi} = 0$, we obtain the gyrokinetic equation
\begin{align} \label{gyrokinetic-star}
 \frac{\partial F_0}{\partial \bar{x}} =0 \text{.}
\end{align}
Therefore, the lowest order distribution function is independent of $\bar{x}$, $F_0 = F_0(t, y_{\star}, z, \mu, U, \sigma_{\parallel}  )$.
Moreover, the dependence of the distribution function on $z$ and $t$ is unimportant, as argued in chapter \ref{chap-traj}.
Hence, I do not consider $z$ and $t$ dependence in the ion distribution function, $F_0 = F_0(y_{\star}, \mu, U, \sigma_{\parallel} )$.

Using that $F_0 = 0$ for $U_{\perp} \rightarrow \infty$ and $v_y = \Omega \left( \bar{x} - x \right) \sim \sqrt{U_{\perp}}$, the limit $\bar{x} \rightarrow \infty$ is equivalent to the limit $x \rightarrow \infty$. 
Hence, the distribution function at the magnetic presheath entrance is
\begin{align} \label{F0-xbar-infinity-star}
\lim_{\bar{x} \rightarrow \infty} F_0 = \lim_{x \rightarrow \infty} F_0 = F_{\text{cl}} \left( y_{\star}, \mu, U, \sigma_{\parallel} \right) \text{ for } \left\langle \dot{\bar{x}} \right\rangle_{\varphi} < 0 \text{,}
\end{align}
where the subscript ``cl'' stands for closed orbits.
Note that the boundary condition only gives the particles that are drifting into the presheath, $\left\langle \dot{\bar{x}} \right\rangle_{\varphi} < 0$.
The assumption of an electron repelling wall implies that no ion comes back from the wall. 
Thus, the only boundary condition to impose at $x=0$ is that there be no forward moving ions.
The distribution function of ions in the magnetic presheath is therefore
\begin{align} \label{F-solution}
F \simeq F_0 = F_{\text{cl}}\left( y_{\star}, \mu, U, \sigma_{\parallel}  \right) \text{.}
\end{align}
This result is similar to the one obtained in reference \cite{Cohen-Ryutov-1998}, but generalized using $y_{\star}$ in order to account for gradients parallel to the wall.

The ordering (\ref{scale-sep}) implies that the collisional layer only determines the boundary conditions at $x \rightarrow \infty$.
A solution of the collisional layer is thus required to obtain the correct form of $F_{\text{cl}}$.
Alternatively, a drift-kinetic code 
of the scrape-off layer could be used to obtain such a distribution function.

\section{Ion density} \label{sec-iondens}

The total ion density is the sum of the closed orbit contribution, $n_{\text{i,cl}}(x,y)$, and the open orbit contribution, $n_{\text{i,op}}(x,y)$,
\begin{align} \label{ni-general}
n_{\text{i}}(x,y) = n_{\text{i,cl}}(x,y) + n_{\text{i,op}}(x,y) \text{.}
\end{align}
In subsection \ref{subsec-iondens-closed} an expression for $n_{\text{i,cl}}(x,y)$ is obtained, and in subsection \ref{subsec-iondens-closed} an expression for $n_{\text{i,op}}(x,y)$ is obtained.

\subsection{Closed orbit density} \label{subsec-iondens-closed}

In order to obtain the ion density of ions in approximately closed orbits, I first proceed to obtain the domain of phase space that allows for periodic solutions to lowest order in $\alpha \sim \delta \ll 1$.
In references \cite{Cohen-Ryutov-1998, Holland-Fried-Morales-1993, Gerver-Parker-Theilhaber-1990, Parks-Lippmann-1994}, analytical expressions for the domain of allowed closed orbits are found with various assumptions. 
My approach is similar to the one presented in reference \cite{Gerver-Parker-Theilhaber-1990}, but includes the velocity dimension parallel to the magnetic field, weak gradients parallel to the wall, and allows for type II orbits.
Type II orbits were not studied in reference \cite{Gerver-Parker-Theilhaber-1990} because the authors were studying the different problem of a plasma with a magnetic field exactly parallel to a wall, which is thus ion-repelling. 

\begin{figure}
\centering
\includegraphics[width=0.45\textwidth]{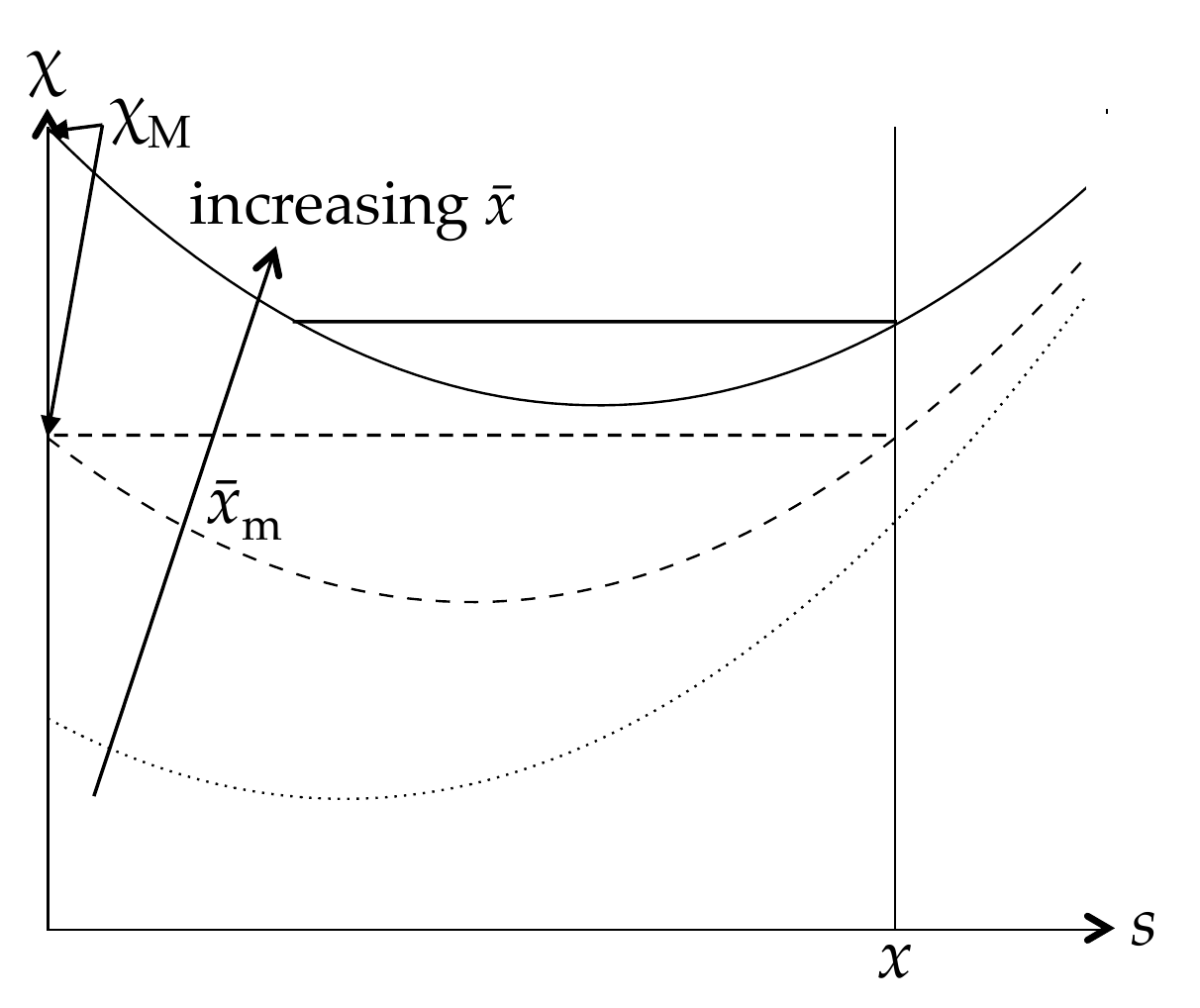}
\includegraphics[width=0.45\textwidth]{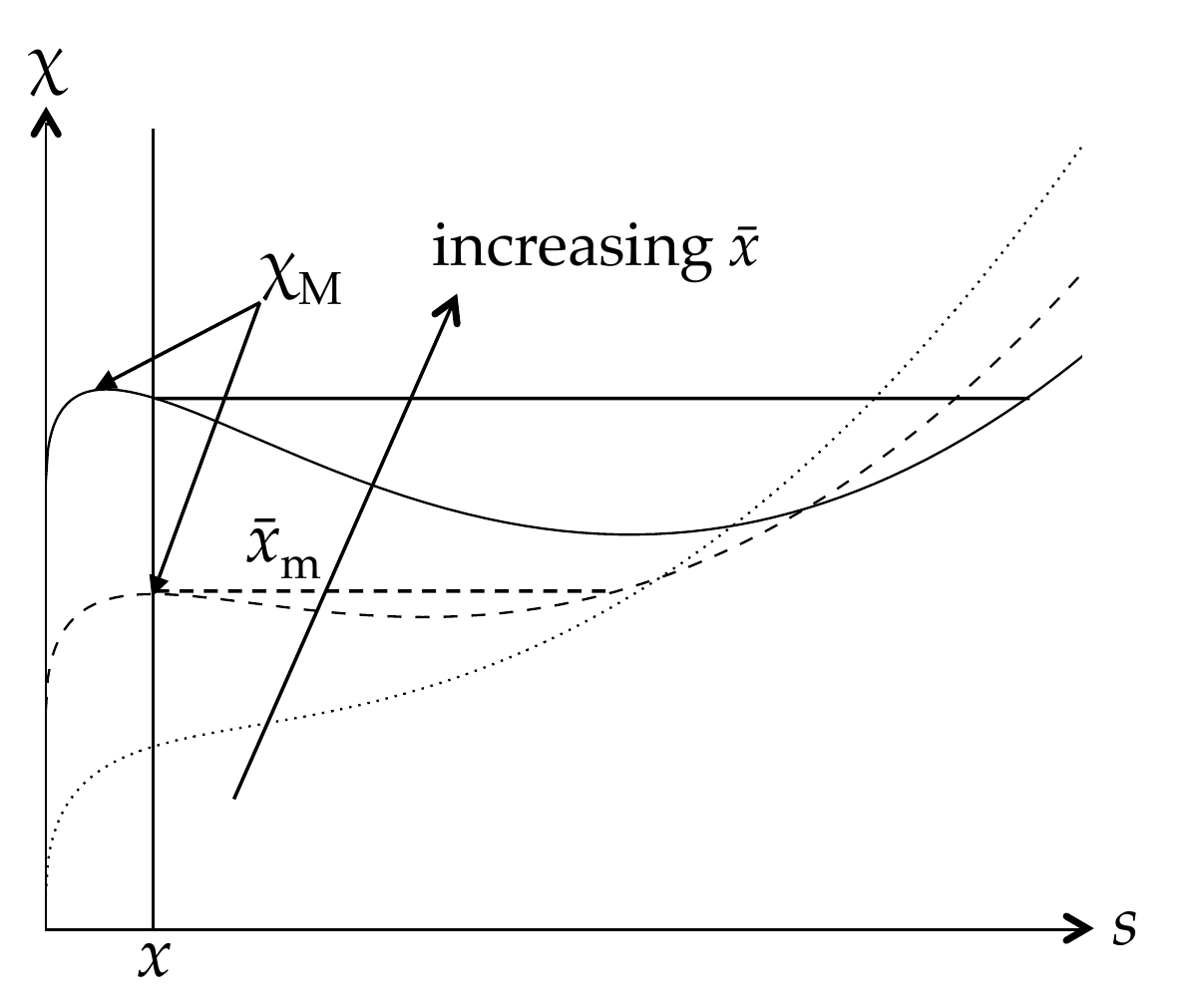}
\caption[Minimum orbit position for a closed orbit]{Two sets of effective potential curves, each set with three different curves $\chi \left(s, \bar{x}, y \right)$ plotted as a function of $s$ that correspond to a different value of the orbit position $\bar{x}$. On the left, type I effective potential curves with a local maximum near the wall, which arises if the electric field is sufficiently strong there, are shown. On the right, type II effective potential curves without a maximum are shown. The solid curves are associated to orbits with position $\bar{x} > \bar{x}_{\text{m}} \left( x, y \right)$. Dashed curves correspond to orbit position $\bar{x}_{\text{m}} \left( x, y \right)$ which is associated with the presence of only one closed orbit that passes through $x$, while dotted ones correspond to orbits at position $\bar{x} < \bar{x}_{\text{m}} \left( x, y \right)$, which are all open if they are to cross point $x$. The horizontal lines are associated with the minimum perpendicular energy required for a closed orbit to lie at position $\bar{x}$ and cross the point $x$, equal to $\chi \left( x, \bar{x}, y \right)$. The effective potential maximum $\chi_{\text{M}} \left( \bar{x}, y \right)$, marked for the closed (and semi-closed) orbit curves.}
\label{figure-xbarmin-condition}
\end{figure}

In this section, I sometimes denote the effective potential $\chi$ as a function of the variable $s$ in order to distinguish the position $x$ at which the ion density is evaluated from the positions $s$ that a particle occupies in its lowest order orbit. 
Obviously $x$ is just one of the many possible values that $s$ can take. 
Electric fields outside the magnetic presheath are weak in our ordering, so that $\partial \phi / \partial s \left( s \rightarrow \infty , y\right) \simeq 0$. The effective potential $\chi$ must therefore be unbounded at infinity for finite $\bar{x}$, 
\begin{align} \label{eff-pot-bound}
\frac{\partial \chi}{\partial s} \left( s \rightarrow \infty , \bar{x}, y \right) \simeq \Omega^2 \left( s - \bar{x} \right) > 0 \text{,}
\end{align}
leading always to a bounce point for sufficiently large $s$. Therefore, in order to have a closed orbit crossing the position $x$ at which we are calculating the integral, the effective potential must be larger than its value at $x$ for some value of $s$ between the particle position $s = x$ and the wall at $s = 0$,
\begin{align}
\chi \left(s; y, \bar{x} \right) > \chi \left( x; y,  \bar{x}\right) \text{ for some or all } s \in \left[0, x\right) \text{.}
\end{align} 
Explicitly, this is (after dividing through by $\Omega^2$)
\begin{align}
\frac{1}{2} \left( s - \bar{x} \right)^2 + \frac{\phi \left( s, y \right)}{\Omega B} > \frac{1}{2}  \left( x - \bar{x} \right)^2 + \frac{\phi \left( x, y \right)}{\Omega B} \text{,}
\end{align}
which reduces to
\begin{align}
\bar{x} \left( x - s \right) > \frac{1}{2} \left( x^2 - s^2 \right) + \frac{\phi \left( x, y \right) - \phi \left( s, y \right)}{\Omega B} \text{.}
\end{align}
This leads to the closed orbit condition
\begin{align} \label{condition-closed}
\bar{x} > g\left( s; x, y \right) \equiv  \frac{1}{2}\left( x + s \right) + \frac{\phi \left( x, y \right) - \phi \left( s, y \right)}{\Omega B \left( x - s \right)} \text{ for some or all } s \in \left[0, x\right) \text{.}
\end{align}
The minimum value of $\bar{x}$, $\bar{x}_{\text{m}} \left( x, y \right)$, that satisifies this condition is obtained by minimizing the function $g\left( s; x, y \right)$ over the interval $\left[0, x\right)$,
\begin{align} \label{xbarm}
\bar{x}_{\text{m}} \left( x, y \right) = \min_{s \in \left[0, x \right)} g\left(s, x, y\right) \text{.}
\end{align}
Note that, from (\ref{condition-closed}), $g\left( s; x, y\right) > x/2$ because $s < x$ and the electrostatic potential $\phi$ is increasing with the distance to the (negatively charged) wall. 
This implies that $\bar{x}_{\text{m}} \left( x , y \right) > x/2$. 
Figure \ref{figure-xbarmin-condition} shows examples of type I and II effective potential curves with values of $\bar{x}$ smaller than, equal to and larger than $\bar{x}_{\text{m}}$.

Closed orbits exist for $\bar{x} > \bar{x}_{\text{m}} \left( x, y \right)$.
However, any orbit which has perpendicular energy $U_{\perp} > \chi_{\text{M}} \left( y, \bar{x} \right)$ does not have a periodic solution to lowest order (with its trajectory intersecting the wall). 
Hence, the distribution function of approximately closed orbits is expected to be non-zero only for $\bar{x} > \bar{x}_{\text{m}}(x,y)$ and $U_{\perp} < \chi_{\text{M}}(y, \bar{x})$, and is therefore given by
\begin{align} \label{fclosed}
f_{\text{cl}} (x, y, v_x, v_y, v_z) \simeq F_{\text{cl}} \left( y_{\star}, \mu  , U, \sigma_{\parallel} \right)  \Theta \left( \bar{x} - \bar{x}_{\text{m}}(x, y) \right) \Theta \left( \chi_{\text{M}} (y,\bar{x}) - U_{\perp} \right)    \text{,}
\end{align}
where $\Theta$ is the Heaviside step function,
\begin{align} \label{Heaviside}
\Theta ( y ) = \begin{cases} 1 & \text{ for } y \geqslant 0 \text{,} \\
0 & \text{ for } y < 0 \text{.}
\end{cases}
\end{align}

The density of ions crossing position $(x,y)$ in approximately closed orbits is an integral in velocity space of the distribution function (\ref{fclosed}),
\begin{align} \label{ion-density-f}
n_{\text{i,cl}} (x, y) = \int  f_{\text{cl}} (x, y, \vec{v})  d^3v   \text{.}
\end{align}
I proceed to change variables in the ion density integral (\ref{ion-density-f}) from $( v_x , v_y , v_z )$ to $( \bar{x}, U_{\perp}, U )$ while holding the ion position co-ordinates $x$ and $y$ fixed. 
Using (\ref{vx-Uperp-xbar-x}), (\ref{vy-xbar-x}) and (\ref{vz-U-Uperp}), the Jacobian is $\left| \partial \left( v_x, v_y, v_z \right) / \partial \left( \bar{x}, U_{\perp}, U \right) \right| = \Omega / \left| v_x v_z \right|$, which can be expressed as
\begin{align}
\left| \frac{\partial \left( v_x , v_y , v_z\right)}{\partial \left( \bar{x}, U_{\perp}, U \right)} \right| = \frac{\Omega}{\sqrt{2\left( U_{\perp} - \chi \left( x; y, \bar{x} \right) \right)} \sqrt{2\left( U - U_{\perp} \right)}} \text{.}
\end{align}
Therefore, the ion density integral becomes
\begin{align} \label{ni-closed-general}
n_{\text{i,cl}} = \sum_{\sigma_{\parallel} = \pm 1} \int_{\bar{x}_m \left( x , y \right) }^{\infty} d\bar{x} \int_{\chi\left( x; y, \bar{x} \right)}^{\chi_{\text{M}} \left( y, \bar{x} \right)}  \frac{2\Omega dU_{\perp}}{\sqrt{2\left( U_{\perp} - \chi \left( x; y, \bar{x} \right) \right)}} \int_{U_{\perp}}^{\infty} \frac{F_{\text{cl}} \left(y_{\star}, \mu, U, \sigma_{\parallel} \right)}{\sqrt{2\left( U - U_{\perp}\right)}} dU \text{,}
\end{align}
where the summation over the two possible values $\sigma_x = \pm 1$ has simplified to a factor of 2 because $F_{\text{cl}}$ is independent of $\varphi$. 
If the distribution function were gyrophase dependent, the summation over $\sigma_x$ would be necessary and (\ref{varphi-def}) would be used in order to obtain $\varphi \left(x, y, \bar{x}, U_{\perp}, \sigma_x \right)$ at each integration point.
The limits ensure that we integrate over the phase space domain in which closed orbits are allowed.

It is worth noting that $n_{\text{i,cl}} (0,y) = 0$, because for type I orbits $\chi_{\text{M}} (y,\bar{x}) = \chi (0, y, \bar{x} )$ while for type II orbits $x=0 < x_{\text{M}} $. 
The fact that $n_{\text{i,cl}} (0,y) = 0$ means that we cannot naively impose quasineutrality with only the approximately closed orbit contribution to the ion density. 
An attempt to impose $Zn_{\text{i,cl}} (0,y) = n_{\text{e}} \left( 0,y \right)$ leads to $n_{\text{e}} \left( 0,y \right) = n_{e\infty} \exp\left( e\phi ( 0,y ) / T_{\text{e}} \right) = 0$ and therefore $\phi ( 0,y ) = - \infty$. 
This is an unphysical result which stems from the fact that we have not kept the dominant contribution to the ion density at (and near) the wall, which comes from ions in open orbits.
This contribution is calculated in the next section.

\subsection{Open orbit density} 
\label{subsec-iondens-open}

Consider an ion at position $(x,y)$ in an open orbit, when $U_{\perp } = \chi_{\text{M}} \left( y, \bar{x} \right) + D$ and $D$ lies in the range (\ref{D-range}). 
For such an ion to exist, the integral $\Delta_{\text{M}}$ defined in equation (\ref{DeltaM}) \emph{must} be positive.
The ion transitioned from being in a closed orbit to being in an open orbit a time $\sim \delta t_{\text{M}} $ before the instant in time considered. 
At this time, the orbit position differed from $\bar{x}$ by $O\left(\alpha \Omega \delta t_{\text{M}} \rho_{\text{i}} \right) $, which is small. 
To lowest order, the ion conserved its adiabatic invariant up to the point where $U_{\perp} = \chi_{\text{M}} (y,\bar{x})$. 
Using $U_{\perp} \simeq \chi_{\text{M}} \left( y, \bar{x} \right)$, the adiabatic invariant of the ion was $\mu  \left( y, \bar{x}, \chi_{\text{M}}  \right) + O\left( \alpha \Omega \delta t_{\text{M}} v_{\text{t,i}} \rho_i \right)$. 
Hence, the distribution function is $F_{\text{cl}} \left( y_{\star}(y, \chi_{\text{M}}, U, \sigma_{\parallel}), \mu \left(y, \bar{x}, \chi_{\text{M}} \right) , U , \sigma_{\parallel} \right) $ to lowest order, which is independent of the value of $D$ \cite{Cary-1986, Neishtadt-1987}.
I proceed to obtain the values of $\bar{x}$ (related to $v_y$) and $U$ (related to $v_z$) for which the distribution function is non-zero at fixed $x$ and $y$.
I also obtain the small range of values of $v_x$ for which the open orbit distribution function is non-zero, at fixed values of $x$, $y$, $\bar{x}$ and $U$.

\begin{figure}
\centering
\includegraphics[width= 0.8\textwidth]{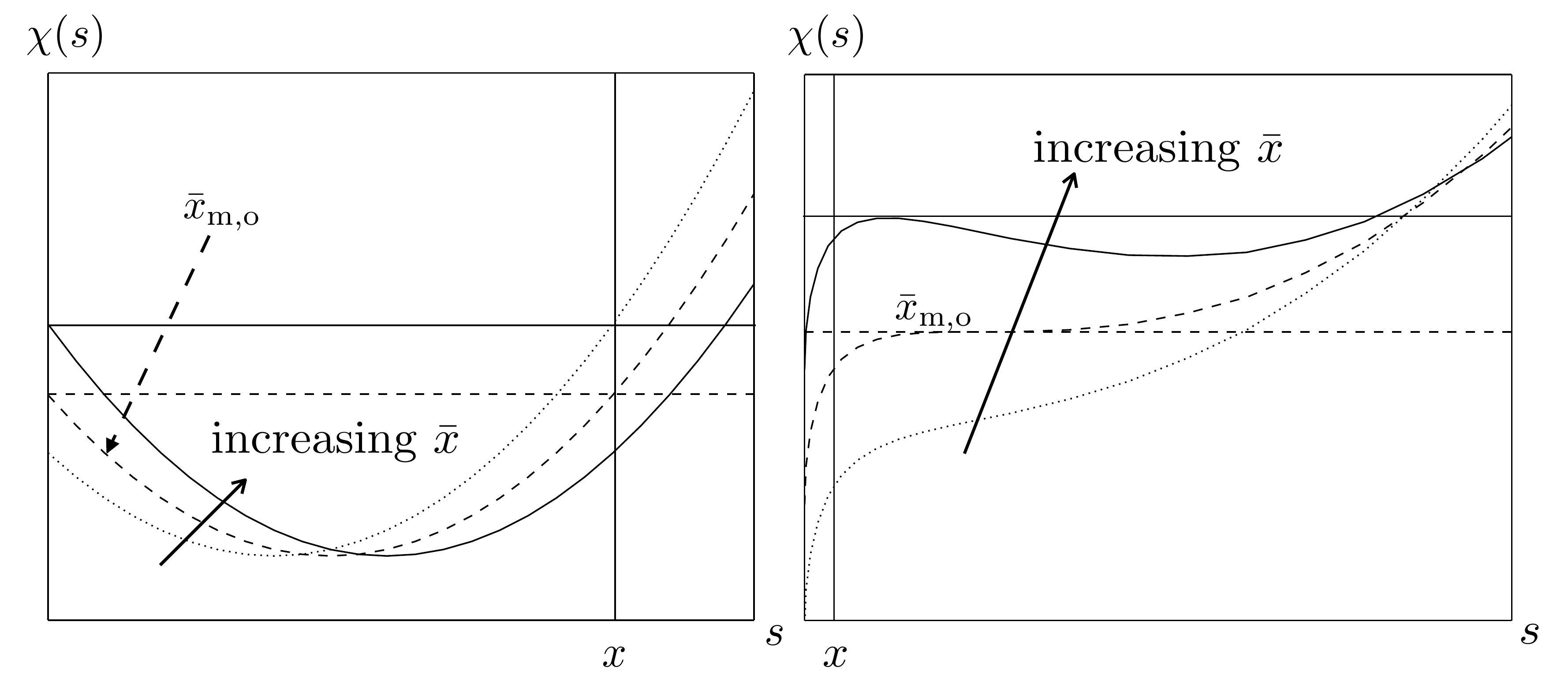}
\caption[Minimum orbit position for an open orbit]{Type I and II effective potential curves are shown on the left and right respectively, for some fixed value of $y$ (not shown). The dashed curves correspond to an orbit position $\bar{x} = \bar{x}_{\text{m,o}} \left( x,y \right) $, which is the minimum value of $\bar{x}$ above which open orbits crossing the position $x$ (vertical line) exist. The solid effective potential curves are the ones corresponding to $\bar{x} > \bar{x}_{\text{m,o}} (x)$. The horizontal lines correspond to $U_{\perp} = \chi_{\text{M}}$, which is the lowest order perpendicular energy of an ion in an open orbit. The dotted curves correspond to $\bar{x} < \bar{x}_{\text{m,o}} \left( x \right)$: no open orbits crossing position $x$ exist for such values of $\bar{x}$ because there are no closed orbits at $s \geqslant x$.
}
\label{fig-effpot-open}
\end{figure}

For an ion in an open orbit to be at position $x$, the range of possible values of $\bar{x}$ (to lowest order) is determined by two constraints. A time $\sim \delta t_{\text{M}}$ before being in an open orbit, the ion must have been in an approximately closed orbit whose existence depends on the presence of an effective potential minimum. 
Hence, a stationary point must exist, which implies that $\bar{x} > \bar{x}_{\text{c}}$ is necessary. 
Moreover, we require that $x< x_{\text{t,M}}$.
For $x< x_{\text{c}} $, it is impossible for an ion to be in the region $x>x_{\text{t,M}}$ because $x_{\text{c}} \leqslant x_{\text{m}} \leqslant x_{\text{t,M}} $, therefore $\bar{x} > \bar{x}_{\text{c}}(y)$ is the necessary and sufficient condition for an open orbit crossing position $x$ in this case. 
For $x> x_{\text{c}}$, we use the fact that $x_{\text{M}} < x_{\text{c}}$ to conclude that the ion must be in the region $x_{\text{M}} < x < x_{\text{t,M}}$; the criterion for an open orbit crossing position $x$ is therefore identical to that of a closed orbit crossing position $x$, $\bar{x} > \bar{x}_{\text{m}} (x,y)$.
Therefore, the condition for an ion in an open orbit to be present at position $x$ is $\bar{x} > \bar{x}_{\text{m,o}} \left( x,y \right)$, where
\begin{align} \label{xbarm-open}
\bar{x}_{\text{m,o}} (x,y) = \begin{cases} 
\bar{x}_{\text{c}}(y)  & \text{ for } x < x_{\text{c}}(y)  \\
\bar{x}_{\text{m}} (x,y) &  \text{ for } x \geqslant  x_{\text{c}}(y) \text{.}  
\end{cases}
\end{align}
Two examples of how the constraint $\bar{x} > \bar{x}_{\text{m,o}} \left( x,y \right)$ arises are shown in Figure \ref{fig-effpot-open}.
This constraint is valid to lowest order in $\alpha \Omega \delta t_{\text{M}}$. 

The ion's total energy has to be larger than the effective potential maximum, $U > \chi_{\text{M}} \left( y, \bar{x} \right)$, and the $z$ component of the velocity is approximated by $ V_{\parallel} \left( \chi_{\text{M}} \left( \bar{x} \right) , U \right) $. 
In order to relate values of $v_y$ and $v_z$ to lowest order values of $\bar{x}$ and $U$ for ions in open orbits, in what follows I refer extensively to equations (\ref{xbar-def}) and
\begin{align} \label{U-open}
U = \chi_{\text{M}} \left( y, \bar{x} \right) + \frac{1}{2} v_z^2 + O\left( \alpha v_{\text{t,i}}^2 \right) \text{,}
\end{align} 
where the latter equation is obtained by rearranging the equation $v_z \simeq V_{\parallel} \left( \chi_{\text{M}} \left( y, \bar{x} \right) , U \right) $.

The velocity component $v_x$ lies in the range (\ref{vx-range}), which is obtained from the range of allowed values of $D$ for given values of $x$, $y$, $\bar{x}$, and $U$. For the evaluation of the distribution function and density of ions in open orbits, the value of $\Delta v_x$ is crucial because  at a given $x$, $\bar{x}$ and $U$ it gives the \emph{small} range of values of $v_x$ in which the distribution function is non-zero.
The \emph{exact} value of the maximum and minimum $v_x$ only needs to be known to lowest order.
Hence, we can shift $V_{x+} \left( x, y, \bar{x}, U\right)$ by a small amount provided we preserve the same value of $\Delta v_x$.
With this in mind, we proceed to obtain simpler expressions for $V_{x+} \left( x, y, \bar{x}, U\right)$ and $\Delta v_x$. 
Two regions must be distinguished: $|x-x_{\text{M}}| \sim \rho_{\text{i}}$ where $\chi_{\text{M}}  - \chi  \sim v_{\text{t,i}}^2$, and $|x-x_{\text{M}}| \sim \alpha^p \rho_{\text{i}}$ where $\chi_{\text{M}}  - \chi  \sim \alpha v_{\text{t,i}}^2$ (with $p$ defined in equation (\ref{p-def})).

In the region $|x - x_{\text{M}} | \sim \rho_i$, we have
\begin{align} \label{outer-ordering-2}
\alpha^{1+p} v_{\text{t,i}}^2 \ll \Delta_{\text{M}}  \sim \Delta_+  \sim \alpha v_{\text{t,i}}^2 \ll \chi_{\text{M}}  - \chi  \sim v_{\text{t,i}}^2 \text{.}
\end{align}
By using equations (\ref{Vx+}) and (\ref{Deltavx}), the ordering (\ref{outer-ordering-2}) leads to
\begin{align} \label{outer-ordering-1}
 \Delta v_x \sim \alpha  v_{\text{t,i}} \ll  V_{x+}  \sim  v_{\text{t,i}} \text{.} 
 \end{align}
Neglecting the term $\Delta_+ \sim \alpha v_{\text{t,i}}^2$ in the square root of equation (\ref{Vx+}) for $V_{x+}$, we obtain 
\begin{align}  \label{Vx+-simpler}
V_{x+} \left( x; \bar{x}, U \right) & = \sqrt{ 2 \left(   \chi_{\text{M}} \left( \bar{x} \right) - \chi \left( x, \bar{x} \right) \right)  + O \left( \alpha v_{\text{t,i}}^2 \right)  } \nonumber \\  
&  = V_x \left( x; \bar{x}, \chi_{\text{M}} \left( \bar{x} \right) \right) +  O\left( \alpha v_{\text{t,i}}  \right)   \text{.}
\end{align}
I expand the terms $\Delta_{\text{M}}$ and $\Delta_+$ out of the square root in equation (\ref{Deltavx}) for $\Delta v_x$ using the ordering (\ref{outer-ordering-2}), and thus obtain
\begin{align} \label{Deltavx-simplifying}
\Delta v_x  = & \left[  \sqrt{ 2 \left(  \Delta_{\text{M}} \left( y, \bar{x}, U \right) - \Delta_+ \left( x; y, \bar{x}, U \right)  + \chi_{\text{M}} \left( y, \bar{x} \right) - \chi \left( x; y,  \bar{x} \right) + O\left( \alpha^{1+p} v_{\text{t,i}}^2 \right)  \right)  } \right.  \nonumber \\ & \left.  - \sqrt{ 2 \left(  -\Delta_+ \left( x; y, \bar{x}, U \right) + \chi_{\text{M}} \left( y, \bar{x} \right)  - \chi \left( x; y, \bar{x} \right) + O\left( \alpha^{1+p} v_{\text{t,i}}^2 \right) \right)  } \right]  \nonumber \\
= & \frac{  \Delta_{\text{M}} \left( y, \bar{x}, U \right)  }{ \sqrt{2\left( \chi_{\text{M}}(y, \bar{x}) - \chi(x,y, \bar{x}) \right) } } \left( 1 + O\left( \alpha^{p}  \right) \right) \text{.}
\end{align} 
Note that the terms proportional to $\Delta_+$ have cancelled to first order, and the error in the last line of (\ref{Deltavx-simplifying}) comes from the $O\left( \alpha^{1+p} v_{\text{t,i}}^2 \right)$ error in the evaluation of $D$ (see equation (\ref{D-range})). 
For convenience, I re-express (\ref{Deltavx-simplifying}) in the form
\begin{align} \label{Deltavx-simpler}
\Delta v_x  = & \left[  \sqrt{ 2 \left(  \Delta_{\text{M}} \left( y,\bar{x}, U \right) + \chi_{\text{M}} \left( y, \bar{x} \right) - \chi \left( x,  y, \bar{x} \right) \right)  } \right.  \nonumber \\ & \left.  - \sqrt{ 2 \left(  \chi_{\text{M}} \left( y, \bar{x} \right) - \chi \left( x, y, \bar{x} \right) \right)  } \right] \left( 1 +  O\left( \alpha^{p}  \right) \right) \text{.}
\end{align} 

I proceed to show that equations (\ref{Vx+-simpler}) and (\ref{Deltavx-simpler}) are also valid in the region $|x-x_{\text{M}}| \sim \alpha^p \rho_{\text{i}}$.
In this region, the scalings
\begin{align} \label{inner-ordering}
 \Delta_+  \lesssim \alpha^{1+p} v_{\text{t,i}}^2  \ll \Delta_{\text{M}} \sim \chi_{\text{M}}  - \chi  \sim \alpha v_{\text{t,i}}^2 \text{}
 \end{align}
 hold.
From equations (\ref{Vx+}), (\ref{Deltavx}) and (\ref{inner-ordering}) we have
\begin{align} \label{inner-ordering-1}
 \Delta v_x \sim  V_{x+}  \sim  \alpha^{1/2} v_{\text{t,i}} \text{.} 
 \end{align}
The term $\Delta_+$ in the ordering (\ref{inner-ordering}) is small because the range of integration in equation (\ref{Delta+}) is small. 
Importantly, the $O\left( \alpha^{1+p} v_{\text{t,i}}^2 \right) $ error in the evaluation of $D$ is larger than (or comparable to) $\Delta_+$.
Hence, the term $\Delta_+ $ is negligible in equations (\ref{Vx+}) and (\ref{Deltavx}), and equations (\ref{Vx+-simpler}) and (\ref{Deltavx-simpler}) are valid in the region $|x-x_{\text{M}}| \sim \alpha^p \rho_{\text{i}}$.

\begin{figure}[h]
\centering
\includegraphics[width=0.75\textwidth]{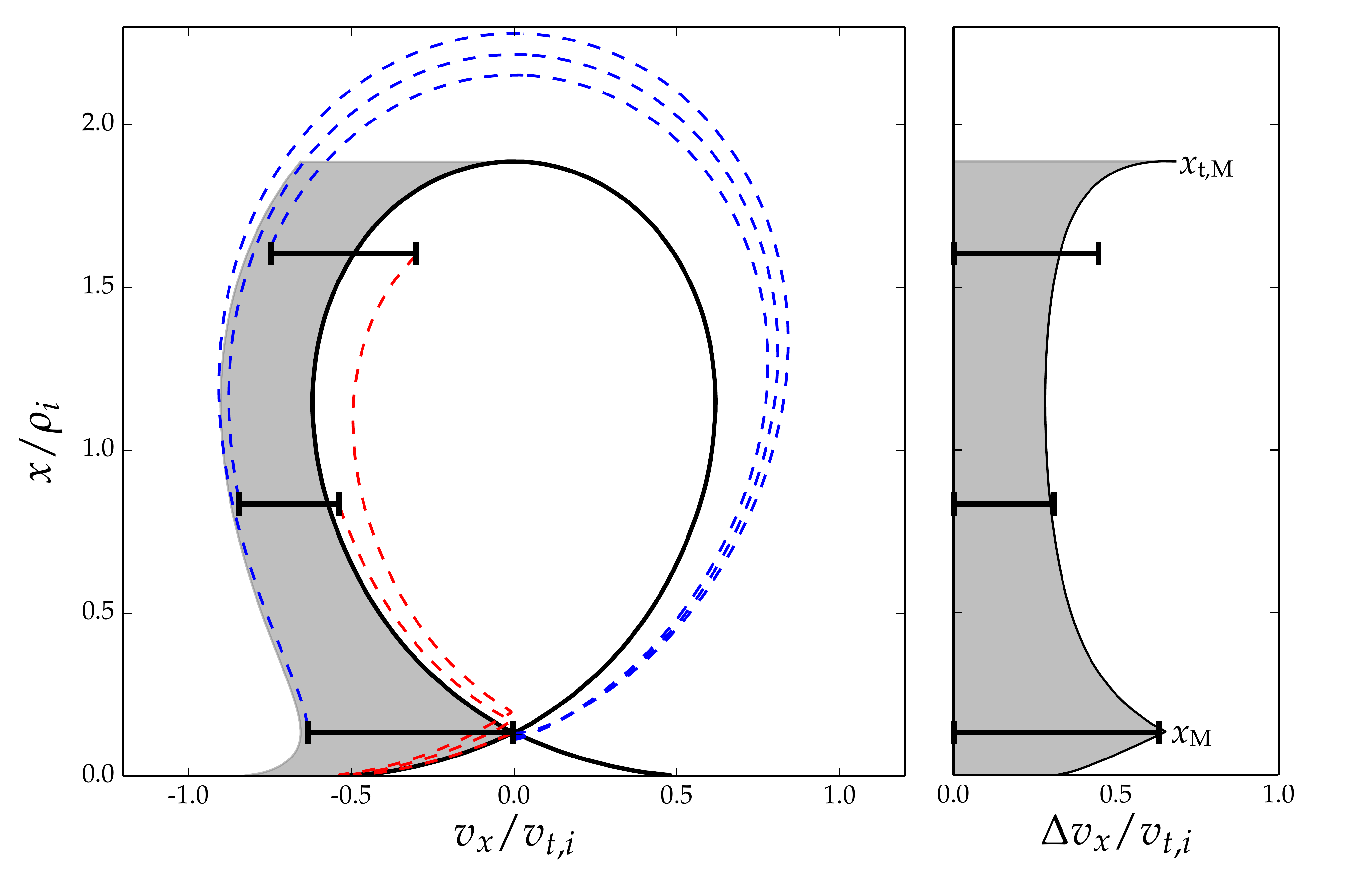}
\caption[Phase space ion trajectories and accuracy of the range of velocities of an open orbit]{
The left diagram shows ion trajectories (dashed lines) for $\alpha = 0.02$, obtained using equations (\ref{x-EOM-exact}), (\ref{vx-Uperp-xbar-x}) and (\ref{xbardot})-(\ref{Uperpdot}) with the numerical electrostatic potential solution presented in chapter \ref{chap-KMPS} (with no $y$-dependence). 
At a given time, the trajectories have $\bar{x} = 1.6 \rho_{\text{i}}$ and $U - \chi_{\text{M}}  =  v_{\text{t,i}}^2$ at three different positions (marked with thick black lines). Blue lines are past ion trajectories chosen to have the largest value of $U_{\perp}$ for which a bottom bounce point exists. Red lines are future ion trajectories chosen to have the smallest value of $U_{\perp}$ for which the ion crosses the effective potential maximum $x_{\text{M}}$ and reaches the wall. The thick black lines connect the red and blue trajectories at the three positions. Thus, they measure the difference between the maximum and minimum $v_x$ of the open orbits. The shaded region on the left is $- V_x \left( x; \bar{x}, \chi_{\text{M}} \right) - \Delta v_x  <  v_x <  - V_x \left( x, \bar{x}, \chi_{\text{M}}  \right)$. On the right diagram, the difference between the maximum and minimum velocities of the open orbits at the three values of $x$ is compared to the width of the shaded region, given by $\Delta v_x$. 
}
\label{fig-Deltavx}
\end{figure}

In the above discussion I neglected the factor of $2\pi$ in the scaling $\Delta_{\text{M}} \sim 2\pi \alpha v_{\text{t,i}}^2$ of equation (\ref{DeltaM}). 
When this factor is included, equation (\ref{Deltavx-simpler}) gives the scaling 
\begin{align} \label{Deltavx-size}
2\pi \alpha v_{\text{t,i}} \lesssim \Delta v_x \lesssim \sqrt{2\pi \alpha } v_{\text{t,i}} \text{,}
\end{align} 
where $\Delta v_x \sim \sqrt{2\pi \alpha } v_{\text{t,i}}$ holds in the neighbourhood of the effective potential maximum $x_{\text{M}}$, while $\Delta v_x \sim 2\pi \alpha  v_{\text{t,i}}$ holds almost everywhere else.
The behaviour of $\Delta v_x$ as a function of $x$ is shown in Figure~\ref{fig-Deltavx}.
Note that there is a small region near the top bounce point that satisfies $|x-x_{\text{t,M}}| \sim \alpha \rho_{\text{i}} $ in which $\Delta v_x \sim \sqrt{2\pi \alpha } v_{\text{t,i}}$. 
In this region, equations (\ref{Vx+-simpler}) and (\ref{Deltavx-simpler}) are not valid because $ \Delta_+  \sim \Delta_{\text{M}} \sim \chi_{\text{M}}  - \chi  \sim \alpha v_{\text{t,i}}^2$ and thus $\Delta_+$ cannot be neglected.
However, after equation (\ref{ni-open-order}) I will argue that the contribution to the ion density due to this region is small.
Recall that $\Delta v_x$, calculated from equation (\ref{Deltavx-simpler}), should be equal to the difference between the maximum and minimum velocity that an open orbit with a given $y$, $\bar{x}$ and $U$ can have.
Indeed, from Figure \ref{fig-Deltavx} (which does not include $y$-dependence) we see that $\Delta v_x$ is a good approximation to the range of allowed velocities at two out of three positions shown, and is a bad approximation only at the position close to $x_{\text{t,M}}$. 

The range of velocities in (\ref{vx-range}) reduces, using equations (\ref{Vx+-simpler}) and (\ref{Deltavx-simpler}), to 
\begin{align} \label{vx-range-simpler}
- V_x \left( x, y, \bar{x}, \chi_{\text{M}}  \right) - \Delta v_x  <  v_x <  - V_x \left( x, y, \bar{x}, \chi_{\text{M}} \right)  \text{.}
\end{align}
Note the major simplification: equations (\ref{Vx+-simpler}) and (\ref{Deltavx-simpler}), and therefore the range (\ref{vx-range-simpler}), are independent of $\Delta_+$. 
Equation (\ref{vx-range-simpler}) gives the range of values of $v_x$ for which the distribution function of open orbits is non-zero. 
Using this range in $v_x$ and $\bar{x} > \bar{x}_{\text{m,o}}(x,y)$, we have
\begin{align} \label{fopen}
f_{\text{op}} (x, y, v_x, v_y, v_z) \simeq & F_{\text{cl}} \left( y_{\star}(y,V_{\parallel}), \mu\left( y, \bar{x}, \chi_{\text{M}}  \right) , U, \sigma_{\parallel}  \right) \Theta \left( \bar{x} - \bar{x}_{\text{m,o}}(x,y) \right)   \nonumber  \\
 & \times \hat{\Pi} \left(  v_x ,  - V_x \left( x, y, \bar{x}, \chi_{\text{M}} \right) - \Delta v_x  , - V_x \left( x, y, \bar{x}, \chi_{\text{M}}\right)  \right) \text{,}
\end{align}
where we defined the top-hat function $\hat{\Pi}\left(r, l_1, l_2\right)$ as
\begin{align}
\hat{\Pi} \left( r, l_1, l_2 \right) = 
\begin{cases}
1 & \text{ if } l_1 \leqslant r < l_2 \text{,} \\
0 & \text{ else.} 
\end{cases}
\end{align}
In equation (\ref{fopen}) we can use (\ref{xbar-def}) and (\ref{U-open}) to re-express $\bar{x}$ and $U$ in terms of $x$, $v_y$ and $v_z$. 
The subscript ``op'' stands for ``open''.

The density of ions in open orbits is an integral of the distribution function in velocity space at fixed $x$, hence
\begin{align}
n_{\text{i,op}}  \left( x \right) = \int f_{\text{op}} \left( x, y, \vec{v} \right) d^3v
 \text{.}
\end{align}
Changing variables in the integral using equations (\ref{xbar-def}) and (\ref{U-open}) we get
\begin{align} \label{ni-open-general}
n_{\text{i,op}} \left( x, y \right) = \int_{\bar{x}_{\text{m,o}}(x,y)}^{\infty} \Omega d\bar{x} \int_{\chi_{\text{M}} \left( y, \bar{x}  \right) }^{\infty} \frac{F_{\text{cl}} \left( y_{\star}(y, v_z), \mu (y,\bar{x}, \chi_{\text{M}}  ), U, \sigma_{\parallel} \right) }{\sqrt{2\left( U - \chi_{\text{M}}  \right)}}  \Delta v_x  dU  \text{.}
\end{align}
The relative error in equation (\ref{ni-open-general}) is $O(\alpha^p)$.

From equations (\ref{Deltavx-size}) and (\ref{ni-open-general}), the characteristic size of the open orbit density is
\begin{align} \label{ni-open-order}
\alpha   n_{\text{e}\infty}  \lesssim n_{i,\text{op}} \lesssim \alpha^{1/2} n_{\text{e}\infty} \text{.}
\end{align}
The ordering $ n_{i,\text{op}}  \sim \alpha^{1/2} n_{\text{e}\infty} $ is valid for $x \lesssim \alpha \rho_{\text{i}}$ only if there is a sufficiently large number of type I orbits, that is, $\bar{x}_{\text{m,I}} \sim \rho_{\text{i}}$ (see Figure \ref{fig-typesofcurves}). 
Type I effective potential curves have $x_{\text{M}} = 0$ by definition, so all type I ion orbits must cross the effective potential maximum at the same position $x=0$, with a range of values of $v_x$ given by $\Delta v_x \sim \alpha^{1/2} v_{\text{t,i}}$.
For type II orbits, the open orbit density is always $ n_{i,\text{op}} \sim \alpha n_{\infty} $ because ions with different values of $\bar{x}$ cross the effective potential maximum at different locations $x_{\text{M}} $.
At some position $x$, there is a small range of values of $x_{\text{M}}$ (and therefore of $\bar{x}$), given by $|x-x_{\text{M}}| \sim \alpha^{1/2} \rho_{\text{i}}$, in which $\Delta v_x\sim \alpha^{1/2} v_{\text{t,i}}$. 
Multiplying the factor $\alpha^{1/2}$ from the range of values of $x_{\text{M}}$ by the factor $\alpha^{1/2}$ from the size of $\Delta v_x$ gives a contribution of order $\alpha n_{\text{e}\infty}$ to the ion density from ions in the region $|x-x_{\text{M}}| \sim \alpha^{1/2} \rho_{\text{i}}$. 
Physically, the ions approach the wall more slowly near the effective potential maximum (where $v_x$ is smaller), leading to a larger number of ions in this region due to flux conservation.
However, ions in type II orbits slow down at different locations depending on their orbit position $\bar{x}$.
Thus, there is not a \emph{single} location where the ions in type II orbits accumulate.
Therefore, their contribution to the density has the same characteristic size at all values of $x$.
Conversely, ions in type I orbits are all slowly crossing the effective potential maximum at the same position $x=0$, and therefore their contribution to the density at $x=0$ is larger.
Despite the fact that $\Delta v_x \sim \alpha^{1/2} v_{\text{t,i}}$ near $x_{\text{t,M}} $, the contribution to the density from ions in this region is of order $\alpha^{3/2} n_{\text{e}\infty} $ because the ions must be very close to $x_{\text{t,M}}$ for $\Delta v_x$ to be large, that is, $|x-x_{\text{t,M}}| \sim \alpha \rho_{\text{i}}$. 
Consequently, the fact that $\Delta v_x$ is a bad approximation to the range of values of $v_x$ near $x_{\text{t,M}}$ (Figure \ref{fig-Deltavx}) is unimportant.

\chapter{Electron-repelling magnetic presheath}
\label{chap-KMPS}

The previous chapter provides the equations from which the ion distribution function and density can be obtained across the magnetic presheath \text{if} the electrostatic potential $\phi(x,y)$ is known.
However, the electrostatic potential is not known a priori, but has to be determined from the quasineutrality equation. 
This chapter is devoted to solving the quasineutrality equation.

I assume a strongly electron-repelling wall and thus an adiabatic electron model.
The conditions under which this assumption is justified are explained in section \ref{sec-KMPS-Boltzmann}. 
Setting the ion and electron density equal to each other gives the quasineutrality equation, as shown in section \ref{sec-KMPS-quasi}.
At the end of section \ref{sec-KMPS-quasi}, I take the limit $\delta \ll \alpha$ in the magnetic presheath, effectively setting $\delta = 0$ and thus ignoring turbulent gradients parallel to the wall for the remainder of this thesis.
In section \ref{sec-KMPS-Chodura}, I derive a condition that must be satisfied for the quasineutrality equation to have a steady-state solution.
This condition is the kinetic generalization of the well-known Chodura condition (to lowest order in $\alpha$).
In section \ref{sec-KMPS-Bohm}, I prove that the solution of my kinetic model satisfies the kinetic Bohm condition with the equality sign, and obtain the expected behaviour of the electrostatic potential near $x=0$.
In section \ref{sec-KMPS-nummethod} I explain the numerical method and iteration procedure used to obtain the solution to the magnetic presheath quasineutrality equation.
I conclude this chapter by presenting numerical results in section \ref{sec-KMPS-results}, discussing how the angle $\alpha$ affects the electrostatic potential and ion flow profiles, as well as the ion distribution function entering the Debye sheath.

\section{Boltzmann electrons} \label{sec-KMPS-Boltzmann}

\begin{figure}
\centering
\includegraphics[width=0.47\textwidth]{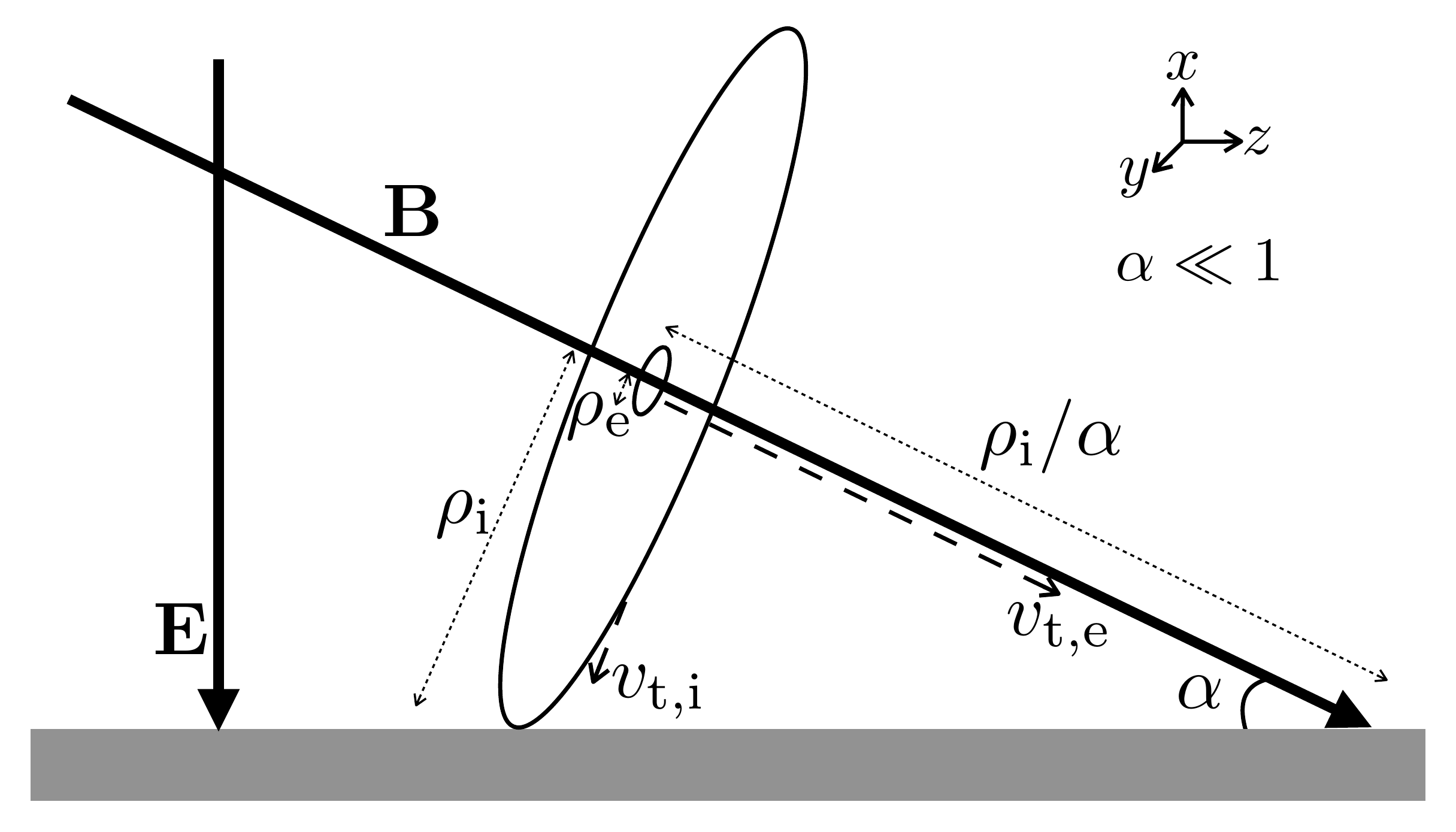}
\caption[Condition for adiabatic electrons (general ion temperature)]{An ion (large) and electron (small) gyro-orbit are shown schematically an ion gyroradius $\rho_{\text{i}}$ away from the wall (grey horizontal surface). 
The ion reaches the wall during a gyro-orbit, taking a time $\rho_{\text{i}} / v_{\text{t,i}}$, where $v_{\text{t,i}}$ is the ion thermal velocity.
The electron gyroradius is much smaller, $\rho_{\text{e}} \ll \rho_{\text{i}}$.
The electron reaches the wall by travelling a distance $\rho_{\text{i}}/\alpha$ parallel to the magnetic field at its thermal velocity $v_{\text{t,e}}$, taking a time $\rho_{\text{i}}/(\alpha v_{\text{t,e}}) = (v_{\text{t,i}}/(\alpha v_{\text{t,e}}))  \times (\rho_{\text{i}}/v_{\text{t,i}})$.
The $x$, $y$ and $z$ directions are shown relative to the magnetic and electric fields $\vec{B}$ and $\vec{E}$, respectively.
}
\label{fig-geometry-elrep}
\end{figure}

Consider a magnetized plasma in contact with a wall without a potential difference between them and with a magnetic field making a small angle $\alpha $ with the wall.
Ions and electrons are expected to travel at characteristic velocities equal to their thermal speeds.
The ion thermal speed is defined in (\ref{v-order}) and the electron thermal speed is
\begin{align} \label{vte}
v_{t,e} = \sqrt{\frac{2T_{\text{e}} }{m_{\text{e}}}} \text{.}
\end{align}
A distance $\rho_{\text{i}}$ from the wall, many ions reach the wall during their Larmor orbit.
On the other hand, the typically much faster electrons ($v_{\text{t,i}} \ll v_{\text{t,e}}$) have to travel along the magnetic field because they are tied much more closely to the magnetic field lines than the ions ($\rho_{\text{e}} \ll \rho_{\text{i}}$), as shown in Figure \ref{fig-geometry-elrep}. 
Hence, electrons travel a distance longer by a factor of $1/\alpha$ compared to ions, and they reach the wall in a time equal to $ v_{\text{t,i}}/(\alpha v_{\text{t,e}}) $ multiplied by the time taken by ions.
The time taken by an electron to reach the wall is much smaller than the time taken by an ion if $ v_{\text{t,i}}/(\alpha v_{\text{t,e}}) \ll 1$, which implies
\begin{align}  \label{ordering-angle}
\alpha \gg  \sqrt{ \frac{ m_{\text{e}} T_{\text{i}}}{ m_{\text{i}} T_{\text{e}} } } \text{.}
\end{align}
For $T_{\text{i}} \sim T_{\text{e}}$, we have $\alpha \gg \sqrt{m_{\text{e}}/ m_{\text{i}}} \sim 0.02 \left( \simeq 1^{\circ} \right) $ for Deuterium ions.

If the ordering (\ref{ordering-angle}) is satisfied, the plasma will lose a much larger number of electrons than ions and a potential difference that repels most of the electrons will be set up between the plasma and the wall.
In current tokamaks, the angle $\alpha$ usually lies in the range $3^{\circ}-11^{\circ}$ ($\alpha \sim 0.05-0.2$ in radians) and the ordering (\ref{ordering-angle}) is approximately satisfied. 
If (\ref{ordering-angle}) is not well satisfied, a kinetic electron model is necessary to solve for the magnetic presheath and Debye sheath.
In ITER, it is expected that $\alpha$ will be around $2^{\circ}$ ($0.03$ radians) \cite{Pitts-2009}.
Note that the presence of an electron current to the wall reduces the need for the wall to repel electrons, and thus large electron currents can increase the right hand side of \ref{ordering-angle}.
The electron-repelling assumption can therefore be violated at larger values of $\alpha$ than estimated above.
Large electron currents are measured close to divertor or limiter targets in the near Scrape-Off-Layer \cite{Loizu-2017}.
This discussion highlights the need for a kinetic electron model in future studies of the sheath and presheath.

If (\ref{ordering-angle}) is satisfied (and there are no large electron currents to the wall), the electron density in the magnetic presheath is expected to be well-approximated by a Boltzmann distribution,
\begin{align} \label{ne-general}
n_{\text{e}}(x,y) = n_{\text{e}\infty} \left( y \right) \exp{\left(\frac{e\left( \phi \left(x, y \right) - \phi_{\infty} \left( y \right) \right)}{T_e }\right)} \text{.}
\end{align}
In (\ref{ne-general}), $n_{\text{e}\infty} \left(y \right)$ and $\phi_{\infty} \left( y \right)$ are the electron density and the electrostatic potential at the magnetic presheath entrance ($x \rightarrow \infty$). 
The ordering (\ref{ordering-angle}) implies that the truncation of the Maxwellian electron distribution function, due to the high energy electrons reaching the wall instead of being reflected, can be ignored.
Such truncation applies to the high-energy tail of the electron distribution function and leads to a negligible correction, of order $n_{\text{e},\infty}\sqrt{m_{\text{e}} T_{\text{i}}/m_{\text{i}}} T_{\text{e}}$, to the Boltzmann distribution (\ref{ne-general}) \cite{Ingold-1972}. 

\section{Quasineutrality equation} 
\label{sec-KMPS-quasi}

The quasineutrality equation is
\begin{align} \label{quasineutrality-general-star}
& n_{\text{e}\infty} \left( y \right) \exp{\left(\frac{e\left( \phi \left( x, y \right) - \phi_{\infty} \left( y \right) \right)}{T_e }\right)} = n_{\text{i}} (x,y)  \text{.}
\end{align}
The solution of this equation is the electrostatic potential $\phi \left( x , y \right)$ across a magnetic presheath with turbulent gradients. 
Solving the magnetic presheath numerically amounts to finding a potential for which (\ref{quasineutrality-general-star}) is satisfied.
As previously proposed (without $y$ dependence) by Cohen and Ryutov \cite{Cohen-Ryutov-1998}, one can solve the quasineutrality equation (\ref{quasineutrality-general-star}) by means of an iterative procedure. 
In order to solve for the self-consistent electrostatic potential in the magnetic presheath, the electrostatic potential profile at $x \rightarrow \infty$, $\phi_{\infty}(y)$, must be known as a boundary condition. 
This is $\phi_{\infty} \left( y \right) = \phi \left( x \rightarrow \infty, y \right)$. 
For some guessed potential $\phi \left( x, y \right)$, the integral on the right hand side of (\ref{quasineutrality-general-star}) can be computed numerically. 
If the initial guess can be corrected to a new guess such that the difference between the right hand side and the left hand side of (\ref{quasineutrality-general-star}) is reduced, this procedure can be iterated until convergence.

I implemented such an iteration procedure to solve the simpler problem with no gradients in the $y$ direction ($\delta = 0$), as will be discussed in section \ref{sec-KMPS-nummethod}. 
\emph{I henceforth set $\delta = 0$ and thus neglect the $y_{\star}$ and $y$ dependences from (\ref{quasineutrality-general-star})}.
The electron density is
\begin{align} \label{ne}
n_{\text{e}}(x) = n_{\text{e}\infty} \exp{\left(\frac{e \phi \left( x \right) }{T_e }\right)} \text{,}
\end{align}
where $n_{\text{e}\infty}$ is a constant and I set $\phi_{\infty} = 0$.
The total ion density is 
\begin{align} \label{ni}
n_{\text{i}} \left( x \right) = n_{\text{i,cl}} (x) + n_{\text{i,op}} \left( x \right) \text{.}
\end{align}
The ion distribution function simplifies to $F(\mu, U)$ and thus the closed orbit density is
\begin{align} \label{ni-closed}
n_{\text{i,cl}} (x) =  \int_{\bar{x}_m \left( x  \right) }^{\infty} d\bar{x} \int_{\chi\left( x; \bar{x} \right)}^{\chi_{\text{M}} \left( \bar{x} \right)}  \frac{2\Omega dU_{\perp}}{\sqrt{2\left( U_{\perp} - \chi \left( x; \bar{x} \right) \right)}} \int_{U_{\perp}}^{\infty} \frac{F_{\infty} \left( \mu, U \right)}{\sqrt{2\left( U - U_{\perp}\right)}} dU \text{.}
\end{align}
I have removed the summation over $\sigma_{\parallel}$ because $\sigma_{\parallel} = + 1 $ is the only allowed value at $\bar{x} \rightarrow \infty$ when the turbulent $\vec{E} \times \vec{B}$ drift is not present (because $\dot{\bar{x}} = - \sigma_{\parallel} \alpha V_{\parallel} (U_{\perp}, U )$ must be negative at $x \rightarrow \infty$).
The open orbit density is
\begin{align} \label{ni-open}
n_{\text{i,op}} \left( x \right) = \int_{\bar{x}_{\text{m,o}}(x)}^{\infty} \Omega d\bar{x} \int_{\chi_{\text{M}} \left(  \bar{x}  \right) }^{\infty} \frac{F_{\infty} \left(  \mu (\bar{x}, \chi_{\text{M}}  ), U \right) }{\sqrt{2\left( U - \chi_{\text{M}}  \right)}}  \Delta v_x  dU \left[ 1 + O  \left( \alpha^{p} \right) \right]  \text{.}
\end{align}
The quasineutrality equation thus becomes 
\begin{align} \label{quasineutrality}
n_{\text{e}\infty} \exp \left( \frac{e\phi\left( x \right) }{T_{\text{e}}} \right) = Zn_{\text{i}} \left( x \right)  \text{.}
\end{align}

Note that removing $y$ dependence in equation (\ref{DeltaM}) leads to
\begin{align} \label{DeltaM-noy}
\Delta_{\text{M}} \left( \bar{x}, U \right) = \alpha V_{\parallel} \left( \chi_{\text{M}} \left(  \bar{x} \right) , U \right) I \left( \bar{x} \right) \text{,}
\end{align}
where I defined the ``open orbit integral'' 
\begin{align} \label{open-integral}
I \left( \bar{x}, U \right) = 2 \Omega^2 \int_{x_{\text{M}}}^{x_{\text{t,M}}} \frac{ x-x_{\text{M}}  }{ V_x \left( x, \bar{x}, \chi_{\text{M}} \left(  \bar{x} \right) \right)  } ds = 2\pi \frac{d}{d\bar{x}} \left( \mu \left( \bar{x}, \chi_{\text{M}}(\bar{x}) \right) \right)   \text{.}
\end{align}
The last equality of (\ref{open-integral}) is proved as follows.
Taking the total derivative of $\mu$ with respect to $\bar{x}$ while fixing $U_{\perp} = \chi_{\text{M}} (\bar{x})$ gives
\begin{align}
\frac{d}{d\bar{x}} \left( \mu \left( \bar{x}, \chi_{\text{M}}(\bar{x}) \right) \right)  = \frac{\partial \mu}{\partial U_{\perp} } (\bar{x}, \chi_{\text{M}} )  \frac{d \chi_{\text{M}} }{d \bar{x}} + \frac{\partial \mu}{\partial \bar{x} }  (\bar{x}, \chi_{\text{M}} )  \text{.}
\end{align}
Using equations (\ref{dmudxbar}) and (\ref{dmudUperp}) without $y$ dependence, and using $ \partial \chi_{\text{M}} /  \partial \bar{x}   = \Omega^2 \left( \bar{x} - x_{\text{M}} \right) $, one obtains
 \begin{align} \label{dmudxbar-open}
\frac{d}{d\bar{x}} \left( \mu \left( \bar{x}, \chi_{\text{M}}(\bar{x}) \right) \right) =  \frac{\Omega^2}{\pi} \int_{x_{\text{M}}}^{x_{\text{t,M}}}  \frac{ x- x_{\text{M}} }{V_x \left(x, \bar{x}, \chi_{\text{M}} \right)} dx \text{.}
\end{align}
Hence, the last equality of (\ref{open-integral}) follows.


Throughout this work I assume a single ion species. 
However, the quasineutrality equation can be generalized to include more than one ion species by adding the density integral of the additional species to equation (\ref{quasineutrality-general-star}) or its simplified version (\ref{quasineutrality}). 
Generalizing to a system with more than one ion species is needed to account for the presence of Deuterium and Tritium isotopes in roughly equal amounts near the divertor targets of future fusion devices.

\section{Magnetic presheath entrance}
\label{sec-KMPS-Chodura}

Expanding the quasineutrality equation (\ref{quasineutrality}) near the magnetic presheath entrance $x\rightarrow \infty$ leads to:
\begin{itemize}
\item a solvability condition for the distribution function at the magnetic presheath entrance, with which I choose a realistic boundary condition for the ion distribution function at $x\rightarrow \infty$;
\item the form of the electrostatic potential near $x\rightarrow \infty$, which is needed to determine the potential above a certain value of $x$ in my numerical scheme.
\end{itemize}

At sufficiently large values of $x$, the electrostatic potential must be small, such that
\begin{align} \label{hatphismall}
\hat{\phi} = \frac{e\left| \phi\left(x\right) \right|}{T_{\text{e}}} \ll 1 \text{.}
\end{align}
Here I also assume that the length scale of changes in the electrostatic potential is very large at sufficiently large $x$, such that
\begin{align} \label{epsilonsmall}
\epsilon = \frac{\rho_{\text{i}} \phi'\left( x\right)}{\phi\left( x \right)} \ll 1 \text{.}
\end{align}
Assumption (\ref{epsilonsmall}) is not the most general one, as $\epsilon$ can be of order unity, but it is useful because it is correct for the boundary condition at $x\rightarrow \infty$ chosen in Section \ref{sec-KMPS-results}. 
In general, $\hat{\phi } \lesssim \epsilon^2 \lesssim 1 $, but here I take the more constrained limit
\begin{align} \label{epssmallerdelta}
\hat{\phi } \lesssim \epsilon^2 \ll 1 \text{.}
 \end{align}

For $x \rightarrow \infty$, the open orbit density is higher order in $\alpha$ than the closed orbit density. 
Moreover, if the distribution function is exponentially decaying with energy, like the one I use in section \ref{sec-KMPS-results}, the open orbit density near $x \rightarrow \infty$ is exponentially small because only very large orbits with very large energies can extend all the way from the wall $x=0$ to points with large $x$. 
Using that $ n_{i,\text{open}} (x) \simeq 0$ for large $x$, the closed orbit density is obtained by expanding the near-circular ion orbits about circular orbits, as shown in Appendix \ref{app-quasi-expansion}, to obtain
\begin{align} \label{niclosedinfty}
n_{i,\text{closed}} \left( x \right)   = & \left( 1 +  \frac{\phi''(x)}{\Omega B}  \right) \int_{-\pi}^{\pi} d\varphi \int_0^{\infty} \Omega d\mu \left\lbrace \int_{\Omega \mu}^{\infty} \frac{F_{\text{cl}}(\mu, U') }{\sqrt{2\left(U- \Omega \mu \right)}} dU  \right. \nonumber \\
& \left.  - \sqrt{2\delta U_{\perp}} F_{\text{cl}}(\mu, \Omega \mu ) - \delta U_{\perp}  \int_{\Omega \mu}^{\infty} \frac{\partial F_{\text{cl}} \left(\mu, U \right) / \partial U  }{\sqrt{2\left( U -  \Omega  \mu \right)}} dU   \right.  \nonumber \\ 
&  \left.  +  \frac{1}{3} \left( 2\delta U_{\perp} \right)^{3/2} \frac{ \partial F_{\text{cl}} }{\partial U } (\mu, \Omega \mu ) + \frac{1}{2} \delta U_{\perp}^2 \int_{\Omega \mu}^{\infty}  \frac{\partial^2 F_{\text{cl}}(\mu, U)/ \partial U^2 }{\sqrt{2\left( U - \Omega \mu \right)}} dU  \right\rbrace  \nonumber \\  
& + O\left( \hat{\phi} \epsilon^3 n_{\text{e}\infty}, \hat{\phi}^2 \epsilon^2 n_{\text{e}\infty}, \hat{\phi}^{5/2} n_{\text{e}\infty} \right)  \text{,}
\end{align}
where
\begin{align} \label{deltaUperp}
\delta U_{\perp} = -\frac{\Omega \phi\left( x \right) }{B} + \frac{\Omega \phi'\left( x \right) }{B} \sqrt{\frac{2\mu}{\Omega} } \cos \varphi -  \frac{\mu \phi'' \left( x \right) }{2B} \left( 1 + 2 \cos^2 \varphi \right) \nonumber  \\
+ O \left( \hat{\phi} \epsilon^3 v_{t,i}^2, \hat{\phi}^2 \epsilon^2 v_{t,i}^2 \right) \text{.}
\end{align}
Note that equations (\ref{niclosedinfty}) and (\ref{deltaUperp}) are derived to lowest order in $\alpha \ll 1$. The quantity $\delta U_{\perp}$ is defined so that $U_{\perp} = \Omega \mu - \delta U_{\perp}$, and therefore captures the difference between $U_{\perp}$ and $\Omega \mu$ as the ion travels into the magnetic presheath. Outside of the magnetic presheath, at $x \rightarrow \infty$, ion orbits are circular and $U_{\perp} = \Omega \mu $ (using $\phi(\infty) = \phi'(\infty) = \phi''(\infty) = 0$). 

The electron density in (\ref{ne}) is expanded in $\hat{\phi} \ll 1$ for $x\rightarrow \infty$,
\begin{align} \label{neinfty}
n_{\text{e}} \left( x \right) = n_{\text{e}\infty} + n_{\text{e}\infty} \frac{e\phi(x)}{T_{\text{e}}} + \frac{1}{2}n_{\text{e}\infty}\left( \frac{e\phi(x)}{T_{\text{e}}} \right)^2 + O\left( \hat{\phi}^3 n_{\text{e}\infty} \right) \text{.}
\end{align}
Substituting (\ref{niclosedinfty}) and (\ref{neinfty}) in (\ref{quasineutrality}), and using that $n_{\text{i,op}}\left( x \right) = 0$, I obtain the quasineutrality equation expanded in $\hat{\phi}$ and $\epsilon$,
\begin{align} \label{quasineutrality-expanded-full}
& n_{\text{e}\infty} + n_{\text{e}\infty} \frac{e\phi(x)}{T_{\text{e}}} + \frac{1}{2}n_{\text{e}\infty}\left( \frac{e\phi(x)}{T_{\text{e}}} \right)^2 = Z \left( 1 +  \frac{\phi''(x)}{ \Omega B }  \right) \int_{-\pi}^{\pi} d\varphi \int_0^{\infty} \Omega d\mu \nonumber \\
& \times \left\lbrace \int_{\Omega \mu}^{\infty} \frac{F_{\text{cl}}(\mu, U) }{\sqrt{2\left(U-\mu \Omega \right)}} dU - \sqrt{2\delta U_{\perp}} F_{\text{cl}}(\mu, \mu \Omega) - \delta U_{\perp}  \int_{\Omega \mu}^{\infty} \frac{\partial F_{\text{cl}}(\mu, U)/ \partial U }{\sqrt{2\left( U - \mu \Omega \right)}} dU  \right. \nonumber \\ & \left. +  \frac{1}{3} \left( 2\delta U_{\perp} \right)^{3/2} \frac{\partial F_{\text{cl}}}{\partial U} (\mu, \mu \Omega)   + \frac{1}{2} \delta U_{\perp}^2 \int_{\Omega \mu}^{\infty}  \frac{\partial^2 F_{\text{cl}} (\mu, U)/ \partial U^2 }{\sqrt{2\left( U - \mu \Omega \right)}} dU    \right\rbrace 
\nonumber  \\ & + O\left( \hat{\phi} \epsilon^3 n_{\text{e}\infty}, \hat{\phi}^2 \epsilon^2 n_{\text{e}\infty}, \hat{\phi}^{5/2} n_{\text{e}\infty}  \right)    \text{.}
\end{align}
To zeroth order in $\hat{\phi}$, equation (\ref{quasineutrality-expanded-full}) gives
\begin{align} \label{quasineutrality-expanded-eps0}
Z \int_{-\pi}^{\pi} d\varphi \int_0^{\infty} \Omega d\mu   \int_{\Omega \mu}^{\infty} \frac{F_{\text{cl}}(\mu, U)}{\sqrt{2\left(U - \Omega \mu \right)}} dU  = n_{\text{e}\infty} \text{.}
\end{align}
This is the quasineutrality equation evaluated exactly at $x\rightarrow \infty$.
The next order correction to (\ref{quasineutrality-expanded-eps0}) is a term of order $\hat{\phi}^{1/2}$, giving
\begin{align} \label{quasineutrality-eps12}
- Z \int_{-\pi}^{\pi} d\varphi \int_0^{\infty}  \Omega d\mu \sqrt{2\delta U_{\perp}}  F_{\text{cl}}(\mu, \Omega \mu) = 0 \text{.}
\end{align}
The distribution function $F_{\text{cl}}\left(\mu, U \right)$ is non-negative, and hence the integral in (\ref{quasineutrality-eps12}) is zero only if $F_{\text{cl}}(\mu, \Omega \mu ) = 0$ for all possible values of $\mu$. 
One expects this for an electron-repelling sheath where no ions come back from the magnetic presheath, so $f_{\infty} \left( v_{x}, v_{y}, v_{z} \right) =0$ at $v_{z} < 0$ and therefore $F_{\text{cl}} \left( \mu, \Omega \mu \right) = f_{\infty} \left( v_{x}, v_{y},  0 \right) = 0$.

To next order, $O(\hat{\phi})$, I collect all terms in (\ref{quasineutrality-expanded-full}) which are proportional to $\phi\left( x\right)$ or its derivatives. 
Integrating by parts\footnote{Re-expressing integrals of the form on the left hand side of (\ref{re-express-Chodura}) to the form on the right hand side using integration by parts is common in the community. 
However, the form on the left hand side is more general, as has been pointed out by Riemann in the context of the Debye sheath \cite{Riemann-review}.} 
and using $F_{\text{cl}}\left( \mu, \Omega \mu \right) = 0$, I find the result
\begin{align} \label{re-express-Chodura}
\int_{\mu \Omega}^{\infty} \frac{\partial F_{\text{cl}} (\mu, U) / \partial U }{\sqrt{2\left( U - \mu \Omega \right)}} dU  = \int_{\mu \Omega}^{\infty} \frac{F_{\text{cl}}(\mu, U) }{\left( 2\left( U - \mu \Omega \right) \right)^{3/2}} dU \text{.}
\end{align}
With this result, the order $\hat{\phi}$ piece of (\ref{quasineutrality-expanded-full}) is, keeping terms up to $O\left( \hat{\phi} \epsilon^2 \right)$,
\begin{align} \label{phiinftyk1}
\phi''\left( x \right) = k_1 \phi \left( x \right) + O\left( \hat{\phi} \epsilon^3 \right) \text{,}
\end{align}
where I define $k_1$, a quantity with dimensions of $\left(1/ \text{length}\right)^2$, as 
\begin{align} \label{k1}
k_1 =  \frac{\Omega^2 }{v_{\text{B}}^2 } \frac{ n_{\text{e}\infty} -  2\pi Z v_{\text{B}}^2 \int_0^{\infty} \Omega d\mu  \int_{\mu \Omega }^{\infty}  \frac{F_{\text{cl}}(\mu, U) dU}{\left(2\left( U - \mu \Omega \right) \right)^{3/2}}  }{ n_{\text{e}\infty} + 2\pi Z \int_0^{\infty} \Omega^2  \mu d\mu \int_{\mu \Omega}^{\infty}  \frac{F_{\text{cl}}(\mu, U) dU}{\left(2\left( U - \mu \Omega \right) \right)^{3/2}}} \text{.}
\end{align}
From equation (\ref{phiinftyk1}) and using the boundary condition $\phi =0$ at $x\rightarrow \infty$, I find $\phi \propto \exp\left( - \sqrt{k_1} x \right)$. Consequently, $\sqrt{\left| k_1 \right|} \rho_{\text{i}} \sim \epsilon$, and assumption (\ref{epsilonsmall}) is true only if $k_1$, defined in equation (\ref{k1}), is sufficiently small. 
If this is not the case, I expect $\phi \propto \exp\left( - \lambda x \right)$, but the value of $\lambda$ would have to be determined by carrying out a more general expansion of the quasineutrality equation in $\hat{\phi} \ll 1$ with $\epsilon \sim 1$.

In order to impose that $\phi \left( \infty \right) = 0$, I require a non-oscillating potential profile at $x \rightarrow \infty$, which gives $k_1 \geqslant 0$ as a solvability condition. The numerator of $k_1$ determines the sign of $k_1$ because the denominator is always positive.
Hence, I obtain the condition
\begin{align} \label{solvability}
   2\pi Z v_{\text{B}}^2 \int_0^{\infty} \Omega d\mu \int_{\mu \Omega }^{\infty}  \frac{F_{\text{cl}} (\mu, U)  dU}{\left(2\left( U - \mu \Omega \right) \right)^{3/2}}  \leqslant  n_{\text{e}\infty}\text{,}
\end{align}
where the Bohm velocity $v_{\text{B}}$ is defined in equation (\ref{vB}). 
The equation
\begin{align} \label{changevar-infty}
  2\pi  \int_0^{\infty} \Omega d\mu \int_{\mu \Omega }^{\infty}  \frac{F_{\text{cl}} (\mu, U)  h\left( \mu, U  \right)  }{ \sqrt{ 2\left( U - \mu \Omega \right)  } }   dU   = \int  f_{\infty} \left( \vec{v} \right) h \left( \frac{ v_x^2 + v_y^2}{2\Omega}, \frac{v_x^2 + v_y^2+v_z^2}{2} \right) d^3v \text{,}
\end{align}
is valid for any function $h$ and is obtained using the fact that $\mu = (v_x^2 + v_y^2 )/2\Omega$ and $U=(v_x^2 + v_y^2 +v_z^2)/2$ at $x\rightarrow \infty$ (shown in Appendix \ref{subapp-mu-expansion}).
I can use equation (\ref{changevar-infty}) to re-express the solvability condition as
\begin{align} \label{kinetic-Chodura}
 Zv_{\text{B}}^2 \int  \frac{f_{\infty} \left( \vec{v} \right)}{v_z^2} d^3v \leqslant n_{\text{e}\infty}  \text{.}
\end{align}
The solvability condition (\ref{kinetic-Chodura}) generalizes Chodura's condition for the magnetic presheath entrance \cite{Chodura-1982} to include the effect of kinetic ions at small $\alpha$.\footnote{It has been speculated that a kinetic generalization of Chodura's condition should be satisfied at the magnetic presheath entrance \cite{Daube-Riemann-1999}; however, the existing derivations \cite{Sato-1994, Claassen-Gerhauser-1996-kineticChodura} make some simplifying assumptions.}
In chapter \ref{chap-Tdep}, I show that the cold ion limit of the generalized condition recovers the cold ion limit of Chodura's original condition to lowest order in $\alpha$. 

It is believed that solvability conditions such as (\ref{solvability}) are usually satisfied marginally \cite{Riemann-review}. This means that equation (\ref{solvability}) is expected to hold in the equality form, which justifies considering $k_1 =0$ and hence justifies my initial assumption that $\epsilon \ll 1$. 
When $k_1 =0$, terms of size $\hat{\phi}^{3/2}$ in the expansion of quasineutrality become important. From considering terms of this order in (\ref{quasineutrality-expanded-full}), I obtain
\begin{align} \label{phi''phi32}
\phi''\left( x \right) = - k_{3/2} \left[ - \phi \left( x \right)\right]^{3/2} \text{,}
\end{align}
where $k_{3/2}$ 
is given by
\begin{align} \label{k32}
 k_{3/2} =  \sqrt{ \frac{e}{T_{\text{e}}} } \left( \frac{\Omega}{v_{\text{B}}} \right)^2  \frac{ \frac{2\sqrt{2}}{3} 2\pi \int_0^{\infty} \Omega v_{\text{B}}^3   \frac{  \partial F_{\text{cl}} }{ \partial U  } (\mu, \Omega \mu)   d\mu }{ n_{\text{e}\infty} + 2\pi Z\int_0^{\infty} \Omega^2  \mu d\mu \int_{\Omega \mu}^{\infty}  \frac{F_{\text{cl}}(\mu, U)}{\left(2\left( U - \Omega \mu \right)\right)^{3/2}}  dU } \geqslant 0 \text{.}
\end{align}
The numerator of (\ref{k32}) is positive because $ F_{\text{cl}}(\mu, U ) = 0 $ for $U \leqslant \Omega \mu$ and hence $\partial F_{\text{cl}} (\mu, \Omega \mu ) / \partial U \geqslant 0$ for all values of $\mu$. Moreover, both terms in the denominator of (\ref{k32}) are explicitly positive, so the inequality in (\ref{k32}) follows. The case $ k_{3/2}  =0 $ only arises if $ \partial F_{\text{cl}} (\mu, \Omega \mu) / \partial U = 0 $ for all $\mu$. Note that this condition implies $(1/v_z)\partial f_{\infty}\left( v_{x}, v_{y}, 0 \right) / \partial v_{z} =0$ for all values of $v_{x}$ and $v_{y}$, which corresponds to a very flat ion distribution function near $v_{z} = 0$. 
One example of such a flat ion distribution function is the one used in section \ref{sec-Tdep-cold} of chapter \ref{chap-Tdep}.

Equation (\ref{phi''phi32}) is solved by multiplying by $\phi' \left( x \right)$ and then integrating once using the boundary condition $\phi' \left( x \right) =0 $ when $\phi \left( x \right) = 0$ to get
\begin{align}
 \phi' \left( x \right)^2  =  \frac{4k_{3/2}}{5}  \left[-\phi \left( x \right) \right]^{5/2}  \text{.}
\end{align}
Taking the square root and integrating again, the potential profile is
\begin{align} \label{phisol1}
\phi  \left( x \right) = - \frac{400}{k_{3/2}^2} \frac{1}{(x+C_{3/2})^4} \text{,}
\end{align}
where $C_{3/2}$ is an integration constant. Equation (\ref{phisol1}) implies that $\epsilon \sim \hat{\phi}^{1/4} \gg \hat{\phi}$. 
The boundary condition that I use to obtain the numerical results (in Section \ref{sec-KMPS-results}) has $k_{3/2} \neq 0$, so equation (\ref{phisol1}) is the form of the electrostatic potential to which I must match my numerical solution at large $x$.

If $\partial F \left( \mu, \Omega \mu \right) / \partial U =0$, then $k_{3/2}  = 0$ and I must go to higher order in $\hat{\phi}$ to solve for the electrostatic potential at large $x$. 
For $\partial F_{\text{cl}} \left( \mu, \Omega \mu \right) / \partial U =0$, one can integrate by parts twice the term with $\partial^2 F\left( \mu, \Omega \mu \right) / \partial U^2 $ to get
\begin{align}
\int_{\Omega\mu}^{\infty} \frac{\partial^2 F_{\text{cl}}\left( \mu, U \right) / \partial U^2 }{\sqrt{2\left( U -\Omega \mu \right)}} dU = 3\int_{\Omega\mu}^{\infty} \frac{F_{\text{cl}}\left( \mu, U \right) }{\left( 2\left( U -\Omega \mu \right) \right)^{5/2}} dU  \text{.}
\end{align}
Balancing the term of order $\hat{\phi} \epsilon^2$ with terms of order $\hat{\phi}^2$ in (\ref{quasineutrality-expanded-full}), I get
\begin{align} \label{phi''phi2}
\phi''\left( x \right) = - k_2 \left[ \phi \left( x \right) \right]^2 \text{,}
\end{align}
where $k_2$ is given by
\begin{align} \label{k2}
 k_2  =   \frac{\Omega^2 e}{2v_{\text{B}}^2 T_{\text{e}}}  \frac{  6\pi Z v_{\text{B}}^4 \int_0^{\infty}  \Omega d\mu \int_{\Omega \mu}^{\infty}  \frac{F_{\text{cl}}(\mu, U) }{\left( 2\left( U - \Omega \mu \right) \right)^{5/2} } dU  -  n_{\text{e}\infty} }{  n_{\text{e}\infty}  +  2\pi Z \int_0^{\infty} \Omega^2 \mu d\mu \int_{\Omega \mu}^{\infty}  \frac{F_{\text{cl}}(\mu, U)}{\left(2\left(U-\Omega \mu \right) \right)^{3/2}} dU   } > 0 \text{.}
\end{align}
Both terms in the denominator of (\ref{k2}) are positive, therefore the inequality on the right hand side of (\ref{k2}) is the result of the numerator being positive, which is demonstrated in Appendix \ref{app-k2>0} using the fact that (\ref{kinetic-Chodura}) is satisfied with the equality sign. Equation (\ref{phi''phi2}) is solved in the same way as equation (\ref{phi''phi32}), and the result is
\begin{align} \label{phisol2}
\phi \left( x \right) = - \frac{6}{k_2} \frac{1}{(x+C_2)^2} \text{,}
\end{align}
where $C_2$ is an integration constant.
The fact that $k_2$ is positive and $k_2 \rho_{\text{i}}^2 T_{e}/e \sim 1$ implies that I do not need to carry out the expansion of (\ref{quasineutrality-expanded-full}) any further, because the order $\hat{\phi}^2$ term is guaranteed to be non-zero if the solvability condition (\ref{kinetic-Chodura}) is marginally satisfied. Hence, $\epsilon \gtrsim \hat{\phi}^{1/2}$ as stated in equation (\ref{epssmallerdelta}).

\section{Debye sheath entrance}
\label{sec-KMPS-Bohm}

By expanding the quasineutrality equation near the Debye sheath entrance, $x=0$, I conclude that
\begin{itemize}
\item that the self-consistent solution of the system gives an ion distribution function at $x=0$ that marginally satisfies the kinetic Bohm condition, with which I can check the numerically calculated distribution function;
\item the self-consistent form of the potential near $x=0$, with which I choose a suitable numerical discretization for the system.
\end{itemize}  

I define the normalized electrostatic potential relative to $x=0$,
\begin{align}
\delta \hat{\phi} = \frac{ e \delta \phi }{T_{\text{e}} }  = \frac{ e }{T_{\text{e}} } \left( \phi ( x ) - \phi ( 0) \right) \ll 1 \text{.}
 \end{align}
Each term of the quasineutrality equation (\ref{quasineutrality}) is expanded in $\delta \hat{\phi} \ll 1$ separately, order by order.
Denoting the electron density at $x=0$ as $n_{\text{e}0}$, such that
\begin{align}
n_{\text{e}0} = n_{\text{e}\infty} \exp \left( \frac{e \phi(0) }{T_{\text{e}}} \right) \text{,}
\end{align}
the electron density near $x=0$ is
\begin{align} \label{ne-near0}
n_{\text{e}} (x) = n_{\text{e}0} \exp \left( \frac{e\delta \phi }{T_{\text{e}}} \right) \text{.}
\end{align}
Using the result $n_{\text{i,cl}} (0) = 0$ and equation (\ref{quasineutrality}) gives
\begin{align} \label{quasi-0}
n_{\text{e}0} = Zn_{\text{i,op}} (0) =  \int_{\bar{x}_{\text{m,o}}(0)}^{\infty}  \Omega d\bar{x}  \int_{\chi_{\text{M}} (\bar{x} )}^{\infty}  \frac{ F_{\text{cl}} ( \mu (\bar{x}, \chi_{\text{M}}) , U ) }{\sqrt{2\left( U - \chi_{\text{M}} (\bar{x}) \right)}} \Delta v_{x0} dU     \text{,}
\end{align}
where 
\begin{align} \label{Deltavx-0}
\Delta v_{x0} = \Delta v_x \rvert_{x=0} = \sqrt{ 2\left( \Delta_{\text{M}} (\bar{x}, U ) + \chi_{\text{M}} (\bar{x} ) - \chi (0, \bar{x} ) \right) } - \sqrt{2\left( \chi_{\text{M}} (\bar{x} ) - \chi (0, \bar{x} )  \right)} \text{.}
\end{align}
Subtracting equation (\ref{quasi-0}) from equation (\ref{quasineutrality}), I obtain the perturbed quasineutrality equation near $x=0$,
\begin{align} \label{quasi-perturbed}
n_{\text{e}} (x) - n_{\text{e}0} = Z \left( n_{\text{i,cl}} (x) +  n_{\text{i,op}}(x) - n_{\text{i,op}}(0) \right) \text{.}
\end{align} 

The value of $\bar{x}$, $\bar{x}_{\text{m,I}}$, above which the orbits become type I becomes infinitely large in the magnetic presheath. 
Therefore, type I orbits are absent. 
To prove this result, I first assume the more general scenario in which both type I and type II orbits are present, with $\phi'(0)$ being finite, and calculate the dominant contribution to equation (\ref{quasi-perturbed}).

I proceed to obtain the term $n_{\text{i,cl}}(x)$ in equation (\ref{quasi-perturbed}) to leading order.
Firstly, I observe that a \emph{closed} orbit near $x=0$ must lie at a position $x$ such that $0 \leqslant x_{\text{M}} \leqslant x$, with $\chi ( x, \bar{x} ) \simeq \chi_{\text{M}} ( \bar{x} ) $. 
Remembering that for a closed orbit the perpendicular energy lies in the range $\chi ( x, \bar{x} ) \leqslant U_{\perp} \leqslant \chi_{\text{M}} ( \bar{x} ) $, I can take the integral over $U_{\perp}$ in (\ref{ni-closed}) by approximating
\begin{align}
F_{\text{cl}}  \left( \mu ( \bar{x}, U_{\perp} ), U \right) \simeq F_{\text{cl}}  \left( \mu ( \bar{x}, \chi_{\text{M}} (\bar{x}) ), U \right)  \text{}
\end{align}
and $ \sqrt{2\left( U- U_{\perp} \right)} \simeq  \sqrt{2\left( U- \chi_{\text{M}} (\bar{x} ) \right)}$.
With these approximations, the integral (\ref{ni-closed}) becomes
\begin{align} \label{ni-closed-expanded}
n_{\text{i,cl}} (x) \simeq & ~  2\int_{\bar{x}_{\text{m}} (x) }^{\infty} \Omega \sqrt{2 \left(\chi_{\text{M}} (\bar{x}) - \chi (x, \bar{x}) \right) } d\bar{x} \nonumber \\
&~  \times \int_{\chi_{\text{M}} (\bar{x})}^{\infty} \frac{F_{\text{cl}} \left(\mu \left(\bar{x}, \chi_{\text{M}} \left(\bar{x}\right) \right), U \right) }{\sqrt{2\left( U - \chi_{\text{M}} (\bar{x}) \right) }}  dU  \text{.}
\end{align}
The contributions to $n_{\text{i,cl}}(x)$ of type I and type II closed orbits have different sizes. 
Introducing the small quantity
\begin{align} \label{deltachi}
\delta \chi = \chi(0, \bar{x} ) - \chi(x, \bar{x}) \simeq - \frac{\Omega \delta \phi }{B} + \Omega^2 \bar{x} x  \text{,}
\end{align}
where I neglected the term proportional to $x^2$, the closed orbit density (\ref{ni-closed-expanded}) is dominated by type I closed orbits (which have $\chi_{\text{M}} (\bar{x}) = \chi (0, \bar{x})$), whose leading order density is given by
\begin{align} \label{ni-closed-expanded-lowest}
n_{\text{i,cl}} (x) \simeq  2\int_{\bar{x}_{\text{m,I}}}^{\infty} \Omega \sqrt{2 \delta \chi } d\bar{x} \int_{\chi_{\text{M}}}^{\infty} \frac{F_{\text{cl}} \left(\mu \left(\bar{x}, \chi_{\text{M}} \left(\bar{x}\right) \right), U \right) }{\sqrt{2\left( U - \chi_{\text{M}} (\bar{x}) \right) }} dU  \text{.}
\end{align}
The reason for neglecting the contribution to the density of type II closed orbits is that the contribution from ions with $x_{\text{M}} > 0$ is smaller, as shown explicitly in Appendix \ref{app-notypeIIclosed}.

I now obtain $n_{\text{i,op}} (x) - n_{\text{i,op}} (0)$ to leading order. 
Using equations (\ref{ni-open}) and (\ref{xbarm-open}), the perturbed open orbit density is
\begin{align} \label{ni-open-perturbed-1}
n_{\text{i,op}} (x) - n_{\text{i,op}} (0) & \simeq \int_{\bar{x}_{\text{c}}}^{\infty} \Omega d\bar{x} \int_{\chi_{\text{M}}(\bar{x})}^{\infty} \frac{ F_{\text{cl}} ( \mu, U ) }{\sqrt{2\left( U - \chi_{\text{M}} (\bar{x}) \right)}} \left[ \Delta v_x - \Delta v_{x0} \right] dU \nonumber \\
& - \int_{\bar{x}_{\text{c}}}^{\bar{x}_{\text{m,o}}(x)} \Omega d\bar{x} \int_{\chi_{\text{M}}(\bar{x})}^{\infty} \frac{ F_{\text{cl}} ( \mu, U ) }{\sqrt{2\left( U - \chi_{\text{M}} (\bar{x}) \right)}} \Delta v_{x0}  dU
\end{align}
where 
\begin{align} \label{Deltavx-Deltavx0}
\Delta v_x - \Delta v_{x0}  & =  \sqrt{ 2\left( \Delta_{\text{M}} (\bar{x}, U ) + \chi_{\text{M}} (\bar{x} ) - \chi (0, \bar{x} ) + \delta \chi \right) } \nonumber \\
& - \sqrt{2\left( \chi_{\text{M}} (\bar{x} ) - \chi (0, \bar{x} ) + \delta \chi \right)}  -  \sqrt{ 2\left( \Delta_{\text{M}} (\bar{x}, U ) + \chi_{\text{M}} (\bar{x} ) - \chi (0, \bar{x} ) \right) } \nonumber  \\
& + \sqrt{2\left( \chi_{\text{M}} (\bar{x} ) - \chi (0, \bar{x} ) \right)}  \text{.}
\end{align}
The second term in (\ref{ni-open-perturbed-1}) is zero if type II orbits are present ($x_{\text{c}} > 0$) because, from equation (\ref{xbarm-open}), $\bar{x}_{\text{m,o}}(x) = \bar{x}_{\text{c}}$ for $ x < x_{\text{c}} \neq 0$.
If no type II orbits are present ($x_{\text{c}} = 0$), equation (\ref{xbarm-open}) gives $\bar{x}_{\text{m,o}}(x) = \bar{x}_{\text{m}} (x)$ and, from equation (\ref{xbarm}), we expect the variation in $\bar{x}_{\text{m}} (x)$ to be linear in $x$ and $\delta \phi$.
For type I orbits, $\chi_{\text{M}} (\bar{x} ) = \chi (0, \bar{x} )$.
Then, the second term in equation (\ref{Deltavx-Deltavx0}) is of order $\sqrt{\delta \hat{\phi}}$, the fourth term is zero, and the first and third terms together cancel to lowest order leaving a piece of order $\delta \hat{\phi}$. 
For $\chi_{\text{M}} (\bar{x} ) > \chi (0, \bar{x} )$, no term as large as $\sqrt{\delta \hat{\phi}}$ appears.
Therefore, the dominant contribution to $\Delta v_x - \Delta v_{x0}$ is of order $\sqrt{\delta \hat{\phi}}$.
Type II open orbits have $\chi_{\text{M}}(\bar{x}) \neq \chi(0, \bar{x})$, and hence they contribute at most an order $\delta \hat{\phi}$ piece to $\Delta v_x - \Delta v_{x0}$%
\footnote{Some type II open orbits have $\chi_{\text{M}}(\bar{x}) - \chi(0, \bar{x}) \sim \delta \chi$, such that the second term in equation (\ref{Deltavx-Deltavx0}) is $\sqrt{ \chi''_{\text{M}}  x_{\text{M}}^2 + 2\delta \chi } \sim  \sqrt{\delta\hat{\phi}} v_{\text{t,i}}$. 
The range of values of $x_{\text{M}}$ for type II orbits which satisfy $\Delta v_x \sim \sqrt{\delta \hat{\phi}} v_{\text{t,i}}$ is small, $x_{\text{M}} \sim \sqrt{ \delta \hat{\phi } } \rho_{\text{i}} $.
Hence, integrating over such type II orbits results in a contribution to the density of the order of $\delta \hat{\phi} n_{\text{e} \infty}$.}%
.
The minimum value of $\bar{x}$ for which type I open orbits are present near $x=0$ is approximately $\bar{x}_{\text{m,I}}$, giving
\begin{align} \label{ni-open-near0-1/2}
n_{\text{i,op}} (x) - n_{\text{i,op}} (0) \simeq - \int_{\bar{x}_{\text{m,I}}}^{\infty} \sqrt{2 \delta \chi}  \Omega d\bar{x}  \int_{\chi_{\text{M}} (\bar{x} )}^{\infty}  \frac{ F_{\text{cl}}  \left( \mu \left( \bar{x}, \chi_{\text{M}}(\bar{x}) \right), U \right)  }{\sqrt{2\left( U - \chi_{\text{M}} (\bar{x}) \right)}} dU      \text{.}
\end{align}

From equation (\ref{ne-near0}), one sees that there is no term in the expansion of the electron density that has a size $\sqrt{\delta \hat{\phi}}$.
Hence, the dominant terms in the perturbed quasineutrality equation (\ref{quasi-perturbed}) for small $x$ are obtained by adding equations (\ref{ni-closed-expanded-lowest}) and (\ref{ni-open-near0-1/2}) and setting the sum to zero,
\begin{align} \label{quasi-near0-leading}
0 =  Z\int_{\bar{x}_{\text{m,I}}}^{\infty} \sqrt{2 \delta \chi}  \Omega d\bar{x}  \int_{\chi_{\text{M}} (\bar{x} )}^{\infty}  \frac{ F_{\text{cl}} \left( \mu \left( \bar{x}, \chi_{\text{M}}(\bar{x}) \right), U \right) }{\sqrt{2\left( U - \chi_{\text{M}}(\bar{x}) \right)}} dU  \text{.} 
\end{align}
The right hand side of equation (\ref{quasi-near0-leading}) vanishes only if $\bar{x}_{\text{m,I}} \rightarrow \infty$, which from equation (\ref{xbarmI}) implies a divergent electric field at $x=0$, $\phi'(0) \rightarrow \infty$.
The fact that $\bar{x}_{\text{m,I}} \rightarrow \infty$ means that only type II orbits are present in the magnetic presheath and $n_{\text{i,cl}}(x)$ is exponentially small, as argued in Appendix \ref{app-notypeIIclosed}. 
Therefore, I consider $n_{\text{i,cl}} (x) \simeq 0$ in equation (\ref{quasi-perturbed}) and focus on the perturbed open orbit density $n_{\text{i,op}}(x) - n_{\text{i,op}}(0)$.

With type I orbits absent, the effective potential maximum lies at $x_{\text{M}} \neq 0$, giving $\chi_{\text{M}}(\bar{x}) \neq \chi(0, \bar{x})$. 
Taking $\bar{x} \rightarrow \infty$ corresponds to $x_{\text{M}} \rightarrow 0$, leading to
\begin{align} \label{xbar-infty-limit}
\lim_{\bar{x} \rightarrow \infty} \chi_{\text{M}}(\bar{x}) = \chi(0, \bar{x}) \simeq \frac{1}{2} \Omega^2 \bar{x}^2  \text{.}
\end{align}
If the distribution function $F_{\text{cl}}$ decays exponentially at large energies, it is exponentially small in the region of the integral where $\chi_{\text{M}}(\bar{x}) - \chi(0, \bar{x}) \sim \delta \chi$ (which corresponds to $\bar{x}$ being large). This is because, according to equation (\ref{xbar-infty-limit}), $U_{\perp} \simeq \chi_{\text{M}}(\bar{x})$ is very large in that region. 
As a consequence, $\delta \chi \ll \chi_{\text{M}}(\bar{x}) - \chi(0, \bar{x}) $ for values of $\bar{x}$ where the distribution function is not exponentially small.
When $\delta \chi \ll \chi_{\text{M}} (\bar{x}) - \chi(0, \bar{x} ) $, both terms in equation (\ref{Deltavx-Deltavx0}) can be Taylor expanded to obtain
\begin{align} \label{Deltavx-expanded}
\Delta v_x - \Delta v_{x0} = - \Delta \left[\frac{1}{ v_{x0}} \right]  \delta \chi  + \frac{1}{2} \Delta \left[ \frac{1}{v_{x0}^3} \right]  \delta \chi^2 \text{,}
\end{align}
where I introduced the positive quantities
\begin{align}
\Delta \left[ \frac{1}{v_{x0}} \right]  =  \frac{1}{ \sqrt{2\left( \chi_{\text{M}} (\bar{x} ) - \chi (0, \bar{x} )  \right) } }  - \frac{1}{  \sqrt{ 2\left( \Delta_{\text{M}} (\bar{x}, U ) + \chi_{\text{M}} (\bar{x} ) - \chi (0, \bar{x} ) \right) }} \text{,}
\end{align}
and 
\begin{align}
\Delta \left[ \frac{1}{v_{x0}^3} \right]  =  \frac{1}{ \left[ 2\left( \chi_{\text{M}} (\bar{x} ) - \chi (0, \bar{x} ) \right) \right]^{3/2} } - \frac{1}{ \left[ 2\left( \Delta_{\text{M}} (\bar{x}, U ) + \chi_{\text{M}} (\bar{x} ) - \chi (0, \bar{x} ) \right) \right]^{3/2} }\text{.}
\end{align} 
I expand the open orbit density (\ref{ni-open}) using equation (\ref{Deltavx-expanded}) for the expansion of $\Delta v_x$, obtaining
\begin{align} \label{ni-open-perturbed}
n_{\text{i,op}} (x) - n_{\text{i,op}} (0) \simeq  & - \int_{\bar{x}_{\text{c}}}^{\infty}  \delta \chi  \Omega d\bar{x}  \int_{\chi_{\text{M}} (\bar{x} )}^{\infty}  \frac{ F_{\text{cl}} ( \mu(\bar{x}, \chi_{\text{M}}), U ) }{\sqrt{2\left( U - \chi_{\text{M}} (\bar{x}) \right)}} \Delta \left[ \frac{1}{v_{x0}} \right]  dU \nonumber \\ 
& + \frac{1}{2}\int_{\bar{x}_{\text{c}}}^{\infty}  \delta \chi^2  \Omega d\bar{x}  \int_{\chi_{\text{M}} (\bar{x} )}^{\infty}  \frac{ F_{\text{cl}} ( \mu(\bar{x}, \chi_{\text{M}}), U ) }{\sqrt{2\left( U - \chi_{\text{M}} (\bar{x}) \right)}} \Delta \left[ \frac{1}{v_{x0}^3} \right]  dU
\text{.}  
\end{align}

Expanding the electron density (\ref{ne-near0}), I get 
\begin{align} \label{ne-perturbed}
n_{\text{e}} (x) - n_{\text{e}0} \simeq \frac{e\delta \phi}{T_{\text{e}}} + \frac{1}{2} \left( \frac{e\delta \phi}{T_{\text{e}}}  \right)^2 \text{.}
\end{align}
The perturbed quasineutrality equation (\ref{quasi-perturbed}), to order $\delta \hat{\phi}$, then implies that
\begin{align} \label{quasi-near0-1}
n_{\text{e}0} \frac{e\delta \phi }{T_{\text{e}}} =   \frac{\Omega \delta \phi }{B}Z\int_{\bar{x}_{\text{c}}}^{\infty}  \Omega d\bar{x}  \int_{\chi_{\text{M}} (\bar{x} )}^{\infty}  \frac{ F_{\text{cl}} \left( \mu\left(\bar{x}, \chi_{\text{M}}(\bar{x}) \right), U \right)  }{\sqrt{2\left( U - \chi_{\text{M}}(\bar{x})\right)}} \Delta \left[ \frac{1}{v_x} \right]  dU \nonumber \\
 -  x  \Omega Z\int_{\bar{x}_{\text{c}}}^{\infty} \Omega^2 \bar{x}   d\bar{x}  \int_{\chi_{\text{M}} (\bar{x} )}^{\infty}  \frac{ F_{\text{cl}} \left( \mu\left(\bar{x}, \chi_{\text{M}}(\bar{x}) \right), U \right) }{\sqrt{2\left( U - \chi_{\text{M}}(\bar{x})\right)}} \Delta \left[ \frac{1}{v_x} \right]  dU  \text{.} 
\end{align}
This can be rearranged to obtain
\begin{align} \label{deltaphi-1}
 \delta \phi = \phi(x) - \phi(0) = \frac{x}{q_1} \text{,}
\end{align}
where $q_1$ is given by
\begin{align} 
q_1 = \frac{e}{\Omega T_{\text{e}}} \frac{  Zv_{\text{B}}^2\int_{\bar{x}_{\text{c}}}^{\infty}   \Omega d\bar{x}  \int_{\chi_{\text{M}} (\bar{x} )}^{\infty}  \frac{ F_{\text{cl}} \left( \mu\left(\bar{x}, \chi_{\text{M}}(\bar{x}) \right), U \right)  }{\sqrt{2\left( U - \chi_{\text{M}}(\bar{x})\right)}} \Delta \left[ \frac{1}{v_x} \right]  dU -  n_{e0}  }{   Z\int_{\bar{x}_{\text{c}} }^{\infty}  \Omega^2 \bar{x} d\bar{x}  \int_{\chi_{\text{M}} (\bar{x} )}^{\infty}  \frac{ F_{\text{cl}} \left( \mu\left(\bar{x}, \chi_{\text{M}}(\bar{x}) \right), U \right)  }{\sqrt{2\left( U - \chi_{\text{M}}(\bar{x})\right)}} \Delta \left[ \frac{1}{v_x} \right] dU  } \text{.}
\end{align}
Equation (\ref{deltaphi-1}) implies that $\phi'(0) = q_1^{-1}$. The magnetic presheath \emph{is driven} towards $q_1 = 0$ because $\phi'(0) \rightarrow \infty$ is required from equation (\ref{quasi-near0-leading}) and the discussion following it. 
Hence, the numerator of $q_1$ must be zero,
\begin{equation} \label{Bohm-not-obvious}
 Zv_{\text{B}}^2\int_{\bar{x}_{\text{c}}}^{\infty}   \Omega d\bar{x}  \int_{\chi_{\text{M}} (\bar{x} )}^{\infty}  \frac{ F_{\text{cl}} ( \mu (\bar{x}, \chi_{\text{M}} ), U ) }{\sqrt{2\left( U - \chi_{\text{M}}\right)}} \Delta \left[ \frac{1}{v_{x0}} \right]  dU = n_{e0}  \text{.} 
\end{equation} 

I proceed to show that equation (\ref{Bohm-not-obvious}) is equivalent to the marginal form of the kinetic Bohm condition \cite{Harrison-Thompson-1959, Cavaliere-1965, Riemann-review},
\begin{align} \label{Bohm-kinetic-marginal}
Zv_{\text{B}}^2 \int \frac{f_0 ( \vec{v} ) }{v_{x}^2}  d^3v  =  n_{e0}  \text{.}
\end{align}
From (\ref{fopen}), the distribution function at $x=0$ is
\begin{align} \label{f0}
f_{0} (\vec{v})  & = f_{\text{open}} (0, \vec{v})   \nonumber  \\
& \simeq  F_{\text{cl}} \left( \mu \left( \bar{x}, \chi_{\text{M}} \left( \bar{x} \right) \right) , U \right)   \hat{\Pi} \left(  v_x ,  - V_x \left( 0, \bar{x}, \chi_{\text{M}} \right) - \Delta v_{x0}  , - V_x \left( 0, \bar{x}, \chi_{\text{M}} \right)  \right) \text{.}
\end{align}
Using the definition (\ref{f0}) and the change of variables $(\bar{x}, U) \rightarrow (v_y, v_z)$ (equations (\ref{xbar-def}) and (\ref{U-open})) at $x=0$, I can re-express the integral in (\ref{Bohm-not-obvious}) to obtain 
\begin{align} 
 \int_{\bar{x}_{\text{c}}}^{\infty}   \Omega d\bar{x} &  \int_{\chi_{\text{M}} (\bar{x} )}^{\infty}  \frac{  F_{\text{cl}} \left( \mu \left( \bar{x}, \chi_{\text{M}} \left( \bar{x} \right) \right) , U \right)   }{\sqrt{2\left( U - \chi_{\text{M}} (\bar{x}) \right)}} \Delta \left[ \frac{1}{v_{x0}} \right]  dU \nonumber \\
& = \int_{\bar{x}_{\text{c}}}^{\infty}   \Omega d\bar{x}  \int_{\chi_{\text{M}} (\bar{x} )}^{\infty}  \frac{ F_{\text{cl}} \left( \mu \left( \bar{x}, \chi_{\text{M}} \left( \bar{x} \right) \right) , U \right) }{\sqrt{2\left( U - \chi_{\text{M}} (\bar{x}) \right)}} dU \nonumber \\
& ~ \times \int_{-\infty}^{0}  \frac{1}{v_x^2} \hat{\Pi} \left(  v_x ,  - V_x \left( 0, \bar{x}, \chi_{\text{M}} \right) - \Delta v_{x0}  , - V_x \left( 0, \bar{x}, \chi_{\text{M}} \right)  \right) dv_x \nonumber \\
 & = \int \frac{f_0 ( \vec{v} ) }{v_{x}^2}  d^3v  \text{.}
\end{align}
This shows that equations (\ref{Bohm-not-obvious}) and (\ref{Bohm-kinetic-marginal}) are equivalent. 
Hence, the system is driven to marginally satisfying the kinetic Bohm condition (\ref{Bohm-kinetic-marginal}).

Because $q_1=0$, I must consider terms of size $ \sim \delta \hat{ \phi}^2$ in equation (\ref{quasi-perturbed}).
Using equations (\ref{ni-open-perturbed}) and (\ref{ne-perturbed}), I obtain
\begin{align} \label{quasi-near0-2}
\frac{1}{2} n_{\text{e}0} \left( \frac{e\delta \phi }{T_{\text{e}}} \right)^2 & =  - Zx\int_{\bar{x}_{\text{c}}}^{\infty}  \Omega^3 \bar{x} d\bar{x}  \int_{\chi_{\text{M}} (\bar{x} )}^{\infty}  \frac{ F_{\text{cl}} \left( \mu \left( \bar{x}, \chi_{\text{M}}(\bar{x}) \right), U \right) }{\sqrt{2\left( U - \chi_{\text{M}}(\bar{x})\right)}} \Delta \left[ \frac{1}{v_x} \right] dU \\
& +   \frac{Z\Omega^2 \delta \phi^2 }{2B^2}  \int_{\bar{x}_{\text{c}}}^{\infty}   \Omega d\bar{x}  \int_{\chi_{\text{M}} (\bar{x} )}^{\infty}  \frac{ F_{\text{cl}} \left( \mu \left( \bar{x}, \chi_{\text{M}}(\bar{x}) \right), U \right) }{\sqrt{2\left( U - \chi_{\text{M}} (\bar{x}) \right)}} \Delta \left[ \frac{1}{v_x^{3}} \right] dU  \text{.} 
\end{align}
This leads to
\begin{align} \label{phi-near0-2}
 \delta \phi = \phi ( x ) - \phi ( 0 )  = q_{2}^{-1/2}  x^{1/2} \text{,}
\end{align}
where 
\begin{align} \label{q2-def}
q_{2} = \frac{1}{2} \left( \frac{e}{T_{\text{e}}} \right)^2  \frac{ Z v_{\text{B}}^4 \int_{\bar{x}_{\text{c}}}^{\infty}  \Omega d\bar{x} \int_{\chi_{\text{M}}(\bar{x})}^{\infty} \frac{F_{\text{cl}} \left( \mu \left( \bar{x}, \chi_{\text{M}}(\bar{x}) \right), U \right) }{\sqrt{ 2\left( U - \chi_{\text{M}} ( \bar{x} ) \right)  } } dU \Delta \left[ \frac{1}{v_x^{3}}\right] - n_{\text{e}0}  }{ Z\int_{\bar{x}_{\text{c}}}^{\infty}  \Omega^3 \bar{x} d\bar{x}  \int_{\chi_{\text{M}} (\bar{x} )}^{\infty}  \frac{ F_{\text{cl}} \left( \mu \left( \bar{x}, \chi_{\text{M}}(\bar{x}) \right), U \right) }{\sqrt{2\left( U - \chi_{\text{M}}(\bar{x})\right)}} \Delta \left[ \frac{1}{v_x} \right] dU   } > 0 \text{.}
\end{align}
In Appendix \ref{app-k2>0}, I show that $q_{2}$ is always positive and never small because equation (\ref{Bohm-kinetic-marginal}) is satisfied. Therefore, equation (\ref{phi-near0-2}) is the scaling of the electrostatic potential I expect to observe in my numerical results.

\section{Numerical method} \label{sec-KMPS-nummethod}

I discretize the potential on a grid $x_{\eta}$ (labelled by the index $\eta$)
\begin{align} \label{x-grid}
\frac{x_{\eta}}{\rho_{\text{i}}} = \begin{cases}
\left( 0.05\eta \right)^2 &  \text{ for } 0 \leqslant \eta < 10 \text{,} \\
 0.25 +  0.1\left( \eta - 10 \right) &  \text{ for } 10 \leqslant \eta < \eta_2 = 129 \text{.} 
\end{cases}
\end{align}
I numerically calculate the ion density profile $n_{\text{i}}\left( x_{\eta} \right)$ in the region $0 \leqslant x_{\eta} \leqslant x_{\eta_1} = 6.15 \rho_{\text{i}} $ ($\eta_1 = 69$). 
The domain in $x$ is larger than $[0, x_{\eta_1}]$ because the potential profile in the region $x_{\eta_1} < x \leqslant x_{\eta_2} = 12.15 \rho_{\text{i}}$ is necessary to correctly evaluate the ion density at $ x_{\eta_1} $ and in its neighbourhood. 
The electron density profile $n_{\text{e}} \left( x_{\eta} \right)$ is evaluated by inserting $\phi \left( x_{\eta} \right)$ into equation (\ref{ne}). 
I iterate over electrostatic potential functions $\phi_{\nu} \left( x_{\eta} \right)$, where $\nu$ is an index labelling the iteration number. 
The problem of solving (\ref{quasineutrality}) is equivalent to finding, after $N$ iterations, a $\phi_N \left( x_{\eta}\right)$ for which $n_{\text{e,}N}\left( x_{\eta} \right) \simeq n_{\text{i,}N} \left( x_{\eta} \right)$ in the region $0 \leqslant x \leqslant x_{\eta_1}$. 

\begin{figure}[h] 
\centering
\includegraphics[width=0.47\textwidth]{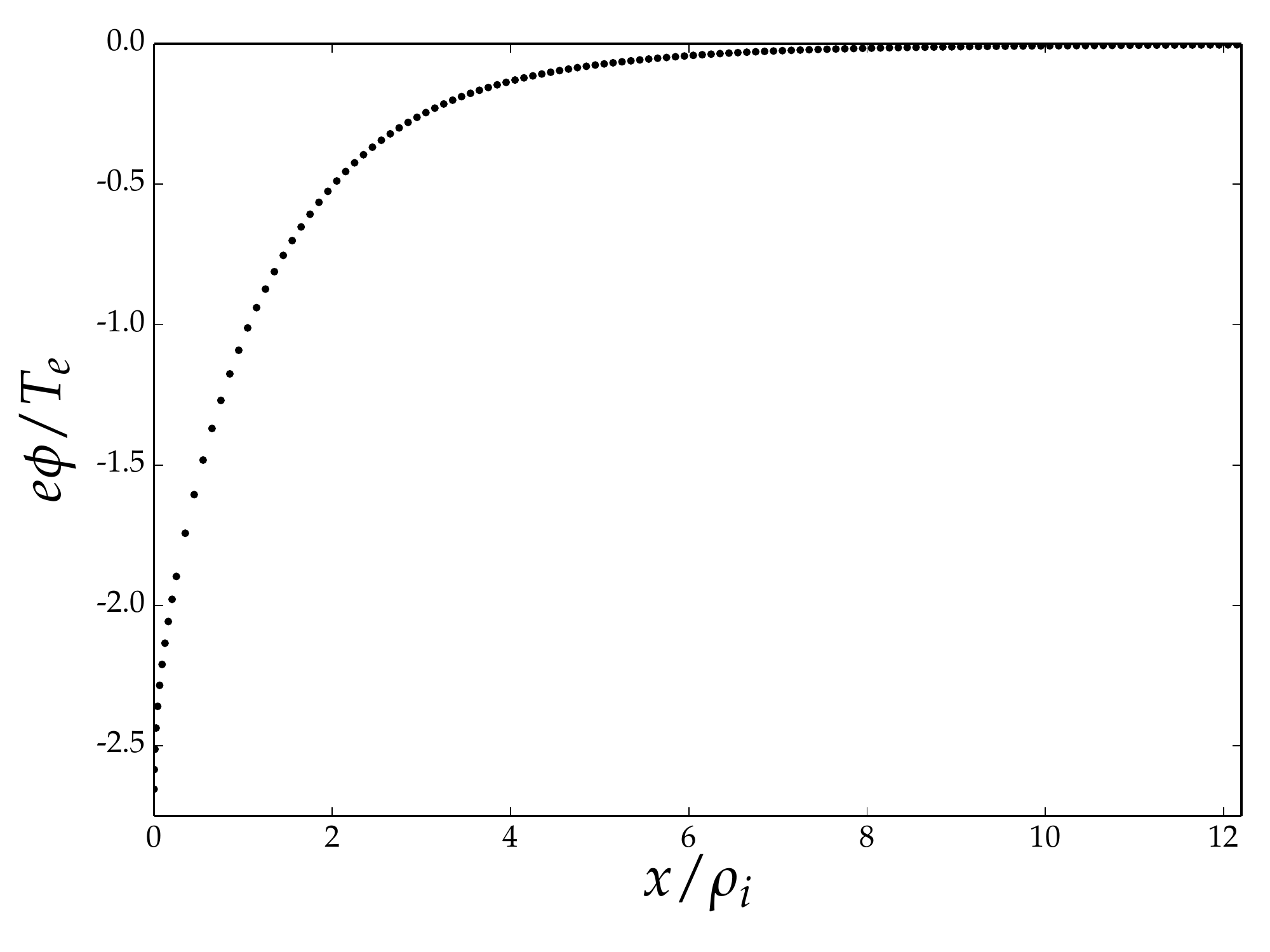}
\includegraphics[width=0.47\textwidth]{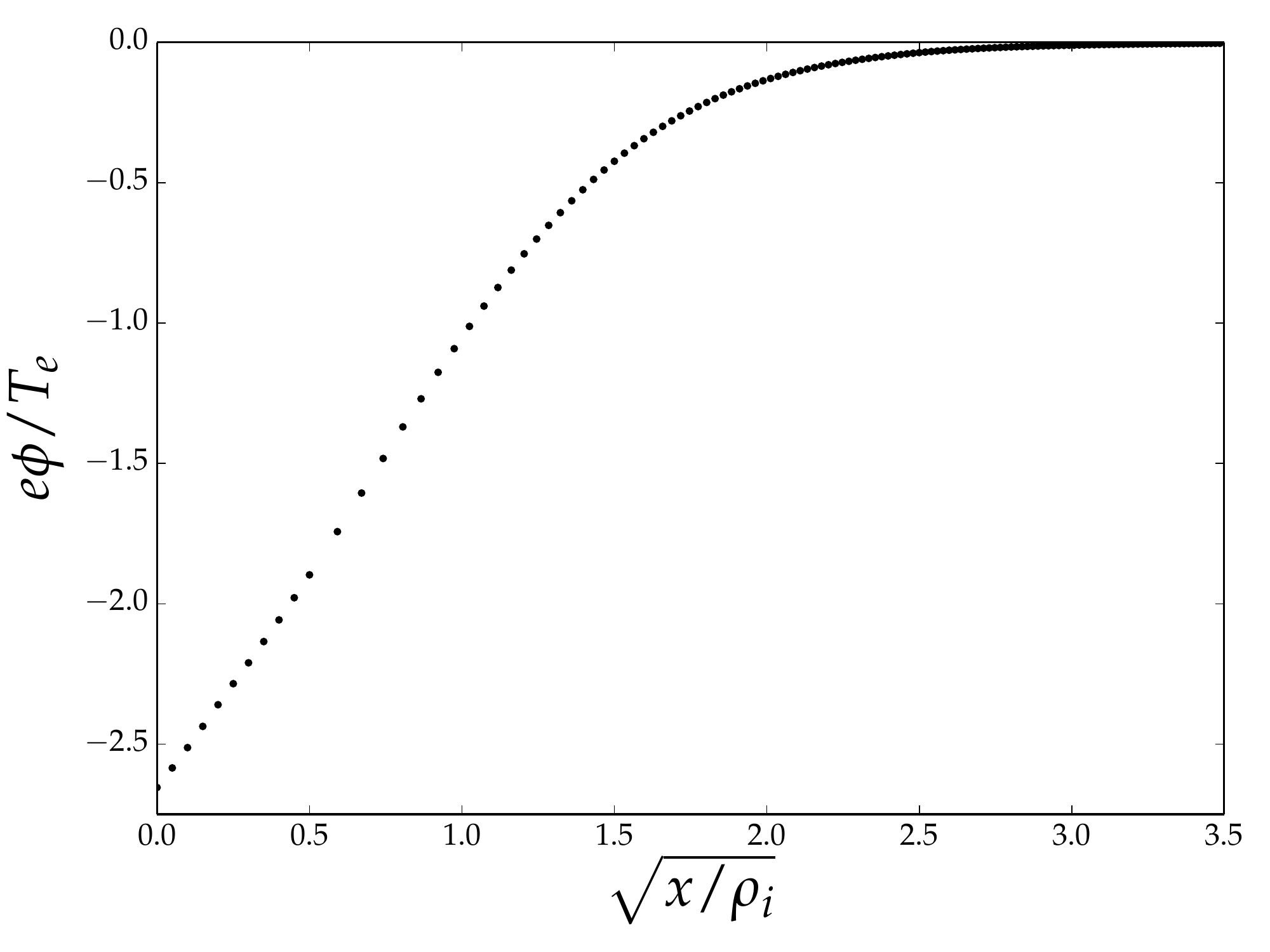}
\caption[Numerical grid of distance from the wall]{An example solution for the electrostatic potential profile (for $\alpha = 0.05$) is plotted on the grid of equation (\ref{x-grid}). Initially $\phi$ increases linearly with $\sqrt{x}$, which justifies my grid choice. }
\label{fig-phires}
\end{figure}

Near $x=0$, the grid (\ref{x-grid}) that I use to discretize all functions of $x$ has evenly spaced values of $\sqrt{x/\rho_{\text{i}}}$ ranging from $0$ to $0.5$ in intervals of $0.05$. The reason for this is that the self-consistent solution of the electrostatic potential is expected to be proportional to $\sqrt{x}$ near $x=0$, as in equation (\ref{phi-near0-2}). This behaviour of the electrostatic potential is captured by my grid as shown in Figure \ref{fig-phires}. For $\sqrt{x/\rho_{\text{i}}}>0.5$, corresponding to $x/\rho_{\text{i}}>0.25$, my grid has evenly spaced values of $x/\rho_{\text{i}}$, ranging from $0.25$ to $12.15$ in intervals of $0.1$.

The density integrals in equations (\ref{ni-closed}) and (\ref{ni-open}) are evaluated numerically at every point $x_{\eta}$ by employing the trapezoidal rule. 
In order to evaluate those integrals, I first evaluate the integrands. I introduce a grid of positions $\bar{x}_{\gamma}$ (labelled with the index $\gamma$),
\begin{align} \label{xbar-grid}
\frac{\bar{x}_{\gamma}}{\rho_{\text{i}}} = 0.01\gamma  \text{ for } 0 \leqslant \gamma < 1200 \text{.} 
\end{align}
Then, I evaluate the function $\chi \left( x_{\eta}, \bar{x}_{\gamma} \right)$ at all possible values of $x_{\eta}$ and $\bar{x}_{\gamma}$. I find the location of the effective potential maximum $x_{\text{M}}$ corresponding to the index $\eta_{\text{M}} \left(\gamma \right)$ that satisfies either
\begin{align} \label{xmax-typeI-numerical}
\chi \left( x_{\eta_{\text{M}} \left(\gamma \right)} , \bar{x}_{\gamma} \right) > \chi \left( x_{\eta_{\text{M}} \left(\gamma \right) + 1} , \bar{x}_{\gamma} \right) & \text{ for } \eta_{\text{M}} \left(\gamma \right) = 0  \text{ (type I) }
\end{align}
or
\begin{align}  \label{xmax-typeII-numerical}
\chi \left( x_{\eta_{\text{M}} \left(\gamma \right)} , \bar{x}_{\gamma} \right) > \chi \left( x_{\eta_{\text{M}} \left(\gamma \right) - 1} , \bar{x}_{\gamma} \right) \nonumber  \\ \text{ and }
\chi \left( x_{\eta_{\text{M}} \left(\gamma \right)} , \bar{x}_{\gamma} \right) > \chi \left( x_{\eta_{\text{M}} \left(\gamma \right) + 1} , \bar{x}_{\gamma} \right) & \text{ for } \eta_{\text{M}} \left(\gamma \right) \geqslant 1 \text{ (type II).}
\end{align}
I also find the location of the effective potential minimum $x_{\text{m}}$ corresponding to the index $\eta_{\text{m}} \left(\gamma \right)$ that satisfies
\begin{align} \label{xmin-numerical}
\chi \left( x_{\eta_{\text{m}} \left(\gamma \right)} , \bar{x}_{\gamma} \right) < \chi \left( x_{\eta_{\text{M}} \left(\gamma \right) - 1} , \bar{x}_{\gamma} \right) \nonumber \\ \text{ and }
\chi \left( x_{\eta_{\text{m}} \left(\gamma \right)} , \bar{x}_{\gamma} \right) < \chi \left( x_{\eta_{\text{M}} \left(\gamma \right) + 1} , \bar{x}_{\gamma} \right) & \text{ for } \eta_{\text{m}} \left(\gamma \right) \geqslant 1 \text{.}
\end{align}

\begin{figure}[h] 
\centering
\includegraphics[width=0.6\textwidth]{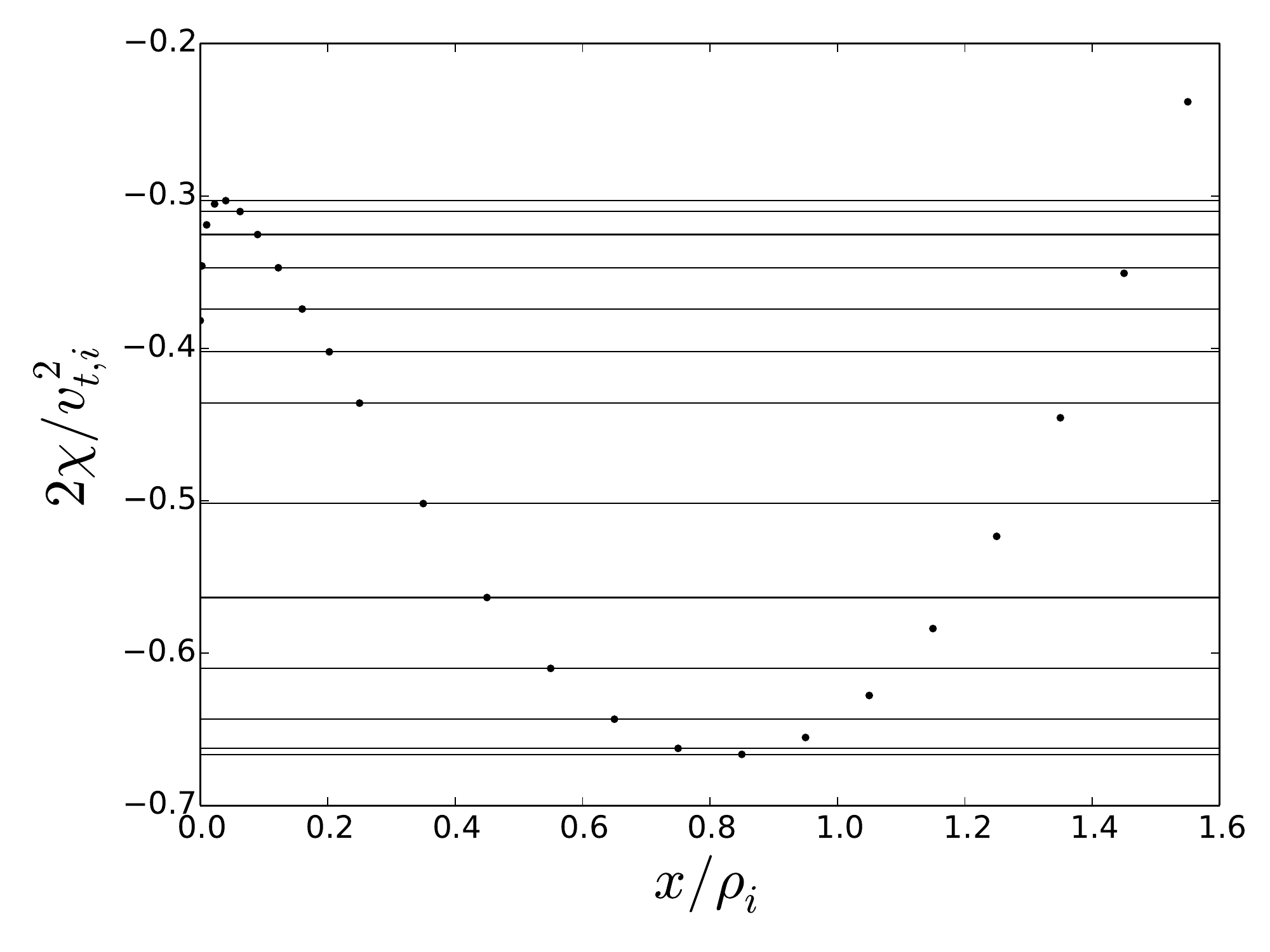}
\caption[Numerical grid of energy]{ The values of $U_{\perp, \gamma \kappa}$ corresponding to different values of $\kappa$ are shown with horizontal lines on top of the effective potential curve $\chi \left( x_{\eta}, \bar{x}_{\gamma} \right)$, for a particular value of $\gamma$. Here, $\kappa$ ranges from $\kappa = 0$ (top line) to $\kappa = 12$ (bottom line). }
\label{fig-Uperpgrid}
\end{figure}

At every value of the orbit parameter $\bar{x}_{\gamma}$, I obtain a grid of possible values of perpendicular energy $U_{\perp, \gamma \kappa}$, indexed with $\gamma$ and $\kappa$, 
\begin{align} \label{Uperp-grid}
U_{\perp, \gamma \kappa} = \chi \left( x_{\kappa + \eta_{\text{M}} \left(\gamma \right) }, \bar{x}_{\gamma} \right) \text{ for } 0 \leqslant \kappa \leqslant   \eta_{\text{m}} \left(\gamma \right) -  \eta_{\text{M}} \left(\gamma \right) \text{.}
\end{align}
This grid is shown in Figure \ref{fig-Uperpgrid}. For all possible $\bar{x}_{\gamma}$ and $U_{\perp,\gamma \kappa}$, I evaluate the adiabatic invariant $\mu \left( \bar{x}_{\gamma}, U_{\perp, \gamma \kappa} \right) $ by performing the integral (\ref{mu-Uperp-xbar}) using the trapezoidal rule. 
Similarly, for all possible values of $\bar{x}_{\gamma}$, I evaluate the integral $I \left( \bar{x}_{\gamma} \right)$ in (\ref{open-integral}) using the trapezoidal rule\footnote{Taking the derivative of $\mu  \left( \bar{x}_{\gamma}, \chi_{\text{M}} (\bar{x}_{\gamma}) \right) $, as in the last equality of (\ref{open-integral}), is faster and gives the same result as the integral (to within a small numerical error).}. 
For all values of $\gamma$ and $\kappa$, the total energy is labelled by the index $\iota$,
\begin{align} \label{U-grid}
\frac{2U_{\gamma\kappa\iota}}{v_{\text{t,i}}^2} = \frac{2U_{\perp, \gamma \kappa}}{v_{\text{t,i}}^2} + (0.2\iota)^2 \text{ for } 0 \leqslant \iota < \iota_{\text{max}} \text{,}
\end{align}
where $\iota_{\text{max}}$ is such that $2U/v_{\text{t,i}}^2 < 15.0$ and $7.5 v_{\text{t,i}}^2$ is a cutoff energy above which the distribution function is essentially zero. 
The distribution function $F_{\text{cl}} (\mu, U )$ must be defined on a grid of values of $2\Omega \mu /v_{\text{t,i}}^2$ and $2U/v_{\text{t,i}}^2$.
This is then bilinearly interpolated at every integration point.
The integrals over $U$ and over $U_{\perp}$ in equations (\ref{ni-closed}) and (\ref{ni-open}) are, for numerical convenience, evaluated over $v_z = \sqrt{2\left( U_{\gamma\kappa\iota} - U_{\perp, \gamma \kappa} \right)} $ and $|v_x | = \sqrt{2\left( U_{\perp, \gamma \kappa} - \chi (x_{\eta}, \bar{x}_{\gamma} )\right) }$, respectively (for this reason $U_{\gamma\kappa\iota}$ is defined such that linear increments in $\iota$ correspond to linear increments in $v_z$). Where necessary, the values of the integrands and of the integration limits of equations (\ref{mu-Uperp-xbar}), (\ref{ni-closed}), (\ref{ni-open}) and (\ref{open-integral}) are found by linear interpolation. 

The iteration scheme to find $\phi(x)$ hinges on imposing
 \begin{align} \label{iterationscheme}
 n_{\text{e,}\nu+1} \left( x_{\eta} \right) = wZn_{\text{i,}\nu} \left( x_{\eta} \right) + \left(1-w\right)n_{\text{e,}\nu} \left( x_{\eta} \right) \text{}
 \end{align}
 at every $(\nu+1)th$ iteration. Here, $w$ is a weight whose value lies in the range $0 < w \leqslant 1$. From (\ref{iterationscheme}), $\phi_{\nu+1} \left( x_{\eta} \right)$ is obtained by inverting the Boltzmann relation for $n_{e,\nu+1} \left( x_{\eta} \right)$, and the new guess for the potential profile is thus obtained for $0 \leqslant \eta \leqslant \eta_1$. 
 For values of $\eta$ in the interval $\eta_1 + 1 \leqslant \eta \leqslant \eta_2$, the electrostatic potential $\phi_{\nu+1}\left(x_{\eta} \right)$ is completed by matching to the appropriate functional form for $\phi \left( x \right)$ at $x \rightarrow \infty$. 
With my choice of distribution function in section \ref{sec-KMPS-results} marginally satisfying the Chodura condition (\ref{solvability}), $\phi(x)$ satisfies equation (\ref{phisol1}) for large $x$. 
The value of $k_{3/2}$ is calculated numerically.
The value of $C_{3/2}$ is obtained by imposing $\phi_{\nu+1} \left( x_{\eta_1} \right) = - 400 k_{3/2}^{-2} (x_{\eta_1}+C_{3/2})^{-4}  $  to get
\begin{align}
C_{3/2} = \sqrt{\frac{20}{k_{3/2}}}  \left[ - \phi_{\nu+1} \left( x_{\eta_1} \right) \right]^{-1/4} - x_{\eta_1} \text{.}
\end{align} 
The new guess for the electrostatic potential is then
\begin{align}
\phi_{\nu+1} \left( x_{\eta} \right) = \begin{cases} \frac{T_{\text{e}}}{  e} \ln \left( w \frac{ Zn_{i\nu} \left( x_{\eta} \right) }{  n_{\text{e}\infty} } + \left(1-w\right) \frac{ n_{e\nu} \left( x_{\eta} \right) }{ n_{\text{e}\infty} } \right) & \text{for } 0 \leqslant \eta \leqslant \eta_1 \text{,} \\ - \frac{ 400 }{  k_{3/2}^2  (x_{\eta}+C_{3/2})^4 }  & \text{for } \eta_1 + 1 \leqslant \eta \leqslant \eta_2 \text{.}
\end{cases} 
\end{align} 
This can be used to evaluate $n_{\text{i,}\nu+1}(x_{\eta})$ in the region $0 \leqslant \eta \leqslant \eta_1$ and continue the iteration. The first potential guess I use is a flat potential profile ($\phi_0 (x_{\eta}) = 0$ for all $\eta$). After $N$ iterations, a numerical solution $\phi_N \left(x_{\eta}\right)$ which satisfies $n_{\text{e,}N} (x_{\eta} ) \simeq n_{\text{i,}N} ( x_{\eta} )$ for all $\eta$ is found. The deviation of $\phi_{\nu} \left( x_{\eta} \right)$ from the exact solution (which satisfies $n_{\text{i}}\left( x_{\eta} \right) = n_{\text{e}}\left( x_{\eta} \right)$) is measured by calculating the quantity
\begin{align}
\tilde{n}_{\nu} \left( x_{\eta} \right) = 1 - \frac{n_{\text{i,}\nu} \left( x_{\eta} \right)}{n_{\text{e,}\nu} \left( x_{\eta} \right)} \text{.}
\end{align} 
Convergence to an acceptable solution is given by the criterion that the root mean square value of $\tilde{n}_{\nu} \left( x_{\eta} \right)$ be less than some error, denoted $\mathcal{E}_{n}$,
\begin{align}\label{convergence-criterion}
\left[  \sum_{\eta=0}^{\eta_1} \frac{1}{\eta_1 + 1} \tilde{n}_{\nu}^2  \left( x_{\eta} \right)  \right]^{1/2} < \mathcal{E}_{n} \text{.}
\end{align} 
For the numerical study of section \ref{sec-KMPS-results}, $\mathcal{E}_{n}$ was set to $0.007$.

\begin{figure}[h] 
\centering
\includegraphics[width=0.6\textwidth]{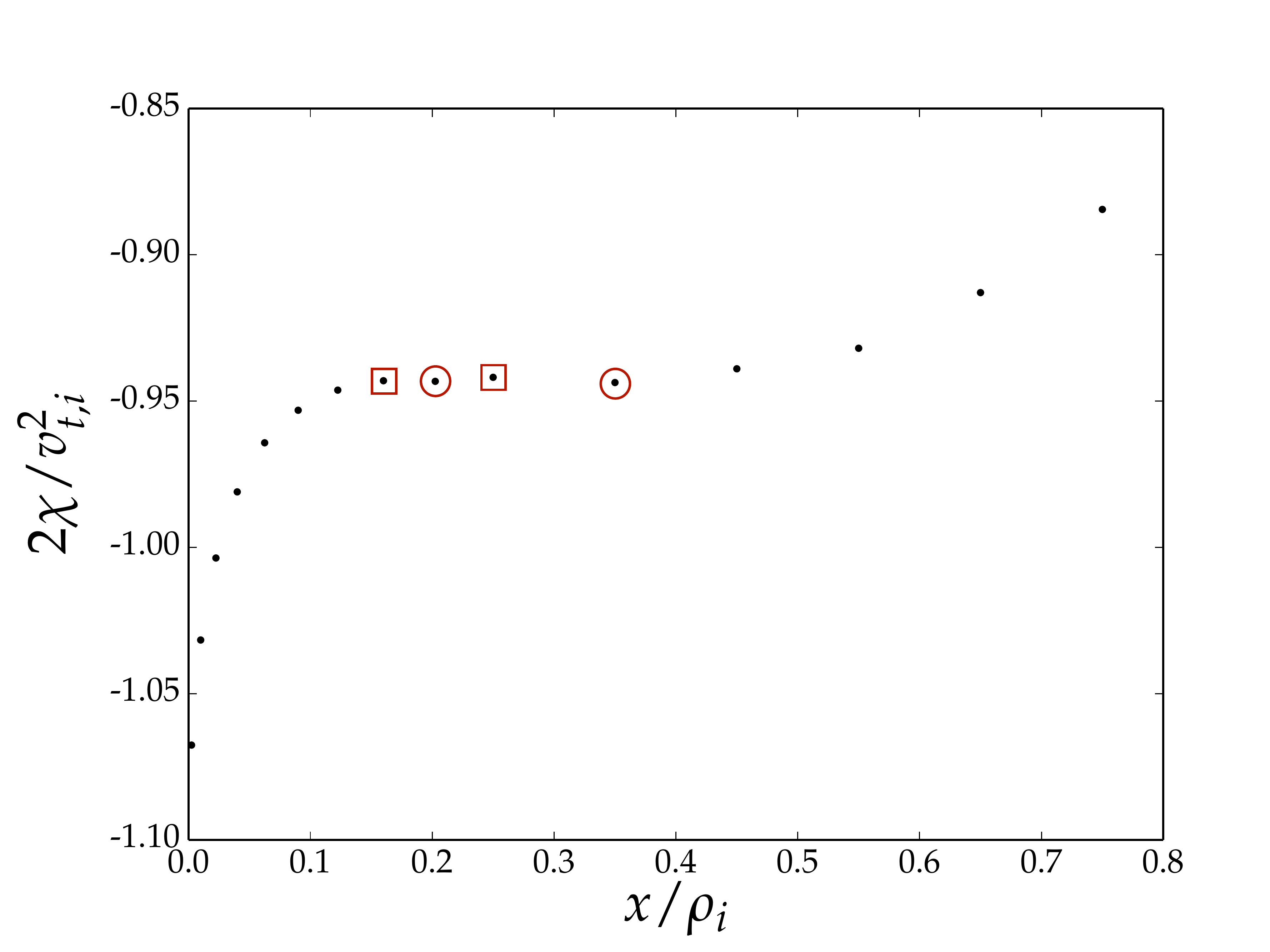}
\caption[Example of failure of the numerical method]{An example of an effective potential $\chi \left( x_{\eta}, \bar{x}_{\gamma} \right) $ in which my algorithm for generating the grid $U_{\perp, \gamma \kappa}$ fails, because it does not take into account the possibility of finding multiple effective potential minima (marked with circles) and maxima (marked with squares) for a given $\gamma$. 
}
\label{fig-effpot-prob}
\end{figure}

The method that I use can give a non-smooth numerical second derivative of the potential $\phi_{\nu} \left( x_{\eta} \right)$.
The numerical noise in the second derivative is problematic because the algorithm fails to take into account the possibility of more than one maximum or minimum of the effective potential existing for some value of $\bar{x}$. 
If at some point during the iteration the function $\phi_{\nu} \left( x_{\eta} \right)$ is such that, for some value of $\gamma$, the function $\chi ( x_{\eta}, \bar{x}_{\gamma} )$ has more than one index $\eta_{\text{M}} (\gamma)$ that satisfies either (\ref{xmax-typeI-numerical}) or (\ref{xmax-typeII-numerical}) (and more than one index $\eta_{\text{m}} (\gamma)$ that satisfies (\ref{xmin-numerical})), a more sophisticated analysis than the one I presented is necessary to obtain the grid of values of $U_{\perp}$.
The appearance of multiple maxima and minima, shown in Figure \ref{fig-effpot-prob}, is due to the numerical second derivative of $\phi \left( x_{\eta} \right)$ having pronounced oscillations, even when $\phi \left( x_{\eta} \right)$ looks smooth to the naked eye.
To avoid the appearance of multiple maxima and minima, in this work I perform a smoothing operation on the second derivative of $\phi_{\nu} \left( x \right)$ (with respect to $\sqrt{x}$) before iteration number $\nu+1$, for a certain number of iterations until the densities obtained using $\phi_{\nu} (x)$ are close to satisfying criterion (\ref{convergence-criterion}). After that, I carry out the last few iterations without smoothing. 
In my iterations, $w=0.6$ when the smoothing operation is performed, while $w=0.2$ when it is not.

The computing time necessary to obtain the solutions that I present in the next section is small.
The number of iterations required for convergence is typically less than 20, and each iteration runs in approximately 3 seconds on a laptop. 
Consequently, the total run time of the code on a laptop is typically less than one minute.
The computing time can be further reduced by using a better initial guess, improving the integration schemes and reducing the number of integration points.
 
From here on, I omit all indices associated with quantities and functions evaluated numerically.

\section{Numerical results}
\label{sec-KMPS-results}

\begin{figure}[h] 
\centering
\includegraphics[width=0.7\textwidth]{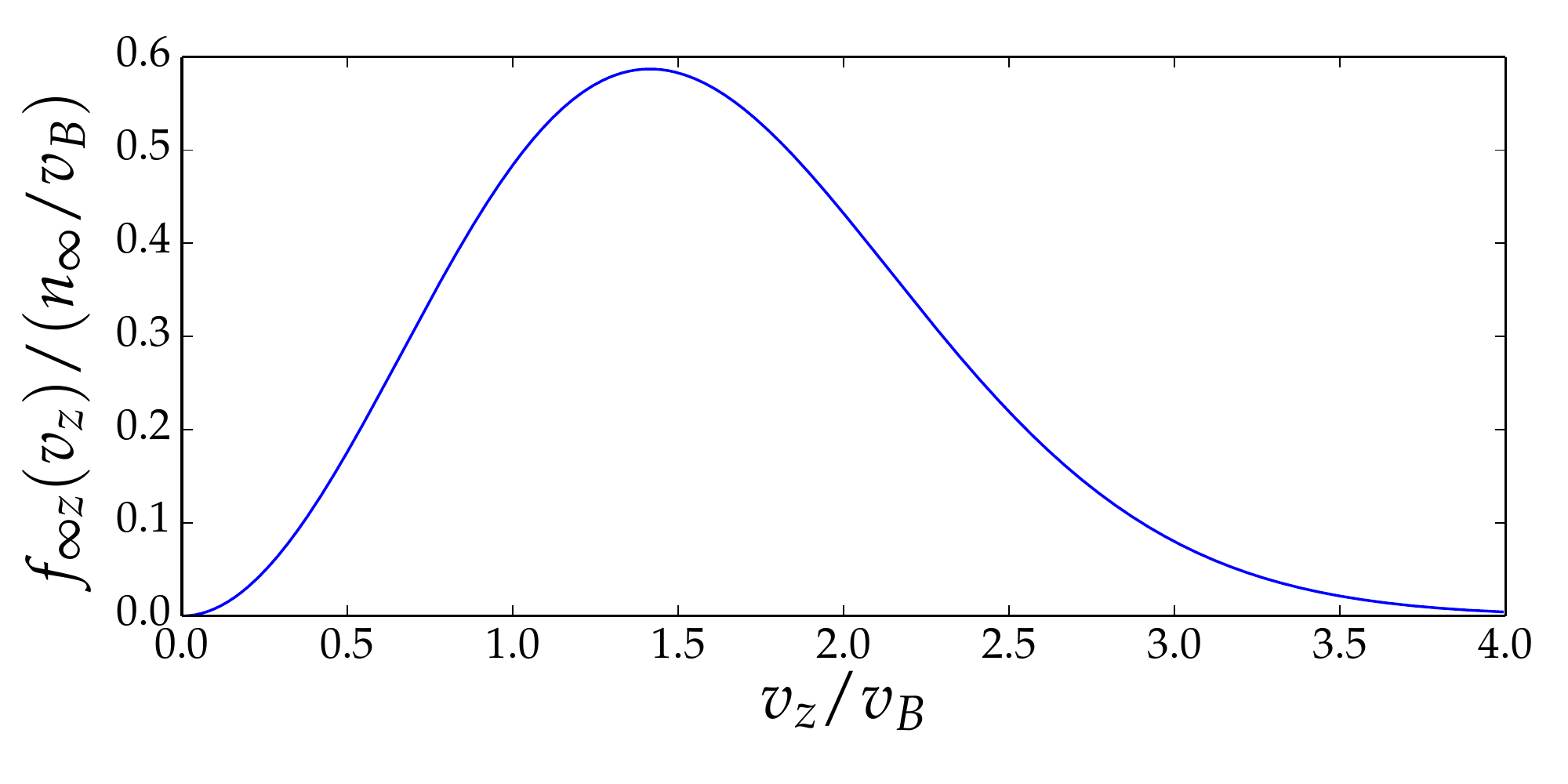}
\caption[Distribution function at magnetic presheath entrance]{The distribution function in (\ref{f-infty}) is shown as a function of the parallel velocity $v_z$ only, $f_{\infty z} \left(v_{z} \right) = \int  \int f_{\infty} \left(\vec{v} \right) dv_x dv_y $. This distribution function marginally satisfies (\ref{solvability}), $\int dv_z  f_{\infty z} \left( v_z \right) / v_z^2 = n_{\text{e}\infty} / v_{\text{B}}^2 $. Its first moment is $u_{z\infty} = \left( 1/n_{\text{e}\infty} \right) \int dv_z  v_z f_{\infty z} \left( v_z \right) \simeq 1.60 v_{\text{B}} $. }
\label{fig-finfinity}
\end{figure}

\begin{figure}[h] 
\centering
\includegraphics[width=0.7\textwidth]{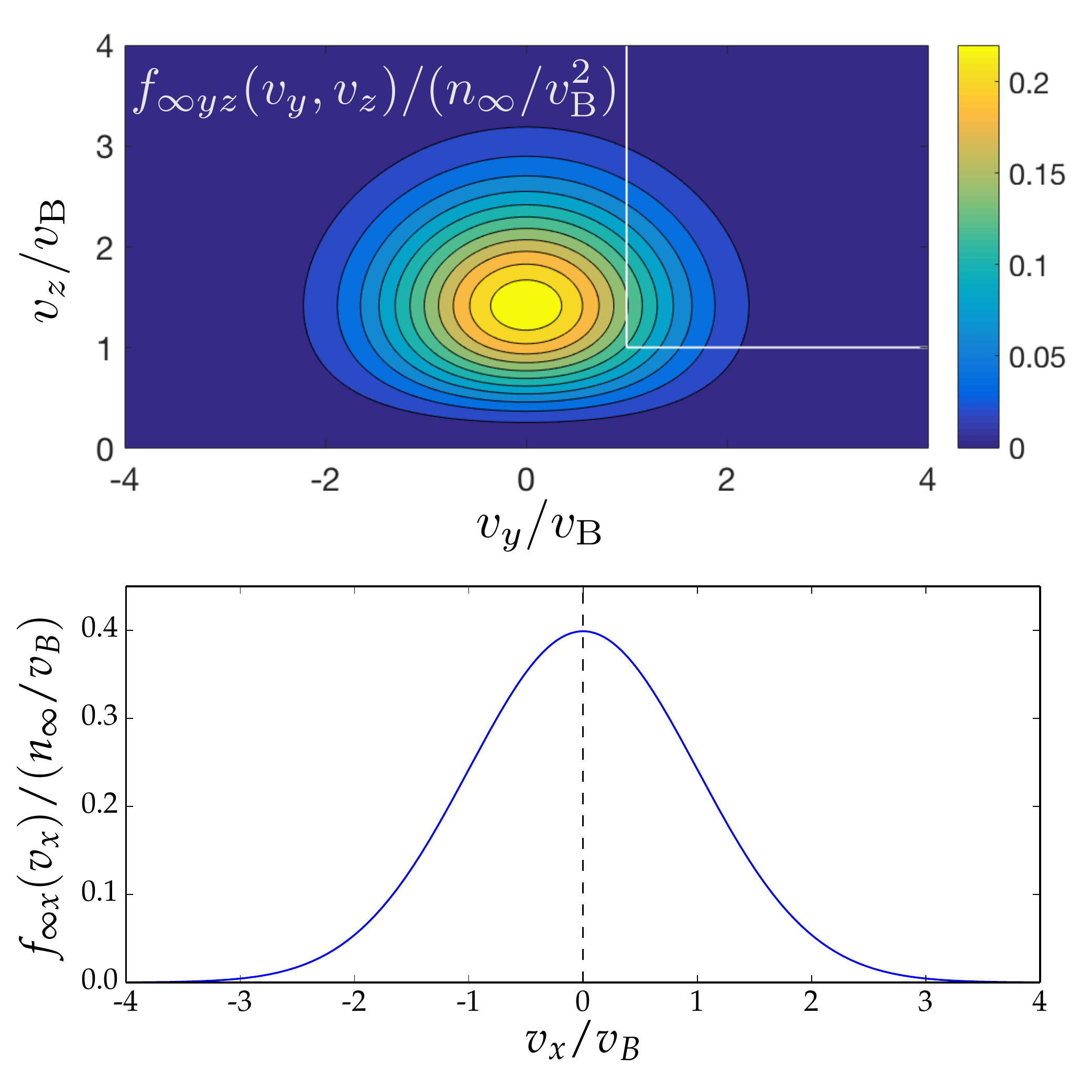}
\caption[Distribution function at magnetic presheath entrance (in 3 dimensions)]{The distribution function (\ref{f-infty}) entering the magnetic presheath is shown as a function of the co-ordinates $\left( v_x, v_y, v_z \right)$. I define $f_{\infty x}(v_x) = \int_{-\infty}^{\infty} dv_y \int_0^{\infty} f_{\infty} (\vec{v} )  dv_z $ and $f_{\infty yz}(v_y, v_z) = \int_{-\infty}^{\infty} f_{\infty} (\vec{v} ) dv_x $. To compare with the distribution function $f_0 \left( \vec{v} \right) $ leaving the magnetic presheath, the box delimited by the white lines and the top right corner in the top diagram has the same size as Figure \ref{fig-f0yz}, and the region to the left of the dashed line in the bottom diagram is the domain of Figure \ref{fig-f0}. }
\label{fig-finfinitycomparison}
\end{figure}

In the following, I take $Z = 1$, thus considering singly charged ions.
Quasineutrality (\ref{quasineutrality}) implies that the ion and electron number densities are equal, and their value at $x \rightarrow \infty$ is denoted $n_{\infty}$.
I assume the following form for the lowest order ion distribution function at the magnetic presheath entrance,
\begin{align} \label{f-infty}
f_{\infty} \left( \vec{v} \right) = \frac{4}{\pi^{3/2}} n_{\text{e}\infty} \left( \frac{m_{\text{i}}}{2T_{\text{e}}} \right)^{5/2} v_z^2 \exp\left( - \frac{m_{\text{i}} | \vec{v} |^2 }{2T_{\text{e}}}  \right) \text{.}
\end{align}
Equation (\ref{f-infty}) satisfies the marginal form of the kinetic Chodura condition (\ref{kinetic-Chodura})
\begin{align} \label{kinetic-Chodura-marginal}
 Zv_{\text{B}}^2 \int  \frac{f_{\infty} \left( \vec{v} \right)}{v_z^2} d^3v = n_{\text{e}\infty}  \text{.}
\end{align}
Changing to variables $\mu$ and $U$, the distribution function (\ref{f-infty}) is
\begin{align} \label{F-numerical-mu-U}
  F_{\text{cl}} \left(\mu, U \right) =  \frac{8}{\pi^{3/2}} n_{\text{e}\infty} \left( \frac{m_{\text{i}}}{2T_{\text{e}}} \right)^{5/2}\left( U - \Omega \mu \right) \exp\left( - \frac{m_{\text{i}} U}{T_{\text{e}}} \right) \Theta \left(v_z\right) \text{,}
\end{align} 
which is constant throughout the magnetic presheath to lowest order in $\alpha$.
This form was used in other studies, for example \cite{Coulette-Manfredi-2016}, and it is plotted in Figures \ref{fig-finfinity} and \ref{fig-finfinitycomparison}. 
Equation (\ref{F-numerical-mu-U}) is used to obtain a discretized version of the distribution function $F_{\text{cl}} (\mu, U )$, defined on a square grid of values of $2\Omega \mu /v_{\text{t,i}}^2$ and $2U/v_{\text{t,i}}^2$ which lie between $0$ and $15.0$ in intervals of $0.05$.
For the distribution function (\ref{F-numerical-mu-U}), I define the ion thermal velocity $v_{\text{t,i}} = \sqrt{2T_{\text{e}}/m_{\text{i}}}$ and the ion gyroradius $\rho_{\text{i}} = v_{\text{t,i}}/ \Omega$. 
The Bohm speed is $v_{\text{B}} = \sqrt{T_{\text{e}}/m_{\text{i}}} = v_{\text{t,i}}/ \sqrt{2}$. 
The distribution function (\ref{f-infty}) marginally satisfies the solvability condition (\ref{kinetic-Chodura}), and the coefficient $k_{3/2}$ can be computed from (\ref{k32}), obtaining
\begin{align} \label{k32-Finfty}
\sqrt{ \frac{T_{\text{e}}}{e} } \left( \frac{v_{\text{B}}}{ \Omega } \right)^2  k_{3/2}  = \frac{8}{3\sqrt{\pi}}  \simeq 1.50  \text{.}
\end{align}
The numerically calculated value of $k_{3/2}$ coincides (to within a numerical error of $2\%$) with equation (\ref{k32-Finfty}). 
The average ion velocity in the $z$ direction at the magnetic presheath entrance is 
\begin{align} \label{finfty-flow}
u_{z\infty}  = \frac{1}{n_{\text{e}\infty}} \int f_{\infty}\left( \vec{v} \right) v_{z}  d^3v = 2\sqrt{\frac{2}{\pi}}v_{\text{B}} \simeq 1.60v_{\text{B}} \text{.}
\end{align} 

\begin{figure}[h] 
\centering
\includegraphics[width=0.6\textwidth]{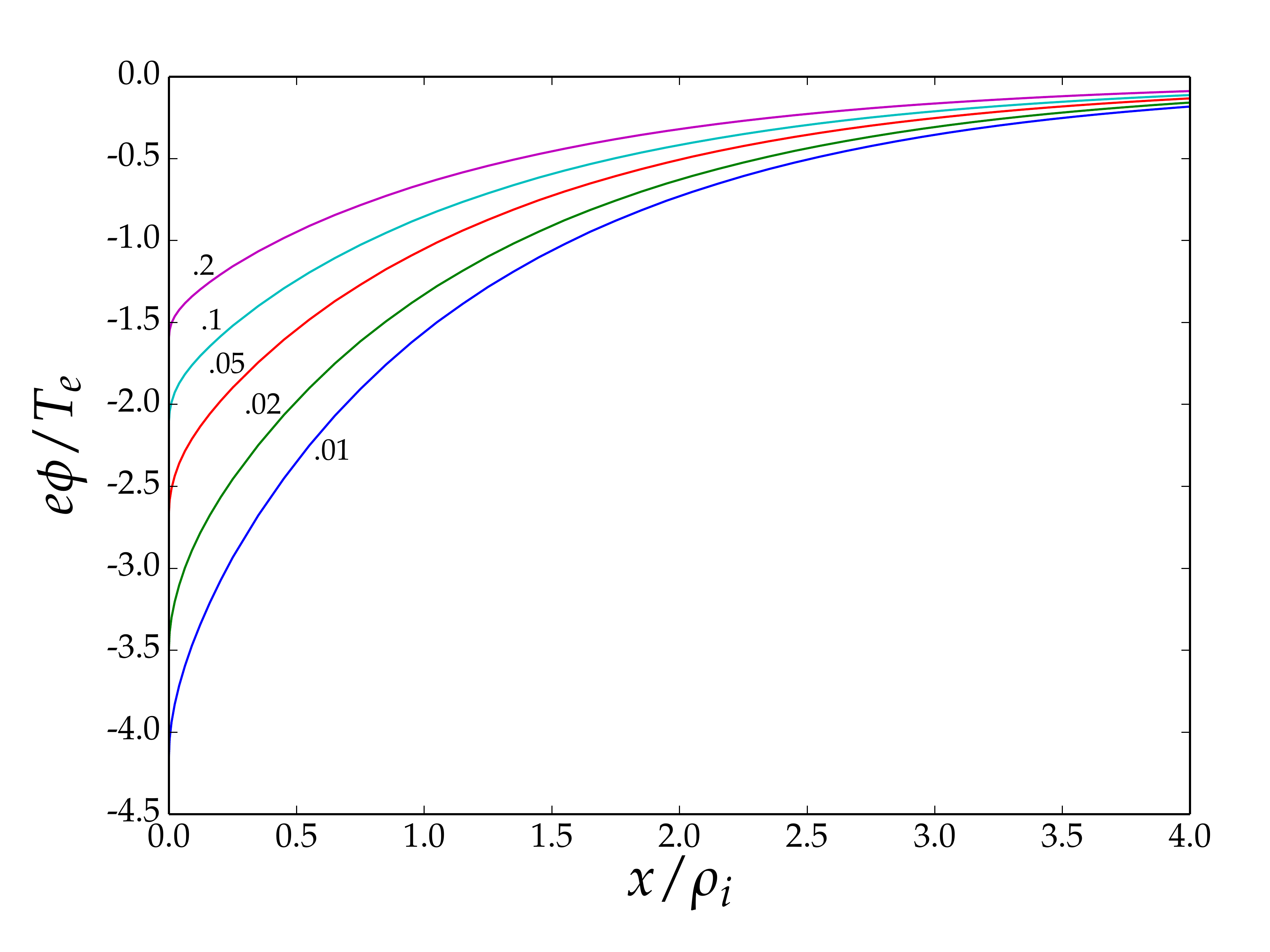}
\caption[Electrostatic potential profiles]{The electrostatic potential profile is plotted for a range of angles $\alpha$, which are indicated next to the corresponding curve. Near $x=0$, $\phi(x) - \phi (0) \propto \sqrt{x} $.}
\label{fig-phiprofile}
\end{figure}
\begin{figure}[h]
\centering
\includegraphics[width=0.47\textwidth]{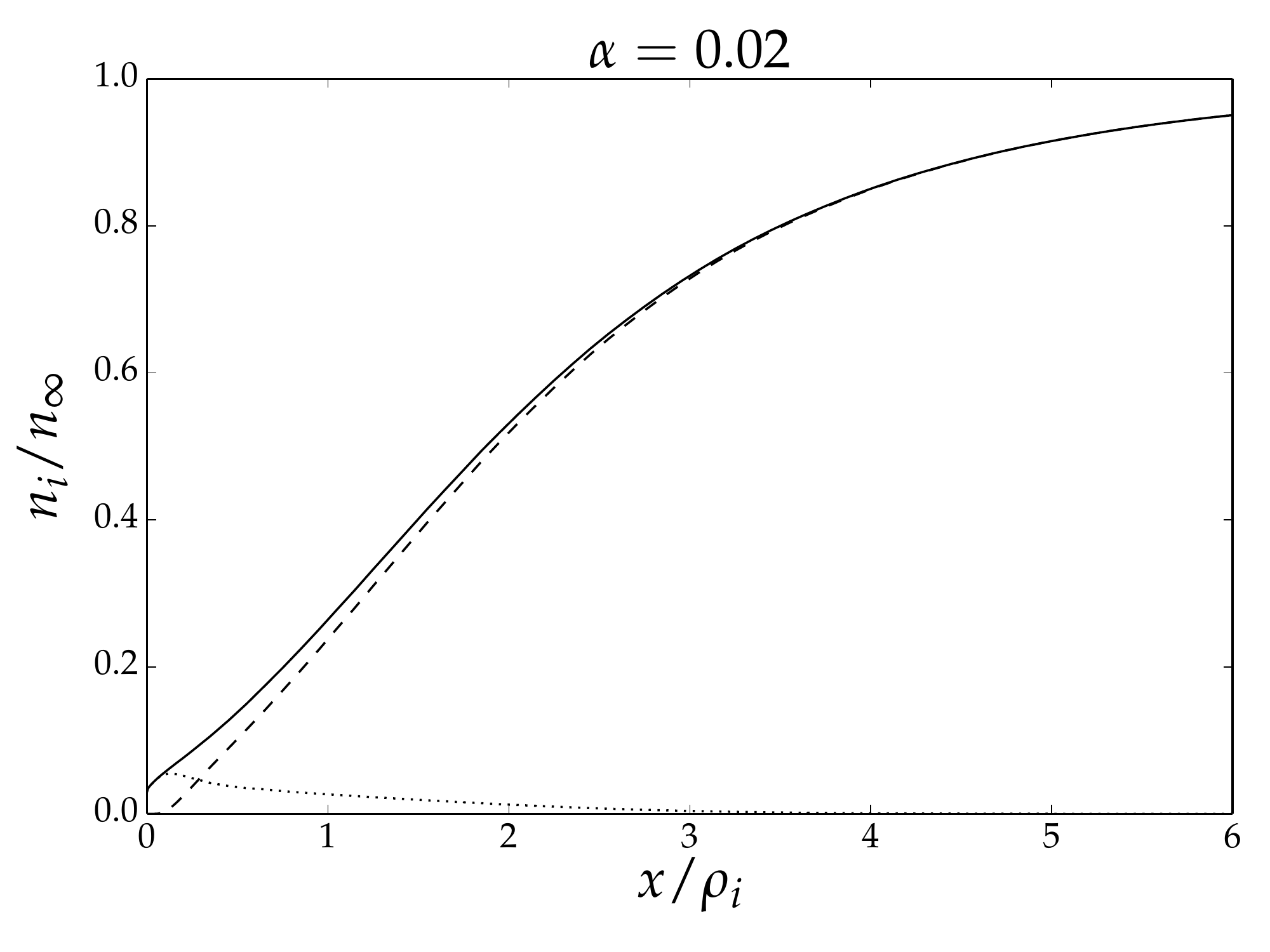}
\includegraphics[width=0.47\textwidth]{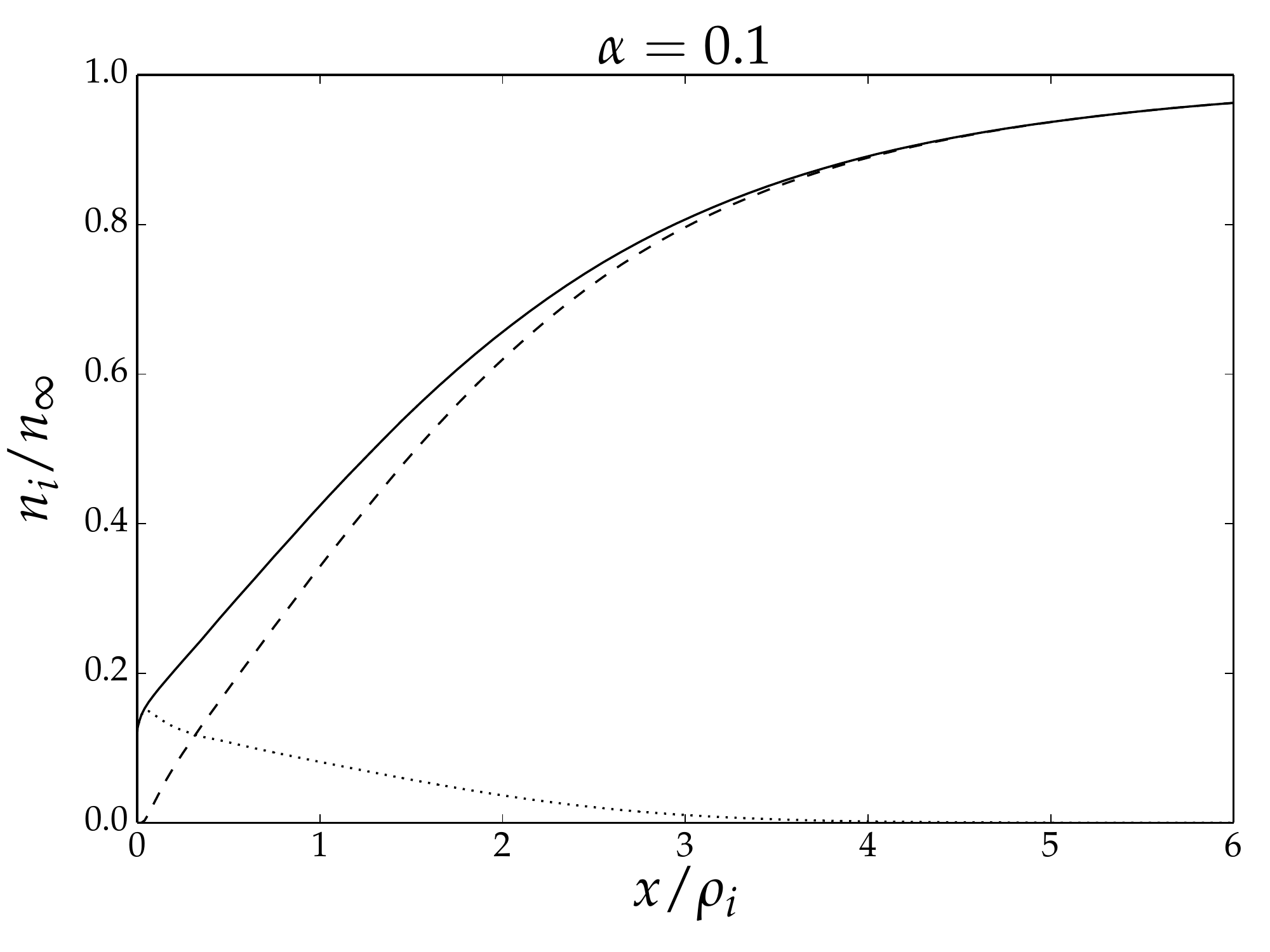}
\caption[Ion density profiles]{The ion density (solid line) for $\alpha = 0.02$ and $\alpha = 0.1$ is shown with the contributions from the closed ion orbits (dashed line) and the open orbits (dotted line) clearly marked. The open orbits clearly dominate in a very small region near $x=0$, there is an overlap region in which the open orbit contribution and the closed orbit contribution have a similar size, while at larger values of $x$ the closed orbit density dominates.}
\label{fig-ni}
\end{figure}
\begin{figure}[h]
\centering
\includegraphics[width=0.6\textwidth]{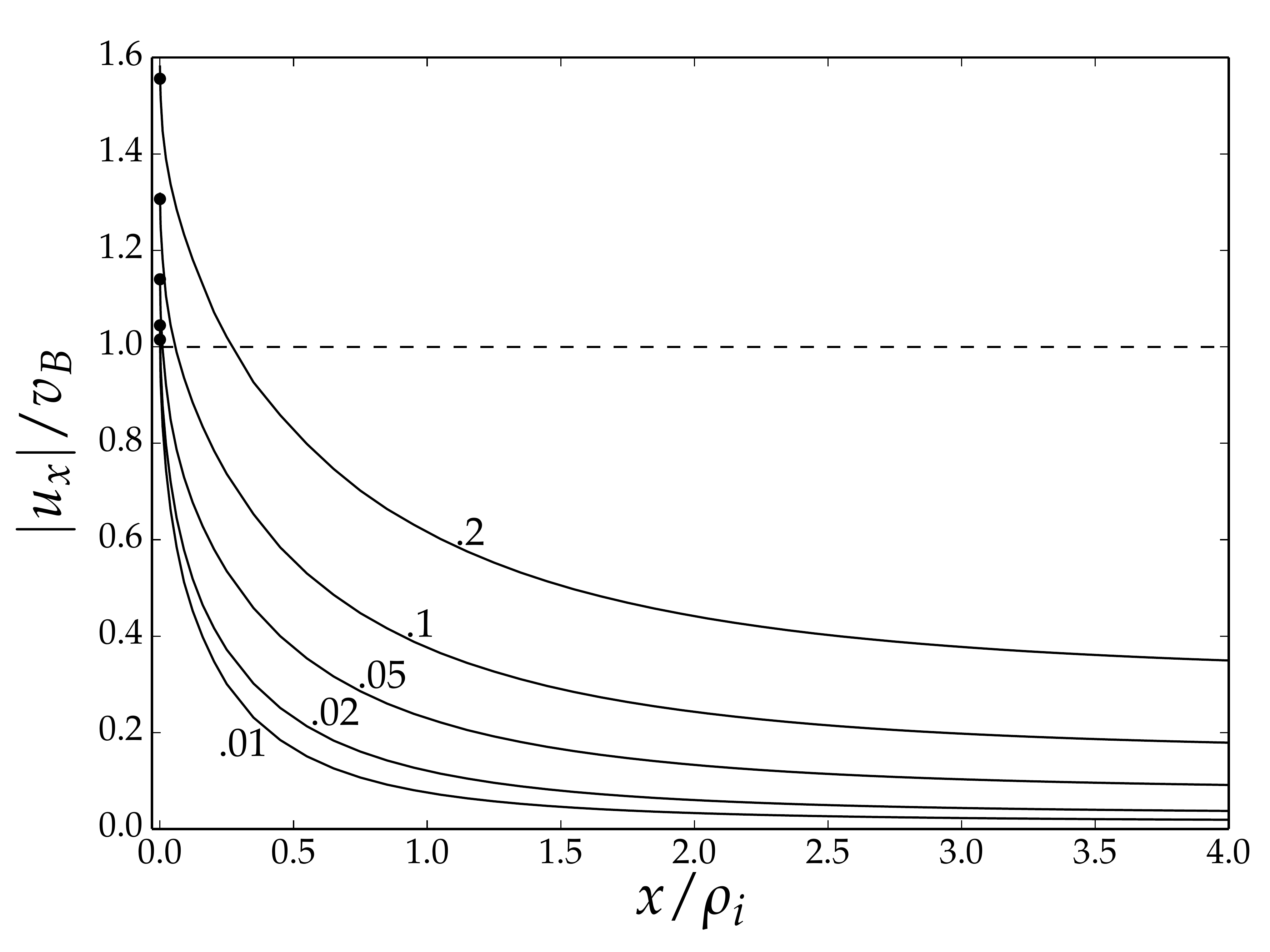}
\caption[Ion fluid velocity profiles]{The average ion velocity in the direction normal to the wall is shown at various angles $\alpha$ (labelled next to the corresponding curve). The flow velocity obtained via the integral (\ref{ux0-integral}) is shown with a black circle at $x=0$, which coincides with the value I calculate from continuity. 
The usual cold ion Bohm limit is indicated by the dashed line $|u_x|/v_{\text{B}} = 1$. The ion flow lies above the cold ion Bohm limit at $x=0$ because the ions are ``warm'' ($T_{\text{i}} \neq 0$). However, at small angles $\alpha \lesssim 0.05$, the ion flow at $x=0$ approaches the cold ion Bohm limit. }
\label{fig-flux}
\end{figure}

The normalized electrostatic potential $e \phi ( x ) / T_{\text{e}}$ is shown in Figure \ref{fig-phiprofile} for a range of angles $\alpha$. 
A general property of the potential curves is that they rise very steeply near $x=0$, with the scaling $\phi(x) - \phi (0) \propto \sqrt{x} / q_2$ in that region (as can be seen explicitly in Figure \ref{fig-phires}). 
I have shown that this behaviour of $\phi(x)$ is expected, and it is connected with the marginal kinetic Bohm condition (\ref{Bohm-kinetic-marginal}) being satisfied. 
The value of $q_2$ that I calculate numerically from the distribution function at $x=0$, using equation (\ref{q2-def}), is consistent with the behaviour of the electrostatic potential near $x=0$.

The ion density profiles for $\alpha=0.02$ and $\alpha = 0.1$ are shown in Figure \ref{fig-ni}. 
The open orbit density can be seen to increase initially and then quickly decrease with distance from the wall. 
This behaviour is consistent with the behaviour of $\Delta v_x$ for type II orbits (see Figure~\ref{fig-Deltavx} and the discussion following equation (\ref{ni-open-order})). 
The open orbit density is clearly the dominant contribution to the density in the neighbourhood of $x=0$, while for large $x$ closed orbits give the largest contribution.
 
The flow velocity of ions across the magnetic presheath is commonly calculated in fluid models. 
Therefore, it is useful to calculate it to compare with previous results.
Here I calculate the flow by using the ion continuity equation.
The ion flux towards the wall across the magnetic presheath must be constant (no ion sources) due to particle conservation,
\begin{align}
\frac{\partial }{\partial x} \left( n_{\text{i}} \left( x \right) u_x \left( x\right) \right) = 0 \text{,}
\end{align} 
where $ u_x \left( x\right) $ is the average velocity of ions in the $x$ direction.
 At the magnetic presheath entrance $x \rightarrow \infty$, the flow towards the wall is obtained from the average over the distribution function of the gyroaveraged motion of ions towards the wall, given by $\dot{\bar{x}}$ (note that, due to distortion of the orbits, this does not remain true across the magnetic presheath). 
 Using equations (\ref{vz-U-Uperp}) and (\ref{xbardot}), the flow in the $z$ direction, $u_{z\infty}$, is related to the flow in the $x$ direction, $u_{x\infty} $, via $u_{x\infty} =   - \alpha u_{z\infty}$. 
 This is equivalent to the boundary condition of flow being parallel to the magnetic field at $x \rightarrow \infty$ \cite{Riemann-1994}.
 The flow $u_{z\infty}$ is obtained as a moment of the incoming distribution function (see equation (\ref{finfty-flow}))
  \begin{align} \label{uzinfty}
 u_{z\infty} = \frac{1}{n_{\text{e}\infty}} \int f_{\infty} ( \vec{v} ) v_{z} d^3v \text{.}
 \end{align}
The flux of ions towards the wall is conserved and therefore given by $ n_{\text{i}} \left( x \right) u_x \left( x\right) = n_{\text{i}\infty} u_{x,\infty} =  - \alpha n_{\text{i}\infty} u_{z\infty} $. 
The average lowest order ion flow velocity towards the wall at a general position $x$ is therefore
 \begin{align} \label{ux-ni}
u_x \left( x\right) \simeq - \frac{\alpha n_{\text{e}\infty} u_{z \infty} }{n_{\text{i}} \left( x \right)} \text{.}
 \end{align}
Using the quasineutrality equation, $n_{\text{i}}(x) = n_{\text{e}}(x) = n_{\text{e}\infty} \exp \left( e\phi(x) / T_{\text{e}} \right)$,
\begin{align} \label{ux}
u_x \left( x \right) \simeq - \alpha u_{z \infty} \exp \left( - \frac{e\phi(x)}{T_{\text{e}} } \right)   \text{.}
 \end{align}
The function (\ref{ux}) evaluated at $x=0$ can be checked, for consistency, against the appropriate integral of the distribution function (\ref{f0}),
\begin{align} \label{ux0-integral}
u_{x0} = \frac{1}{n_{\text{i}} \left( 0 \right) } \int f_{0} \left( \vec{v} \right) v_x d^3v \text{.}
\end{align}
In Figure \ref{fig-flux}, I plot the average ion velocity profile $u_x\left( x\right)$, obtained using equation (\ref{ux}), for a range of angles $\alpha$. The magnetic presheath acceleration turns the ion flow from being (super)sonic in the direction parallel to the magnetic field to being (super)sonic in the $x$ direction normal to the wall. 
At $x=0$, the flow velocity is calculated in an alternative way, by taking the integral of the distribution function as in equation (\ref{ux0-integral}). The value thus obtained is marked on the curves for each value of $\alpha$, and it is consistent with the value obtained by using equation (\ref{ux}).

\begin{figure}[h]
\centering
\includegraphics[width=0.6\textwidth]{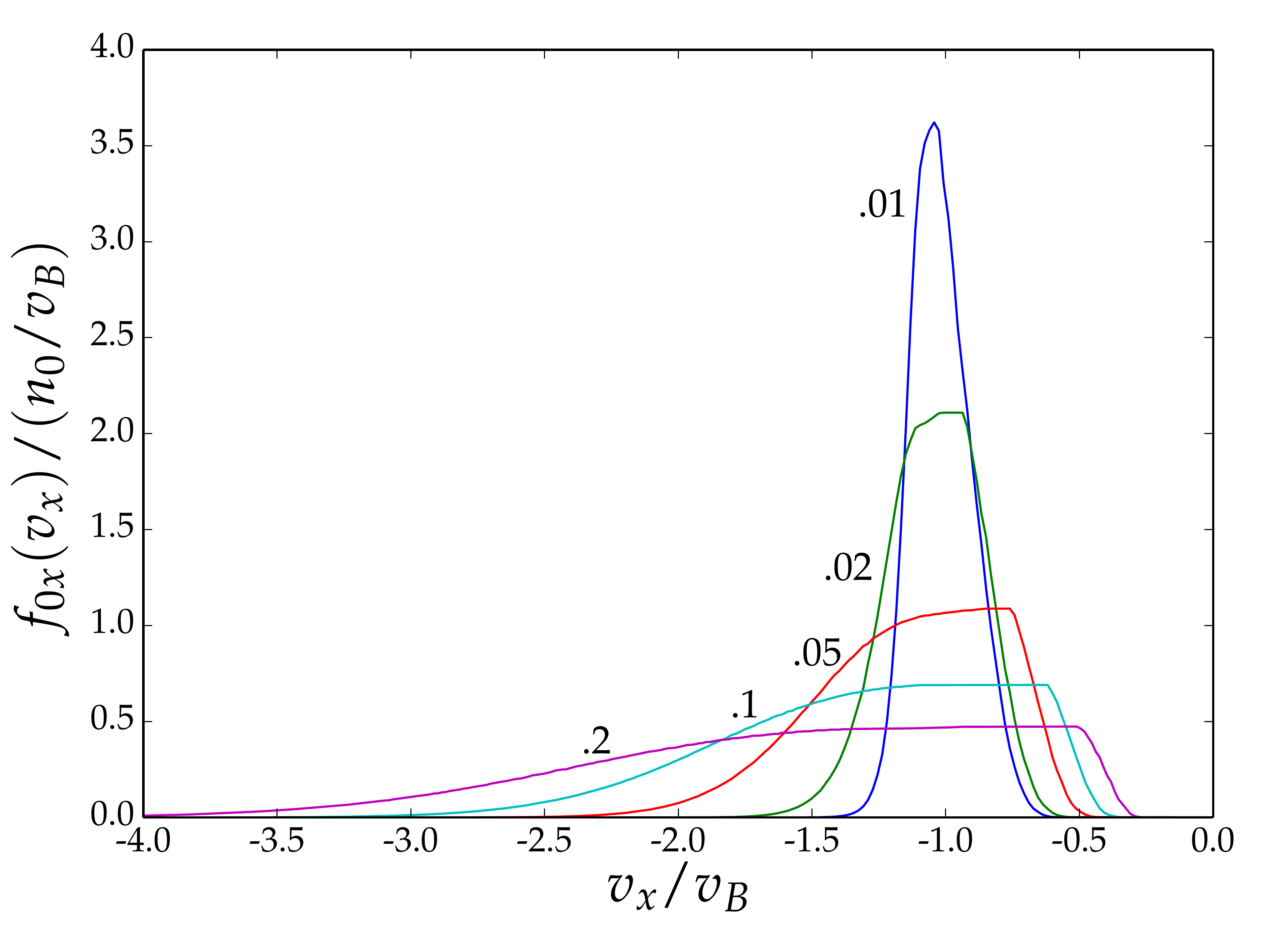}
\caption[Ion distribution function at the Debye sheath entrance (velocity component normal to wall)]{The distribution function $f_{0x} \left(v_x\right) = \int_{-\infty}^{\infty}   dv_y \int_{-\infty}^{\infty}  f_{0}\left( v_x, v_y, v_z \right) dv_z $ for a range of angles $\alpha$, marked next to the corresponding curve. }
\label{fig-f0}
\end{figure}
\begin{figure}[h]
\centering
\includegraphics[width=0.7\textwidth]{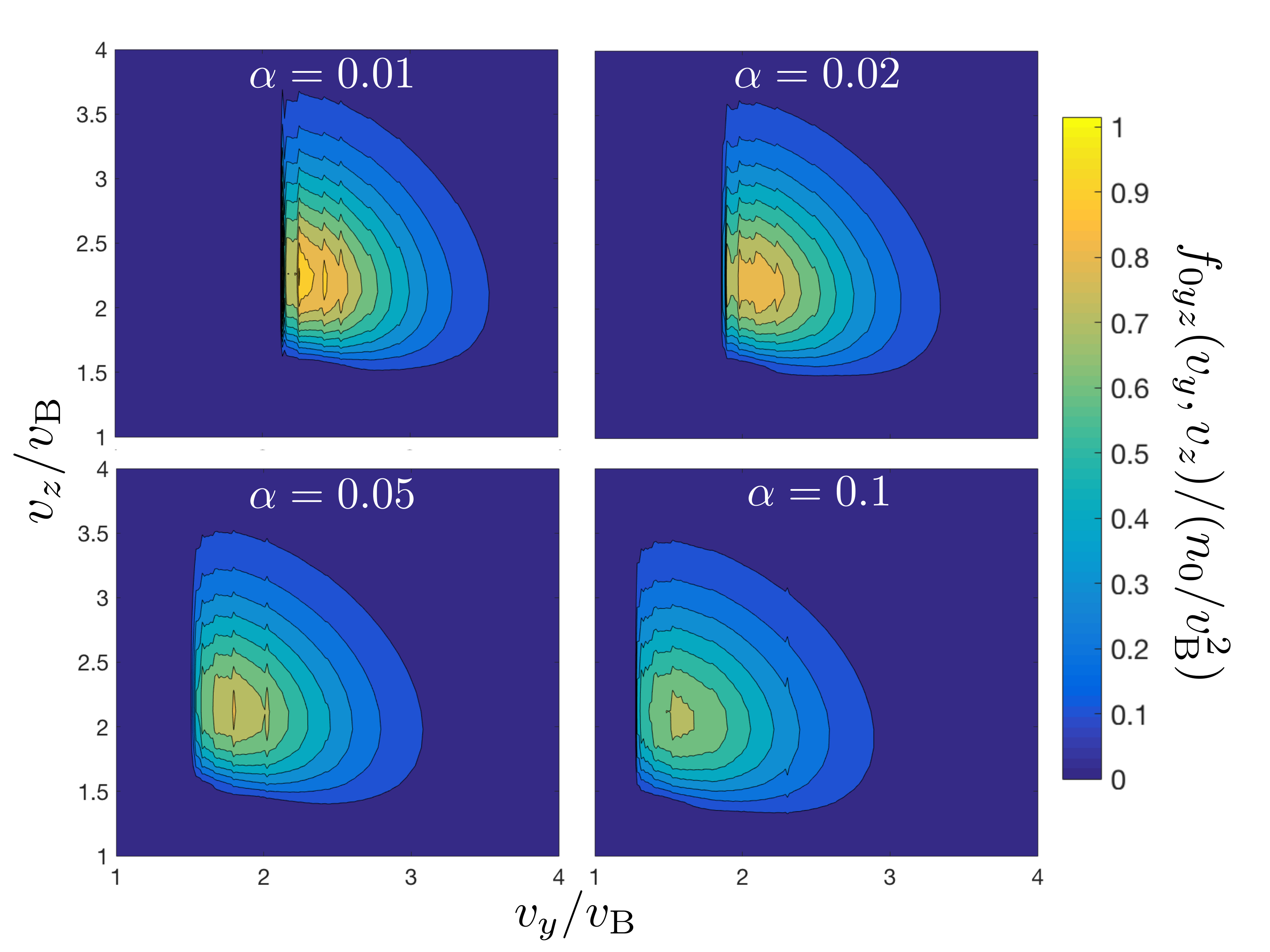}
\caption[Ion distribution function at the Debye sheath entrance (velocity components parallel to wall)]{The distribution function $f_{0yz} \left(v_y, v_z\right)  =  \int_{-\infty}^0   f_{0}\left( v_x, v_y, v_z \right) dv_x $ for a range of angles $\alpha$, marked on each panel. }
\label{fig-f0yz}
\end{figure}
\begin{figure}[h]
\centering
\includegraphics[width=0.7\textwidth]{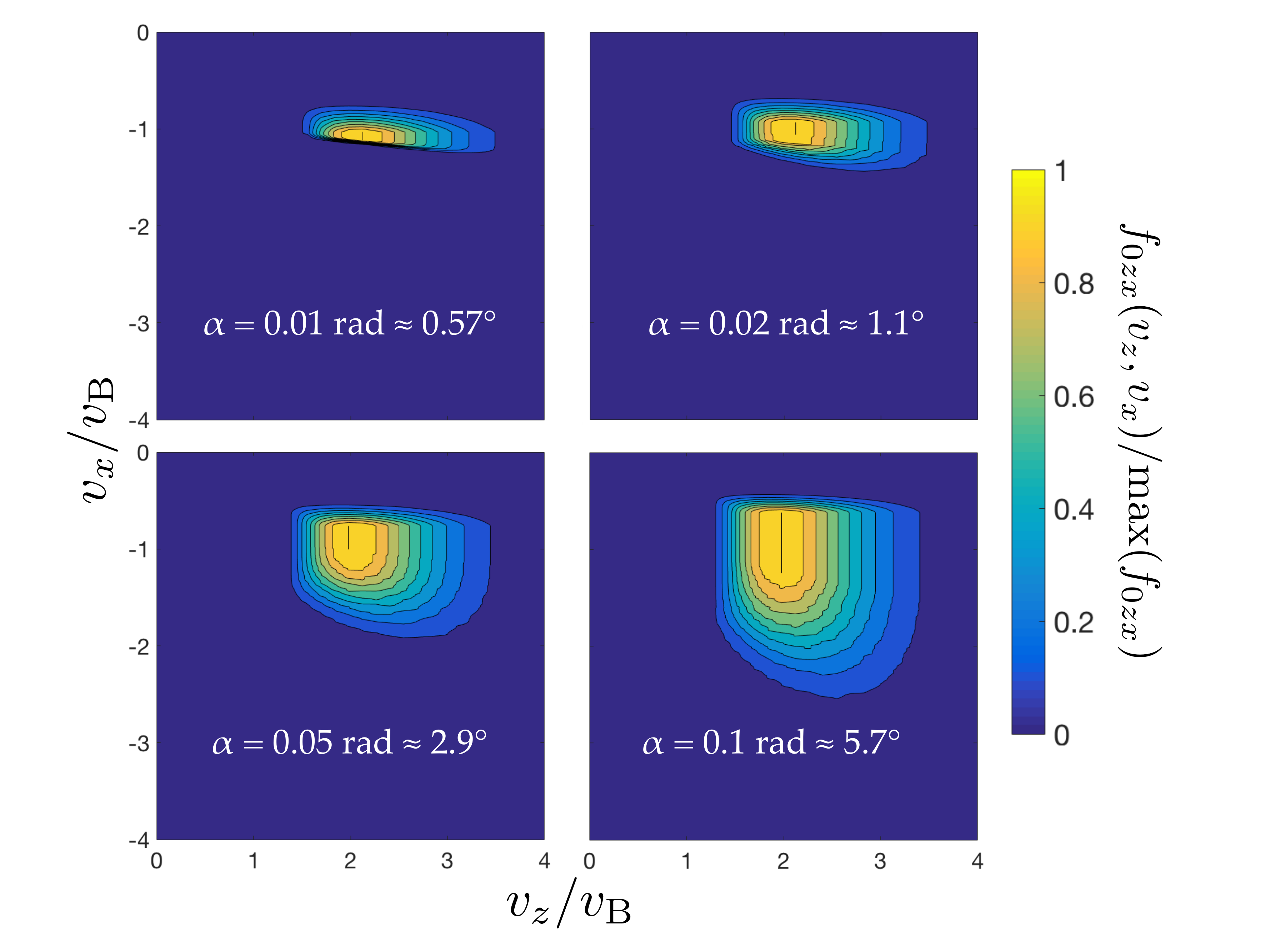}
\caption[Ion distribution function at the Debye sheath entrance (velocity components parallel to the magnetic field and perpendicular to wall)]{The distribution function $f_{0zx} \left(v_z, v_x\right)  =  \int_{0}^{\infty}   f_{0}\left( v_x, v_y, v_z \right) dv_y $ for a range of angles $\alpha$, marked on each panel. 
The distribution of the corresponding velocity components of ions entering the magnetic presheath is the same as the one shown in Figure \ref{fig-finfinitycomparison}, with the axis labelled $v_y$ representing $v_x$.}
\label{fig-f0zx}
\end{figure}

By asymptotic matching, the distribution function in (\ref{f0}) is the distribution function entering the Debye sheath.
In the Debye sheath, electrostatic forces normal to the wall dominate over magnetic forces, hence $v_x$ is the only velocity component that changes significantly \cite{Riemann-review}. 
Therefore, only knowledge of the function 
\begin{align} \label{f0x}
 f_{0x}\left( v_x \right)  = &  \int_{-\infty}^{\infty} dv_y \int_{-\infty}^{\infty}  f_{0}\left( v_x, v_y, v_z \right) dv_z  \nonumber \\
\simeq &  \int_{\bar{x}_{\text{m,o}}}^{\infty} \Omega d\bar{x} \int_{\chi_{\text{M}} \left( \bar{x} \right) }^{\infty} \frac{F_{\text{cl}}\left( \mu \left( \bar{x},  \chi_M \left( \bar{x} \right) \right), U \right) }{V_{\parallel} \left( \chi_{\text{M}} \left( \bar{x} \right) , U \right) }  \nonumber \\
& \times \hat{\Pi} \left(  v_x ,  - V_x \left( 0, \bar{x}, \chi_{\text{M}} \right) - \Delta v_x  , - V_x \left( 0, \bar{x}, \chi_{\text{M}} \right)  \right) dU   \text{}
\end{align}
 is needed to solve for the electrostatic potential in the Debye sheath. 
The distribution $f_{0x} \left( v_x \right) $ is shown in Figure \ref{fig-f0} for a range of angles $\alpha$.  
A general feature of this function is that it is very close to zero near $v_x =0$. 
This is expected from the discussion in Section \ref{sec-KMPS-Bohm}, where I concluded that there is an exponentially small number of ions with small values of $v_x$ if the distribution function $F_{\text{cl}}$ decays exponentially at large energy $U$.
Another pronounced feature of Figure \ref{fig-f0} is that the distribution function becomes narrower with decreasing $\alpha$. 
For the cases $\alpha = 0.01$ and $\alpha = 0.02$, the distribution function is thin, approximately symmetric and centred at the Bohm speed $v_{\text{B}}$. 
For all angles $\alpha$, the marginal form of the kinetic Bohm condition (\ref{Bohm-kinetic-marginal}) is found to be satisfied, as I predicted in Section \ref{sec-KMPS-Bohm}, with an error of $\lesssim 2\%$.
A thin distribution function implies that the distribution function must be centred at the sonic speed. 
If the ions entering the Debye sheath have a narrow velocity distribution, this can be approximated by a Dirac delta function, $f_{0x} \left( v_x \right) \simeq \delta_{\text{Dirac}} \left( v_x - u_{x0} \right) $. 
Substituting this approximation into (\ref{Bohm-kinetic-marginal}), one obtains the ``fluid'' marginal Bohm condition $u_{x0} = v_{\text{B}} $.

The broadening of the distribution function $f_{0x} \left( v_x \right) $ at larger values of $\alpha$ is due to typical values of $\Delta v_{x}$, given in equation (\ref{Deltavx-simpler}), becoming larger. The scaling $\Delta v_x \sim \sqrt{2\pi \alpha} v_{\text{t,i}}$ gives $\Delta v_x \sim v_{\text{t,i}} $ for $\alpha \sim 0.1$. 
The asymptotic expansion relies on $\Delta v_x$ being small, so one might question the validity of my results when $\Delta v_x \sim v_{\text{t,i}}$. 
While it is true that the accuracy of our expansion may to some extent be compromised at such large values of $\Delta v_x$, the broadening of the distribution function is expected to be physical. 

In Figure \ref{fig-f0yz} I show a contour plot of $f_{0yz} \left( v_y, v_z \right) $, which is given by 
\begin{align} \label{f0yz}
f_{0yz}\left( v_y, v_z \right)  =  \int   f_{0}\left( v_x, v_y, v_z \right) dv_x 
 \simeq  F_{\text{cl}} \left( \mu \left( \bar{x},  \chi_M \left( \bar{x} \right) \right), U \right)  \Delta v_x \text{,}
\end{align} 
where (\ref{xbar-def}) and (\ref{U-open}) can be used to re-express $\bar{x}$ and $U$ in terms of $v_y$ and $v_z$ in equation (\ref{f0yz}). Comparing with the distribution function at the magnetic presheath entrance (shown in Figure \ref{fig-finfinitycomparison}), the distribution function at $x=0$ is narrower (it occupies a smaller area in the $v_y-v_z$ plane of phase space) and it has shifted to larger $v_z$ and to very large and positive $v_y$. 
The net motion of the ions in the $y$ direction can be explained by the fact that they acquire very large $\vec{E} \times \vec{B}$ velocities in the magnetic presheath (see Figure \ref{fig-iontraj}).  
A contour plot of the distribution function in the $x-z$ plane (containing the normal to the wall and the magnetic field) is also shown in Figure \ref{fig-f0zx}.
From Figures \ref{fig-f0}, \ref{fig-f0yz} and \ref{fig-f0zx}, I infer that ions entering the Debye sheath travel with a typical speed of $\sim 3v_{\text{B}}$, making an angle of $15-30^{\circ}$ with the plane parallel to the wall.
The ion speed and the angle that the ion trajectory makes to the wall are expected to increase in the Debye sheath as the electric field accelerates ions in the $x$-direction.

 \begin{figure}[h!]
\centering
\includegraphics[width=0.6\textwidth]{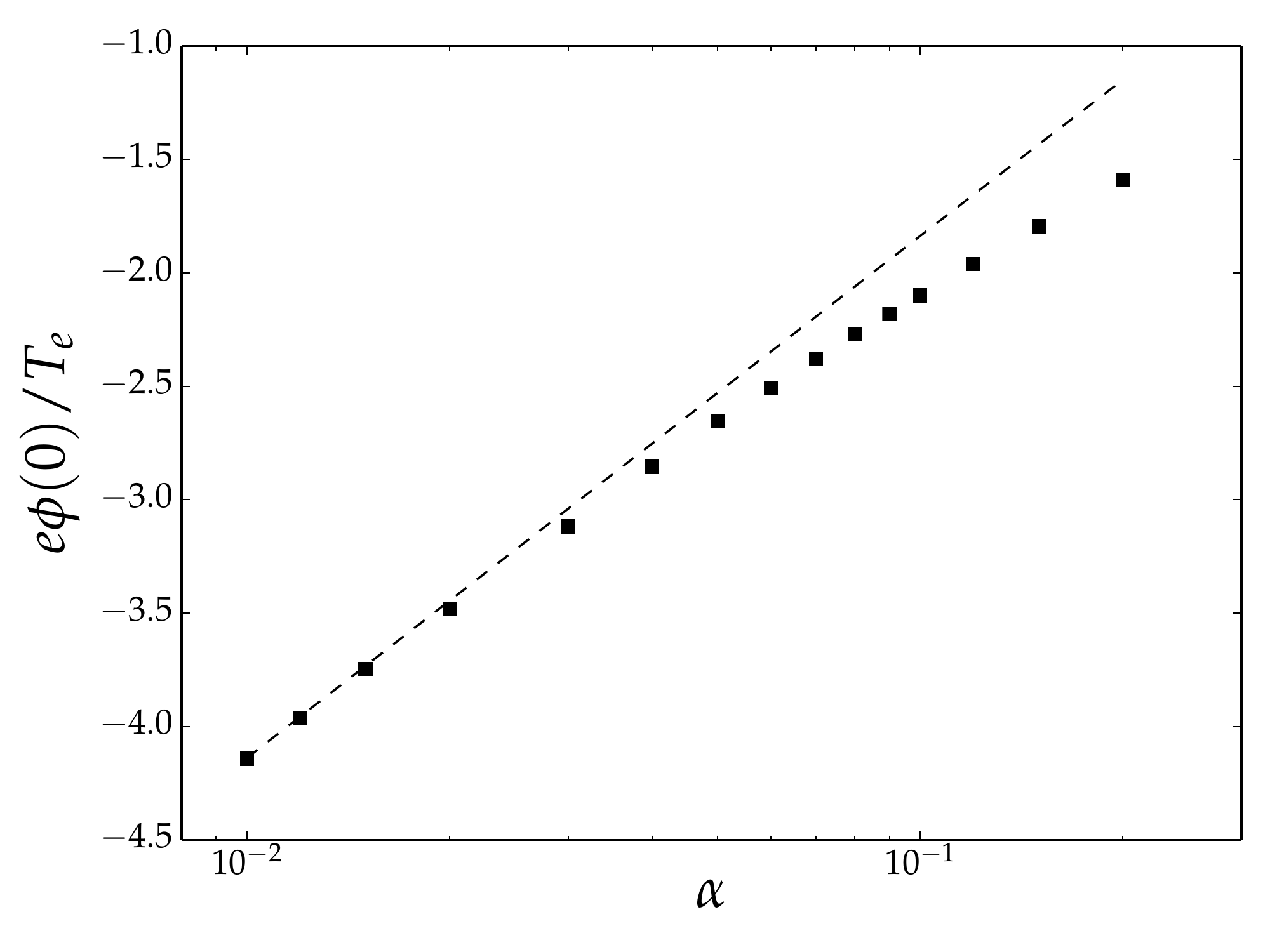}
\caption[Electrostatic potential drop]{The total electrostatic potential drop across the magnetic presheath for a range of angles $\alpha$ is shown with squares.
The dashed line represents the potential drop expected if the ions entering the Debye sheath are cold and the Bohm condition is marginally satisfied, $h(\alpha) = \ln \left(  \alpha u_{z \infty}   / v_{\text{B}} \right) $. For $\alpha \lesssim 0.05$, my results converge to the dashed line.
}
\label{fig-phidrop}
\end{figure}

The electrostatic potential drop across the magnetic presheath is shown in Figure \ref{fig-phidrop}. 
At small angles, $e\phi ( 0 )/T $ converges to the function
\begin{align} \label{phidrop-Bohm}
h \left( \alpha \right)   = \ln \left(  \frac{  \alpha u_{z \infty}  }{ v_{\text{B}} } \right) \text{,}
\end{align}
which is depicted in Figure \ref{fig-phidrop} using a dashed line. 
The reason is the following. 
At $x=0$, the flow into the wall is  $n_{\text{e}\infty} \exp \left( e\phi\left(0\right)/T \right) | u_{x0} | $. Equating this to the flux through $x \rightarrow \infty$, equal to $\alpha n_{\text{e}\infty} u_{z\infty} $, and rearranging, I obtain an expression for the potential drop in terms of the ion flows into and out of the magnetic presheath, 
\begin{align} \label{phidrop}
\frac{e\phi\left(0\right)}{T}   = \ln \left(  \frac{  \alpha u_{z \infty}  }{ | u_{x0} |}   \right) \text{.}
\end{align}
Moreover, I previously found 
that for $\alpha \lesssim 0.05$ the cold ion Bohm condition is almost marginally satisfied, $|u_{x0}| \simeq v_{\text{B}}$, due to the thinness of the distribution function (see Figure \ref{fig-f0}). 
Then, the potential drop across the magnetic presheath can be predicted using equation (\ref{phidrop}) with $u_{x0} = v_{\text{B}}$, which is equation (\ref{phidrop-Bohm}).
Therefore, the potential drop converges to the dashed line in Figure \ref{fig-phidrop}.

\chapter{Ion temperature dependence}
\label{chap-Tdep}

At the end of the previous chapter, a magnetic presheath in which the ion and electron temperatures are similar was studied.
Setting $Z=1$, the magnetic presheath solution depends on: (i) the angle $\alpha$; and (ii) the ion distribution function at the magnetic presheath entrance.

There is an infinite number of possible distribution functions at the magnetic presheath entrance, but only a small subset of them can be used in a realistic study.
The width of the ion distribution function is characterized by the quantity $T_{\text{i}} = m_{\text{i}} v_{\text{t,i}}^2 / 2$.
In this chapter, I study the dependence of the magnetic presheath on the width of the ion distribution function using the parameter
\begin{align}
\tau = \frac{T_{\text{i}}}{T_{\text{e}}} \text{}
\end{align}
by using a prescribed, arbitrary set of distribution functions.

Dependence on the ion temperature at the entrance of the magnetic presheath is important because $\tau$ is measured to be greater than unity near the divertor targets of tokamak plasmas \cite{Mosetto-2015}.
No ions bounce back from the magnetic presheath, and thus we expect the ion distribution function at the entrance of the magnetic presheath to be zero for $v_z < 0$.
That is, the distribution function is not a Maxwellian.
Fluid equations require that the distribution function be sufficiently close to a Maxwellian. 
If the ion temperature is small compared to the electron temperature, $\tau \ll 1$, the ions can be assumed mono-energetic and fluid equations can accurately describe the ion dynamics.
Conversely, if $\tau \gtrsim 1$, the ion distribution function is very far from Maxwellian and the validity of a fluid description becomes questionable.
Therefore, it is important to include the effect of a strongly non-Maxwellian ion distribution function, and the only way to do this is via a kinetic treatment of the ions such as the one carried out in this thesis.

Some of the basic orderings presented in section \ref{sec-traj-orderings} change with $\tau$. 
The electrostatic potential $\phi$ must be ordered as in (\ref{order-pot}) if the wall is electron-repelling.
Ions falling in such a potential gain energies of order $Z e \phi \sim Z T_{\text{e}}$, where $Ze$ is the ion charge. 
At such energies, ions have at least a velocity of the order of the Bohm speed (\ref{vB}).
Hence, considering that the ion's speed must at least be equal to its thermal velocity, I order
\begin{align} \label{cs}
| \vec{v} | \sim c_s = \sqrt{ v_{\text{B}}^2 + \frac{1}{2}v_{\text{t,i}}^2 } = \sqrt{\frac{ZT_e + T_i}{m_i}} \text{.}
\end{align}
Note that $c_{\text{s}} = v_{\text{B}}$ when $\tau = 0$ and $c_{\text{s}} = v_{\text{t,i}} / \sqrt{2}$ when $\tau = \infty$.

The typical size of the magnetic presheath, denoted $d_{\text{mps}}$, also changes with $\tau$.
In order to obtain an estimate for $d_{\text{mps}}$, consider the two limits $\tau \ll 1$ and $\tau \gg 1$ separately. 
When the ion temperature is much smaller than the electron temperature, $\tau \ll 1$, the only way by which ions can acquire the Bohm velocity $v_{\text{B}}$ in the direction normal to the wall --- necessary to satisfy the Bohm condition at the Debye sheath entrance \cite{Riemann-review} --- is if the electric field becomes large enough that it breaks the Larmor orbits \cite{Loizu-2012}. 
From the ordering $| \vec{v} | \sim v_{\text{B}}$ for the ion speed and by balancing the magnetic and electric forces, one obtains $\phi' (x) \sim T_{\text{e}}/ e d_{\text{mps}}  \sim v_{\text{B}} B $, leading to $d_{\text{mps}} \sim v_{\text{B}} / \Omega $. 
When the ion temperature is large, $\tau \gg 1$, the length scale of the magnetic presheath is set by the ion density variation, which cannot have a characteristic length scale smaller than the ion gyroradius, giving $d_{\text{mps}} \sim  \rho_{\text{i}} = v_{\text{t,i}} / \Omega $. 
When $\tau \sim 1$, both arguments are valid since $\rho_{\text{i}} \sim v_{\text{B}}/\Omega$.
Thus, the size of the magnetic presheath is given by the ion sound gyroradius,
\begin{align} \label{rho-s}
d_{\text{mps}} \sim \rho_{\text{s}} = \frac{c_s}{\Omega} \text{.}
\end{align} 

The rest of this chapter is structured as follows.
In section \ref{sec-Tdep-bc}, I describe the set of distribution functions used as boundary conditions.
Subsequently, in section \ref{sec-Tdep-finite} I present numerical results obtained for a range of parameters $\tau$ using the numerical method outlined in chapter \ref{chap-KMPS}. 
The numerical results are consistent with the limits $\tau \rightarrow 0$ and $\tau \rightarrow \infty$, which are studied in sections \ref{sec-Tdep-cold} and  \ref{sec-Tdep-hot} respectively.
In section \ref{sec-Tdep-disc} I discuss the results of this chapter.

\section{Boundary conditions} \label{sec-Tdep-bc}

For different values of the parameter $\tau$, the boundary condition for the ion distribution function at the magnetic presheath entrance is given by
\begin{align} \label{f-infty-Tdep}
f_{\infty} \left( \vec{v} \right) =  \begin{cases}
\mathcal{N}  n_{\infty} \frac{4 v_z^2}{\pi^{3/2} v_{\text{t,i}}^5}   \exp \left( - \frac{ \left| \vec{v} - u v_{\text{t,i}} \hat{\vec{z}} \right|^2 }{v_{\text{t,i}}^2} \right) \Theta \left( v_z \right) & \text{ for } \tau \leqslant 1 \text{,} \\
\mathcal{N}  n_{\infty}  \frac{ 4 v_z^2 }{  \pi^{3/2} v_{\text{t,i}}^3 \left( v_{\text{t,i}}^2 +r v_z^2 \right)} \exp \left( - \frac{\left| \vec{v} \right|^2 }{v_{\text{t,i}}^2} \right)\Theta \left( v_z \right)  & \text{ for } \tau > 1 \text{.}
\end{cases}
\end{align}
The value of the normalization constant $\mathcal{N}$ is (see Appendix \ref{integrals-Tdep})
\begin{align} \label{N-infty}
\mathcal{N}  =  \begin{cases}
\left[  \left( 1 + 2u^2 \right) \left( 1 + \text{erf}(u) \right) + \frac{2u}{\sqrt{\pi}}  \exp(-u^2) \right]^{-1}  & \text{ for } \tau \leqslant 1 \text{,} \\
r^{3/2} \left[ 2\sqrt{r} - 2\sqrt{\pi} \exp\left(\frac{1}{r}\right) \left( 1 - \text{erf} \left( \frac{1}{\sqrt{r}} \right) \right) \right]^{-1}  & \text{ for } \tau > 1 \text{,}
\end{cases}
\end{align}
ensuring that
\begin{align} \label{n-infty}
n_{\infty} = \int f_{\infty} \left( \vec{v} \right) d^3v  \text{.}
\end{align}
At every value of $\tau$, the positive constant $r$ or $u$ is obtained by imposing the marginal form of the kinetic Chodura condition (\ref{kinetic-Chodura}),
\begin{align} \label{kinetic-Chodura-marginal}
\int \frac{ f_{\infty} \left( \vec{v} \right)}{v_z^2} d^3v = \frac{n_{\infty}}{v_{\text{B}}^2}   \text{.}
\end{align}
Hence, $u$ satisfies the equation
\begin{align} \label{r-def}
 \left( 1 + \text{erf}(u) \right) = \tau \left[ \left( 1 + 2u^2 \right) \left( 1 + \text{erf} (u) \right) + \frac{2u}{\sqrt{\pi}} \exp(-u) \right]  \text{,}
\end{align}
and $r$ satisfies
\begin{align} \label{u-def}
r \sqrt{\pi} \exp\left(\frac{1}{r}\right) \left( 1 - \text{erf} \left( \frac{1}{\sqrt{r}} \right) \right)  = \tau \left[  2\sqrt{r} - 2\sqrt{\pi} \exp\left(\frac{1}{r}\right) \left( 1 - \text{erf} \left( \frac{1}{\sqrt{r}} \right) \right) \right] \text{.}
\end{align}
Equations (\ref{r-def}) and (\ref{u-def}) are derived in Appendix \ref{integrals-Tdep}.
The positive constant $u$ can be determined iteratively for a given $\tau \leqslant 1$, and the positive constant $r$ can be determined iteratively for a given $\tau > 1$.
The fluid velocity in the $z$ direction at the magnetic presheath entrance, $u_{z\infty}$, is evaluated in Appendix \ref{integrals-Tdep}, giving
\begin{align} \label{flow-u}
\frac{ u_{z\infty} }{v_{\text{t,i}} } =  \frac{  u \left( 3+ 2u^2\right) \left( 1 + \text{erf} \left( u \right) \right) + \frac{2\exp(-u^2)}{\sqrt{\pi}} \left(1+u^2\right) }{  \left( 1 + 2u^2 \right) \left( 1 + \text{erf}(u) \right) + \frac{2u}{\sqrt{\pi}}  \exp(-u^2) }   & \text{ for } \tau \leqslant 1 \text{.}
\end{align}
and
\begin{align} \label{flow-r}
\frac{ u_{z\infty} }{v_{\text{t,i}} } =  \frac{2}{\sqrt{r} \sqrt{\pi}} \frac{ r - \exp\left(\frac{1}{r} \right)  E_1 \left( \frac{1}{r} \right) }{  2\sqrt{r} - 2\sqrt{\pi} \exp\left(\frac{1}{r}\right) \left( 1 - \text{erf} \left( \frac{1}{\sqrt{r}} \right) \right)  }   & \text{ for } \tau > 1 \text{.}
\end{align}
In equation (\ref{flow-r}), I have introduced the exponential integral,
\begin{align} \label{E1}
E_1(\xi) = \int_{\xi}^{\infty} \frac{\exp(-\eta)}{\eta} d\eta \text{.}
\end{align}

I proceed to write the distribution functions (\ref{f-infty-Tdep}) in the variables $\mu$ and $U$. 
Using the results of Appendix \ref{app-quasi-expansion}, the equations
\begin{align} \label{mu-infty-vx-vy}
\mu = \frac{v_x^2 + v_y^2 }{2\Omega}  
\end{align}
and
\begin{align} \label{U-infty-vx-vy-vz}
U = \Omega \mu + \frac{1}{2} v_z^2 \text{}
\end{align}
are valid at $x \rightarrow \infty$.
From equation (\ref{f-infty-Tdep}), (\ref{mu-infty-vx-vy}) and (\ref{U-infty-vx-vy-vz}), we can write
\begin{align} \label{F-infty}
F \left( \mu, U \right) =  \begin{cases}
\mathcal{N}  n_{\infty} \frac{8\left( U - \Omega \mu \right) }{\pi^{3/2} v_{\text{t,i}}^5}  \exp \left[ - \frac{2}{v_{\text{t,i}}^2}  \left(  \Omega \mu + \left( \sqrt{2\left(U - \Omega \mu \right)}  - u v_{\text{t,i}} \right)^2 \right)  \right]   & \text{ for } \tau \leqslant 1 \text{,} \\
\mathcal{N}  n_{\infty}  \frac{ 8\left( U - \Omega \mu \right) }{  \pi^{3/2} v_{\text{t,i}}^3 \left( v_{\text{t,i}}^2 + 2r \left(U - \Omega \mu \right) \right)} \exp \left( - \frac{2U}{v_{\text{t,i}}^2} \right)   & \text{ for } \tau > 1 \text{.}
\end{cases}
\end{align} 

The boundary conditions in equation (\ref{f-infty-Tdep}) are only one possible set.
The ion distribution function entering the magnetic presheath is unknown in the absence of a kinetic solution in the bulk plasma or in the collisional presheath.

\section{Finite ion temperature} \label{sec-Tdep-finite}

\begin{figure} 
\centering
\includegraphics[width = 0.6\textwidth]{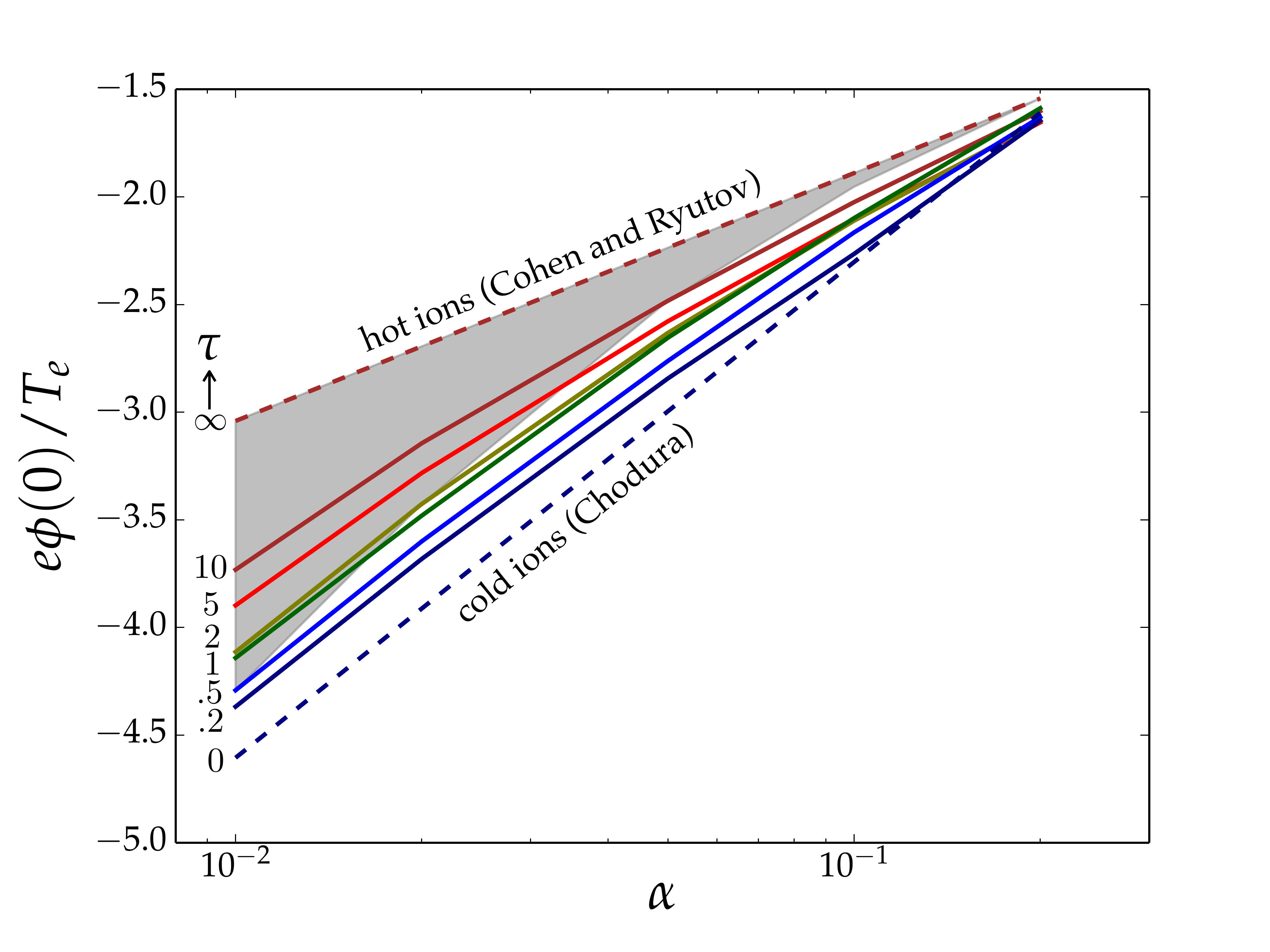} 
\caption[Electrostatic potential drop as a function of $\alpha$ and $\tau$]{The electrostatic potential drop across the magnetic presheath $\phi \left( 0 \right)$ is shown as a function of the angle $\alpha$ and the parameter $\tau$. 
The region where $ \alpha \lesssim \sqrt{\tau} \sqrt{m_e / m_i  } $, and therefore the ordering (\ref{ordering-angle}) breaks down, is shaded. }
\label{fig-phi-Tdep}
\end{figure}

\begin{figure}
\centering
\includegraphics[width=0.6\textwidth]{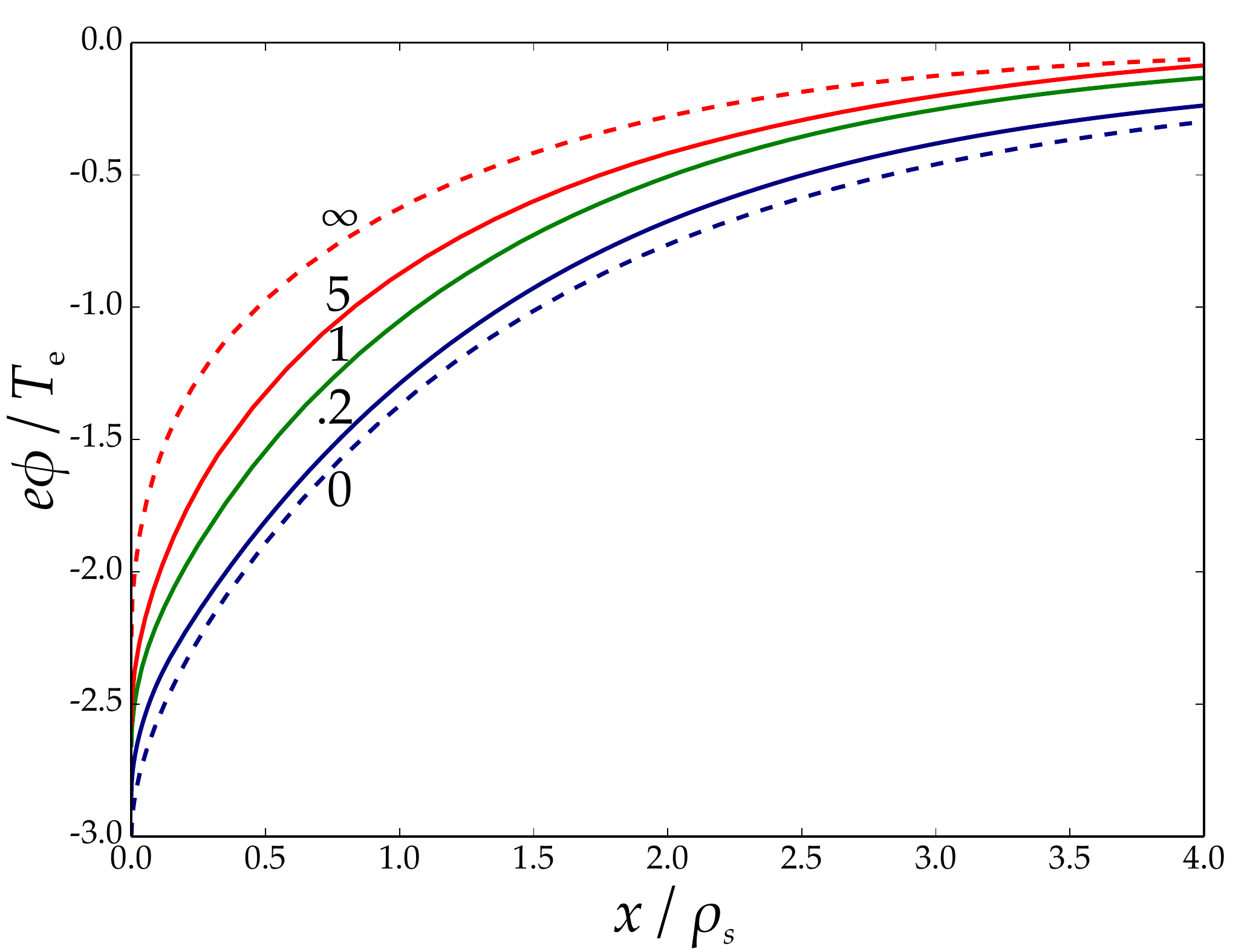} 
\caption[Electrostatic potential profiles for $\alpha = 0.05$ at sevaral values of $\tau$]{Electrostatic potential for $\alpha = 0.05$ at different values of $\tau$, marked on the curves. }
\label{fig-phiprofile-Tdep}
\end{figure}

The electrostatic potential drop across the magnetic presheath is shown in Figure \ref{fig-phi-Tdep} as a function of $\alpha$ and $\tau$, for the range $0.01 \leqslant \alpha \leqslant 0.1$ and $0.2 \leqslant \tau \leqslant 10$.
The numerical results approaching $\tau = 0.2$ and $\tau = 10$ are consistent with the analytical results obtained in the cold ion and cold electron limits, treated in sections \ref{sec-Tdep-cold} and \ref{sec-Tdep-hot}, respectively.
The shaded region is where we expect the assumption of an electron-repelling wall not to be suitable for Deuterium ions, $\alpha \lesssim \sqrt{\tau} \sqrt{m_{\text{e}}/m_{\text{e}}} \sim 0.02 \sqrt{\tau} $.
Considering the unshaded region in Figure \ref{fig-phi-Tdep}, the potential drop with finite ion temperature is up to $10-15\%$ smaller than the cold ion ($\tau =0$) potential drop. 
A kinetic model for electrons, as well as ions, would be necessary to study the transition from the unshaded to the shaded region.
Moreover, in the shaded region, the wall may be an ion-repelling wall and hence it may be that $\phi(0)>0$. 
Therefore, one would also need to relax some of the assumptions that were used to obtain the ion distribution function in chapter \ref{chap-dens}.
Specifically, the backward-travelling ions reflected in the magnetic presheath or Debye sheath must be considered. 
Finally, note that for a sufficiently large ion current towards the wall, the wall remains electron-repelling and the potential drop shown in Figure \ref{fig-phi-Tdep} will be correct for values of $\alpha / \sqrt{\tau}$ smaller than $0.02$.

\begin{figure}
\centering
\includegraphics[width=0.45\textwidth]{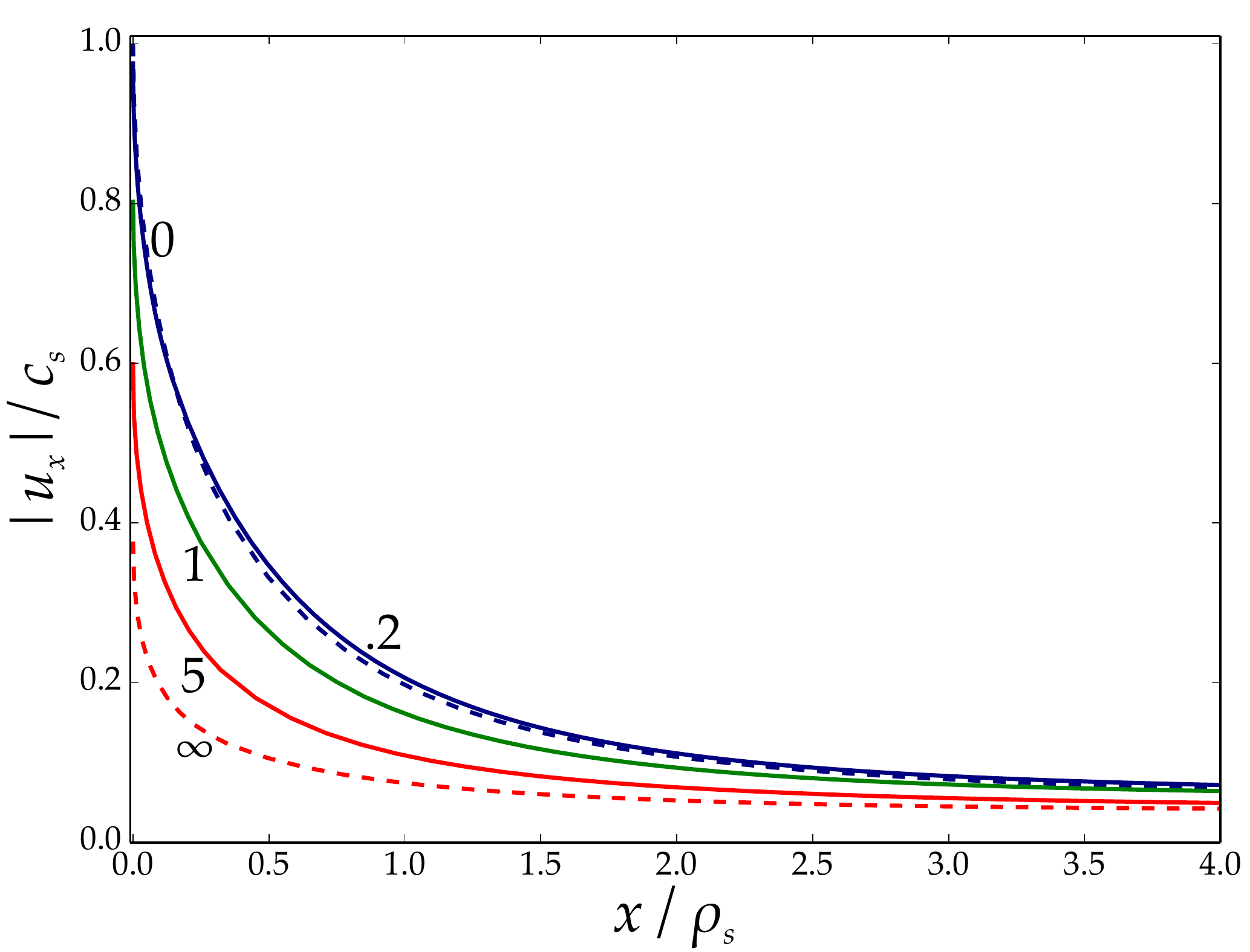}
\includegraphics[width=0.45\textwidth]{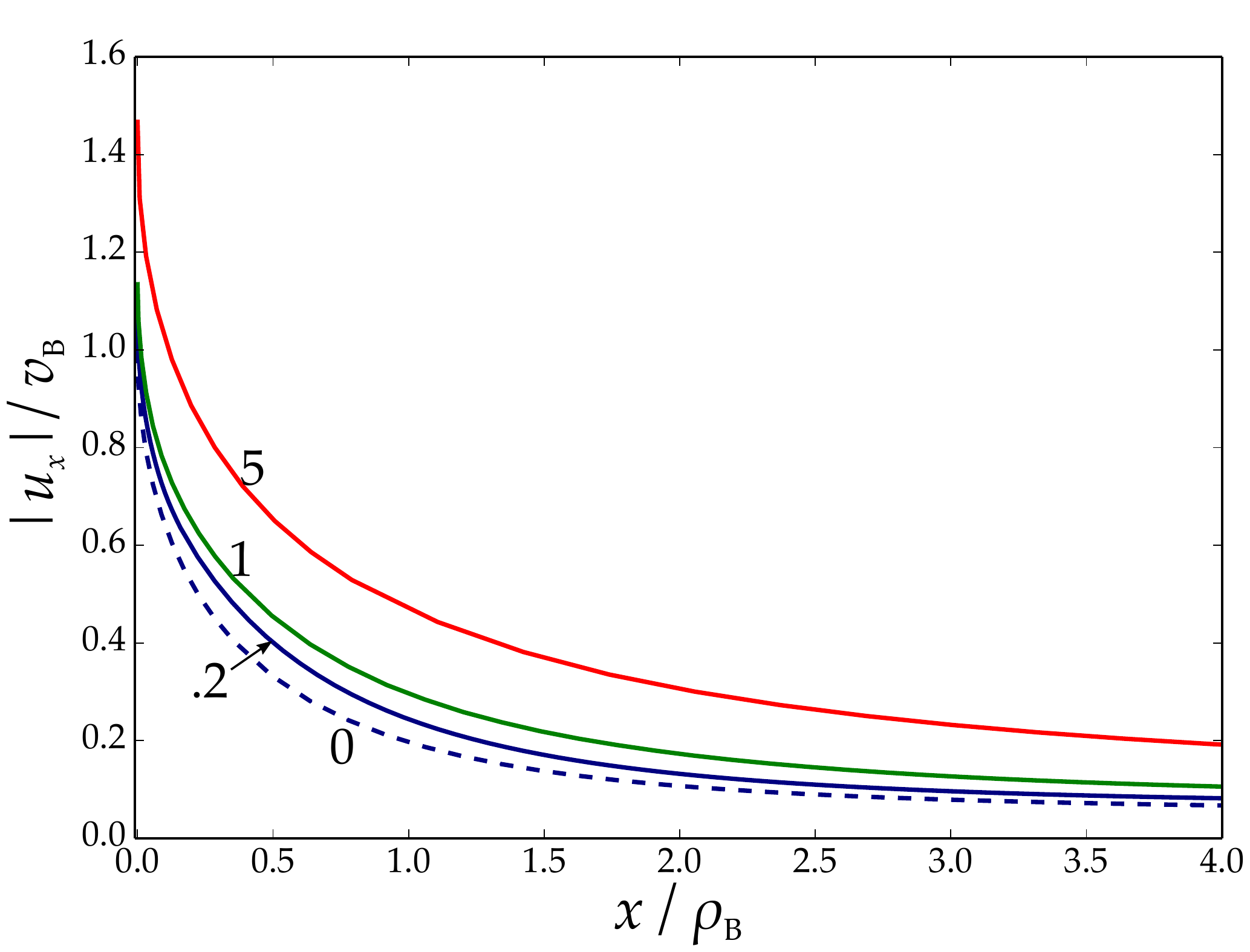} 
\caption[Flow profiles for $\alpha = 0.05$ at different values of $\tau$]{Flow profiles for $\alpha = 0.05$ at different values of $\tau$, marked on the curves. 
The flow speed $|u_x |$ is normalized to $c_s$ on the left and to $v_{\text{B}}$ on the right. 
The distance from the wall is normalized to $\rho_{\text{s}}$ on the left and to $\rho_{\text{B}}$ on the right.  
}
\label{fig-ux-Tdep}
\end{figure}

The electrostatic potential profiles are shown in Figure~\ref{fig-phiprofile-Tdep}. 
The numerical profiles lie in between the known limits $\tau = 0$ and $\tau = \infty$.
The flow profiles are shown in Figure~\ref{fig-ux-Tdep} for different values of $\tau$.
The ion fluid velocity is normalized to the sound speed $c_{\text{s}}$ on the left diagram, and to the Bohm speed $v_{\text{B}}$ on the right diagram.
The numerical profiles are consistent with the low and high temperature limits $\tau = 0$ and $\tau = \infty$. 
The fluid velocity normal to the wall is always simultaneously sub-sonic and super-Bohm, $v_{\text{B}} \leqslant |u_x(0)| \leqslant c_{\text{s}}  $.
At $\tau = 0$, the sound speed and the Bohm speed coincide and
the cold ion Bohm condition, $|u_x(0)| = v_{\text{B}} = c_{\text{s}}$, is satisfied.

\begin{figure}
\centering
\includegraphics[width = 0.7\textwidth]{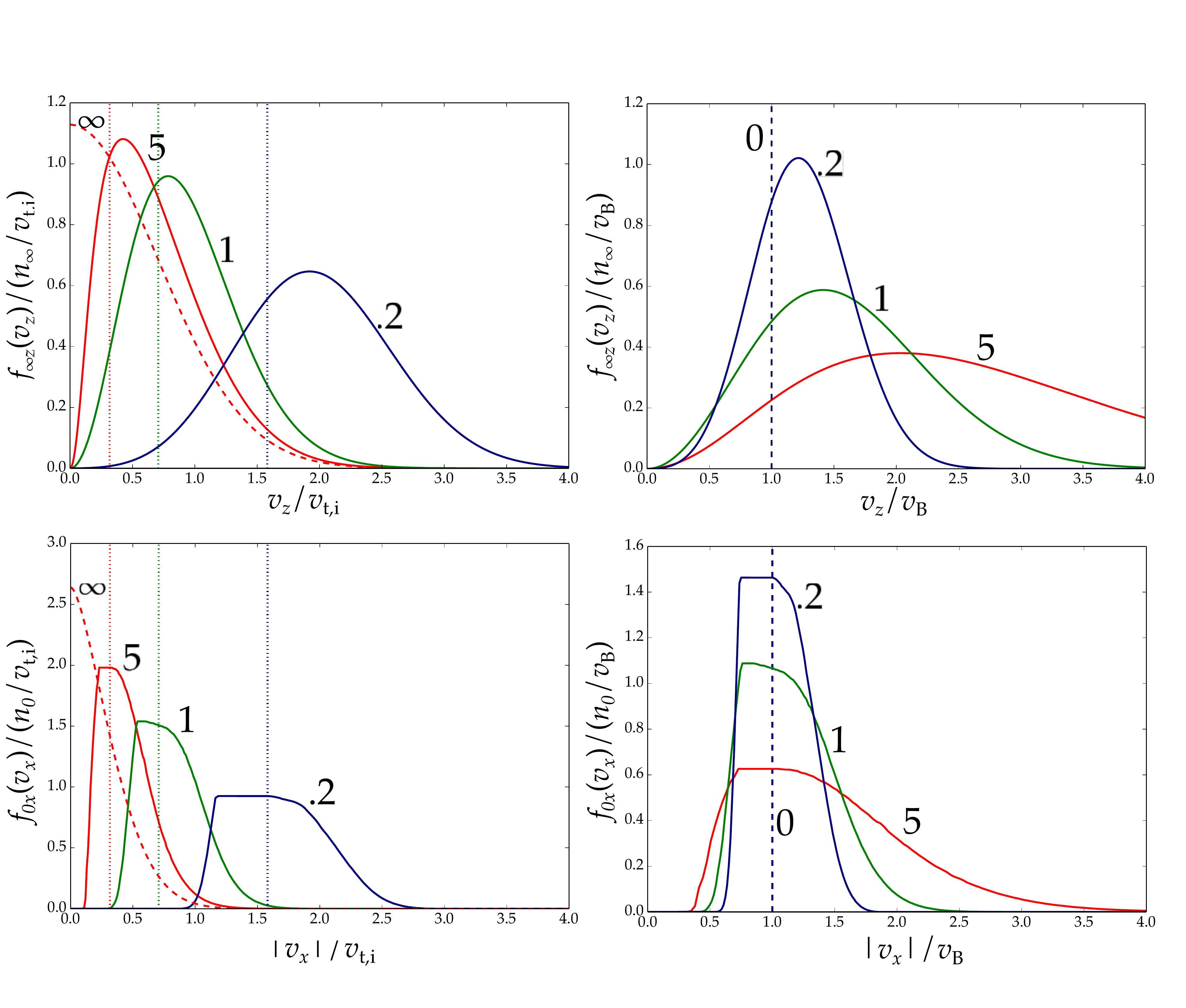} 
\caption[1D distribution function at the wall and at the magnetic presheath entrance]{The distributions of the component $v_z$ of the ion velocity at the magnetic presheath entrance $x \rightarrow \infty$ (top) and the component $v_x$ of the velocity at the Debye sheath entrance $x=0$ (bottom) are shown for $\alpha = 0.05$ 
for three different values of the parameter $\tau$, labelled next to the corresponding curve. 
The velocities are normalized to $v_{\text{t,i}}$ on the left diagrams and to $v_{\text{B}}$ on the right diagrams.
Magnetized ions at the magnetic presheath entrance move parallel to the magnetic field. 
Hence, $v_z$ is responsible for the flow of ions to the wall.
At the Debye sheath entrance, the ion flow towards the wall is determined by $|v_x|$. 
The red dashed lines on the left diagrams are the distribution functions in the limit $\tau \rightarrow \infty$.
The blue vertical dashed lines on the right diagrams are the cold ion distribution functions, $\tau = 0$.
}
\label{fig-f0x}
\end{figure}

For different values of temperature, we plot the functions $f_{0x} (v_x ) $ (defined in equation (\ref{f0x})) and $f_{\infty z} (v_z) = \int_{-\infty}^{\infty} \int_{-\infty}^{\infty} f_{\infty } (\vec{v}) dv_y dv_x$ in Figure~\ref{fig-f0x}.
The kinetic Bohm condition (\ref{Bohm-kinetic-marginal}) is numerically satisfied for all distribution functions.
This is a property of the self-consistent solution of equation (\ref{quasineutrality}), as shown in section \ref{sec-KMPS-Bohm} of chapter \ref{chap-KMPS}.
The distribution $f_{0x} (v_x )$ is found to be narrower than $f_{\infty z} (v_z)$ for all values of $\tau$.

\begin{figure}
\centering
\includegraphics[width = 0.6\textwidth]{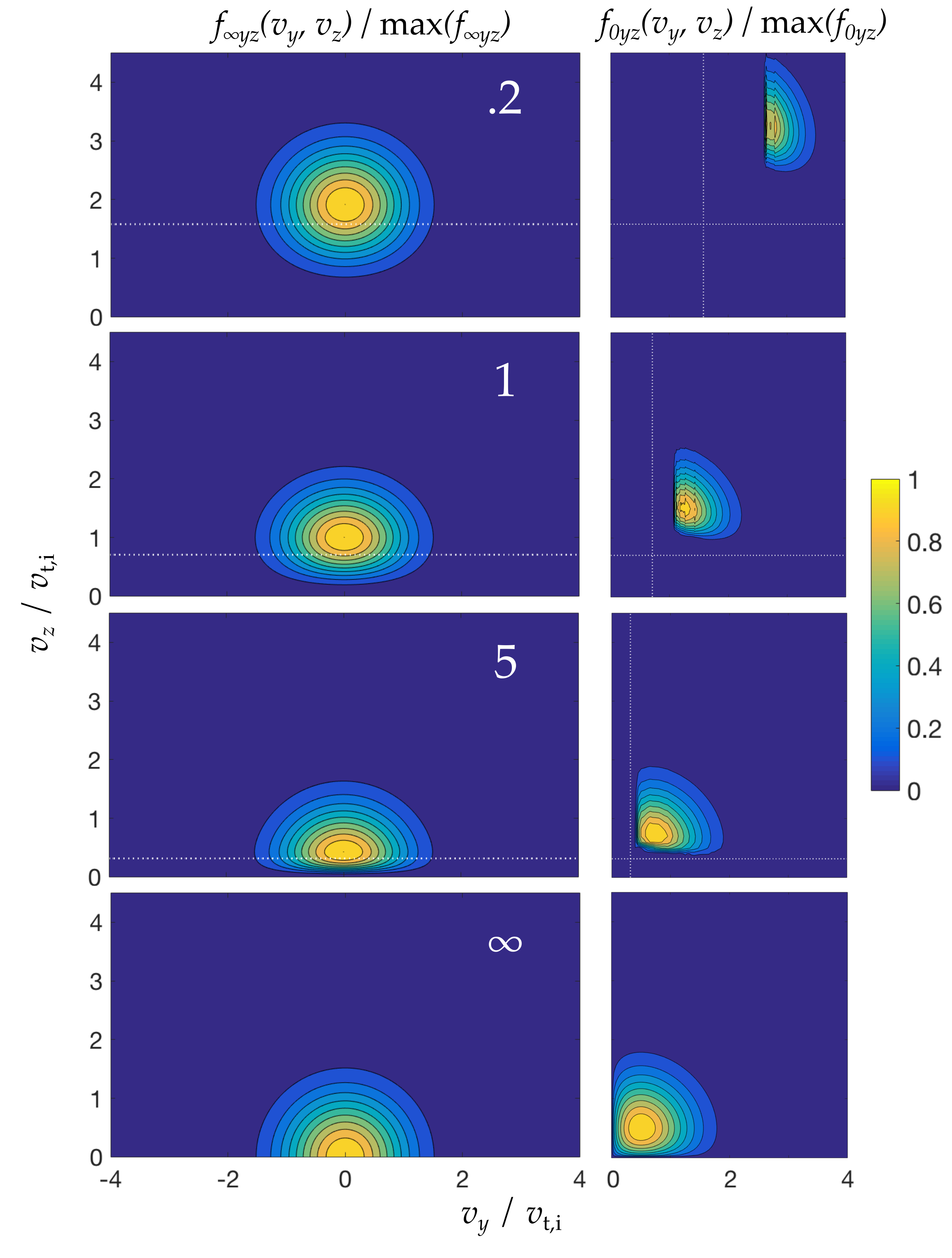} 
\caption[2D distribution function at the Debye and magnetic presheath entrance]{The ion distribution functions $f_{\infty yz}(v_y, v_z)$ (left) and $f_{0yz} (v_y, v_z)$ (right) for $\alpha = 0.05$ and, from top to bottom, for $\tau = 0.2$, $\tau = 1$, $\tau = 5$ and $\tau = \infty$ (see section \ref{sec-Tdep-hot}). 
The Bohm speed $v_{\text{B}}/v_{\text{t,i}} = 1/ \sqrt{2\tau} $ is marked as a horizontal line in all panels, and also as a vertical line on the right panels. 
}
\label{fig-f0yz-Tdep}
\end{figure}

In Figure \ref{fig-f0yz-Tdep}, we plot the functions $f_{\infty yz}(v_y, v_z)$ and $f_{0yz}(v_y, v_z)$. 
The ions have very large tangential velocities at $x=0$ (compared with $x=\infty$) due to the large increase in the $y$-component of the velocity, related to the $\vec{E} \times \vec{B}$ drift acquired by the ion orbit in the magnetic presheath. 

\section{Zero ion temperature limit} \label{sec-Tdep-cold}


When $T_{\text{i}} = 0$, the ion equations of motion become the equations for the ion fluid velocity. 
Letting $\vec{u} = (u_x, u_y, u_z)$ be the ion fluid velocity, the momentum equation for the ion species is, in components,
\begin{align} \label{x-momentum}
u_x u_x' = - \frac{\Omega \phi'}{B} + \Omega u_y \cos \alpha \text{,}
\end{align}
\begin{align} \label{y-momentum}
u_x u_y' = -  \Omega u_x \cos \alpha - \Omega u_z \sin \alpha \text{,}
\end{align}
\begin{align} \label{z-momentum}
u_x u_z' =  \Omega u_y \sin \alpha \text{.}
\end{align}
Here, $'$ indicates differentiation with respect to $x$.
The momentum equations (\ref{x-momentum})-(\ref{z-momentum}) follow from the particle equations of motion (\ref{vx-EOM-exact})-(\ref{vz-EOM-exact}) in section (\ref{subsec-traj-orbitparameters}), by replacing the ion velocity $\vec{v}$ with the ion fluid velocity $\vec{u}$ and using $u_x = \dot{x}$ to write $\dot{\vec{u}} = u_x \vec{u}'$ (thus changing the time derivative of every velocity component to a spatial derivative).
The fluid velocity at $x\rightarrow \infty$ is denoted $\vec{u}_{\infty} $.
Equations (\ref{x-momentum})-(\ref{z-momentum}) are solved using the boundary conditions of a flow that is aligned with the magnetic field,
\begin{align} \label{bc-infty}
\vec{u}_{\infty}  =  \begin{pmatrix} 
u_{x\infty} \\
 u_{y\infty} \\
 u_{z\infty} 
 \end{pmatrix} =
  \begin{pmatrix} 
- u_{\infty} \sin \alpha  \\
 0 \\ u_{\infty} \cos \alpha 
 \end{pmatrix}  \text{.}
\end{align} 
The quasineutrality equation, together with the electron Boltzmann distribution and the ion continuity equation, gives the equation
\begin{align} \label{ux-fluid}
u_x = u_{x\infty} \exp\left( - \frac{e\phi}{T_{\text{e}}} \right) \text{.}
\end{align}
The derivation of (\ref{ux-fluid}) is identical to that of (\ref{ux}), but uses $u_x = u_{x\infty}$ at $x \rightarrow \infty$.
Then, differentiating (\ref{ux-fluid}) and rearranging, one obtains
\begin{align} \label{phi'(ux)}
\phi' = - \frac{T_{\text{e}}}{e} \frac{u_x'}{u_x} \text{.}
\end{align}

I follow Riemann's derivation of a first order differential equation for $u_x$ from equations (\ref{x-momentum})-(\ref{z-momentum}) and (\ref{phi'(ux)}) \cite{Riemann-1994}. 
The original derivation is due to Chodura \cite{Chodura-1982}.
Equation (\ref{x-momentum}) can be rearranged to obtain 
\begin{align} \label{uy-inter}
 u_y = \frac{ \phi'}{B  \cos \alpha  } + \frac{ u_xu'_x }{\Omega \cos \alpha }  \text{.}
\end{align}
By substituting equation (\ref{phi'(ux)}) into equation (\ref{uy-inter}), one obtains
\begin{align} \label{uy(ux)}
 u_y = - \frac{ v_{\text{B}}^2 u'_x }{u_x \Omega \cos \alpha  } + \frac{ u_xu'_x }{\Omega \cos \alpha }  \text{,}
\end{align}
which can be substituted in equation (\ref{z-momentum}) to obtain
\begin{align} \label{uz-inter}
u_z' = \tan \alpha \left( 1- \frac{v_{\text{B}}^2}{u_x^2} \right) u_x' \text{.}
\end{align}
Using the boundary conditions in (\ref{bc-infty}), equation (\ref{uz-inter}) integrates to
\begin{align} \label{uz(ux)}
u_z = u_x \tan \alpha + \frac{ v_{\text{B}}^2 \tan \alpha }{u_x } + \frac{u_{\infty}}{\cos \alpha}  +\frac{ v_{\text{B}}^2 }{u_{\infty}  \cos \alpha } \text{.}
\end{align}

Analogously to the procedure for deriving conservation of energy for a single particle, one can add equations (\ref{x-momentum})-(\ref{z-momentum}) multiplied by $u_x$, $u_y$ and $u_z$ respectively, and divide by $u_x$ again to obtain
\begin{align} \label{energy-equation}
\frac{1}{2} u_{\infty}^2 = \frac{1}{2} u_x^2  + \frac{1}{2} u_y^2 + \frac{1}{2} u_z^2 + \frac{\Omega \phi}{B} \text{,}
\end{align}
where we used $\phi (x) =0$ at $ x \rightarrow \infty$ and the boundary conditions (\ref{bc-infty}). 
Substituting equations (\ref{uy(ux)}) and (\ref{uz(ux)}) into the energy equation (\ref{energy-equation}), one obtains the differential equation
\begin{align} \label{udiff-Riemann}
\frac{ \left( u_x^2 - v_{\text{B}}^2 \right)^2 }{ u_x^2 \Omega^2 \cos^2 \alpha } u_x'^2  = f(u_x )  \text{,}
\end{align}
where 
\begin{align} \label{f-Riemann}
f(u_x ) = u_{\infty}^2 - u_x^2 - 2 v_{\text{B}}^2 \ln \left(  \frac{ u_{\infty} \sin \alpha }{-u_x} \right) -  \left( \tan \alpha \left( u_x + \frac{v_{\text{B}}^2 }{u_x} \right) + \frac{u_{\infty} + \frac{v_{\text{B}}^2}{u_{\infty}}}{\cos \alpha}  \right)^2 \text{.}
\end{align}
Equation (\ref{udiff-Riemann}) is solved by realizing that the singularity occurring when $u_x = - v_{\text{B}}$ corresponds to the Debye sheath entrance, $x=0$ \cite{Chodura-1982, Riemann-1994}.
Then the position $x$ is the definite integral 
\begin{align}
x = \int_{-v_{\text{B}}}^{u_x}  \frac{ \eta^2 - v_{\text{B}}^2  }{ \eta \Omega \cos \alpha } \left[ f(\eta ) \right]^{-1/2} d\eta   \text{.}
\end{align}
This equation can be inverted to obtain $u_x (x)$, and then equation (\ref{ux-fluid}) can be used to obtain $\phi(x)$.

Taking $\alpha \ll 1$, the direction of the fluid velocity at $x \rightarrow \infty$ is, to lowest order, along the $z$-axis.
Hence, $u_{\infty} \simeq u_{z\infty}$.
At the Debye sheath entrance, the Bohm condition is satisfied and thus $u_x \sim v_{\text{B}}$ close to $x=0$.
Therefore, the size of $u_x$ changes from $\sim \alpha v_{\text{B}}$ to $\sim v_{\text{B}}$ from $ x\rightarrow \infty$ to $x=0$.
For $u_x  \sim \alpha  v_{\text{B}}$, the lowest order terms in equations (\ref{udiff-Riemann}) and (\ref{f-Riemann}) give
\begin{align} \label{udiff-far} 
  \frac{v_{\text{B}}^2}{\Omega^2}  \left( \frac{u_x'}{u_x} \right)^2 & = - \frac{v_{\text{B}}^2}{u_{z\infty}^2} - 2 + 2 \ln \left( \frac{ - u_x }{\alpha u_{z\infty}} \right)   - \frac{\alpha^2 v_{\text{B}}^2  }{u_x^2}  + \frac{2 \alpha v_{\text{B}}^2}{-u_x u_{z\infty} } +  \frac{2 \alpha u_{z\infty} }{-u_x}     \text{.} 
 \end{align}
 For $u_x \sim v_{\text{B}}$, the lowest order terms in  (\ref{udiff-Riemann}) and (\ref{f-Riemann}) give
  \begin{align} \label{udiff-near} 
 \frac{  \left( v_B^2 - u_x^2 \right)^2 }{ v_{\text{B}}^2 u_x^2 } \frac{ u_x'^2 }{\Omega^2 } & =   - \frac{ u_x^2 }{v_{\text{B}}^2 }  - \frac{v_{\text{B}}^2}{u_{z\infty}^2} -2    +  2\ln \left( \frac{-u_x}{ \alpha u_{z\infty}  } \right)     \text{.} 
\end{align}
Equations (\ref{udiff-far}) and (\ref{udiff-near}) both result in the same equation to lowest order in $\alpha$ in the limit $\alpha \ll u_x / v_{\text{B}} \ll 1$,
\begin{align} \label{udiff-int} 
 \frac{ v_B^2 }{ u_x^2  } \frac{ u_x'^2 }{\Omega^2 }  =  - \frac{v_{\text{B}}^2}{u_{z\infty}^2} -2 + 2\ln \left( \frac{-u_x}{ \alpha u_{z\infty}  } \right)    \text{.} 
\end{align}
A lowest order equation in $\alpha$ valid for all values of $u_x$ is
\begin{align} \label{udiff-uni-marginal-general}
 \frac{  \left( v_B^2 - u_x^2 \right)^2 }{ v_{\text{B}}^2 u_x^2 } \frac{ u_x'^2 }{\Omega^2 }  = f_{\alpha} (u_x )     \text{,}
\end{align}
with
\begin{align} \label{f-alpha}
f_{\alpha} (u_x ) =  - \frac{ \left( u_x + \alpha u_{z\infty} \right)^2}{ v_{\text{B}}^2} - \frac{v_{\text{B}}^2}{u_{z\infty}^2} - 2 + 2 \ln \left( \frac{ - u_x }{\alpha u_{z\infty}} \right)   - \frac{\alpha^2 v_{\text{B}}^2  }{u_x^2}   + \frac{2 \alpha v_{\text{B}}^2}{-u_{z\infty} u_x }  + \frac{2 \alpha u_{z\infty}}{-u_x } \text{.}
\end{align}
Equations (\ref{udiff-uni-marginal-general}) and (\ref{f-alpha}) are obtained by expanding (\ref{udiff-Riemann}) and (\ref{f-Riemann}) in $\alpha$ and keeping enough lowest order terms such that the conditions $f_{\alpha}(-\alpha u_{z\infty}) = f_{\alpha}'(-\alpha u_{z\infty}) = 0$ are satisfied.
These conditions correspond to the electrostatic potential and ion flow tending to a constant value at $x \rightarrow \infty$ \cite{Riemann-1994}.

By imposing the equality form of the fluid Chodura condition, $u_{z\infty} = v_{\text{B}}$, in (\ref{udiff-uni-marginal-general}) and (\ref{f-alpha}), one obtains
\begin{align} \label{udiff-uni-marginal}
 \frac{  \left( v_B^2 - u_x^2 \right)^2 }{ v_{\text{B}}^2 u_x^2 } \frac{ u_x'^2 }{\Omega^2 }  = - \frac{ \left( u_x + \alpha v_{\text{B}} \right)^2}{ v_{\text{B}}^2} - 3 + 2 \ln \left( \frac{ - u_x }{\alpha v_{\text{B}}} \right)   - \frac{\alpha^2 v_{\text{B}}^2  }{u_x^2}  - \frac{4 \alpha v_{\text{B}}}{u_x }   \text{.}
\end{align}
The solution of (\ref{udiff-uni-marginal}) is 
\begin{align} \label{x-ux-lowest}
x = \int_{-v_{\text{B}}}^{u_x}  \frac{ \eta^2 - v_{\text{B}}^2  }{ \eta \Omega } \left[ - \frac{ \left( \eta + \alpha v_{\text{B}} \right)^2}{ v_{\text{B}}^2} - 3 + 2 \ln \left( \frac{ - \eta }{\alpha v_{\text{B}}} \right)   - \frac{\alpha^2 v_{\text{B}}^2  }{\eta^2}  - \frac{4 \alpha v_{\text{B}}}{\eta } \right]^{-1/2} d\eta   \text{.}
\end{align}
The ion flow $u_x$ is obtained by inverting (\ref{x-ux-lowest}).
The electrostatic potential profile is then obtained by using the equation relating flow and potential, (\ref{ux}),
\begin{align}
\phi (x) = - \frac{T_{\text{e}}}{e} \ln \left( \frac{- u_x (x)}{\alpha v_{\text{B}} } \right) \text{.}
\end{align}
With the flow solution to (\ref{udiff-uni-marginal}) marginally satisfying Bohm's condition at the Debye sheath entrance (\cite{Riemann-review, Geraldini-2018}), $u_{x}(0) = - v_{\text{B}}$, the potential drop is 
\begin{align} \label{phidrop-tau0}
 \phi (0)  = - \frac{T_{\text{e}}}{e }  \ln \left( \frac{1}{ \alpha } \right) \text{.}
\end{align}

\begin{figure}
\centering
\includegraphics[width = 0.6\textwidth]{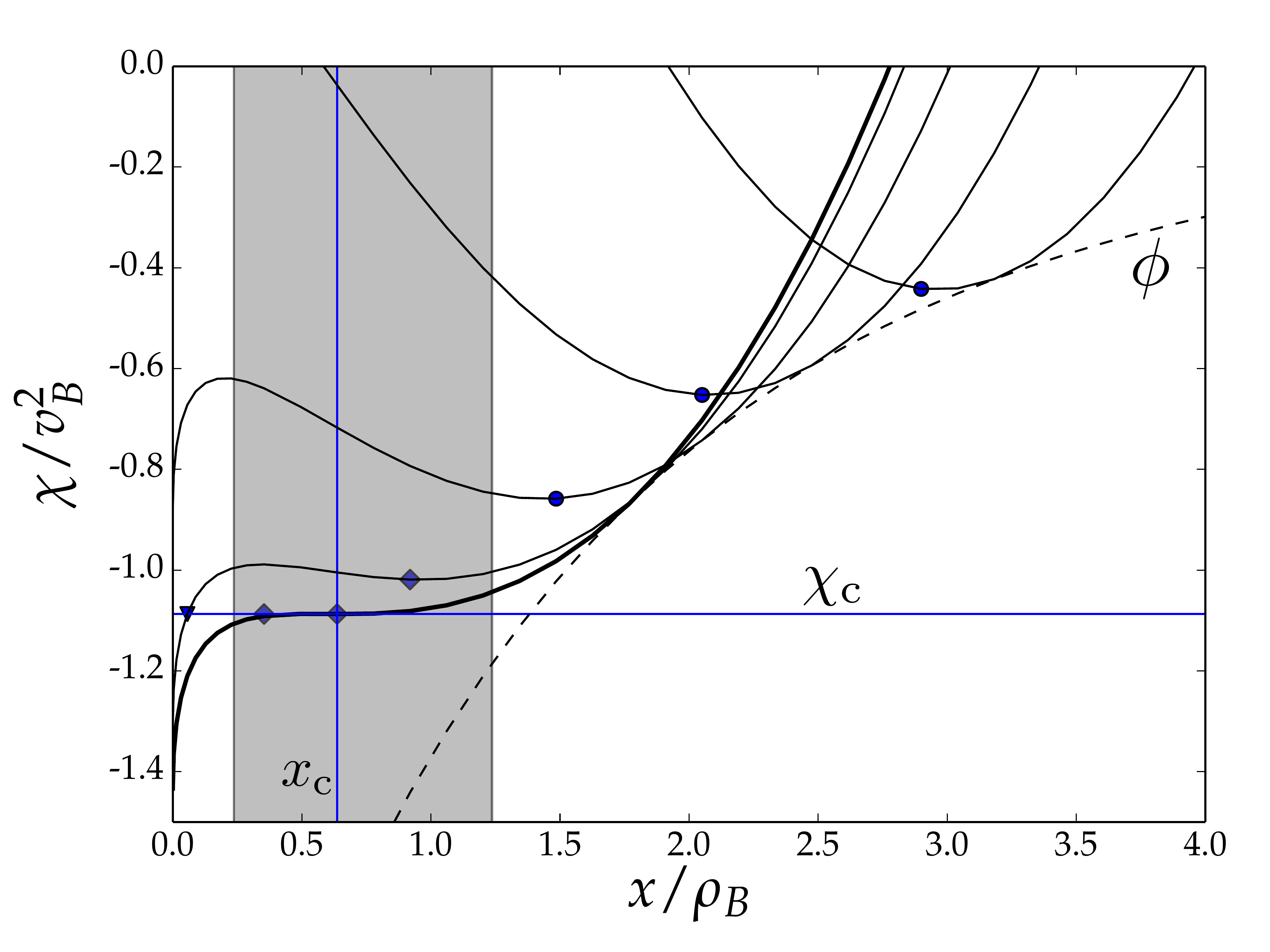}
\caption[Effective potential curves of small-temperature ions]{Some effective potential curves $\chi(x, \bar{x})$, corresponding to the electrostatic potential profile $\phi(x)$ (dashed line) that solves the magnetic presheath for $\tau = 0$ and $\alpha = 0.05$, are shown for a range of values of $\bar{x}$ (where the dashed line intersects the solid lines, $\chi(x, \bar{x}) = \phi(x)$ hence $ x = \bar{x}$). 
The position of an ion with $\mu = 0$ corresponding to each value of $\bar{x}$ is shown by a marker: (i) a circle if the ion is in a closed orbit to the right of the shaded region, $x \gg x_{\text{c}}$; (ii) a triangle if the ion is in an open orbit to the left of the shaded region, $x \ll x_{\text{c}}$; (iii) a diamond if the ion is in the shaded region, $x\sim x_{\text{c}}$, transitioning from a closed to an open orbit.
In the shaded transition one finds $\chi''(x, \bar{x}) \simeq  0$ and $\chi'''(x, \bar{x}) = \phi'''(x, \bar{x}) \simeq 0$.
}
\label{fig-chiChodura}
\end{figure}

In section \ref{sec-KMPS-Bohm}, it was shown that there cannot be any type I orbits, and hence $\phi(x) - \phi(0) \propto \sqrt{x}$ for $x \rightarrow 0$.
As a result, the critical point $x_{\text{c}}$ (defined in equation (\ref{xbarc})) must be non-zero, and there exists a value of orbit position $\bar{x}_c$ for which the effective potential $\chi(x, \bar{x}_c)$ has a point of inflection with zero derivative at $x_c$.
For each $\bar{x}$, closed orbits with $\mu \sim \tau v_{\text{B}}^2 / \Omega $ must lie in a region surrounding the effective potential minimum, and the size of this region is small for $\tau \ll 1$ and $x \gg x_{\text{c}}$.
Since minima of the effective potential only exist for $x > x_{\text{c}}$, most closed orbits must lie in the region $x \gg x_{\text{c}}$, while only open orbits reach the region $x\ll  x_{\text{c}}$, as shown schematically in Figure~\ref{fig-chiChodura}.
In Figure~\ref{fig-chiChodura}, a set of effective potential curves corresponding to the solution of the fluid equation (\ref{x-ux-lowest}) for $\alpha = 0.05$ are shown.
It can be seen that the effective potential curves corresponding to minima near $x_{\text{c}}$ are very flat.
Hence, for $\tau \ll 1$, the size of closed orbits approaching $x_{\text{c}}$ from large $x$ grows and can become of the order of the system scale length, $\rho_{\text{B}}$. 

From this discussion, the quasineutrality equation (\ref{quasineutrality}) evaluated for $\tau \ll 1$ leads to three regions of interest:
$x\gg x_{\text{c}}$, $x \sim x_{\text{c}}$ and $x\ll x_{\text{c}}$.
In subsection \ref{subsec-cold-kinetic-closed}, I show that the quasineutrality equation (\ref{quasineutrality}) for $x \gg x_{\text{c}}$ reduces to equation (\ref{udiff-far}), which is the fluid differential equation obtained for $u_x \sim \alpha v_{\text{B}}$.
In subsection \ref{subsec-cold-kinetic-open}, I show that the quasineutrality equation for $x \ll x_{\text{c}}$ reduces to the form in equation (\ref{udiff-near}), which is the fluid differential equation obtained for $u_x \sim  v_{\text{B}}$.
In subsection \ref{subsec-cold-kinetic-intermediate}, I discuss the intermediate region $x \sim x_{\text{c}}$. 

\subsection{Closed orbit region ($x\gg x_{\text{c}}$)} \label{subsec-cold-kinetic-closed}

At large values of $x$ the ion density is dominated by closed orbits.
For $\tau \ll 1$, the distribution function (\ref{f-infty-Tdep}) is clustered in a thin region a few $v_{t,i}$ from $v_z = u v_{\text{t,i}} \simeq  u_{z\infty} \gg v_{\text{t,i}}$.
In terms of $\mu$ and $U$, equation (\ref{F-infty}) reduces to a very thin Maxwellian for $\tau \rightarrow 0$,
\begin{align} \label{f-infty-smalltau}
F \left( \mu, U \right) = \frac{1}{2\pi^{3/2}\tau^{3/2}} \left( \frac{m_{\text{i}}}{2\tau T_{\text{e}}} \right)^{3/2} & \exp \left( - \frac{m_{\text{i}} \Omega \mu}{2 \tau T_{\text{e}} } \right) \nonumber \\ 
& \times \exp\left(  - \frac{ m_{\text{i}} \left( \sqrt{2\left( U - \Omega \mu \right)} -  u_{z\infty} \right)^2 }{2 \tau T_{\text{e}} } \right)    \text{.} 
\end{align}
For most of this section, we consider the general case $u_{z\infty} \geqslant v_{\text{B}}$, but the numerical results are all based on equation (\ref{F-infty}) and hence $u_{z\infty} = v_{\text{B}}$.
Expanding the square root in the exponential around $\sqrt{ 2\left( U - \Omega \mu \right) } = u_{z\infty}$ gives
\begin{align}
\sqrt{ 2\left( U - \Omega \mu \right) } = \sqrt{ u_{z\infty}^2 + 2\left( U - \Omega\mu - \frac{1}{2}u_{z\infty}^2 \right) } \simeq u_{z\infty} \left (1 + \frac{ \left( U - \Omega \mu - \frac{1}{2} u_{z\infty}^2 \right) }{u_{z\infty}^2} \right) \text{.}
\end{align}
Hence, the distribution function is
\begin{align} \label{f-infty-smalltau-2}
F \left( \mu, U \right)  =  \left( \frac{m_{\text{i}}}{2\tau T_{\text{e}}  \pi } \right)^{3/2}  \exp \left( - \frac{m_{\text{i}} \Omega \mu}{ \tau T_{\text{e}} } \right) \exp\left(  - \frac{ m_{\text{i}} \left( U - \Omega \mu  -  \frac{1}{2} u_{z\infty}^2 \right)^2 }{2 \tau T_{\text{e}} u_{z\infty}^2} \right)    \text{.} 
\end{align}
Considering that $e\phi \sim T_{\text{e}}$, I order $\Omega \mu \sim U \sim T_{\text{e}} / m_{\text{i}}$, thus when taking $\tau \rightarrow 0$ equation (\ref{f-infty-smalltau}) is equivalent to the product of two Dirac delta functions
\begin{align} \label{F-cold}
F (\mu, U ) =  \frac{u_{z\infty}}{2\pi \Omega} \delta (  \mu ) \delta \left( U  - \frac{1}{2} u_{z\infty}^2 \right) \text{.}
 \end{align}

For $\tau \ll 1$ and $x\gg x_{\text{c}}$, the particle is confined close to the minimum, $x\simeq x_{\text{m}}$, and the effective potential looks like a parabola locally near the minimum.
Hence, 
\begin{align} \label{chi-parabola}
\chi(x, \bar{x}) - \chi_{\text{m}} (\bar{x}) \simeq \frac{1}{2} \chi'' (x) \left( x - x_{\text{m}} \right)^2 \text{.}
\end{align}
Using equation (\ref{stationary-points}), we have that 
\begin{align} \label{chi-minimum}
\Omega^2 \left( \bar{x} - x_{\text{m}} \right) = \frac{ \Omega \phi'( x_{\text{m}} ) }{ B } \text{.}
\end{align}
From $x\simeq x_{\text{m}}$, one finds $U_{\perp}   \simeq \chi \left( x, \bar{x} \right) \simeq \chi_{\text{m}} \left( \bar{x} \right) $, and using equation (\ref{chi-minimum}), it follows that
\begin{align}
U_{\perp}    \simeq \frac{1}{2} \left( \frac{  \phi'\left(x \right) }{B} \right)^2 +  \frac{ \Omega  \phi \left(x\right) }{B}  \text{.}
\end{align}
The adiabatic invariant is
\begin{align}  \label{mu-cold-1}
\mu  \simeq  \frac{\sqrt{2\left( U_{\perp} - \chi_{\text{m}} (\bar{x}) \right) }}{\pi} \int_{x_{\text{b}}}^{ x_{\text{t}} } \sqrt{  1 - \frac{\chi''(x_{\text{m}})  \left( x - x_{\text{m}} \right)^2 }{ 2\left( U_{\perp} - \chi (x_{\text{m}})  \right) }  } dx \text{,}
\end{align}
with $x_{\text{b}} = x_{\text{m}} - \sqrt{ 2\left( U_{\perp} - \chi_{\text{m}} (\bar{x}) \right) / \chi''(x_{\text{m}}) }$ and $x_{\text{t}} = x_{\text{m}} + \sqrt{ 2\left( U_{\perp} - \chi_{\text{m}} (\bar{x}) \right) / \chi''(x_{\text{m}}) }$.
Upon using $\chi''(x_{\text{m}}) \simeq \chi''(x)$, equation (\ref{mu-cold-1}) reduces to
\begin{align} \label{mu-Uperp-cold}
\mu \simeq \frac{U_{\perp} -\chi_{\text{m}}(\bar{x})}{\sqrt{ \chi''(x) } }  \text{.}
\end{align}
Inserting the distribution function of equation (\ref{f-infty-smalltau}) into the closed orbit integral (\ref{ni-closed}) and changing from $U_{\perp}$ to $\mu$ using equation (\ref{mu-Uperp-cold}) gives
\begin{align} \label{ni-closed-cold-1}
n_{\text{i,cl}}(x) =  \int_{\bar{x}_{\text{m}}(x)}^{\infty} \Omega d\bar{x} \int_0^{ \infty } \frac{  2  \sqrt{ \chi''(x) } d\mu}{\sqrt{2\left(\sqrt{ \chi''(x) } \mu + \chi_{\text{m}}(\bar{x}) - \chi (x, \bar{x}) \right)}} \nonumber \\ \times \int_{\Omega \mu}^{\infty} \frac{ F(\mu, U ) dU  }{\sqrt{2\left( U - \frac{1}{2} \left( \frac{ \phi' }{B} \right)^2 - \frac{\Omega \phi}{B} \right)}}  \text{.}
\end{align}
The upper limit of integration in $\mu$ is allowed to be $\infty$ for $x  \gg x_{\text{c}} $ because $F (\mu, U)$ is exponentially small for orbits with $\mu \gg \tau v_{\text{B}}^2 / \Omega $.

To take the integral in equation (\ref{ni-closed-cold-1}), I change variable from $\bar{x}$ to $x_{\text{m}}$.
From (\ref{chi-minimum}) one obtains the relation 
\begin{align} \label{dxbardxm}
\frac{\partial \bar{x} }{\partial x_{\text{m}} } = \frac{\chi''(x_{\text{m}})}{\Omega^2} \simeq 1 + \frac{\phi''(x)}{\Omega B} \text{.}
\end{align}
Using (\ref{chi-parabola}) and (\ref{dxbardxm}), equation (\ref{ni-closed-cold-1}) becomes
\begin{align} \label{ni-closed-cold-2}
n_{\text{i,cl}}(x) = \frac{\chi''(x) }{\Omega^2} \int_0^{\infty}   d\mu  \int_{\Omega\mu}^{\infty} \frac{ F \left( \mu, U \right) dU  }{\sqrt{2\left( U - \frac{1}{2} \left( \frac{ \phi'(x) }{B} \right)^2 - \frac{\Omega \phi (x)}{B} \right)}}   \nonumber \\  \times    \int_{x-\frac{ \sqrt{2\mu}}{(\chi''(x))^{1/4}}}^{x+\frac{ \sqrt{2\mu} }{(\chi''(x))^{1/4}}}  \frac{2  \sqrt{ \chi''(x) } dx_{\text{m}}}{\sqrt{2\left(\sqrt{ \chi''(x) } \mu - \frac{1}{2} \chi'' (x) \left( x - x_{\text{m}} \right)^2  \right)}} \text{.}
\end{align}
Note that the order of integration was changed in order to obtain (\ref{ni-closed-cold-2}).
The rightmost integral evaluates to $2\pi$, and thus equation (\ref{ni-closed-cold-2}) becomes
\begin{align} \label{ni-closed-cold-3}
n_{\text{i,cl}}(x) = \frac{2\pi\chi''(x) }{\Omega^2} \int_0^{\infty}   d\mu  \int_{\Omega\mu}^{\infty} \frac{ F \left( \mu, U \right) dU  }{\sqrt{2\left( U - \frac{1}{2} \left( \frac{ \phi'(x) }{B} \right)^2 - \frac{\Omega \phi (x)}{B} \right)}}   \text{.}
\end{align}
Equation (\ref{ni-closed-cold-2}) is valid for $x \gg x_{\text{c}} $.
To lowest order in $\tau \ll 1$, one can substitute equation (\ref{F-cold}) for the distribution function to obtain
\begin{align}
n_{\text{i,cl}}(x) =    \frac{n_{\infty} u_{z\infty} \chi'' (x) }{ \Omega}   \int_0^{\infty}  \delta (\mu ) d\mu    \int_{\Omega \mu}^{\infty} \frac{ \delta \left( U - \frac{1}{2} u_{z\infty}^2 \right) }{\sqrt{2\left( U - \frac{1}{2} \left( \frac{ \phi'(x) }{B} \right)^2 - \frac{ \Omega \phi(x) }{B}  \right)}}    dU \text{.}
\end{align}
Taking the straightforward integrals over Dirac delta functions, the density of closed orbits is
\begin{align} \label{ni-closed-cold}
n_{\text{i,cl}}(x) = \frac{n_{\infty} u_{z\infty}  \left( 1 + \frac{\phi''(x) }{\Omega B } \right) }{\sqrt{u_{z\infty}^2 - \left( \frac{\phi'(x) }{B} \right)^2 - \frac{2\Omega \phi (x)}{B}}} \text{.}
\end{align}
Note that equation (\ref{ni-closed-cold}) satisfies $n_{\text{i,cl}}(x) \rightarrow 0$ for $x \rightarrow x_{\text{c}}$, since $1 + \phi'' (x_{\text{c}}) / \Omega B= 0 $.

The quasineutrality equation (\ref{quasineutrality}) for cold ions in the region $x \gg x_{\text{c}}$ is, using $n_{\text{i,op}}(x) = 0$ and equation (\ref{ni-closed-cold}),
\begin{align} \label{quasi-cold-1}
\exp \left( \frac{e\phi  }{T_{\text{e}}} \right) = \frac{u_{z\infty} \left( 1 + \frac{\phi'' }{\Omega B } \right) }{\sqrt{u_{\infty}^2 - \left( \frac{\phi' }{B} \right)^2 - \frac{2\Omega \phi }{B}}} \text{.}
\end{align}
Multiplying equation (\ref{quasi-cold-1}) by $\Omega \phi'/B$, integrating once and using the boundary condition $\phi = \phi' = 0$ at $x \rightarrow \infty$ I obtain
\begin{align}
v_{\text{B}}^2  \exp \left( \frac{e\phi }{T_{\text{e}}} \right)  = v_{\text{B}}^2 + u_{z\infty}^2 - u \sqrt{ u_{z\infty}^2- \left( \frac{\phi' }{B} \right)^2 - \frac{2\Omega \phi }{B}}   \text{,}
\end{align}
which can be rearranged to
\begin{align}
\left( \frac{\phi' }{B} \right)^2 = u_{z\infty}^2  - \frac{2\Omega \phi }{B} - \frac{ v_{\text{B}}^4 }{ u_{z\infty}^2} \left( 1 + \frac{u_{z\infty}^2}{v_{\text{B}}^2 }  - \exp\left(\frac{e\phi}{T_{\text{e}}} \right) \right)^2 \text{.}
\end{align}
Using the relation (\ref{ux}) between fluid velocity and potential leads to equation (\ref{udiff-far}).
Hence, Chodura's differential equation is recovered for $x \gg x_{\text{c}}$.

\subsection{Open orbit region ($x\ll x_{\text{c}}$)} 
\label{subsec-cold-kinetic-open}

I proceed to evaluate the quasineutrality equation (\ref{quasineutrality}) in the region $x \ll x_{\text{c}} $.
In order to do so, I calculate the open orbit density for $\tau \ll 1$. 
While an ion is in an open orbit, it falls down the effective potential well in the region $x \ll x_{\text{c}} $. 
The perpendicular energy of the open orbit remains constant to lowest order in $\alpha$, taking the value it had at the effective potential maximum.
From Figure \ref{fig-chiChodura}, the perpendicular energy is given, to lowest order in $\alpha$ and $\tau$, by
\begin{align}
U_{\perp} \simeq \chi_{\text{M}} \simeq \chi_{\text{c}} = \chi (x_{\text{c}}, \bar{x}_{\text{c}} ) \text{.}
\end{align}
For $x \ll x_{\text{c}} $, one can use $\Delta_{\text{M}} \sim \alpha v_{\text{t,i}}^2 $ and $\chi_{\text{c}} - \chi (x, \bar{x}_{\text{c}}) \sim v_{\text{B}}^2 $ to show that $\Delta_{\text{M}} \ll \chi_{\text{c}} - \chi$.
Then, equation (\ref{Deltavx-simpler}) is expanded to give
\begin{align} \label{vxopen-expanded}
\Delta v_x \simeq \frac{ \Delta_{\text{M}}  }{\sqrt{ 2\left( \chi_{\text{c}} -  \frac{1}{2} \Omega^2 ( x - \bar{x}_{\text{c}} )^2 - \frac{ \Omega \phi(x)}{B} \right) }  } \text{.}
\end{align}
In what follows, I denote $d\mu / d\bar{x}|_{\text{open}} = d \mu (\bar{x},\chi_{\text{M}} (\bar{x}))/ d\bar{x}$.
By changing variable from $\bar{x}$ to $\mu$ and substituting (\ref{vxopen-expanded}), the integral (\ref{ni-open}) simplifies to
\begin{align} \label{ni-op-int}
n_{\text{i,op}} = \frac{1}{\sqrt{ 2\left( \chi_{\text{c}} -  \frac{1}{2} \Omega^2 ( x - \bar{x}_{\text{c}} )^2 - \frac{ \Omega \phi(x)}{B} \right) }   }    \int_{0}^{\infty}  \left. \frac{ d\mu}{d\bar{x}}  \right\rvert_{\text{open}}^{-1}   \Omega d\mu  \nonumber \\ \times \int_{ \Omega \mu }^{\infty} \frac{ F(\mu, U ) \Delta_{\text{M}}  dU}{\sqrt{2\left( U - \chi_{\text{c}} \right) }} 
\text{.}  
\end{align}
To evaluate (\ref{ni-op-int}), one uses the relation
\begin{align} \label{DeltaM-cold}
\Delta_{\text{M}} \simeq 2\pi \alpha \sqrt{2\left( U - \chi_{\text{c}}\right) } \left. \frac{d\mu}{d\bar{x}} \right\rvert_{\text{open}} \text{,}
\end{align}
obtained using (\ref{DeltaM-noy}) and (\ref{open-integral}).
Inserting equation (\ref{DeltaM-cold}) into (\ref{ni-op-int}) gives
\begin{align} \label{ni-op-int-2}
n_{\text{i,op}} = \frac{2\pi\alpha}{\sqrt{ 2\left( \chi_{\text{c}} -  \frac{1}{2} \Omega^2 ( x - \bar{x}_{\text{c}} )^2 - \frac{ \Omega \phi(x)}{B} \right) }   }    \int_{0}^{\infty}   \Omega d\mu  \int_{ \Omega \mu }^{\infty} F(\mu, U ) dU
\text{.}  
\end{align}
Using (\ref{F-cold}) for the distribution function, the density of open orbits is
\begin{align} \label{ni-open-cold}
n_{\text{i,op}}(x) = \frac{  \alpha  n_{\infty} u_{z\infty} }{\sqrt{ 2\left( \chi_{\text{c}} - \frac{1}{2} \Omega^2 ( x - \bar{x}_{\text{c}} )^2 - \frac{\Omega \phi}{B} \right) }} \text{,}
\end{align}
valid for $x \ll x_{\text{c}} $.
For $ x \rightarrow x_{\text{c}}$, the open orbit density diverges, $n_{\text{i,op}} (x) \rightarrow \infty$, while we previously found that the closed orbit density goes to zero at the same point, $n_{\text{i,cl}} (x) \rightarrow 0$. 
Both estimates were derived by considering positions far from the critical point and are thus not valid in the region $x \sim x_{\text{c}}$.

Imposing quasineutrality (\ref{quasineutrality}) in the region $x \ll x_{\text{c}} $ using $n_{\text{i,cl}}(x) = 0$ and equation (\ref{ni-open-cold}), one obtains
\begin{align} \label{quasi-open}
\exp\left(\frac{ e\phi}{T_{\text{e}}}\right)  =\frac{ \alpha u_{z\infty} }{ \sqrt{ 2\chi_{\text{c}} - \Omega^2 ( x - \bar{x}_{\text{c}} )^2 - \frac{2 \Omega \phi}{B} }} \text{.}
\end{align}
The constants $\chi_{\text{c}}$ and $\bar{x}_{\text{c}}$ are to be determined, because they depend on the form of the solution $\phi$. 
Rearranging equation (\ref{quasi-open}) gives
\begin{align} \label{open-rearranged}
 \Omega^2 ( x - \bar{x}_{\text{c}} )^2 = 2\chi_{\text{c}}  -   \frac{2\Omega \phi}{B} - \alpha^2  u_{z\infty}^2 \exp\left(-\frac{ 2e\phi}{T_{\text{e}}}\right)   \text{,}
\end{align}
which is differentiated to obtain
\begin{align} \label{open-rearranged-differentiated}
\Omega^2 ( x - \bar{x}_{\text{c}} ) =  \frac{ e \phi' }{T_{\text{e}}} \left( \alpha^2  u_{z\infty}^2  \exp\left(-\frac{ 2e\phi}{T_{\text{e}}}\right)   - v_{\text{B}}^2 \right)    \text{.}
\end{align}
Hence, inserting equation (\ref{open-rearranged-differentiated}) into (\ref{open-rearranged}), the constant $\bar{x}_{\text{c}}$ is eliminated and the first order differential equation
\begin{align}
\left( \frac{ e \phi' }{\Omega T_{\text{e}}} \right)^2 \left( \alpha^2  u_{z\infty}^2  \exp\left(- \frac{ 2e\phi}{T_{\text{e}}}\right)   - v_{\text{B}}^2 \right)^2   = 2\chi_{\text{c}} -  \frac{2 \Omega\phi}{B} - \alpha^2 u_{z\infty}^2 \exp\left(-\frac{ 2e\phi}{T_{\text{e}}}\right)  \text{}
\end{align}
is obtained.
Using the equation relating fluid velocity to electrostatic potential, (\ref{ux}), gives a differential equation for $u_x$,
\begin{align} \label{udiff-open}
 \frac{  \left( v_{\text{B}}^2 - u_x^2 \right)^2 }{ u_x^2 v_{\text{B}}^2 } \frac{ u_x'^2 }{\Omega^2 }   =   \frac{ 2\chi_{\text{c}}}{v_{\text{B}}^2}  - \frac{ u_x^2 }{v_{\text{B}}^2} +   2\ln \left( \frac{ -u_x }{\alpha u_{z\infty} } \right) \text{.}
\end{align}

%
Note that (\ref{udiff-open}) has the same form as (\ref{udiff-near}), but with an additional constant of integration to be determined.
The singularity in (\ref{udiff-open}) at $u_x = - v_{\text{B}}$ implies that the system will reach the Bohm velocity with an infinite gradient at the Debye sheath entrance, $x=0$ \cite{Riemann-review}.
Hence, we expect that, to lowest order in $\alpha$ and $\tau$, the distribution of the ion velocity normal to the wall at $x=0$ is
\begin{align} \label{f0xcold}
f_{0x} (v_x) = \alpha n_{\infty} \delta \left(v_x + v_{\text{B}} \right) \text{,} 
\end{align}
shown as a vertical dashed line in the bottom-right panel of Figure \ref{fig-f0x}.

\subsection{Intermediate region ($x\sim x_{\text{c}}$)} \label{subsec-cold-kinetic-intermediate}


The equations derived in the two previous subsections assumed that $x$ was a point sufficiently far away from the critical point $x_{\text{c}}$. 
In what follows, I speculate that the self-consistent electrostatic potential must be a parabola in the region $x\sim x_{\text{c}}$, and also argue that my kinetic model requires non-zero $\tau$ (and therefore requires kinetic effects) in order to have a solution in this intermediate region. 
Figure \ref{fig-chiChodura} shows effective potentials corresponding to the solution of the fluid model (equation (\ref{x-ux-lowest}), valid for $\tau = 0$; for $\bar{x} = \bar{x}_{\text{c}}$ there is a large intermediate region near $x_{\text{c}}$ where the effective potential is flat over a length scale of $\rho_{\text{B}}$, and this behaviour is connected with the electrostatic potential being close to parabolic for $x\sim x_{\text{c}}$.

The closed orbit and the open orbit equations both have the form
\begin{align} \label{udiff-intermediate-1 }
 \frac{  v_B^2 }{ \Omega^2 } \frac{ u_x'^2 }{u_x^2 }  =  C  + 2 \ln \left( \frac{ - u_x }{\alpha u_{z\infty}} \right)   \text{,}
\end{align}
when $\alpha \ll u_x / v_{\text{B}} \ll 1$ is taken.
Suppose that equation (\ref{udiff-intermediate-1 }) describes the dynamics in the region $x\sim x_{\text{c}}$.
Using (\ref{ux}) to re-express (\ref{udiff-intermediate-1 }) as a differential equation for the electrostatic potential, one obtains
\begin{align} \label{phiprime-par}
 \rho_B \frac{ e\phi' }{T_{\text{e}}} = \sqrt{ C  - \frac{2e\phi}{T_{\text{e}}} } \text{.}
\end{align}
The solution to equation (\ref{phiprime-par}) is the parabola
\begin{align} \label{phipar-Te}
\frac{ e\phi }{T_{\text{e}}} =  \frac{1}{2}C  - \frac{ \left( x-x_0 \right)^2}{2\rho_{\text{B}}^2}  \text{,}
\end{align}
which can be re-expressed as
\begin{align} \label{phipar-v2}
\frac{ Ze\phi }{m_{\text{i}}} =  \frac{1}{2} C v_{\text{B}}^2 -  \frac{1}{2} \Omega^2  \left( x-x_0 \right)^2  \text{.}
\end{align}
The effective potential for ions reaching $x\sim x_{\text{c}}$ is thus
\begin{align}
\chi (x; \bar{x} ) = \frac{ 1 }{2} \Omega^2 \left( x - \bar{x}
 \right)^2  -  \frac{1}{2} \Omega^2  \left( x-x_0 \right)^2  +  \frac{1}{2} C v_{\text{B}}^2  \text{.}
\end{align}
The effective potential of cold ions reaching $x_{\text{c}}$ must be such that $\chi'(x; \bar{x}) = 0$ to lowest order in $\alpha$ and $\tau$, because these ions are in closed orbits that eventually become open, and closed orbits satisfy $U_{\perp} = \chi_{\text{m}} (\bar{x})$ to lowest order in $\tau$.
Hence, $x_0 = \bar{x}$ and the effective potential is constant.
The value of $\bar{x}$ for which $\chi'(x_{\text{c}}; \bar{x}) = \chi''(x_{\text{c}}) = 0$ is $\bar{x}_{\text{c}}$, as proved in section \ref{subsec-traj-effpottypes}.
Hence, $ x_0 = \bar{x}_{\text{c}} $ and the effective potential is equal to $ \chi (x_{\text{c}}; \bar{x}_{\text{c}}) = \chi_{\text{c}}$ to lowest order in $\alpha$ and $\tau$,
\begin{align}
\chi (x; \bar{x} ) \simeq  \frac{1}{2} C v_{\text{B}}^2 = \chi_{\text{c}} \text{.}
\end{align}
From the closed orbit quasineutrality equation for $x\gg x_{\text{c}}$, $C = - v_{\text{B}}^2 / u_{z\infty}^2 - 2 $.
Hence, $\chi_{\text{c}} = - v_{\text{B}}^2 / u_{z\infty}^2 - 2$, and the open orbit solution for $x\ll x_{\text{c}}$ coincides with equation (\ref{udiff-near}), which is the expansion of Chodura's differential equation for $u_x \sim v_{\text{B}}$.

Note that this discussion relies on the supposition that the solution be a parabola in the intermediate region. 
Moreover, the exact values of $x_{\text{c}}$ and $\bar{x}_{\text{c}}$ remain undetermined, and depend on the higher order corrections to equation (\ref{udiff-intermediate-1 }).
Hence, the electrostatic potential solution in the intermediate region depends on higher order corrections in $\alpha$ and $\tau$.
My model is formulated to lowest order in $\alpha$, and thus I expect that non-zero value of $\tau$ is required in order to obtain a solution in the intermediate region.
Hence, my numerical scheme relies on the presence of kinetic effects, although it can be seen that my numerical solutions qualitatively approach the cold ion limit when $\tau$ is small.

In this subsection, I have speculated (without proof) that the form of the solution to the quasineutrality equation (\ref{quasineutrality}) in the intermediate region is a parabola when $\tau$ is small.
I have thus argued that the electrostatic potential solution of (\ref{quasineutrality}) relies on $\tau \neq 0$, and is not necessarily equivalent to the electrostatic potential solution of the fluid model (valid for $\tau =0$).
Nonetheless, the numerical profiles for $\tau = 0.2$ approach the solutions for $\tau = 0$, lending support to the accuracy of the kinetic model in the intermediate region for small $\tau$.
Moreover, note that I have proved in \ref{subsec-cold-kinetic-closed} and \ref{subsec-cold-kinetic-open} that the kinetic model recovers the correct ion flow profile --- and electrostatic potential profile --- for the regions $ x \gg x_{\text{c}}$ and $x \ll x_{\text{c}}$.
It is therefore unlikely that the flow and electrostatic potential in the intermediate region should tend to a profile that is too different from the fluid profile in the region $x \sim x_{\text{c}}$.

\section{Infinite ion temperature limit} \label{sec-Tdep-hot}

In the hot ion limit $\tau \rightarrow \infty$, the distribution function of equation (\ref{f-infty-Tdep}) must have the value $r \rightarrow \infty$ in order to satisfy the marginal form of the kinetic Chodura condition (\ref{kinetic-Chodura-marginal}), and therefore tends to a half-Maxwellian
\begin{align} \label{F-hot}
F = 2 n_{\infty} \left( \frac{ m_i }{2\pi T_i} \right)^{3/2} \exp\left( -\frac{ m_{\text{i}} U }{ T_{\text{i}}} \right)  \text{.}
\end{align}
This distribution function was used in \cite{Cohen-Ryutov-1998} to study a magnetic presheath where the electrons are cold. 
The limit $\tau \rightarrow \infty$ is inconsistent with the ordering (\ref{ordering-angle}) for finite values of $m_{\text{i}}/m_{\text{e}}$.
The inconsistency arises because, at large values of $\tau$, we would expect the sheath-presheath system to be ion-repelling.
However, the system may stay electron-repelling if the value of $m_i / m_e$ is sufficiently large that $\tau \ll \alpha^2 m_{\text{i}}/ m_{\text{e}}$ can be satisfied.
Therefore, the limit $1 \ll \tau \ll \alpha^2 m_{\text{i}} / m_{\text{e}} $ is not inconsistent (although it is unrealistic at large values of $\tau$ for typical values of $m_{\text{i}}/ m_{\text{e}}$).

For $\tau \rightarrow \infty$, ion orbits are undistorted by the presheath potential drop necessary to repel most of the cold electrons. 
We expect $e\phi(x)/T_{\text{e}} \sim 1$, and therefore the ion flow and density can be computed using $e\phi(x)/T_{\text{i}} = (1/\tau) e\phi(x)/T_{\text{e}} = 0$ across the magnetic presheath. 
The effective potential is a parabola with its minimum at $x_{\text{m}} = \bar{x}$,
\begin{align}
\chi(x, \bar{x}) = \frac{1}{2} \Omega^2 \left(x-\bar{x} \right)^2 \text{.}
\end{align}
This is a type I effective potential.
Hence,
\begin{align}
\chi_{\text{M}} (\bar{x} ) = \chi (0, \bar{x} ) = \frac{1}{2} \Omega^2 \bar{x}^2 \text{.}
\end{align}
In chapter \ref{chap-KMPS}, I found that the self-consistent solution for the electrostatic potential leads to the presence of type II orbits only.
Hence, for $\tau \rightarrow \infty$ there must be an infinitely small region near $x=0$ in which the type II behaviour emerges.
I neglect this region and proceed with the analysis.
The minimum value of $\bar{x}$ necessary for an ion at position $x$ to be in a closed orbit or an open orbit is, using equations (\ref{xbarm}) and (\ref{xbarm-open}) with $\phi(x)=0$,
\begin{align} \label{xbarm-open-flatphi}
\bar{x}_{\text{m,o}} \left( x \right) = \bar{x}_{\text{m}} \left( x \right) = \frac{1}{2} x \text{.}
\end{align}
Moreover, the adiabatic invariant is $ \mu = U_{\perp}/\Omega $.

Inserting the distribution function (\ref{F-hot}) into equation (\ref{ni-closed}), the closed orbit density is 
\begin{align}
n_{\text{i,cl}}(x) = 2 n_{\infty} \left( \frac{ m_i }{2\pi T_i} \right)^{3/2}  \int_{x/2}^{\infty} \Omega d\bar{x} \int_{\frac{1}{2} \Omega^2 \left(x-\bar{x} \right)^2}^{\frac{1}{2}\Omega^2 \bar{x}^2 } \frac{ 2 dU_{\perp} }{\sqrt{2\left(U_{\perp} - \chi (x) \right)}} \int_{\Omega\mu}^{\infty}  \frac{ \exp \left(  - \frac{m_{\text{i}} U}{T_{\text{i}} }  \right) dU }{\sqrt{2\left( U - \Omega \mu \right)}}  \text{.}
\end{align}
Changing variables to $\tilde{ v }_y =  \left(  \bar{x} - x \right)/\rho_{\text{i}} $, $\tilde{U}_{\perp} = m_{\text{i}} \left(  U_{\perp} - \frac{1}{2} \Omega^2 (x-\bar{x})^2 \right) / T_{\text{i}}$ and $\tilde{U} = m_{\text{i}} \left(U - \Omega \mu \right) / T_{\text{i}}$ gives
\begin{align}
n_{\text{i,cl}}(x) = \frac{ n_{\infty} }{ \pi^{3/2} }  \int_{ - \frac{1}{2}\frac{x}{\rho_{\text{i}}} }^{\infty} d\tilde{v}_y \exp (-  \tilde{ v }_y^2)  \int_{0}^{ \frac{x}{\rho_{\text{i}}} \left( 2\tilde{v}_y + \frac{x}{\rho_{\text{i}}} \right) }  \tilde{U}_{\perp}^{-1/2} \exp (-\tilde{U}_{\perp}) d\tilde{U}_{\perp} \nonumber \\
\times  \int_{0}^{\infty}  \tilde{U}^{-1/2} \exp(-\tilde{U}) d\tilde{U}  \text{.}
\end{align}
Evaluating the integral over $\tilde{U}$ and the integral over $\tilde{U}_{\perp}$ leads to
\begin{align} \label{ni-closed-coldelectrons}
n_{\text{i,cl}}(x) = \frac{n_{\infty}}{\sqrt{\pi}}  \int_{-\frac{1}{2}\frac{x}{\rho_{\text{i}}}}^{\infty}  \exp \left(- \tilde{v}_y^2 \right) \text{erf}  \left( \sqrt{  \frac{x}{\rho_{\text{i}}} \left( 2\tilde{ v }_y + \frac{x}{\rho_{\text{i}}}\right) } \right) d\tilde{v}_y \text{.}
\end{align}

The density of open orbits is given by
\begin{align} \label{ni-open-coldelectrons-1}
n_{\text{i,op}} (x)  = & \int_{\frac{1}{2}x }^{\infty}  \Omega d\bar{x}  \int_{\frac{1}{2} \Omega^2 \bar{x}^2 }^{\infty} \frac{ F\left( \mu, U \right) }{\sqrt{2\left( U - \chi_{\text{M}}  \right) }} \nonumber \\
 & \times \left( \sqrt{2\left( \chi_{\text{M}}  - \chi \left( x, \bar{x} \right) + \Delta_{\text{M}} \right) } - \sqrt{2\left( \chi_{\text{M}}  - \chi \left( x, \bar{x} \right)  \right) } \right) dU \text{.}
\end{align}
The adiabatic invariant of ions with $U_{\perp} = \chi_{\text{M}} = \chi ( 0, \bar{x} )= \Omega^2 \bar{x}^2 / 2 $ is given by
\begin{align}
\mu =\frac{1}{2} \Omega \bar{x}^2 \text{.}
\end{align}
Using equation (\ref{DeltaM-noy}), we obtain 
 \begin{align}
\Delta_{\text{M}} = 2 \alpha \pi \Omega \bar{x} \sqrt{2\left( U - \frac{1}{2} \Omega^2 \bar{x}^2   \right) }  \text{.}
\end{align}
Then, using the dimensionless integration variables $\tilde{v}_z = \sqrt{ m_{\text{i}} \left( U - \Omega^2 \bar{x}^2/2 \right) / T_{\text{i}} }$ and $\tilde{\bar{x}} = \bar{x}/\rho_{\text{i}}$, equation (\ref{ni-open-coldelectrons-1}) reduces to
\begin{align} \label{ni-open-coldelectrons}
n_{\text{i,op}} (x) = & \frac{2}{\pi^{3/2}} \int_{\frac{x}{2\rho_{\text{i}}} }^{\infty}   \exp \left( - \tilde{\bar{x}}^2 \right) d\tilde{\bar{x}} \int_{ 0 }^{\infty} \exp\left(-\tilde{v}_{z}^2 \right)  \nonumber \\
 & \times  \left( \sqrt{ \frac{x}{\rho_{\text{i}}} \left( 2\tilde{ \bar{x} } - \frac{x}{\rho_{\text{i}}} \right)  + 4 \alpha \pi \tilde{ \bar{x} } \tilde{v}_z  } - \sqrt{ \frac{x}{\rho_{\text{i}}} \left( 2\tilde{\bar{x}} - \frac{x}{\rho_{\text{i}}} \right) } \right) d\tilde{v}_z   \text{,}
\end{align}
which does not simplify further for $x \neq 0$.
At $x=0$, equation (\ref{ni-open-coldelectrons}) evaluates to
\begin{align} \label{ni-open-coldelectrons-x=0}
n_{i,\text{op}} (0)  
= n_{\infty} \frac{  \Gamma^2 \left( 3/4 \right) }{\pi } \sqrt{\alpha } \text{.}
\end{align}

The ion density profile in the hot ion limit is, according to (\ref{ni}) the sum of equations (\ref{ni-closed-coldelectrons}) and (\ref{ni-open-coldelectrons}).
The potential profile is obtained by imposing quasineutrality and inverting the Boltzmann relation for the electron density, to find
\begin{align} \label{phi-hot}
\frac{ e\phi (x) }{T_{\text{e}}} =  \ln \left( \frac{ n_{\text{i}}(x) }{n_{\infty}} \right) \text{.}
\end{align}
The potential drop across the magnetic presheath can be calculated by using $n_{\text{i,cl}}(0) = 0$ (from equation (\ref{ni-closed-coldelectrons})) and equation (\ref{ni-open-coldelectrons-x=0}),
\begin{align} \label{phidrop-tauinf}
\frac{ e\phi (x) }{T_{\text{e}}} =  \ln \left(  \frac{  \Gamma^2 \left( 3/4 \right) }{\pi } \sqrt{\alpha }  \right) \simeq  \ln \left( 0.48 \sqrt{\alpha }  \right) \text{.}
\end{align}
Equation (\ref{phidrop-tauinf}) is the dashed line marked $\tau = \infty$ in Figure \ref{fig-phi-Tdep}.
The electrostatic potential profile obtained using equation (\ref{phi-hot}) is the red dashed curve (marked $\tau = \infty$) in Figure \ref{fig-phiprofile-Tdep}.
The dashed curve in Figure \ref{fig-ux-Tdep} is obtained by inserting $\phi(x)$ into equation (\ref{ux}) for the ion fluid velocity $u_x$.

Inserting the distribution function (\ref{F-hot}) and the value of $\bar{x}_{\text{m,o}}$ in (\ref{xbarm-open-flatphi}) into equation (\ref{f0x}), the distribution of the ion velocity component perpendicular to the wall at $x=0$ is 
\begin{align} \label{f0x-hot}
f_{0x}(v_x ) = \frac{n_{\infty}}{v_{\text{t,i}}\pi} \Theta (v_x) \int_0^{\infty} \exp \left( - \tilde{\bar{x}}^2 \right) \left[ 1 - \text{erf} \left( \frac{v_x^2 }{4\pi \alpha \tilde{\bar{x}} v_{\text{t,i}}^2} \right) \right] d\tilde{\bar{x}}  
\end{align}
Inserting the distribution function (\ref{F-hot}) into equation (\ref{f0yz}), the distribution of the ion velocity components parallel to the wall at $x=0$ is 
\begin{align} \label{f0yz-hot}
f_{0yz}(v_y, v_z ) = \frac{4\sqrt{\alpha} n_{\infty}}{\pi} \frac{  \sqrt{v_y v_z } }{v_{\text{t,i}}^3} \exp \left( - \frac{ v_y^2 + v_z^2  }{v_{\text{t,i}}^2} \right)  \Theta (v_y) \Theta(v_z) \text{.}
\end{align}
Equation (\ref{f0x-hot}) is the dashed curve on the bottom-left panel in Figure \ref{fig-f0x}, while equation (\ref{f0yz-hot}) is the bottom right panel in Figure \ref{fig-f0yz-Tdep}.

\section{Discussion} 
\label{sec-Tdep-disc}

In this chapter I have studied the dependence of an electron-repelling magnetic presheath on ion temperature using the kinetic model described in this thesis.
The electrostatic potential drop across the magnetic presheath deviates from the fluid one $\tau = 0$ by up to approximately $15\%$ for $\alpha / \sqrt{\tau} \gtrsim \sqrt{m_{\text{e}}/ m_{\text{i}}} \sim 0.02$ (assuming Deuterium ions), a regime in which the wall is electron-repelling.
For very large values of $\tau$, the potential drop is up to a factor of $30\%$ smaller than for $\tau = 0$, but the wall is likely to not be electron-repelling, which would change the potential.
In order to study the transition between an electron- and an ion-repelling wall ($\alpha / \sqrt{\tau} \sim \sqrt{m_{\text{e}} / m_{\text{i}} }$), a kinetic treatment of both ions and electrons is essential.

I have argued that the numerical profiles of potential and fluid velocity obtained with the kinetic model are consistent with the small and large $\tau$ limit. 
The limit $\tau \ll 1$ is Chodura's fluid model.
For $\tau \ll 1$, my kinetic treatment analytically recovers the potential drop predicted from fluid theory (equation (\ref{phidrop-tau0})), and recovers the fluid potential and flow profiles for $x \gg x_{\text{c}}$ and for $x\ll x_{\text{c}}$.
In the intermediate region $x \sim x_{\text{c}}$, I have speculated that the solution to my equations must become close to a parabola when $\tau$ is small, but have argued that $\tau$ must be non-zero in order to obtain a solution to the kinetic model, and that such a solution may not be equal to the solution to the fluid model to lowest order in $\tau$ and $\alpha$.
The numerical results for $\tau = 0.2$, however, lend support to the view that the solution to the kinetic model approaches the solution to the fluid model when $\tau$ is small.


For large ion temperatures, $\tau \gtrsim 5$, the velocity component normal to the wall at the Debye sheath entrance is reduced by the small angle $\alpha$ for two reasons.
Firstly, there are a large number of slow ions with $|v_x| \simeq v_{\text{B}} \ll v_{\text{t,i}}$, and secondly, the velocity spread of the distribution function is $\sqrt{\alpha} v_{\text{t,i}}$ (see Figure \ref{fig-f0x} and equation (\ref{f0x-hot})).
The tangential velocity of a typical ion remains roughly of the same size $v_y \sim v_z \sim v_{\text{t,i}}$, and therefore the angle between the ion trajectory and the wall is shallow at the Debye sheath entrance.
There are several ions (in the peak of the distribution function $f_{0x}$) whose trajectory makes an angle with the wall of size $1/\sqrt{\tau}$.
Most ions, in the tail of the distribution function, make an angle with the wall of size $\sqrt{\alpha}$.
For $\tau \lesssim 1$, the typical size of all the velocity components is $v_{\text{B}}$ and thus the angle between the ion trajectory and the wall is of order unity (for $\alpha = 0.05$ and $\tau = 0$ this angle is approximately $30^{\circ}$). %
Hence, an ion is expected to impinge on the wall at an angle whose size is small when $\tau \gg 1$ and order unity when $\tau \lesssim 1$.

It has been experimentally shown that $\tau \gtrsim 1$ near the divertor targets of tokamaks \cite{Mosetto-2015}. 
The kinetic treatment of this thesis has introduced the necessary ingredient to tackle the presence of ions at a temperature comparable to or larger than the electron temperature in the magnetic presheath.
A kinetic treatment of the electrons will also be necessary to study systems in which $\alpha / \sqrt{\tau} \lesssim \sqrt{m_{\text{e}}/ m_{\text{i}}}$.

\chapter{Conclusions}
\label{chap-conc}

In this thesis I presented and solved a model of the magnetic presheath.
The ordering (\ref{scale-sep}) was used to assume that the magnetic presheath is collisionless and quasineutral.
The ion trajectories were discussed in detail in chapter~\ref{chap-traj}, where a treatment that relies on an asymptotic expansion in the small angle $\alpha$ between magnetic field and wall was presented.
Weak turbulent gradients were assumed in the $y$ direction perpendicular to the magnetic field and parallel to the wall.
These gradients give rise to a turbulent electric field component which is smaller than the component normal to the wall by a factor $\delta \sim \alpha \ll 1$.
Given an electrostatic potential profile $\phi(x,y)$, the ion density (\ref{ni-general}) is obtained in chapter~\ref{chap-dens}.
The density includes the contribution (\ref{ni-closed-general}) of approximately closed orbits and the crucial contribution (\ref{ni-open-general}) of open orbits.
The continuity equation is used to obtain the fluid velocity from the ion density via equation (\ref{ux-ni}).

By making the simplifying assumption that the wall is strongly electron-repelling, in chapter~\ref{chap-KMPS} I have taken a Maxwell-Boltzmann response (\ref{ne-general}) for the electron density in the magnetic presheath.
The self-consistent electrostatic potential is the solution of the quasineutrality equation (\ref{quasineutrality-general-star}).
Taking $\delta = 0$ and thus assuming no gradients in the directions parallel to the wall, the electrostatic potential that solves the simplified quasineutrality equation (\ref{quasineutrality}) can be determined by the iteration scheme presented in section \ref{sec-KMPS-nummethod}.
The ion distribution function at the magnetic presheath entrance was shown to have to satisfy a kinetic generalization of the Chodura condition, equation (\ref{kinetic-Chodura}).
The ion distribution function at the Debye sheath entrance is obtained from the analysis of chapter \ref{chap-dens}, and is given in equation (\ref{f0}).
It is shown both analytically and numerically that this distribution function satisfies the kinetic Bohm condition.

The kinetic treatment of ions in the magnetic presheath is generalizable to an arbitrary number of ion species, and also applicable to impurities, and can thus be used to predict impurity trajectories and the effect of impurities.
The ion distribution function (\ref{f0}) is important in order to predict the amount of sputtering off the wall, since each ion has a sputtering yield which depends on the ion's species, kinetic energy and angle of incidence with the surface \cite{Eckstein-1993-sputtering}.
Sputtering by impurities usually has a lower kinetic energy threshold than sputtering by the main ion species \cite{Mellet-2016}, and therefore it is especially important to predict the velocity distribution of impurities reaching the wall.
In the limit of small impurity density, the impurity ions can be considered as a trace in the plasma and their distribution function at the Debye sheath entrance can be obtained from their distribution function at the magnetic presheath entrance. 

The effect of finite ion temperature on an electron-repelling magnetic presheath was studied in chapter~\ref{chap-Tdep}, where the distribution function at the magnetic presheath entrance was varied with the parameter $\tau = T_{\text{i}}/ T_{\text{e}}$.
The marginally satisfied kinetic Chodura condition (\ref{solvability}) was assumed.
All the numerical results were shown to be consistent with the zero and infinite ion temperature limits.

The inclusion of kinetic electrons is an imporant next step that must be taken to understand the physics of plasma-wall interaction near the divertor targets of a fusion device. 
Kinetic electrons are expected to change the behaviour of the magnetic presheath.
Firstly, if the parameter $\alpha/ \sqrt{\tau}$ is small enough, a strongly electron-repelling wall may not be a good approximation (see equation (\ref{ordering-angle})).
In this case, the number of electrons reaching the wall would be too large to ignore in the electron distribution, and an appropriate treatment of the whole electron distribution function would have to be retained.
Plasma in contact with a wall that is not strongly electron-repelling can only be studied with a kinetic ion model such as the one described in this thesis.
This is because fluid ion models can only be justified for zero ion temperature, and therefore are --- from equation (\ref{ordering-angle}) --- only consistent with an electron-repelling wall.

The ion kinetic treatment presented in this thesis is expected to be applicable to electrons in the Debye sheath if $\rho_{\text{e}} \sim \lambda_{\text{D}}$, a regime relevant for the boundary of fusion devices.
Hence, an accurate and realistic solution of the Debye sheath and magnetic presheath with kinetic electrons may be obtained using the tools presented in this thesis.
For $ \alpha / \sqrt{\tau} \lesssim \sqrt{m_{\text{e}} / m_{\text{i}} }$ the inclusion of kinetic electrons is expected to dramatically change the potential drop across the sheath and presheath under ambipolar conditions, as well as the relationship between current and potential drop.
With a full kinetic treatment, the ion and electron distribution functions at the wall could be obtained.
The ion distribution function is important to predict sputtering off the wall, while the electron distribution function can be used to predict the amount of secondary electron emission.

In summary, this thesis has presented a powerful asymptotic theory of charged particle trajectories a gyroradius away from a wall, on which a uniform magnetic field impinges at a small angle $\alpha$.
Applying this theory to ions, a strongly electron-repelling magnetic presheath was solved for several values of $\alpha$ and for several values of the ratio of the ion and electron temperatures.
The numerical solutions typically require less than one minute to compute on a laptop, making the numerical scheme described in this thesis computationally cheap.
The kinetic framework can be generalized to include multiple ion species and impurities, and can be applied to electrons in the Debye sheath.
The numerical and analytical studies presented in this thesis generalize known magnetic presheath results, and demonstrate the potential applicability of the asymptotic theory to more complex problems such as a Debye sheath with $\rho_{\text{e}} \sim \lambda_{\text{D}}$, a magnetic presheath affected by turbulence ($\delta \sim \alpha$) and small angles ($\alpha \sim \sqrt{\tau m_{\text{e}} / m_{\text{i}} }$).

\cleardoublepage

\appendix

\chapter{Typical widths of the plasma-wall boundary layers in a tokamak} 
\label{appendix:widths}

In this Appendix, I estimate the typical typical width of the different plasma-wall boundary layers (see Figure \ref{figure-boundary-layers}).
In order to do so, I take data from reference \cite{Militello-Fundamenski-2011}, which contains a comparison of some important parameters between different tokamaks. 
I use JET data because it is the most relevant for fusion. 
From Tables 1 to 5 of \cite{Militello-Fundamenski-2011}, the magnetic field is
\begin{align}
B \sim 2 \text{ T,}
\end{align}
the electron and ion temperatures are (by taking $T_{\text{e}} \sim T_{\text{i}} \sim T$),
\begin{align}
T \sim 50 \text{ eV} = 8 \times 10^{-18} \text{ J} \text{,}
\end{align}
and the electron density is,
\begin{align} \label{estimate-density}
n_{\text{e}} \sim 10^{19} \text{ m}^{-3} \text{.}
\end{align}
For a deuterium plasma, the ion mass is $m_{\text{i}} \sim 3 \times 10^{-27} \text{ kg}$, which leads to the estimate
\begin{align}
v_{\text{t,i}} = \sqrt{\frac{2T_{\text{i}}}{m_{\text{i}}}} \sim 7 \times 10^{4} \text{ ms}^{-1}
\end{align}
for the ion thermal velocity.
The ion gyrofrequency is estimated to be
\begin{align}
\Omega = \frac{eB}{m_{\text{i}}} \sim  10^8 \text{ s}^{-1}  \text{,}
\end{align}
which leads to the estimate for the ion gyroradius
\begin{align}
\rho_{\text{i}} \sim \frac{v_{\text{t,i}}}{\Omega} \sim 7 \times 10^{-4} \text{ m} = 0.7 \text{ mm} \text{.}
\end{align}
The Debye length is (with $\epsilon_0 = 8.85 \times 10^{-12} \text{ Fm}^{-1}$)
\begin{align}
\lambda_{\text{D}} = \sqrt{\frac{\epsilon_0 T_{\text{e}}}{n_{\text{e}}e^2}} \sim  2 \times 10^{-5} \text{ m} = 0.02 \text{ mm} \text{.}
\end{align}

The ion mean free path is estimated for the two dominant collision processes occurring close to the divertor target: Coulomb collisions and charge exchange. 
For Coulomb collisions, I determine the frequency of ion-ion collisions as (from the NRL Plasma Formulary \cite{NRL-plasma-formulary})
\begin{align} \label{estimate-Coulomb-collisions}
\nu_{\text{ii}} \sim \frac{4\sqrt{\pi}}{3} \frac{e^4n_{\text{i}} \ln \Lambda}{\left(4\pi \epsilon_0 \right)^2 m_{\text{i}}^{1/2} T_{\text{i}}^{3/2} } \sim  2 \times 10^{4} \text{ s}^{-1}\text{,} 
\end{align}
where $\ln \Lambda \sim 15$ is the Coulomb logarithm for ion-ion collisions.
Therefore, the collisional mean free path is
\begin{align} \label{estimate-meanfreepath}
\lambda_{\text{mfp,ii}} \sim \frac{v_{\text{t,i}}}{\nu_{\text{ii}}} \sim  4 \text{ m.} 
\end{align}
I evaluate $\alpha \lambda_{\text{mfp,ii}}$ for $\alpha \sim 0.1$ to obtain
\begin{align}
\alpha \lambda_{\text{mfp,ii}} \sim 0.4 \text{ m.}
\end{align}
For charge exchange I use the value of $ 5 \times 10^{-14} \text{ m}^3\text{s}^{-1}$ (extracted from reference \cite{Havlickova-Fundamenski-2013}) as an approximate rate coefficient for the reaction at $T_{\text{i}} \sim 50 \text{ eV}$. 
I then multiply this by an estimate of the neutral density, $n_{\text{n}} \sim 10^{18} \text{ m}^{-3}$, in order to obtain the charge exchange collision frequency
\begin{align}
\nu_{\text{cx}} \sim 5 \times 10^4 \text{ s}^{-1}\text{.}
\end{align}
From this collision frequency, we obtain a mean free path that is slightly smaller than the Coulomb collision one,
\begin{align} \label{estimate-meanfreepath-2}
\lambda_{\text{mfp,cx}} \sim 1 \text{ m.}
\end{align}
The estimate for the width of the collisional layer becomes
\begin{align}
\alpha \lambda_{\text{mfp,cx}} \sim 0.1 \text{ m.}
\end{align}

It is worth making a few comments on these numbers and the scalings associated with them. 
The scaling $\rho_{\text{i}} / \lambda_{\text{D}} \sim 40 \gg 1$ which implies a quasineutral magnetic presheath is robust. 
The dependence on density and temperature of the ion gyroradius and Debye length is weak, with $\rho_{\text{i}} \propto \sqrt{T_{\text{i}}}$ and $\lambda_{\text{D}} \propto \sqrt{T_{\text{e}}/n_{\text{e}}}$, so the error associated with both these estimates is small. If one of $T_{\text{i}}$, $T_{\text{e}}$ or $n_{\text{e}}$ is wrong by a factor of $10$, the corresponding estimate for $\rho_{\text{i}}$ or $\lambda_{\text{D}}$ will be wrong by a factor of $\sim 3$ only. 

The discussion about the length of the collisional layer is more complex. 
I assumed Coulomb or charge exchange collisions in order to make the estimates (\ref{estimate-meanfreepath}) and (\ref{estimate-meanfreepath-2}), which resulted in $\alpha \lambda_{\text{mfp}} / \rho_{\text{i}} \sim 100 \gg 1$ using the slightly more conservative charge exchange estimate. 
This seems to favour a collisionless model for the magnetic presheath and a scale separation between the magnetic and collisional layers. 
However, note that $\lambda_{\text{mfp,ii}} \propto T_{\text{i}}^2$, so that if the ion temperature were smaller than the estimated temperature by a factor of $10$, this separation of scales would no longer be valid due to collisions becoming important in the magnetic presheath. 
Moreover, the charge exchange frequency depends linearly on the neutral density close to divertor targets, which we estimated crudely. 
These estimates warrant caution towards the idea of a completely collisionless magnetic presheath and the separation into the three layers of Figure \ref{figure-boundary-layers}, even though it is still a physically motivated and theoretically attractive way to model the plasma-wall boundary.

\chapter{Orderings in the context of a tokamak} 
\label{appendix:turbulence}

The orderings I used for the length scales parallel to the wall in the magnetic presheath and the characteristic timescale are applicable to tokamaks, as I show in this Appendix.
Figure \ref{fig-tokamakgeometry} shows the plasma-wall boundary in the context of the tokamak geometry.
In Appendix \ref{subapp-lengthscales-turbulence}, I recover the orderings (\ref{xyz-order}) and (\ref{t-dep}) of Section 2 by developing an ordering for the turbulent structures in the SOL. 
In Appendix \ref{subapp-parallelgradients-turbulence}, I calculate the characteristic steady-state gradients parallel to the wall in the magnetic presheath by projecting the SOL width onto the $y$ and $z$ directions parallel to the wall. 
The steady-state lengths are longer than, or of the same order as, the characteristic lengths parallel to the wall due to turbulent structures, as expected.

\begin{figure}[h]
\centering
\includegraphics[width=0.8\textwidth]{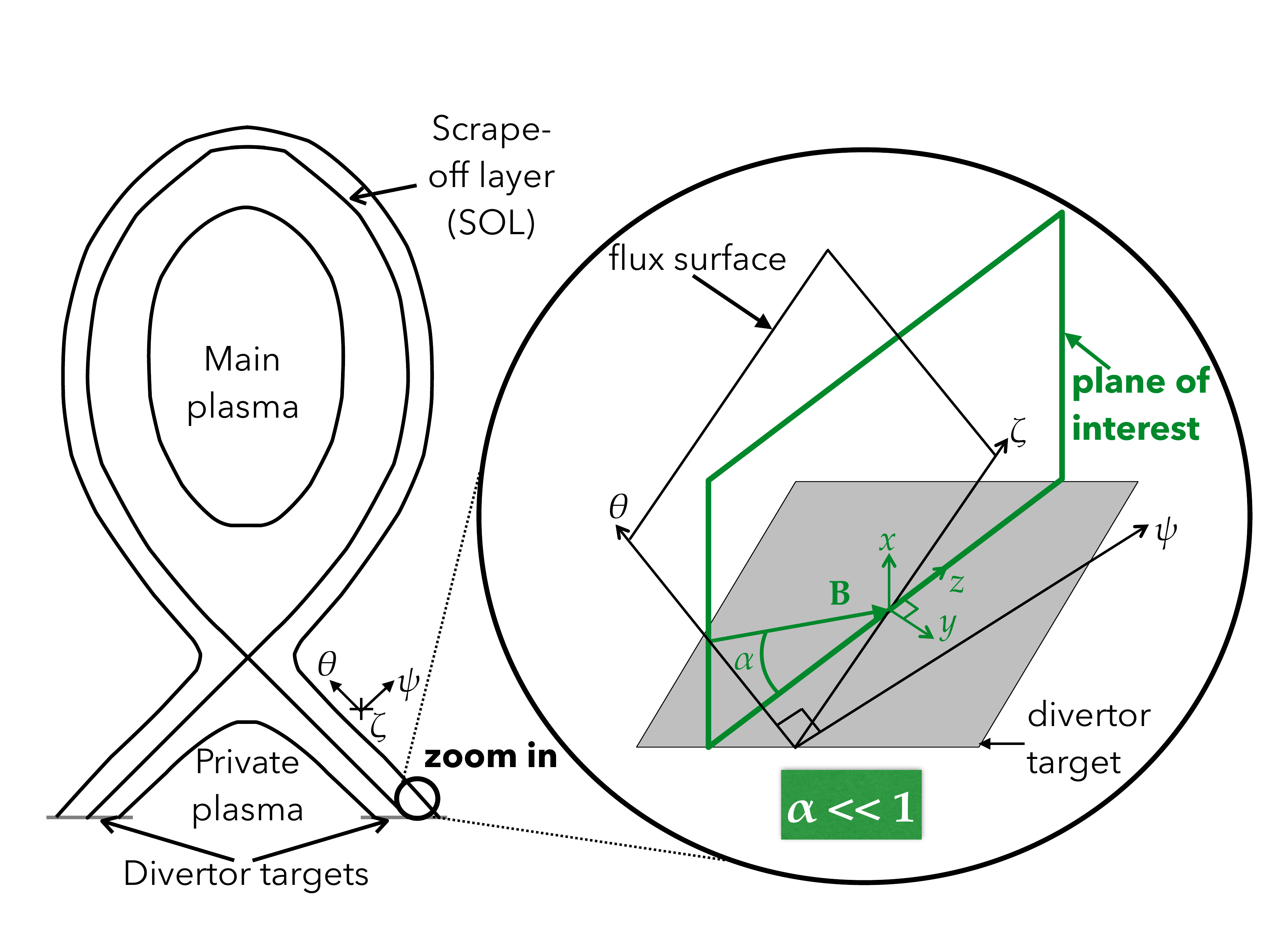}
\caption[Plasma-wall boundary geometry in the context of a tokamak]{On the left, a cartoon of the flux surface contours in the poloidal plane of a typical tokamak, with different plasma regions labelled. The region close to one of the two divertor targets (grey horizontal lines) is enlarged and shown in 3 dimensions on the right. Here, the cartesian coordinate axes $(x,y,z)$ used as a basis in this thesis are shown in green. The divertor target (also referred to as the wall) is the grey surface S, and two planes are shown cutting through it: the flux surface containing the field line $\vec{B}$ (green arrow) and the toroidal direction $\zeta$, and the plane perpendicular to $y$ (framed in green) containing the field line and the normal to the wall.
In both drawings we have identified, in the region near the divertor target, the local poloidal and toroidal axes $\theta$ and $\zeta$ respectively, and the axis locally normal to the flux surface $\psi$. 
On the right, we have labelled the minimum angle $\alpha$ between the field and the wall.}
\label{fig-tokamakgeometry}
\end{figure}

\section{Turbulence}
\label{subapp-lengthscales-turbulence}

The size of the turbulent structures is assumed of order $l \sim \rho_{\text{i}} / \delta$ in any direction perpendicular to field lines, with $\delta \ll 1$. I proceed to estimate the parallel length $l_{\parallel}$ and turnover time $t_{\text{turn}}$ associated with such structures.
In the perpendicular direction, over a characteristic turbulent timescale, I assumed that plasma travels a distance $l$. 
The distance it travels in the parallel direction is larger than $l$ by the factor by which the typical velocity along the field line, the thermal velocity $v_{\text{t,i}}$, is larger than the cross-field one. 
I order the cross-field velocity of plasma in the SOL the same as the $\vec{E} \times \vec{B}$ drift I expect turbulence to produce, $\sim \delta v_{\text{t,i}}$. 
I can thus assume that turbulent structures have a size
$l_{\parallel} = l / \delta$
parallel to the magnetic field.
\begin{figure}[h]
\centering
\includegraphics[scale=0.4]{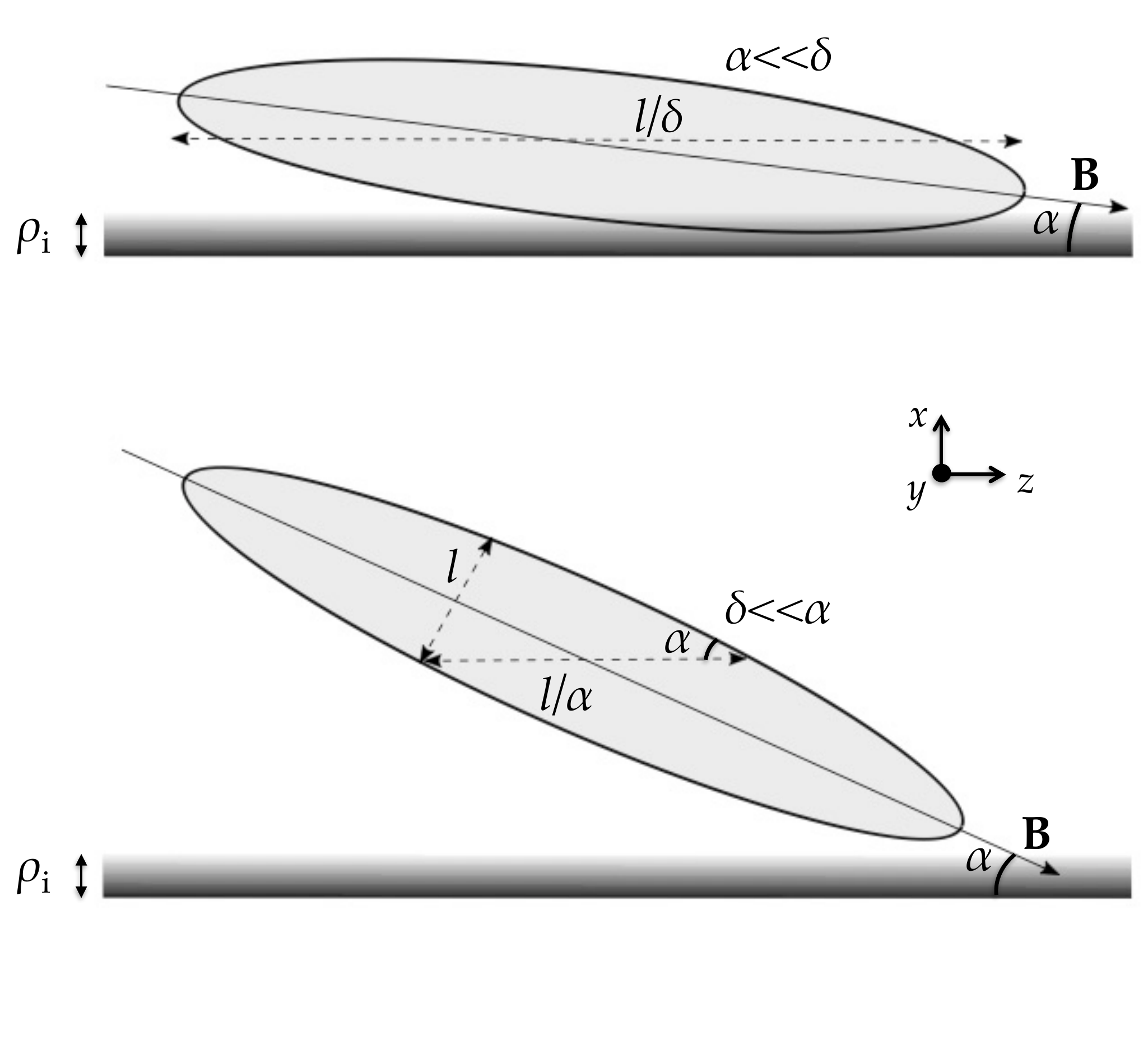}
\caption[Turbulence orderings]{Turbulent structures in the SOL, as they approach  the magnetic presheath (thin shaded region of thickness $\sim \rho_{\text{i}}$), are shown here. The elongation of these structures is by a factor $1/\delta$, which comes from the characteristic size of perpendicular velocities compared to parallel ones. Two cases (i) $\alpha \ll \delta$ and (ii) $\delta \ll \alpha$ are shown. In (i), the size of the turbulent structure in the $z$ direction is determined by the length of the turbulent structure parallel to the field line, $l/\delta$. In (ii), it is determined by the length of a cut across the eddy, $l/\alpha$.}
\label{figure-turbulence-size}
\end{figure}

 I refer to turbulent scale lenghs in the $z$ direction as $l_z$. For $\alpha \ll \delta$, it should be clear from Figure \ref{figure-turbulence-size} that gradients in the $z$-direction in the magnetic presheath, arising due to the turbulent structures impinging on the wall, are set by $l/ \delta$, so that $l / l_z \sim \delta$. On the other hand, when $\delta \ll \alpha$ the length scale in the $z$-direction is set by the horizontal cut across the eddy shown in the lower picture, of length $l/ \alpha$, so that $l / l_z \sim \alpha$. Therefore, $l_z \sim \min \left( l / \alpha , l / \delta \right) $. By ordering $z \sim l_z$, I obtain the ordering of (\ref{xyz-order}).
  
  The turnover time of turbulence is obtained from the characteristic length and velocity scales associated with the turbulence,
$t_{\text{turn}} \sim l / \delta v_{\text{t,i}} \sim 1 / \delta ^2 \Omega \text{.}$
 This leads to an estimate for the characteristic frequency of changes within turbulent structures, using $\partial / \partial t \sim 1 / t_{\text{turn}}$, from which the ordering of (\ref{t-dep}) follows.

The gradients in the $x$ direction outside of the magnetic presheath are determined by the cross-field size $l$ of turbulent structures, but they get larger as the magnetic presheath is approached (its characteristic thickness is $\rho_{\text{i}}$). Pictorially, this can be viewed as a squeezing that the turbulent structures undergo in the direction normal to the wall as they approach it. However, the discussion on the characteristic lengths \emph{parallel to the wall} in the magnetic presheath is unaffected, because these scales are inherited from the boundary conditions at the magnetic presheath entrance ($x \rightarrow \infty$).

\section{Steady-state scrape-off layer width}
\label{subapp-parallelgradients-turbulence}

\begin{figure}[h]
\centering
\includegraphics[scale=0.4]{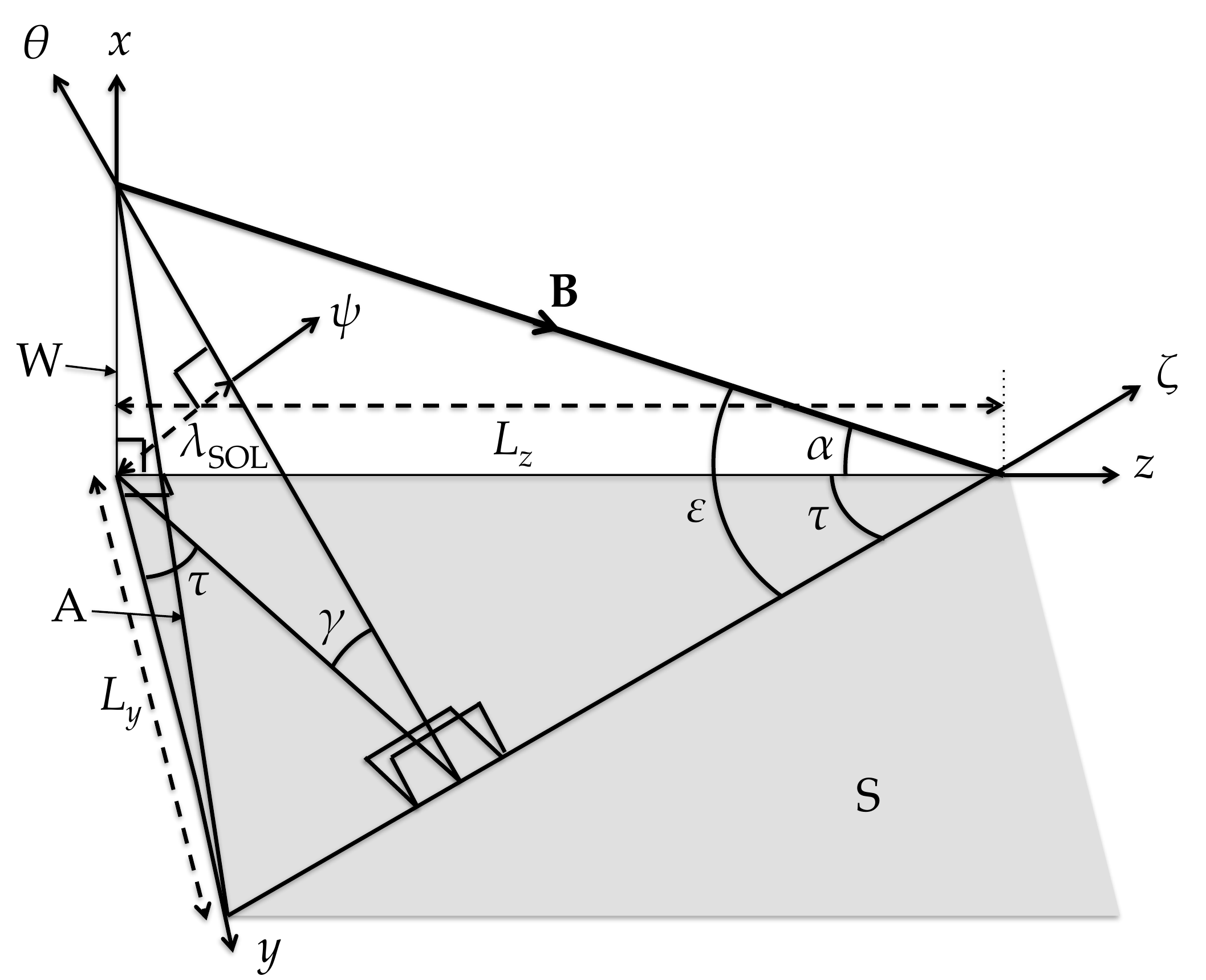}
\caption[Length scales in magnetic presheath]{A schematic that shows how the SOL thickness $\lambda_{\text{SOL}}$, measured in the flux coordinate direction $\psi$, is projected to lengths $L_y$ and $L_z$ in the directions parallel to the wall. 
The plane A is the flux surface, W is the plane perpendicular to $y$ (framed green in Figure \ref{fig-tokamakgeometry}) and S is the divertor target.
The following angles are shown: $\gamma$, the angle between the flux surface A and the divertor target S; $\varepsilon$, the angle between the toroidal direction $\zeta$ and the magnetic field line $\vec{B}$; $\tau$, the angle between the $z$ and $\zeta$ directions.}
\label{diagram-SOL-thick}
\end{figure}

One expects gradients in the $y$ and $z$ directions due to the steady-state SOL width because the flux coordinate direction $\psi$ has components in the $y$ and $z$ directions (see Figure \ref{diagram-SOL-thick}).\footnote{The $x$ direction also has a component in the $\psi$ direction, but the scale of the SOL width $\lambda_{\text{SOL}}$ is large compared to the magnetic presheath scale $\rho_{\text{i}}$, so this does not matter.}
Note that, in a typical tokamak, the SOL width $\lambda_{\text{SOL}}$ is of the order of the width of turbulent structures, $\lambda_{\text{SOL}} \sim l \sim 10 \text{ mm}$ \cite{Carralero-2015}.
One can calculate the projection in the $y$ and $z$ directions of the SOL width, and thus obtain an estimate for the characteristic steady-state scale lengths in those directions, $L_{y}$ and $L_{z}$. 
These must be greater than or equal to the turbulent scales in those directions, $L_y \gtrsim l$ and $L_z \gtrsim l_z \sim \min \left( l/\alpha , l / \delta \right)$.

In Figure \ref{diagram-SOL-thick}, the angle $\varepsilon$ is related to the ratio of the poloidal component of the magnetic field to the toroidal component of the magnetic field, $\tan \varepsilon = \left| B_{\theta} / B_{\zeta} \right| $. The ratio $B_{\theta} / B_{\zeta}$ is usually small in tokamaks, so $\varepsilon \ll 1$. The angle $\gamma$ is the angle between the flux surface and the divertor target. From Figure \ref{diagram-SOL-thick}, I obtain an expression relating $\alpha$ to $\varepsilon$ and $\gamma$, 
\begin{align}
\sin \alpha = \sin\varepsilon \sin\gamma \text{.}\end{align}
In order to achieve $\alpha \ll 1$ it is sufficient to have $\varepsilon \ll 1 $, which is valid for most tokamaks. However, the divertor target inclination in the poloidal plane is a free design parameter, and the flux surface geometry can be controlled with the external magnets. Therefore, the angle $\gamma$ between the divertor target and the flux surface is often also made small, $\gamma \lesssim 1$, in order to make $\alpha$ even smaller. I order $\alpha$ with respect to $\varepsilon$ as $\alpha \lesssim \varepsilon$.

In what follows, it will be convenient to use the angle $\tau$, shown in Figure \ref{diagram-SOL-thick}. I express $\tau$ in terms of $\varepsilon$ and $\gamma$, 
\begin{align} \label{tau-epsilon-gamma}
\tan\tau = \tan \varepsilon \cos\gamma \text{.}
\end{align}
I proceed to express the length scales $L_y$ and $L_z$ in terms of the SOL width $\lambda_{\text{SOL}}$ and the angles $\varepsilon$ and $\gamma$. Projecting the SOL width $\lambda_{\text{SOL}}$ onto the $z$-axis and using (\ref{tau-epsilon-gamma}), I obtain 
\begin{align} \label{Lz}
L_z = \frac{\lambda_{\text{SOL}}}{ \sin \gamma \sin \tau }  = \frac{ \sqrt{1 + \tan^2\varepsilon \cos^2 \gamma }}{ \sin \gamma \cos \gamma \tan \varepsilon }  \lambda_{\text{SOL}} \sim \frac{1}{\varepsilon \sin\gamma} \lambda_{\text{SOL}} \gg \lambda_{\text{SOL}}  \text{.}
\end{align}
The presence of $\sin \gamma$ in the denominator of (\ref{Lz}) implies that $L_z \sim \lambda_{\text{SOL}} / \alpha \gtrsim l_z \sim \min \left( l/ \alpha , l / \delta \right) $. The SOL width projected in the $y$ direction is, using (\ref{tau-epsilon-gamma}), 
\begin{align}
L_y = \frac{\lambda_{\text{SOL}}}{\cos \tau \sin \gamma} = \frac{\sqrt{1 + \tan^2 \varepsilon \cos^2 \gamma}}{\sin\gamma} \lambda_{\text{SOL}} \sim \frac{\lambda_{\text{SOL}}}{\sin\gamma} \gtrsim \lambda_{\text{SOL}} \text{.}
\end{align}
This also implies $L_y \gtrsim l$. The ratio of the two steady-state length scales is
\begin{align}
\frac{L_y}{L_z} = \tan \varepsilon \cos \gamma \sim \varepsilon \text{.}
\end{align}
This means that $L_z$ is much larger than $L_y$, by a factor $\sim 1 / \varepsilon$.


\chapter{How large can the parallel current be?}
\label{appendix:largejz}

In this appendix, I consider the validity of our equations when a large current $\vec{j}^L$ is driven parallel to the magnetic field through the plasma in the magnetic presheath. 
In Section 2, I ordered the plasma currents using the particle drifts, and the relationship between current components obtained using Maxwell's equations. 
This means that the ordering I obtained there, $j_z^D \sim \alpha n_{\text{i}} e v_{\text{t,i}} \sim \delta n_{\text{i}} e v_{\text{t,i}} $ is consistent with the piece of the parallel current that flows through the plasma in response to the currents due to the perpendicular drifts. 
This parallel current is present to satisfy $\nabla \cdot \vec{j}^D = 0$ and maintain charge neutrality, that may otherwise be broken by the divergence of the perpendicular current. 
In this Appendix, I show that that the orderings are consistent with a larger parallel current, $j^L \gg \delta e n_{\text{i}} v_{\text{t,i}}$, provided that the size of this current does not become too large, at which point neglecting the plasma produced magnetic fields and the induced electric fields would no longer be a valid assumption. 

The large parallel current $\vec{j}^L$ has components $j^L_z = j^L \cos \alpha \sim j^L$, $j_x^L = - j^L \sin \alpha  \sim \alpha j^L $ and $j_y^L = 0$. 
This parallel current must satisfy $\nabla \cdot \vec{j}^L = 0$. 
Using (\ref{xyz-order}) I find $\partial j^L_x / \partial x \sim \alpha j^L / \rho_{\text{i}}$ and $\partial j^L_z / \partial z \sim \delta \alpha j^L / \rho_{\text{i}} \sim \delta^2 j^L / \rho_{\text{i}}$. 
Thus, $\nabla \cdot \vec{j}_L = 0$ requires $\partial j^L / \partial x = 0$, and to lowest order $\vec{j}^L$ is not affected by the magnetic presheath.
Therefore, the length scale in the $x$ direction of the large parallel current and the magnetic and electric fields associated with it is larger than the magnetic presheath scale $\rho_{\text{i}}$. 
Balancing $\partial j_x^L / \partial x \sim \partial j_z^L / \partial z$ leads to ordering 
\begin{align}
\frac{\partial}{\partial x} \sim \frac{\delta}{\alpha l} \sim \frac{1}{l} \text{,}
\end{align}
so the appropriate length scale in the $x$ direction (for this Appendix only) is the perpendicular turbulent scale $l$.

The magnetic field produced in the plasma by $\vec{j}^L$ is denoted $\vec{B}^{p'}$ and is determined by Amp\`ere's law (\ref{Ampere}), with $\vec{j}^L$ instead of $\vec{j}^D$ and $\vec{B}^{p'}$ instead of $\vec{B}^{p}$. Taking the $y$ component of (\ref{Ampere}) with $j_y^L = 0$, I have $\partial B_z^{p'} / \partial x \sim \partial B_x^{p'} / \partial z$, from which I obtain $B_z^{p'} \sim \alpha B_x^{p'}$. Considering the long length scales in the $z$ direction, this implies that $\partial B_z^{p'} / \partial z$ must be subdominant in $\nabla \cdot \vec{B}^{p'} = 0$, so $\partial B_x^{p'} / \partial x \sim \partial B_y^{p'} / \partial y $. This implies that $B_x^{p'} \sim B_y^{p'}$. The $x$ and $z$ components of Amp\`ere's law determine $B_y^{p'} \sim B_x^{p'} \sim \mu_0 l j^L $. 
Collecting the orderings for the components of the magnetic field produced by a large parallel current $\vec{j}_{\text{L}}$, I have
\begin{align} \label{ordering-Bx-By-Bz-largejz}
B_z^{p'} \sim \alpha B^{p'} \ll B_x^{p'} \sim B_y^{p'} \sim B^{p'} \sim \mu_0 l j^L \sim \frac{ j^L }{\delta en_{\text{i}}v_{\text{t,i}}} \beta  B^c \text{,}
\end{align}
where in the rightmost equation I used $\beta B^c \sim B^p \sim \mu_0 l \delta n_{\text{i}} e v_{\text{t,i}} $ inferred from (\ref{j-orderings}) and (\ref{ordering-beta}).

As explained in section \ref{sec-traj-orderings}, in order to neglect $\vec{B}^{p'}$, I require each component of it to be negligible compared to either the respective component or the smallest retained component of the constant external magnetic field $\vec{B}^c$. The strongest constraint is obtained by the neglect of $B_x^{p'}$ and $B_y^{p'}$ compared to $B_x^c \sim \alpha B^c$. This is $B^{p'} \ll \alpha B^c$, which leads to
\begin{align} \label{ordering-large-current}
j^L \ll  \frac{\alpha}{\beta}  \delta e n_{\text{i}} v_{\text{t,i}} \text{.}
\end{align}

The large parallel current $\vec{j}^L$ is consistent with an electrostatic electric field provided that each component of the electric field $\vec{E}^{p'}$ induced by $\vec{B}^{p'}$ is negligible compared to either the same component or the smallest retained component of $-\nabla \phi$. 
From (\ref{t-dep}), (\ref{ordering-Bx-By-Bz-largejz}) and the length scale orderings of this section (which are $x \sim y \sim l \ll z \sim l / \delta \sim l / \alpha$), one can order the components of the induction equation (\ref{induction}), with $\vec{E}^{p'}$ instead of $\vec{E}^{p}$ and $\vec{B}^{p'}$ instead of $\vec{B}^p$. I obtain (recalling that $\rho_{\text{i}} \Omega = v_{\text{t,i}}$)
\begin{align}
E_x^{p'} \sim E_y^{p'} \sim \alpha E^{p'} \ll E_z^{p'} \sim E^{p'} \sim \delta v_{\text{t,i}} B^{p'} \text{.}
\end{align}
Therefore, I find that the strongest constraint on the electrostatic approximation is $E^{p'}_z \sim \delta v_{\text{t,i}} B^{p'} \ll \delta v_{\text{t,i}} B^c$ and leads to $B^{p'} \ll B^c$. 
This is a weaker condition than the one needed to neglect the magnetic field $\vec{B}^{p'}$; hence, the electrostatic approximation is justified when (\ref{ordering-large-current}) holds. 
Note that (\ref{ordering-large-current}) allows for a large parallel current in the magnetic presheath because $\alpha \gg \beta$, as pointed out in Section 2. 
If currents larger than (\ref{ordering-large-current}) were present in the magnetic presheath, one would have to consider the magnetic fields produced by them in our equations (and in extreme cases also the induced electric fields). For example, currents larger than (\ref{ordering-large-current}) would be large enough to change the angle $\alpha$ between the magnetic field lines and the wall.

\chapter{Adiabatic invariant} 
\label{appendix:mucalc}

I proceed to prove that the lowest order adiabatic invariant has the form in (\ref{mu-Uperp-xbar}), starting from the definition
\begin{align} \label{adiabatic-definition}
\mu = \frac{1}{2\pi m_{\text{i}}}\oint  \vec{\tilde{p}} \cdot \vec{d\tilde{r}} \text{,}
\end{align}
where $\vec{\tilde{p}}$ and $\vec{\tilde{r}} = (\tilde{x}, \tilde{y}, \tilde{z})$ are the canonical momentum and the position vector of the charged particle in the system where $\alpha = \delta = 0$ and in the frame where the parallel motion and $\vec{E}\times \vec{B}$ drift of an ion are zero.
The electric field in this frame is given by $\vec{\tilde{E}} = \vec{E} - \left\langle \vec{E} \right\rangle_{\varphi}$, whereas the magnetic field is unchanged provided we are in the non-relativistic limit. 

From (\ref{mag-vec-pot}), the magnetic vector potential can be expressed as (using $\alpha = 0$).
\begin{align}
\vec{A} = \begin{pmatrix}
0 \\
\tilde{x}B \\ 0
\end{pmatrix} \text{.}
\end{align}
The canonical momentum is $\vec{\tilde{p}} = m_{\text{i}} \vec{\tilde{v}} + Ze\vec{A}$, where $\vec{\tilde{v}} = \vec{v} - \left\langle \vec{v} \right\rangle_{\varphi} $.
Using (\ref{vx-Uperp-xbar-x}) and (\ref{gyroaverage}), the component $v_x$ of the velocity gyroaverages to zero.
The gyroaverage of $v_y$ is (\ref{V-ExB}) and the gyroaverage of $v_z$ is (\ref{vz-U-Uperp}), leading to
\begin{align} \label{v-gyroaveraged}
\left\langle \vec{v} \right\rangle_{\varphi} =
\begin{pmatrix}
0 \\
V_{\vec{E}\times\vec{B}} (\bar{x}, U_{\perp}) \\
\sigma_{\parallel} V_{\parallel} \left( U_{\perp}, U \right) \end{pmatrix} \text{.}
\end{align}
Note that there is no periodicity in the $z$ direction, and thus $\tilde{v}_z = 0$.
The position $\vec{\tilde{r}}$ is obtained by integrating $\vec{\tilde{v}}$ in time (which, due to periodicity, is equivalent to integrating in gyrophase and dividing by $\overline{\Omega}$). 
For simplicity, I choose $\vec{\tilde{r}}$ such that $\left\langle \vec{\tilde{r}} \right\rangle_{\varphi} = 0$. 
Using $\vec{d\tilde{r}} = \vec{\tilde{v}} d\varphi / \overline{\Omega} $, the expression for the adiabatic invariant in (\ref{adiabatic-definition}) becomes
\begin{align}
\mu = \frac{1}{2\pi \overline{\Omega}} \oint \left( \tilde{v}^2 + \Omega \tilde{x} \tilde{v}_y \right) d \varphi \text{.}
\end{align}
From (\ref{vy-xbar-x}) I obtain $\tilde{x} = - \tilde{v}_y / \Omega$ which leads to (using $\tilde{v}_x = v_x$ and $\tilde{v}_z = 0$)
\begin{align} \label{inv2}
\mu = \frac{1}{2\pi \overline{\Omega}} \oint v_x^2 ~d\varphi \text{.}
\end{align}
Using (\ref{vx-Uperp-xbar-x}) and the definition of the gyroaverage in (\ref{gyroaverage}), equation (\ref{inv2}) reduces to the form in (\ref{mu-Uperp-xbar}).

\chapter{Ion density at the magnetic presheath entrance}  
\label{app-quasi-expansion}

Here I derive equation (\ref{niclosedinfty}) in the following steps. 
First, in section \ref{subapp-mu-expansion}, I expand the adiabatic invariant (\ref{mu-Uperp-xbar}) as a function of $\bar{x}$ and $U_{\perp}$ for small electrostatic potential, $e\phi (x) / T_{\text{e}} \ll 1$, and small gradients of the electrostatic potential, $\epsilon = \rho_{\text{i}} \phi'(x) / \phi(x) \ll 1 $.
Then, in section \ref{subapp-varphi-expansion} I expand equation (\ref{varphi-def}) to obtain an expression for $\bar{x}$ as a function of $\varphi$, $x$ and $\mu$. 
I also obtain an expression for $U_{\perp}$ as a function of $\varphi$, $x$ and $\mu$. 
Then, by making the change of variables $(x, \bar{x}, U_{\perp}, U)  \rightarrow (x, \varphi, \mu, U) $, I obtain an expression for the ion density in section \ref{subapp-ni-expansion}.
Finally, this expression for the ion density is carefully expanded in section \ref{subapp-ni-finalexpansion}. 
The results of this appendix are valid to lowest order in $\alpha$. 

\section{Adiabatic invariant expansion} \label{subapp-mu-expansion}

I proceed to derive an expression for $\mu$ as a function of $\bar{x}$ and $U_{\perp}$ by expanding equation (\ref{mu-Uperp-xbar}) near $x\rightarrow \infty$, where $e\phi (x) / T_{\text{e}} \ll 1$. In addition, I assume that the length scale of changes in the electrostatic potential is much larger than the ion gyroradius $\rho_{\text{i}}$, defining the small parameter $\epsilon$ of equation (\ref{epsilonsmall}). I first expand the expression inside the square root of equation (\ref{vx-Uperp-xbar-x}) around $x=\bar{x}$ to second order in $\epsilon$, obtaining
\begin{align}
v_x = \sigma_x V_x\left(x, \bar{x}, U_{\perp} \right) =  \sigma_x \sqrt{2} \left[  U_{\perp} - \frac{1}{2}\Omega^2 \left(x-\bar{x} \right)^2 - \frac{\Omega  \phi\left(\bar{x}\right)}{B} \right.  \nonumber \\  \left. -  \frac{\Omega  \phi'\left(\bar{x}\right)}{B} \left( x - \bar{x} \right) -   \frac{\Omega  \phi''\left(\bar{x}\right)}{2B} \left( x - \bar{x} \right)^2  + O \left( \epsilon^3 \hat{\phi} v_{\text{t,i}}^2  \right) \right]^{1/2} \text{.}
\end{align}
Completing the square in the square root and dropping small terms gives
\begin{align} \label{Vx-infty}
V_x\left(x, \bar{x}, U_{\perp} \right) = & A \Omega \sqrt{1 + \frac{\phi''\left( \bar{x}\right)}{\Omega B}} \nonumber
\\ & \times \sqrt{ 1 - \frac{1}{A^2} \left[ x -\bar{x} + \frac{ \phi'\left( \bar{x} \right)}{ \Omega B}  \right]^2 + O \left(  \hat{\phi} \epsilon^3, \hat{\phi}^2 \epsilon^2 \right)  }  \text{,}
\end{align}
where I have defined the orbit amplitude
\begin{align} \label{A-def}
A = \frac{1}{\Omega}  \left( 1 + \frac{\phi''(\bar{x})}{\Omega B} \right)^{-1/2} \sqrt{ 2U_{\perp} - \frac{ 2\Omega \phi\left( \bar{x} \right) }{B}    }  \text{.}
\end{align}
The bounce points of the closed orbit are obtained by solving $V_x \left( x, \bar{x}, U_{\perp}  \right) = 0$, leading to
\begin{align} \label{xb-infty}
x_{\text{b}} = \bar{x} - \frac{\phi'\left( \bar{x} \right)}{ \Omega B}  - A \text{,}
\end{align}
\begin{align} \label{xt-infty}
x_{\text{t}} = \bar{x} - \frac{\phi'\left( \bar{x} \right)}{ \Omega B}  + A \text{.}
\end{align}
By substituting (\ref{Vx-infty}) into equation (\ref{Omegabar-def}) and using (\ref{xb-infty}) and (\ref{xt-infty}) for the integration limits, I find
\begin{align} \label{varphi-explicit}
\frac{\pi}{\overline{\Omega}} = & \int_{x_{\text{b}}}^{x_{\text{t}}} \left( A \Omega  \sqrt{1 + \frac{\phi''\left( \bar{x}\right)}{\Omega B}}   \sqrt{ 1 - \frac{1}{A^2} \left[ x -\bar{x} + \frac{\phi'\left( \bar{x} \right)}{ \Omega B}  \right]^2  } \right)^{-1} dx  \nonumber \\ & + O \left( \frac{\hat{\phi}  \epsilon^3}{\Omega} , \frac{\hat{\phi}^2  \epsilon^2}{\Omega} \right)  \text{,}
\end{align}
which leads to the modified gyrofrequency
\begin{align} \label{Omegabar-infty}
\overline{\Omega} =  \Omega \sqrt{1 + \frac{\phi''\left( \bar{x}\right)}{\Omega B}} + O \left( \hat{\phi}  \epsilon^3 \Omega ,  \hat{\phi}^2  \epsilon^2 \Omega  \right)   =  \Omega \left( 1 + \frac{\phi''\left( \bar{x}\right)}{2\Omega B} +  O \left( \hat{\phi}  \epsilon^3 ,  \hat{\phi}^2  \epsilon^2  \right) \right) \text{.}
\end{align}
We exploit (\ref{Omegabar-infty}) to simplify equation (\ref{Vx-infty}),
\begin{align} \label{Vx-infty-simpler}
V_x\left(x, \bar{x}, U_{\perp} \right) =   \overline{\Omega} A \sqrt{ 1 - \frac{1}{A^2} \left[ x -\bar{x} + \frac{\phi'\left( \bar{x} \right)}{\Omega B}\right]^2 + O \left( \hat{\phi} \epsilon^3 ,  \hat{\phi}^2 \epsilon^2   \right)  }  \text{.}
\end{align}
By inserting (\ref{Vx-infty-simpler}) into expression (\ref{mu-Uperp-xbar}) for the adiabatic invariant, I find
\begin{align}
\mu = \frac{1}{\pi} \int_{x_{\text{b}}}^{x_{\text{t}}} \overline{\Omega} A \sqrt{ 1 - \frac{1}{A^2} \left[ x -\bar{x} + \frac{ \phi'\left( \bar{x} \right)}{ \Omega B}  \right]^2  } dx + O \left( \hat{\phi }  \epsilon^3 \frac{v_{\text{t,i}}^2}{\Omega} , \hat{\phi }^2  \epsilon^2 \frac{v_{\text{t,i}}^2}{\Omega}  \right) \text{,}
\end{align}
which evaluates to
\begin{align} \label{mu-Omega}
\mu & = \frac{1}{2}\overline{\Omega} A^2 + O \left(  \hat{\phi} \epsilon^3 \rho_{\text{i}}  v_{\text{t,i}},  \hat{\phi}^2 \epsilon^2 \rho_{\text{i}}  v_{\text{t,i}} \right)  \nonumber \\
& =  \frac{1}{\overline{\Omega}} \left( U_{\perp} - \frac{ \Omega \phi\left( \bar{x} \right) }{B}     \right)  + O \left(  \hat{\phi} \epsilon^3 \rho_{\text{i}}  v_{\text{t,i}} , \hat{\phi}^2 \epsilon^2 \rho_{\text{i}}  v_{\text{t,i}} \right) \text{.}
\end{align}
Rearranging equation (\ref{mu-Omega}) and using (\ref{Omegabar-infty}), I obtain
\begin{align} \label{mu-expanded1}
U_{\perp}  & = \overline{ \Omega } \mu  +  \frac{\Omega \phi\left( \bar{x} \right) }{ B }  + O \left(  \hat{ \phi } \epsilon^3 \frac{v_{\text{t,i}}^2}{\Omega},  \hat{\phi}^2 \epsilon^2  \frac{v_{\text{t,i}}^2}{\Omega} \right)  \nonumber \\ 
& = \Omega \mu  +   \frac{\Omega \phi\left( \bar{x} \right) }{B } +  \frac{\mu \phi'' \left( \bar{x} \right) }{  2B } + O \left(  \hat{ \phi } \epsilon^3 \frac{v_{\text{t,i}}^2}{\Omega},  \hat{\phi}^2 \epsilon^2  \frac{v_{\text{t,i}}^2}{\Omega} \right)  \text{.}
\end{align}
At $x\rightarrow \infty$, the zeroth order in $\hat{\phi}$ of all the equations in this Appendix is valid exactly. 
Then, I have $\overline{\Omega} = \Omega$ from equation (\ref{Omegabar-infty}) and $\mu = U_{\perp}/\Omega$ from equation (\ref{mu-Omega}). Hence, the equations $2\Omega \mu = v_x^2 + v_y^2$ and $2\left( U - \Omega \mu \right) = v_z^2$ are valid at $x \rightarrow \infty$.
These equations are used to obtain equation (\ref{changevar-infty}) and to obtain $F_{\text{cl}}(\mu, U)$ from $f_{\infty}(\vec{v})$.


\section{Gyrophase expansion} \label{subapp-varphi-expansion}

I require an expression for $U_{\perp}$ as a function of $\mu$, $\varphi $ and $x$. 
To obtain it from equation (\ref{mu-expanded1}), I need an equation for $\bar{x}$ as a function of $\mu$, $\varphi$ and $x$, which I proceed to derive.
I insert equation (\ref{Vx-infty-simpler}) into the definition of the gyrophase $\varphi$, equation (\ref{varphi-def}), and use the top bounce point in (\ref{xt-infty}) as the lower integration limit to obtain
\begin{align} \label{varphi-app}
\varphi = \sigma_x \int^x_{x_{\text{t}}} \left( A  \sqrt{ 1 - \frac{1}{A^2} \left[ s -\bar{x} + \frac{ \phi'\left( \bar{x} \right)}{ \Omega  B}  \right]^2  } \right)^{-1} ds + O \left(  \hat{\phi} \epsilon^3, \hat{\phi}^2 \epsilon^2  \right)   \text{.}
\end{align}
Note that $\varphi >0$ when $\sigma_x = -1$. 
Using equation (\ref{varphi-app}) and $A=\sqrt{2\mu/\overline{\Omega}}$ (from equation (\ref{mu-Omega})), I obtain the relation
\begin{align} \label{x-xbar-varphi}
x -\bar{x} + \frac{\phi'\left( \bar{x} \right)}{ \Omega B} = \sqrt{\frac{2\mu}{\overline{\Omega}}} \cos \varphi + O \left(  \hat{\phi}  \epsilon^3 \rho_{\text{i}},  \hat{\phi}^2 \epsilon^2 \rho_{\text{i}}  \right) \text{.}
\end{align}
Then, we expand equation (\ref{x-xbar-varphi}) around the lowest order $\bar{x} = x - \sqrt{2\mu/\overline{\Omega}} \cos \varphi $ to obtain
\begin{align} \label{xbar-expanded}
\bar{x} & = x - \left( 1 +  \frac{3 \Omega \phi'' (x)}{4B}  \right) \sqrt{ \frac{2\mu}{\Omega} } \cos \varphi + \frac{ \phi' (x)}{\Omega B} + O \left(  \hat{\phi} \epsilon^3 \rho_{\text{i}},  \hat{\phi}^2 \epsilon^2 \rho_{\text{i}} \right)  \text{.}
\end{align}
Similarly, I expand equation (\ref{mu-expanded1}) around $\bar{x} = x - \sqrt{2\mu/\Omega} \cos \varphi $ to obtain
\begin{align} \label{Uperp-expanded}
U_{\perp} = \Omega \mu +\frac{\Omega \phi\left( x \right) }{B} - \frac{\Omega \phi'\left( x \right) }{B} \sqrt{\frac{2\mu}{\Omega} } \cos \varphi + \frac{\mu \phi'' \left( x \right) }{2 B} \left( 1 + 2 \cos^2 \varphi \right) \nonumber  \\
+ O \left( \hat{\phi} \epsilon^3 v_{\text{t,i}}^2, \hat{\phi}^2 \epsilon^2 v_{\text{t,i}}^2 \right) \text{.}
\end{align}
Defining $ \delta U_{\perp} = \Omega \mu - U_{\perp} $, equation (\ref{Uperp-expanded}) leads to equation (\ref{deltaUperp}).

\section{Change of variables in the ion density integral}  \label{subapp-ni-expansion}

\begin{figure} 
\includegraphics[width=0.9\textwidth]{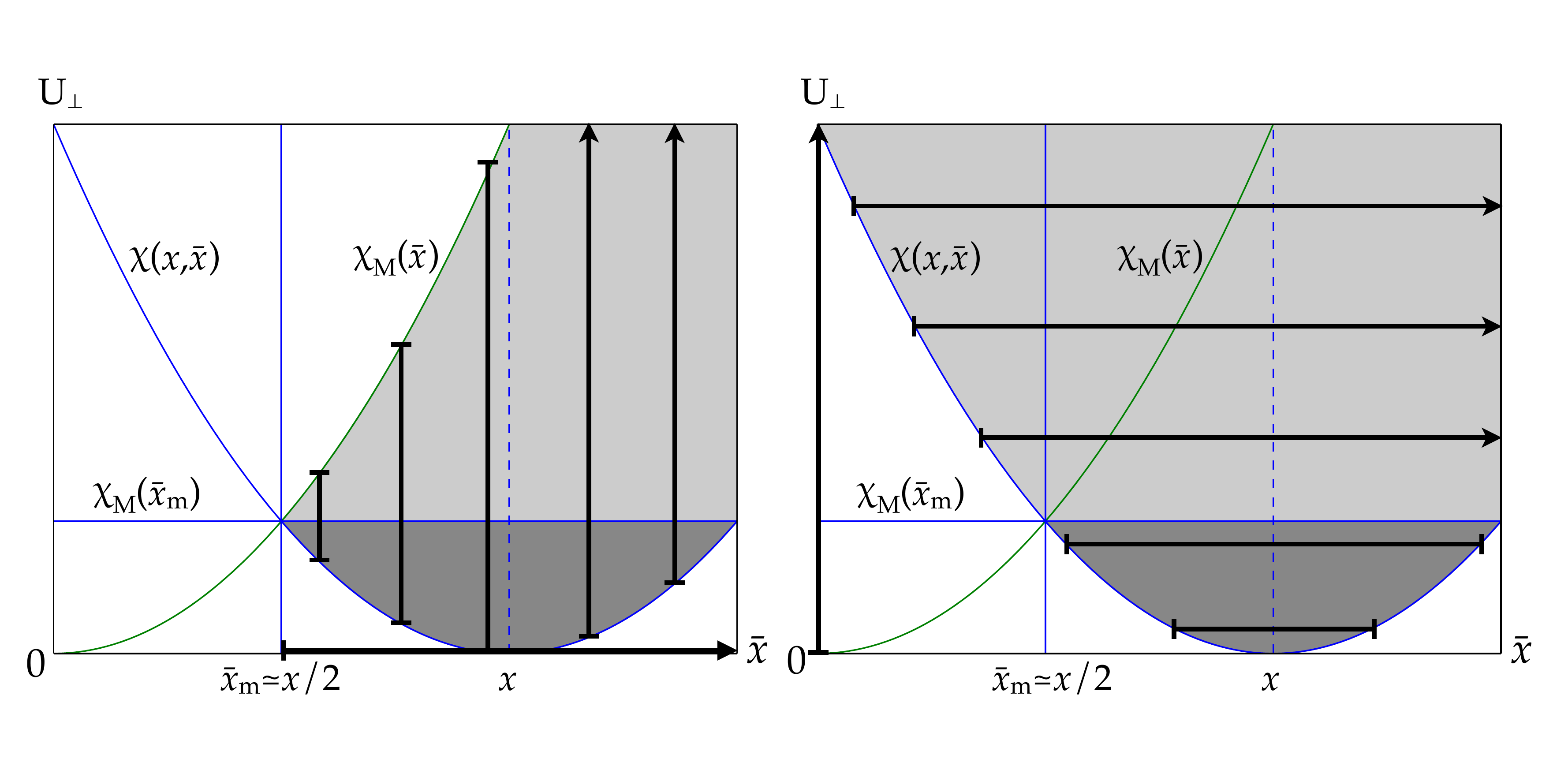}
\caption[Inverting the order of integration in the ion density integral at $x\rightarrow\infty$]{The integration domain in $(\bar{x}, U_{\perp})$ of equation (\ref{ni-closed-change}) consists of both shaded regions on the left hand side drawing. When I exchange the integration order, the integration limits (bold lines) are picked such that the integration domain coincides in the dark grey region but not in the light grey one. The light grey region satisfies $U_{\perp} > \chi_{\text{M}} (x/2) = \Omega^2 x^2 / 8 \gg v_{\text{t,i}}^2 $ near $x\rightarrow\infty$, and at such large energies I expect the distribution function to be exponentially small. Thus, the contribution to the integral from this region of phase space is negligible and the limits of integration of equation (\ref{ni-closed-orderexchange}) are appropriate. }
\label{fig-integrationchange}
\end{figure}

For sufficiently large $x$, the open orbit density is zero and the closed orbit density in (\ref{ni-closed}) is
\begin{align} \label{ni-closed-change}
n_{\text{i,cl}} (x) \simeq \int_{\bar{x}_{\text{m}}(x)}^{\infty} \Omega d\bar{x} \int_{\chi\left( x, \bar{x} \right)}^{\chi_{\text{M}}(\bar{x})} \frac{2dU_{\perp}}{\sqrt{2\left(U_{\perp} - \chi\left(x, \bar{x} \right) \right)}} \int_{U_{\perp}}^{\infty} \frac{F_{\text{cl}} \left( \mu \left( \bar{x}, U_{\perp} \right), U \right) }{\sqrt{2\left( U - U_{\perp} \right)}} dU  \text{.}
\end{align}
The value of $\bar{x}_{\text{m}}(x)$ is given by equation (\ref{xbarm}) evaluated near $x \rightarrow \infty$,
\begin{align}
\bar{x}_{\text{m}} (x) \simeq \frac{1}{2} x \text{,}
\end{align}
The effective potential maximum at large $\bar{x}$ is, from equation (\ref{xbar-infty-limit}), $\chi_{\text{M}} (\bar{x}) \simeq  \Omega^2 \bar{x}^2/2$.
I can exchange the integrals over $\bar{x}$ and $U_{\perp}$ to get 
\begin{align} \label{ni-closed-orderexchange}
n_{\text{i,cl}} (x) \simeq \int_0^{\infty} dU_{\perp} \int_{\bar{x}_{\text{b}}}^{\bar{x}_{\text{t}}} \frac{2\Omega d\bar{x}}{\sqrt{2\left( U_{\perp} - \chi (x, \bar{x}) \right) } } \int_{U_{\perp}}^{\infty} \frac{F_{\text{cl}} \left( \mu \left( \bar{x}, U_{\perp} \right), U \right) }{\sqrt{2\left( U - U_{\perp} \right)}} dU  \text{,}
\end{align}
where 
\begin{align}
\bar{x}_{\text{b}} = x - \frac{1}{\Omega} \sqrt{ 2\left( U_{\perp} - \frac{\Omega \phi(x)}{B} \right) } \text{,}
\end{align}
\begin{align}
\bar{x}_{\text{t}} = x + \frac{1}{\Omega} \sqrt{ 2\left( U_{\perp} - \frac{\Omega \phi(x)}{B} \right) } \text{.}
\end{align}
The change in the integration limits is explained in Figure \ref{fig-integrationchange}. 
Equations (\ref{xbar-expanded}) and (\ref{Uperp-expanded}) can be used to make the change of variables $(x, \bar{x}, U_{\perp}, U ) \rightarrow (x, \varphi, \mu, U ) $ in equation (\ref{ni-closed-orderexchange}).
Using equations (\ref{xbar-expanded}) and (\ref{Uperp-expanded}), the Jacobian
\begin{align}
\left| \frac{ \partial \left( \bar{x}, U_{\perp} \right) }{\partial \left( \varphi, \mu \right) } \right| = \left( 1 + \frac{5\phi''(x)}{4B\Omega} \right) \sqrt{2\Omega \mu} \left| \sin \varphi \right| + O \left(  \hat{\phi} \epsilon^3 v_{\text{t,i}}, \hat{\phi}^2 \epsilon^2 v_{\text{t,i}}   \right) \text{}
\end{align}
and the result
\begin{align}
\frac{1}{\sqrt{2\left( U_{\perp} - \chi (x, \bar{x}) \right)}} = \left( 1 - \frac{\phi''(x)}{4B\Omega } \right) \frac{1}{\sqrt{2\Omega\mu} \left| \sin \varphi \right|} + O\left( \frac{\hat{\phi}\epsilon^3}{v_{\text{t,i}}}, \frac{\hat{\phi}^2\epsilon^2}{v_{\text{t,i}}} \right) \text{,}
\end{align}
I obtain
\begin{align} \label{niclosedinftynotexp}
n_{\text{i,cl}} \left(x \right) = & \left( 1 +  \frac{\phi''(x)}{\Omega B}  \right) \int_{-\pi}^{\pi} d\varphi \int_0^{\infty} \Omega d\mu \int_{\Omega \mu}^{\infty} \frac{ F_{\text{cl}}(\mu, U)  }{\sqrt{2\left( U - \Omega  \mu + \delta U_{\perp} \right)}}   dU \nonumber \\
& + O\left( \hat{\phi} \epsilon^3 n_{\infty},  \hat{\phi}^2 \epsilon^2 n_{\infty} \right) \text{,}
\end{align}
where $\delta U_{\perp}$ is defined in equation (\ref{deltaUperp}). 
 Note that I changed the lower limit of the integral over $U$ from $U_{\perp}$ to $\Omega \mu$ in going from equation (\ref{ni-closed-orderexchange}) to (\ref{niclosedinftynotexp}). The distribution function is zero for $U < \Omega \mu$. Therefore, the integrand is zero in the region $U_{\perp} < U < \Omega \mu $ and $U_{\perp}$ can be replaced by $\Omega \mu$ in the integration limit of the integral in $U$. 

\section{Expansion of the integral over $U$ in equation (\ref{niclosedinftynotexp})} \label{subapp-ni-finalexpansion}

I begin by changing variables from $U$ to $U_{\star} = U - \Omega \mu  + \delta U_{\perp} $,
\begin{align}
\int_{\Omega \mu}^{\infty} \frac{  F_{\text{cl}}(\mu, U) dU}{\sqrt{2\left(U - \Omega\mu + \delta U_{\perp} \right) }}   =  \int_{\delta U_{\perp}}^{\infty} \frac{ F_{\text{cl}}(\mu, U_{\star} +  \Omega \mu - \delta U_{\perp} ) }{\sqrt{2U_{\star}  }}  dU_{\star} \text{.}
\end{align}
Note that $\delta U_{\perp} >0$ for typical values of $\mu$. I Taylor expand the distribution function
\begin{align} \label{Taylor-dist}
& \int_{\delta U_{\perp}}^{\infty} \frac{ F_{\text{cl}}(\mu, U_{\star} + \Omega \mu - \delta U_{\perp} ) }{\sqrt{2U_{\star}  }}  dU_{\star}   =  \int_{\delta U_{\perp}}^{\infty}  \frac{F_{\text{cl}}(\mu, U_{\star} + \Omega \mu )}{\sqrt{2U^{\star}}} dU^{\star} \nonumber \\  
& -  \int_{\delta U_{\perp}}^{\infty} \frac{\delta U_{\perp}}{\sqrt{2U_{\star}}}  \frac{\partial  F_{\text{cl}} }{\partial U} \left( \mu, U_{\star} + \Omega \mu \right)  dU_{\star}  + \frac{1}{2}  \int_{\delta U_{\perp}}^{\infty}  \frac{\delta U_{\perp}^2}{\sqrt{2U_{\star}}}   \frac{ \partial^2 F_{\text{cl}}}{\partial U^2}(\mu, U_{\star} + \Omega \mu )  dU_{\star} + \ldots 
\end{align}
Each of the terms of equation (\ref{Taylor-dist}) can then be split into two separate integrals over $U_{\star}$
\begin{align}
n_{\text{i}} (x) = & \int_{0}^{\infty}  \frac{dU_{\star}}{\sqrt{2U_{\star}}}   \left( F_{\text{cl}}(\mu, U_{\star} + \Omega \mu ) - \delta U_{\perp} \frac{\partial F_{\text{cl}}}{\partial U} (\mu, U_{\star} + \Omega \mu ) \right.  \nonumber \\
& \left.  + \frac{1}{2} \delta U_{\perp}^2 \frac{ \partial^2 F_{\text{cl}}}{\partial U^2} (\mu, U_{\star} + \Omega \mu )  \right) 
 -   \int_{0}^{\delta U_{\perp}}  \frac{dU_{\star}}{\sqrt{2U_{\star}}}   \left(  F_{\text{cl}}(\mu, U_{\star} + \Omega \mu )  \phantom{ \frac{\partial F}{\partial \mu} } \right.  \nonumber \\
& \left.  - \delta U_{\perp} \frac{\partial F_{\text{cl}}}{\partial U} (\mu, U_{\star} + \Omega \mu )    \right) + \ldots \text{.}
\end{align}
Then, for small $\delta U_{\perp}$, I Taylor expand the distribution function near $U_{\star}=0$ in the integrals between $0$ and $\delta U_{\perp}$ (and I neglect terms of order $\delta U_{\perp}^{5/2}$)
\begin{align}
n_{\text{i}} (x) = & \int_{0}^{\infty}  \frac{dU_{\star}}{\sqrt{2U_{\star}}}   \left( F_{\text{cl}}(\mu, U_{\star} + \Omega \mu ) - \frac{\partial F_{\text{cl}}}{\partial U} (\mu, U_{\star} + \Omega \mu )  \delta U_{\perp}  \right. \nonumber \\
& \left. + \frac{1}{2} \frac{\partial^2 F_{\text{cl}}}{\partial U^2} (\mu, U_{\star} + \Omega \mu ) \delta U_{\perp}^2 \right)   -   \int_{0}^{\delta U_{\perp}}  \frac{dU_{\star}}{\sqrt{2U_{\star}}}   \left( F_{\text{cl}}(\mu, \Omega \mu ) \phantom{ \frac{\partial F}{\partial \mu} }  \right. \nonumber \\
& \left. + \left( U_{\star} - \delta U_{\perp}  \right) \frac{\partial F_{\text{cl}}}{\partial U} (\mu, \Omega \mu ) \right)  + \ldots
\end{align}
Carrying out the integrals between $0$ and $\delta U_{\perp}$, I obtain
\begin{align} \label{int-expanded}
n_{\text{i}} (x) = & \int_{0}^{\infty}  \frac{dU_{\star}}{\sqrt{2U_{\star}}}   \left( F_{\text{cl}}(\mu, U_{\star} + \Omega \mu ) - \frac{\partial F_{\text{cl}}}{\partial U} (\mu, U_{\star} + \Omega \mu )  \delta U_{\perp}  \right. \nonumber  \\
& \left. + \frac{1}{2} \frac{\partial^2 F_{\text{cl}}}{\partial U^2} (\mu, U_{\star} + \Omega \mu ) \delta U_{\perp}^2 \right)   -   \sqrt{2\delta U_{\perp}}   F_{\text{cl}}(\mu, \Omega \mu ) \nonumber \\
&  +\frac{1}{3} \left( 2\delta U_{\perp}  \right)^{3/2} \frac{\partial F_{\text{cl}}}{\partial U} (\mu, \Omega \mu ) + \ldots \text{}
\end{align}
Then, inserting (\ref{int-expanded}) into equation (\ref{niclosedinftynotexp}) and changing the dummy integration variable to $U = U_{\star} + \Omega \mu $, one is left with the result of equation (\ref{niclosedinfty}).

\chapter{Proof that $k_2 > 0$ and $q_2>0$} \label{app-k2>0}

In order to show that $k_2 > 0$, I argued in the main text (see discussion below equation (\ref{k2})) that it is sufficient to show that
\begin{align} \label{k2>0equivalent}
 6\pi  \int_0^{\infty} \Omega d\mu  \int_{\Omega \mu}^{\infty}  \frac{F_{\text{cold}}(\mu, U) v_{\text{B}}^4 }{\left( 2\left( U - \Omega \mu \right) \right)^{5/2} } dU - \frac{ n_{e\infty} }{Z}  > 0 \text{.}
\end{align}
Remembering $v_z = \sqrt{2\left( U - \Omega \mu \right)}$ and equation (\ref{changevar-infty}), the integral in the first term can be recast as
\begin{align} \label{Reexpressed}
2\pi \int_0^{\infty} \Omega d\mu \int_0^{\infty}  \frac{F_{\text{cold}}(\mu, U) dU}{\left( 2\left( U - \Omega \mu \right) \right)^{5/2} } =  \int_0^{\infty}  \frac{ f_{\infty z} \left( v_z \right) }{v_z^{4} } dv_{z} \text{,}
\end{align}
where 
\begin{align} \label{Reexpressed-vz}
f_{\infty z} \left( v_z \right) = \int_{-\infty}^{\infty} dv_x  \int_{-\infty}^{\infty}  f_{\infty} \left( \vec{v} \right) dv_y \text{.}
\end{align}

By application of Schwarz's inequality I have the relation
\begin{align} \label{Schwarz-application}
\int_0^{\infty}  \frac{ f_{\infty z} \left( v_z \right) }{v_z^4} dv_z  \int_0^{\infty}  f_{\infty z} \left( v_z \right)  dv_z   \geqslant  \left( \int_0^{\infty}  \frac{ f_{\infty z} \left( v_z \right) }{v_z^{2} }  dv_z \right)^2  \text{.}
\end{align}
From quasineutrality and from the marginal form of the kinetic Chodura condition (\ref{kinetic-Chodura-marginal}) I obtain
\begin{align} \label{Schwarz-quasineutrality}
Z \int_0^{\infty}  f_{\infty z} \left( v_z \right)  dv_z   = n_{e\infty} \text{,}
\end{align}
\begin{align} \label{kinetic-Chodura-marginal-1D}
\int_{0}^{\infty} \frac{f_{\infty z}(v_z) }{v_z^2} = \frac{n_{\infty}}{v_{\text{B}}^2 } \text{.}
\end{align}
Substituting (\ref{kinetic-Chodura-marginal-1D}) and (\ref{Schwarz-quasineutrality}) in (\ref{Schwarz-application}), I obtain 
\begin{align}
Z v_{\text{B}}^4 \int_0^{\infty}  \frac{ f_{\infty z} \left( v_z \right)  }{v_z^4} dv_z   \geqslant  n_{e \infty}  \text{.} 
\end{align}
Re-expressing the left hand side of the inequality in terms of $F(\mu, U)$ and $U$ by using (\ref{Reexpressed}), I obtain  
\begin{align} \label{k2>0-almost}
2\pi \int_0^{\infty} \Omega d\mu \int_{\Omega \mu }^{\infty}  \frac{ F_{\text{cl}}(\mu, U) v_{\text{B}}^4 }{\left( 2\left( U - \Omega \mu \right) \right)^{5/2} } dU \geqslant \frac{n_{e\infty}}{Z}    \text{.}
\end{align}
From (\ref{k2>0-almost}) I see that
\begin{align}
6\pi \int_0^{\infty} \Omega d\mu \int_{\Omega \mu }^{\infty}  \frac{ F_{\text{cl}}(\mu, U) v_{\text{B}}^4  }{\left( 2\left( U - \Omega \mu \right) \right)^{5/2} } dU - \frac{n_{e\infty}}{Z}    \geqslant    \frac{2n_{e\infty}}{Z}       >   0 \text{,}
\end{align}
from which (\ref{k2>0equivalent}) immediately follows. 

This proof can be straightforwardly adapted to show that $q_2 >0$, where $q_2$ is defined in equation (\ref{q2-def}). Again, it suffices to show that the numerator of equation (\ref{q2-def}) is positive,
\begin{align} \label{q2-equivalent}
v_{\text{B}}^4 \int_{\bar{x}_{\text{c}}}^{\infty}   \Omega d\bar{x}  \int_{\chi_{\text{M}} (\bar{x} )}^{\infty}  \frac{  F_{\text{cl}} \left( \mu \left( \bar{x}, \chi_{\text{M}} \left( \bar{x} \right) \right) , U \right)   }{\sqrt{2\left( U - \chi_{\text{M}} (\bar{x}) \right)}} \Delta \left[ \frac{1}{v_{x0}^3} \right]  dU - \frac{n_{\text{e}0}}{Z} > 0 \text{.}
\end{align}
The integral can be re-expressed as
\begin{align} 
 \int_{\bar{x}_{\text{c}}}^{\infty}   \Omega d\bar{x}  & \int_{\chi_{\text{M}} (\bar{x} )}^{\infty}  \frac{  F_{\text{cl}} \left( \mu \left( \bar{x}, \chi_{\text{M}} \left( \bar{x} \right) \right) , U \right)   }{\sqrt{2\left( U - \chi_{\text{M}} (\bar{x}) \right)}} \Delta \left[ \frac{1}{v_{x0}^3} \right]  dU \nonumber \\
& = \int_{\bar{x}_{\text{c}}}^{\infty}   \Omega d\bar{x}  \int_{\chi_{\text{M}} (\bar{x} )}^{\infty}  \frac{ F_{\text{cl}} \left( \mu \left( \bar{x}, \chi_{\text{M}} \left( \bar{x} \right) \right) , U \right) }{\sqrt{2\left( U - \chi_{\text{M}} (\bar{x}) \right)}} dU \nonumber \\
& ~ \times \int_{-\infty}^{0}  \frac{3}{v_x^4} \hat{\Pi} \left(  v_x ,  - V_x \left( 0, \bar{x}, \chi_{\text{M}} \right) - \Delta v_{x0}  , - V_x \left( 0, \bar{x}, \chi_{\text{M}} \right)  \right) dv_x \nonumber \\
 & = 3 \int  \frac{f_0 ( \vec{v} ) }{v_{x}^4}  d^3v 
 \nonumber \\
 & = 3 \int_{-\infty}^0 \frac{f_{0x} ( v_x ) }{v_{x}^4}  dv_x  \text{,}
\end{align}
where $f_{0x}(v_x)$ is defined in equation (\ref{f0x-def}).
The marginal form of Bohm's condition is
\begin{align}
Z v_{\text{B}}^2 \int_{-\infty}^0 \frac{f_{0x} ( v_x ) }{v_{x}^2}  dv_x = n_{\text{e}0}  \text{}
\end{align}
and quasineutrality is
\begin{align}
Z  \int_{-\infty}^0 f_{0x} ( v_x )  dv_x = n_{\text{e}0}  \text{.}
\end{align}
Proceeding in an analogous way to the previous derivation, I conclude that
\begin{align}
v_{\text{B}}^4 \int_{\bar{x}_{\text{c}}}^{\infty}  \Omega d\bar{x} \int_{\chi_{\text{M}}(\bar{x})}^{\infty}  \frac{F_{\text{cl}} \left( \mu (\bar{x}, \chi_{\text{M}}(\bar{x}) , U \right)}{\sqrt{ 2\left( U - \chi_{\text{M}} ( \bar{x} ) \right)  } } \Delta \left[ \frac{1}{v_{x0}^{3}}\right] dU - \frac{n_{\text{e}0}}{Z}   \geqslant    \frac{2n_{\text{e}0}}{Z}       >   0 \text{,}
\end{align}
from which (\ref{q2-equivalent}) immediately follows.

\chapter{Neglecting the contribution of type II closed orbits near $x=0$} \label{app-notypeIIclosed}

The expansion of the closed orbit density near $x=0$ relies on distinguishing type I and type II effective potential curves. 
In section \ref{sec-KMPS-Bohm} I omitted the contribution of closed orbits associated with type II curves, denoted $n_{\text{i,cl,II}} (x) $. 
I proceed to show that this contribution is negligible.

From equation (\ref{ni-closed-expanded}), and using the expansion (\ref{VxII}) of $V_x$ near the stationary maximum $x_{\text{M}}$, I obtain an expression for the contribution to the density near $x=0$ due to ions in approximately closed type II orbits,
\begin{align} \label{ni-closed-near0}
n_{\text{i,cl,II}} (x) \simeq   2\int_{\bar{x}_{\text{m}}(x)}^{\bar{x}_{\text{m,I}}} \Omega \sqrt{| \chi_{\text{M}}'' |}  \left| x - x_{\text{M}} \right|  d\bar{x} 
 \int_{\chi_{\text{M}} (\bar{x})}^{\infty} \frac{ F\left(\mu \left(\bar{x}, \chi_{\text{M} } \right), U \right) }{\sqrt{2\left( U - \chi_{\text{M}}(\bar{x}) \right) }} dU  \text{.}
\end{align}
The upper limit of integration in $\bar{x}$ is $\bar{x}_{\text{m,I}}$, which is the value of $\bar{x}$ above which the effective potential is a type I curve.
It is easier to express the integral in (\ref{ni-closed-near0}) by changing variables from $\bar{x}$ to $x_{\text{M}}$ (for type II curves, $x_{\text{M}}$ depends on the value of $\bar{x}$). 
The Jacobian of this change of variables can be obtained using the equation for a stationary maximum, which is $\chi'(x_{\text{M}}, \bar{x} ) = 0$. Rearranging equation (\ref{stationary-points}) evaluated at the stationary point $x_{\text{M}}$, I deduce
\begin{align} \label{xbar-xM}
\bar{x} = x_{\text{M}} + \frac{\phi' ( x_{\text{M}} )}{\Omega B}  \text{.}
\end{align}
Differentiating this equation with respect to $x_{\text{M}}$, one obtains $\left| d \bar{x} / d x_{\text{M}} \right| =  |\chi_{\text{M}}''| / \Omega^2$. 
Then, the integral (\ref{ni-closed-near0}) can be written in terms of $x_{\text{M}}$. The integration limit $ \bar{x} = \bar{x}_{\text{m,I}}$ corresponds to $x_{\text{M}} = 0$, while the integration limit $\bar{x} = \bar{x}_{\text{m}}(x)$ corresponds to $x_{\text{M}} = x$.

For small $x$, one can Taylor expand the integrand near $\bar{x} = \bar{x}_{\text{m,I}}$ (which corresponds to $x_{\text{M}} = 0$) and retain only the leading order,
\begin{align} \label{ni-closed-near0-expanded}
n_{\text{i,cl,II}} (x)  \simeq & ~  2\int_0^x   ( x - x_{\text{M}} )  \frac{|\chi'' (0, \bar{x}_{\text{m,I}} ) |^{3/2}}{\Omega} d x_{\text{M}} 
  \int_{\chi_{\text{M}} (\bar{x})}^{\infty} \frac{ F_{\text{cl}} \left(\mu \left(\bar{x}_{\text{m,I}}, \chi_{\text{M}} \left(\bar{x}_{\text{m,I}} \right) \right), U \right) }{\sqrt{2\left( U - \chi_{\text{M}} (\bar{x}_{\text{m,I}}) \right) }} dU \nonumber \\
 \simeq & x^2  \frac{|\chi'' (0, \bar{x}_{\text{m,I}} ) |^{3/2}}{\Omega}     
 \int_{\chi_{\text{M}}}^{\infty} \frac{ F_{\text{cl}} \left(\mu \left(\bar{x}_{\text{m,I}}, \chi_{\text{M}} \left(\bar{x}_{\text{m,I}} \right) \right), U \right) }{\sqrt{2\left( U - \chi_{\text{M}} (\bar{x}_{\text{m,I}}) \right) }} dU 
   \text{.}
\end{align}
Hence, the contribution from type II closed orbits near $x=0$ is proportional to $ x^2 $ and therefore subdominant compared to $x$, making it negligible.
In fact, when $\bar{x}_{\text{m,I}} \rightarrow \infty$, I expect the contribution to be exponentially small because $\mu \left( \bar{x}_{\text{m,I}}, \chi_{\text{M}} (\bar{x}_{\text{m,I}}) \right) \rightarrow \infty$ and $F$ is usually exponentially small for $\mu \rightarrow \infty$.

\chapter{Integrals of temperature-dependent distribution functions}
\label{integrals-Tdep}

The distribution functions in (\ref{f-infty}) are normalized according to equation (\ref{N-infty}).
The integrals over $v_y$ and $v_z$ are trivially carried out to obtain the functions
\begin{align} \label{f-inftyz}
f_{\infty z} (v_z) = \int f_{\infty}( \vec{v}) dv_x dv_y =  \begin{cases}
\mathcal{N}  n_{\infty} \frac{4 v_z^2}{\sqrt{\pi} v_{\text{t,i}}^5}   \exp \left( - \frac{ \left( v_z - u v_{\text{t,i}}  \right)^2 }{v_{\text{t,i}}^2} \right) \Theta \left( v_z \right) & \text{ for } \tau \leqslant 1 \text{,} \\
\mathcal{N}  n_{\infty}  \frac{ 4 v_z^2 }{  \sqrt{\pi} v_{\text{t,i}}^3 \left( v_{\text{t,i}}^2 +r v_z^2 \right)} \exp \left( - \frac{v_z^2 }{ v_{\text{t,i}}^2 } \right)\Theta \left( v_z \right)  & \text{ for } \tau > 1 \text{,}
\end{cases} 
\end{align}
All the integrals in this appendix are carried out using the dimensionless variables $\tilde{w}_z = v_z / v_{\text{t,i}} - u$ and $\tilde{v}_z = v_z / v_{\text{t,i}} $.

The normalization condition (\ref{N-infty}) is then
\begin{align} \label{normcondz}
n_{\infty} = \int_0^{\infty} f_{\infty z} (v_z ) dv_z
\end{align}
Applying equation (\ref{normcondz}) for $\tau \leqslant 1$, and changing integration variable to $\tilde{w}_z $ gives 
\begin{align}
n_{\infty} 
& =   \mathcal{N}  n_{\infty} \frac{4 }{\sqrt{\pi} } \int_{-u}^{\infty}  \left( \tilde{w}_z + u \right)^2  \exp \left( -  \tilde{w}_z^2  \right)
\end{align}
Thus,
\begin{align} \label{N-u}
 \frac{4 \mathcal{N}  }{\pi^{1/2} } \int_{-u}^{\infty}  \left( \tilde{w}_z^2  + 2\tilde{w}_z u + u^2 \right)   \exp \left( -  \tilde{w}_z^2  \right) = 1 \text{,}
\end{align}
The integral in equation (\ref{N-u}) evaluates to
\begin{align}
\int_{-u}^{\infty}  \left( \tilde{w}_z^2  + 2\tilde{w}_z u + u^2 \right)   \exp \left( -  \tilde{w}_z^2  \right)  = \frac{\sqrt{\pi}}{4} \left[ \left( 1 + 2u^2 \right) \left( 1 + \text{erf} (u) \right) + 2 u \exp(-u^2) \right]    \text{.}
\end{align}
Hence, equation (\ref{N-infty}) for $\tau \leqslant 1$ follows.

Applying equation (\ref{normcondz}) for $\tau >1$ one finds, after changing variable to $\tilde{v}_z$,
\begin{align} \label{N-r-1}
n_{\infty} 
=  \frac{ 4 \mathcal{N}  n_{\infty} }{  \pi^{1/2}  } \int_0^{\infty}  \frac{ \tilde{v}_z^2  \exp \left( - \tilde{v}_z^2   \right) }{   1 +r \tilde{v}_z^2  }    \text{.}
\end{align}
The last integral in equation (\ref{N-r-1}) is obtained in the following way.
First, one can obtain the integral of the function $\exp(- \tilde{v}_z^2)/(1+r\tilde{v}_z^2)$ (which will be useful when imposing the kinetic Chodura condition (\ref{kinetic-Chodura-marginal}) in the next paragraph).
Re-expressing $1/(1+r\tilde{v}_z^2)$ as a definite integral of the function $\exp\left( -\eta \left( 1+r\tilde{v}_z^2 \right) \right)$ with respect to a dummy variable $\eta$, one has
\begin{align}
\int_0^{\infty} \frac{\exp(-\tilde{v}_z^2) }{1+r\tilde{v}_z^2} dx & = \int_0^{\infty}  \exp(-\eta) d\eta \int_0^{\infty} \exp \left( -\left( 1 + \eta r \right) \tilde{v}_z^2 \right) d\tilde{v}_z  \nonumber \\
& =  \frac{\sqrt{\pi}}{2} \int_0^{\infty} \frac{ \exp(-\eta) }{ \sqrt{\eta r + 1} } d\eta \text{.}
\end{align}
Changing variable to $\xi = \sqrt{ \eta + 1 / r}$ gives
\begin{align} \label{fancy-int-0}
\int_0^{\infty} \frac{\exp(-\tilde{v}_z^2) }{1+r\tilde{v}_z^2} dx & =  \sqrt{\frac{\pi}{r}} \exp \left( \frac{1}{r} \right) \int_{1/\sqrt{r}}^{\infty} \exp(-\xi^2) d\xi   \nonumber \\
&  =  \frac{\pi}{2\sqrt{r}} \exp \left( \frac{1}{r} \right) \left[ 1 - \text{erf} \left( \frac{1}{\sqrt{r}} \right) \right] \text{.}
\end{align}
Then, using the relation
\begin{align}
\int_0^{\infty} \frac{\exp(-\tilde{v}_z^2) }{1+r\tilde{v}_z^2} dx + r\int_0^{\infty} \frac{\tilde{v}_z^2 \exp(-\tilde{v}_z^2) }{1+r\tilde{v}_z^2} dx = \int_0^{\infty} \exp(-\tilde{v}_z^2)   d\tilde{v}_z = \frac{ \sqrt{\pi}}{2}  \text{,}
\end{align}
the integral 
\begin{align} \label{fancy-int-2}
\int_0^{\infty} \frac{\tilde{v}_z^2 \exp(-\tilde{v}_z^2) }{1+r\tilde{v}_z^2} dx = \frac{\sqrt{\pi}}{2r} - \frac{\pi}{2r^{3/2}} \exp \left( \frac{1}{r} \right) \left[ 1 - \text{erf} \left( \frac{1}{\sqrt{r} }\right) \right]
\end{align}
is obtained.
Inserting this integral into (\ref{N-r-1}), one obtains $\mathcal{N}$.

Equation (\ref{kinetic-Chodura-marginal-1D}) is used to obtain the values of the positive constants $u$ and $r$.
For $\tau \leqslant 1$ one inserts the distribution function in (\ref{f-inftyz}) in (\ref{kinetic-Chodura-marginal-1D}) and changes variable to $\tilde{w}_z = v_z / v_{\text{t,i}} - u$ to obtain
\begin{align} \label{Chodura-u-1}
 \frac{ v_{\text{t,i}}^2  }{v_{\text{B}}^2} =  \frac{4  \mathcal{N} }{\pi^{1/2} } \int_{-u}^{\infty}   \exp \left( -  \tilde{w}_z^2  \right) d \tilde{w}  = 2  \mathcal{N}  \left( 1 + \text{erf} \left(  u  \right) \right) \text{.}
\end{align}
Rearranging equation (\ref{Chodura-u-1}) and inserting the value of $\mathcal{N}$, one obtains equation (\ref{u-def}).

For $\tau > 1$, one changes variable to $\tilde{v}_z = v_z / v_{\text{t,i}} $ in the integral (\ref{Chodura-u-1}) to obtain
\begin{align}
 \frac{ v_{\text{t,i}}^2  }{v_{\text{B}}^2} =  \frac{ 4 \mathcal{N}  }{  \pi^{1/2}  } \int_0^{\infty}  \frac{   \exp \left( - \tilde{v}_z^2   \right) }{   1 +r \tilde{v}_z^2  }  \text{.}
\end{align}
Inserting the value of $\mathcal{N}$ and the integral in equation (\ref{fancy-int-0}), one obtains equation (\ref{r-def}).

The ion fluid velocity is evaluated using the equation (\ref{uzinfty})
 \begin{align} \label{uzinfty}
 u_{z\infty} = \frac{1}{n_{\infty}} \int f_{\infty z} ( v_z ) v_{z} dv_z \text{.}
 \end{align}
 For $\tau \leqslant 1$ one has
 \begin{align} \label{uz-u-int}
\frac{  u_{z\infty} }{v_{\text{t,i}}}  & =  \frac{4 \mathcal{N}  n_{\infty} }{\sqrt{\pi}}  \int_0^{\infty} \frac{ v_z^3}{ v_{\text{t,i}}^5 }   \exp \left( - \frac{ \left( v_z - u v_{\text{t,i}}  \right)^2 }{v_{\text{t,i}}^2} \right) dv_z \nonumber \\
 & = \frac{4 \mathcal{N}  n_{\infty} }{\sqrt{\pi}}   \int_{-u}^{\infty}  \left( \tilde{v}_z + u \right)^3   \exp \left( -   \tilde{v}_z^2   \right) d\tilde{v}_z 
 \text{.}
 \end{align}
 The integrals in the final equality of (\ref{uz-u-int}) evaluate to
 \begin{align}
 \int_{-u}^{\infty}  \left( \tilde{v}_z + u \right)^3   \exp \left( -   \tilde{v}_z^2   \right) d\tilde{v}_z  &  =  \int_{-u}^{\infty}  \left( \tilde{v}_z^3 + 3\tilde{v}_z^2 u + 3 \tilde{v}_z u^2 + u^3 \right)  \exp \left( -   \tilde{v}_z^2   \right) d\tilde{v}_z \nonumber \\
 & =  \frac{1}{2} \exp(-u^2) \left( u^2 + 1 \right)  +  \frac{  \sqrt{ \pi } u  }{4}  \left( 3 + 2u^2 \right)\left( 1 + \text{erf} \left( u \right) \right) \text{.}
 \end{align}
 For $\tau > 1$ one has
  \begin{align} \label{uzinfty-r-int}
\frac{  u_{z\infty} }{v_{\text{t,i}}} & = \frac{ 4 \mathcal{N}  n_{\infty} }{\sqrt{\pi} v_{\text{t,i}}} \int_0^{\infty}  \frac{  v_z^3 }{ v_{\text{t,i}}^3 \left( v_{\text{t,i}}^2 +r v_z^2 \right)} \exp \left( - \frac{v_z^2 }{ v_{\text{t,i}}^2 } \right) dv_z \nonumber \\
 & = \frac{4 \mathcal{N}  n_{\infty} }{\sqrt{\pi}} \int_0^{\infty}  \frac{  \tilde{v}_z^3 }{ \left( 1 +r \tilde{v}_z^2 \right)} \exp \left( - \tilde{v}_z^2  \right) d\tilde{v}_z \text{.}
 \end{align}
 The integral in equation (\ref{uzinfty-r-int}) is taken, as before, by expressing $1/(1+rx^2)$ as a definite integral,
 \begin{align}
 \int_0^{\infty}  \frac{  \tilde{v}_z^3 }{ \left( 1 +r \tilde{v}_z^2 \right)} \exp \left( - \tilde{v}_z^2  \right) d\tilde{v}_z & =   \int_0^{\infty} \exp(-\eta) d\eta \int_0^{\infty} \tilde{v}_z^3 \exp \left( - \tilde{v}_z^2 \left( 1 + \eta r\right) \right) d\tilde{v}_z \nonumber \\
 & =   \int_0^{\infty} \frac{ \exp(-\eta) }{ 2 \left( 1 + \eta r \right)^2 } d\eta \nonumber \\
 & =   \frac{1}{2r} - \frac{1}{2r} \int_0^{\infty} \frac{ \exp(-\eta) }{ \left( 1 + \eta r \right) } d\eta \nonumber \\
 & =   \frac{1}{2r} - \frac{\exp(1/r)}{2r^2} \int_{1/r}^{\infty} \frac{ \exp(-\eta') }{ \eta' } d\eta' \nonumber \\
 \end{align}
 where, in the last two steps, I have integrated by parts and then changed integration variable to $\eta' = \eta + 1 / r$.
 Using the definition of the exponential integral in equation (\ref{E1}), one obtains
 \begin{align}
   \int_0^{\infty}  \frac{  \tilde{v}_z^3 }{ \left( 1 +r \tilde{v}_z^2 \right)} \exp \left( - \tilde{v}_z^2  \right) d\tilde{v}_z  = \frac{1}{2r} -  \frac{\exp(1/r)}{2r^2} E_1 \left(\frac{1}{r} \right)  \text{.}
 \end{align}

\cleardoublepage

\backmatter

\bibliographystyle{unsrt}
\bibliography{references}

\end{document}